\documentclass[12pt,letterpaper,english]{article}
\usepackage{microtype}
\usepackage{float}
\usepackage{amsmath} 
\usepackage{amsthm}
\usepackage{mathrsfs}
\usepackage{graphicx} 
\usepackage{caption}
\usepackage{setspace}
\usepackage{amssymb}
\usepackage{multirow}
\usepackage{rotating}
\usepackage{array}
\usepackage{booktabs}
\usepackage{multibib}
\usepackage{xr}
\usepackage{algorithm}
\usepackage{algpseudocode}
\usepackage{color}
\usepackage[small,compact]{titlesec} 
\DeclareMathAlphabet{\mathcalligra}{T1}{calligra}{m}{n}
\usepackage[sectionbib,round]{natbib}
\usepackage{makecell}

\usepackage{times}
\usepackage{abstract}

\usepackage[hang,flushmargin]{footmisc}
\usepackage{fullpage} 
\usepackage{hyperref}
\usepackage[compact]{titlesec}
\usepackage[section]{placeins}
\usepackage{footnote}
\usepackage{epstopdf}
\usepackage{placeins}
\usepackage[toc,page]{appendix}
\usepackage{graphicx}
\usepackage{subcaption}
\usepackage{amsmath} 
\usepackage{amsthm}
\usepackage{graphicx} 
\usepackage{blkarray}
\usepackage{stackengine}
\usepackage{accents}
\usepackage{enumitem}
\usepackage[mathscr]{euscript}
\usepackage{diagbox}

\bibpunct[, ]{(}{)}{;}{a}{}{,}%
\def\bibsep{\smallskipamount}%
\def\newblock{\ }%
\usepackage{etoolbox}
\AtBeginEnvironment{equation}{%
  \setlength{\abovedisplayskip}{4pt}%
  \setlength{\belowdisplayskip}{4pt}%
  \setlength{\abovedisplayshortskip}{4pt}%
  \setlength{\belowdisplayshortskip}{4pt}%
}

\theoremstyle{definition}

\theoremstyle{definition}

\usepackage[utf8]{inputenc}
\usepackage{longtable}
\usepackage{graphicx}
\usepackage{epstopdf}
\usepackage{bbm}
\usepackage{dsfont}
\usepackage{comment}
\usepackage{pgfplots}
\usepackage{xfrac}
\usepackage[flushleft]{threeparttable}
\usepackage[normalem]{ulem} 

\usepackage{pifont}

\usepackage{caption}
\theoremstyle{plain}

\newcommand{\squishlist}{
   \begin{list}{$\bullet$}
    { \setlength{\itemsep}{0pt} \setlength{\parsep}{1pt}
      \setlength{\topsep}{1pt} \setlength{\partopsep}{1pt}
      \setlength{\leftmargin}{1.5em} \setlength{\labelwidth}{1em}
      \setlength{\labelsep}{0.5em} } }

\newcommand{\squishlisttwo}{
   \begin{list}{$\bullet$}
    { \setlength{\itemsep}{0pt} \setlength{\parsep}{0pt}
      \setlength{\topsep}{0pt} \setlength{\partopsep}{0pt}
      \setlength{\leftmargin}{1em} \setlength{\labelwidth}{1.5em}
      \setlength{\labelsep}{0.5em} } }

\newcommand{\squishend}{
    \end{list}  }

\author{Shirsho Biswas, Hema Yoganarasimhan, Haonan Zhang\thanks{Authors contributed equally and names are listed in alphabetical order. E-mails: shirsho@uw.edu, hemay@uw.edu, hzhang96@uw.edu. We thank an anonymous pet supplies retailer for providing the data and for extensive, thoughtful discussions on many aspects of the paper. We would also like to thank Simha Mummalaneni for his detailed comments, which have significantly improved the paper. We also thank the participants of the University of Washington seminar, the Hong Kong University virtual seminar, the 2024 ISMS Marketing Science Conference, and the 2025 China India Insights Conference for their helpful input. The opinions expressed in this paper are the authors' and do not reflect those of the data providers.} \\ University of Washington}

\begin{document}
\title{Channel Adoption Pathways and Post-Adoption Behavior}

\maketitle
\vspace{-0.4in}
\begin{abstract} 
\vspace{-0.4in}
\noindent \singlespacing 

The rapid growth of digital shopping channels has led many traditional retailers to invest in e-commerce websites and mobile apps. While prior research shows that multichannel customers are more valuable, it overlooks how the motive for adopting a new channel shapes post-adoption behavior. Using transaction-level data from a major Brazilian pet supplies retailer, we study offline-only consumers who adopt online shopping through four pathways: organic adoption, the COVID-19 pandemic, Black Friday promotions, and a loyalty program. Using consumer-level panel data and difference-in-differences estimates, we examine how these pathways are associated with post-adoption spending, profitability, and channel usage. We find that all adopters spend more than comparable offline-only consumers, but their post-adoption behavior differs systematically by adoption pathway. Promotion-driven adopters exhibit patterns consistent with forward buying and lower subsequent profitability, whereas COVID adopters display stronger offline persistence consistent with consumer inertia and habit persistence. These findings suggest that managers may benefit from accounting for adoption-pathway heterogeneity when forecasting customer lifetime value and assessing the breakeven and ROI of promotions designed to induce online adoption.

\end{abstract}

\thispagestyle{empty}

\clearpage

\pagenumbering{arabic}
    \setcounter{page}{1}

\section{Introduction}

\subsection*{Background and Motivation}

Online shopping has grown rapidly over the last two decades with the proliferation of computers and smartphones. In response, many grocery and retail chains (e.g., Walmart, Target, Costco), which were historically brick-and-mortar only, now also sell through websites and mobile apps. For traditional retailers, a central question is not only whether online adoption increases consumer value, but also whether consumers who adopt online shopping through different pathways are equally valuable afterward. \citep{goyal_2023, bloomberg_2020}.

Prior research has found that multichannel consumers (those who buy using multiple channels of a retailer) tend to spend more than single-channel consumers; see $\S$\ref{sec:literature}. However, this work largely ignores \emph{why} consumers adopt new channels, implicitly assuming that the benefits from multichannel adoption are the same regardless of the reasons for adoption of new channels. This assumption is unlikely to hold in practice. Some customers adopt online shopping organically because they value the convenience of an additional channel. Others adopt because an external shock pushes them online, or because a firm-driven promotion requires online engagement. These distinct adoption pathways can attract different customers, induce different behaviors around adoption, and generate different post-adoption trajectories.

Understanding this heterogeneity is managerially important. Firms often evaluate the ROI of online adoption campaigns by extrapolating from the behavior of existing multichannel customers or organic adopters. Such extrapolation is valid only if customers induced to adopt by such campaigns behave like organic adopters after adoption. If, instead, promotion-induced or macro shock-induced adopters differ in spend, channel usage, or profitability, then firms could make errors in forecasting customer value, overpay to induce online adoption, and misjudge the ROI of campaigns designed to move customers online. These considerations motivate our research questions.

\subsection*{Research Questions and Theoretical Predictions}

We study two research questions. First, how does adopting a retailer's digital shopping channels affect the purchase behavior of customers who were previously \textit{offline-only} shoppers? Second, conditional on adoption, how does post-adoption behavior differ across organic, macro-shock-driven, and promotion-driven adoption pathways? We focus on three managerially salient outcomes: total spend, profitability, and the share of spending allocated to the offline channel.

Building on the multichannel literature and consumer behavior frameworks (discussed in detail in $\S$\ref{sec:literature} and $\S$\ref{sec:theory}), we hypothesize that customers who adopt a firm's online channels and become multichannel shoppers will, on average, increase their total spending relative to \textit{offline-only customers}. Beyond this average effect, we anticipate substantial heterogeneity across adoption pathways. Consumer inertia and habit theory \citep{verplanken2006interventions,wood2009habitual} suggest that when adoption is induced by external shocks or incentives rather than intrinsic preference, consumers are more likely to revert toward prior routines once the stimulus subsides. Thus, relative to \textit{organic adopters}, externally driven adopters should exhibit a higher post-adoption share of spending on offline channels and slower growth in online channel share. We further anticipate that such differences will be more pronounced for older, habit-driven customers.

For adoption spurred by time-limited promotions, theories of forward buying and demand distortion \citep{neslin1985consumer,hendel2006sales,van2017sales} imply that promotion-driven adopters may temporarily inflate their spending during the promotional period by stocking up, but then reduce spending afterwards. This mechanism can dampen or even offset the long-run spending uplift typically associated with multichannel adoption. Finally, because profit margins generally differ across online and offline channels, heterogeneity in the post-adoption mix of spending across channels is likely to translate into systematic differences in profitability across adopter types, even when their total spend is similar.

In summary, our theoretical framework predicts both an average multichannel effect and systematic pathway heterogeneity. Adopters should spend more and become more profitable than \textit{offline-only customers}, but the magnitude and persistence of these gains should depend on why customers adopted online shopping. Macro-shock-driven adopters may exhibit stronger offline persistence due to inertia, while promotion-driven adopters may show lower post-adoption spend because promotions pull demand forward. These behavioral differences should, in turn, affect profitability through both total spend and channel mix.

\subsection*{Empirical Setting and Methodology}

To test these predictions, we use data from a large pet supplies retailer in Brazil with an extensive network of brick-and-mortar stores and an online presence (website and mobile app). Our data comprise all customer purchases across all channels from 2019 to 2024. During this period, there were three key events: a macro-environmental shock (COVID-19) and two firm-driven promotions designed to incentivize online shopping. This allows us to observe a variety of adoption pathways.

We construct four groups of adopters of online shopping who previously only shopped at the firm's physical stores: (1) \textit{COVID adopters}, who started shopping online during the onset of COVID-19; (2) \textit{Black Friday adopters}, who adopted online shopping to take advantage of deep discounts available online during Black Friday; (3) \textit{Loyalty Program adopters}, who adopted online shopping during the launch of a loyalty program that had to be activated on the retailer’s mobile app; and (4) \textit{organic adopters}, who started shopping online without any identifiable external stimulus. In addition, we consider a control group of non-adopters or \textit{offline-only customers} (consumers who did not adopt online shopping during our observation period).

Using these data, we construct consumer-level monthly panel data to quantify revenue, profitability, and the share of spend in the offline and online channels, and use this panel to answer our research questions. Specifically, we measure the Average Treatment Effect on the Treated (ATT) of online adoption on post-adoption behavior for different adoption pathways using a difference-in-differences framework with individual and time fixed effects. We interpret these estimates as descriptive ATTs with appropriate controls. This estimand is managerially relevant because each pathway reflects a different form of customer self-selection into online adoption. Firms typically cannot force adoption; they can only encourage it through campaigns or observe adoption triggered by external shocks. The relevant managerial object is therefore the post-adoption value of customers who actually adopt through each pathway, rather than a causal ATE for a randomly assigned adoption intervention.

\subsection*{Main Findings, Managerial Implications, and Contributions}

\begin{table}[htp!]
\caption{Summary of Main Findings by Adoption Pathway}
\label{tab:main_result_preview}
\centering
\footnotesize
\renewcommand{\arraystretch}{1.25}
\setlength{\tabcolsep}{4pt}
\begin{tabular}{p{3.2cm}p{4.1cm}p{4.1cm}p{4.1cm}}
\midrule
\makecell[l]{Adoption \\ pathway}
& \textit{COVID adopters}
& \textit{Black Friday adopters}
& \textit{Loyalty Program adopters} \\
\midrule

\makecell[l]{Post-adoption \\ spend}
& Similar to \textit{organic adopters}
& Lower than \textit{organic adopters}
& Lower than \textit{organic adopters} \\
\midrule

\makecell[l]{Buying behavior \\ around adoption}
& No evidence of forward buying
& Forward buying during the promotion, followed by lower post-adoption spending
& Forward buying around adoption, followed by lower post-adoption spending \\
\midrule

\makecell[l]{Post-adoption \\ channel mix}
& Higher offline share and slower movement toward online shopping
& Higher offline share initially, but faster movement toward online shopping over time
& No meaningful difference in offline share relative to \textit{organic adopters} \\
\midrule

\makecell[l]{Post-adoption \\ profitability}
& Higher than \textit{organic adopters}, despite similar spend
& Lower than \textit{organic adopters}
& Lower than \textit{organic adopters} \\
\midrule

\makecell[l]{Main behavioral \\ mechanism}
& Consumer inertia and habit persistence, especially among older customers; higher offline margins raise profitability
& Promotion-induced forward buying reduces later spend; higher offline share does not offset the spending decline
& Promotion-induced forward buying reduces later spend; recurring app-based activation may reinforce online use \\
\midrule

\makecell[l]{Managerial \\ implication}
& Shock-induced adopters can be valuable, but may need targeted reinforcement to develop online habits
& Discount-induced adoption can overstate long-run customer value if managers ignore post-promotion demand pull-forward
& Loyalty-program adoption can create online engagement, but firms should not assume these adopters behave like \textit{organic adopters} \\
\midrule\midrule

\multicolumn{4}{p{15.6cm}}{\scriptsize
\textit{Notes:} This table summarizes the main directional findings and mechanisms from the event-adopter versus organic-adopter analyses. The exact DID estimates, effect sizes, and statistical tests are reported in $\S$\ref{ssec:results_event_organic_adopters}.}
\end{tabular}
\end{table}

Table \ref{tab:main_result_preview} summarizes the main post-adoption differences between each externally induced adopter group and \textit{organic adopters}. The table provides a qualitative summary of the key results and mechanisms; exact DID estimates, effect-size calculations, and statistical tests are reported in $\S$\ref{ssec:results_event_organic_adopters}.

We find three main patterns. First, all adopters spend more and generate higher profits than comparable \textit{offline-only customers}, confirming the average value of multichannel adoption. However, this pooled adopter comparison masks substantial pathway heterogeneity. Second, relative to \textit{organic adopters}, \textit{COVID adopters} exhibit similar post-adoption spend but higher profitability. This profitability advantage is consistent with their greater reliance on the offline channel, where margins are higher in our setting, and with slower movement toward online shopping among older, more habit-driven customers.

Third, promotion-induced adopters behave differently from both \textit{COVID adopters} and \textit{organic adopters}. \textit{Black Friday adopters} and \textit{Loyalty Program adopters} spend less than \textit{organic adopters} after adoption and are also less profitable. Event-study evidence shows that these customers spend more around the adoption period but less afterward, consistent with forward buying and post-promotion demand distortion. The two promotion pathways also differ in channel usage. \textit{Black Friday adopters} initially retain a higher offline share than \textit{organic adopters} but move online faster over time, consistent with their younger profile. In contrast, \textit{Loyalty Program adopters} show no meaningful difference in offline spending share, likely because monthly app-based activation reinforces online engagement.

Taken together, these findings show that the \emph{reason} for online adoption matters. Macro-shock adopters can be more profitable than \textit{organic adopters} when they continue using higher-margin offline channels, whereas promotion-induced adopters can be less valuable because promotional adoption is associated with forward buying and lower subsequent spend. Thus, managers should not assume that customers induced to adopt online shopping through external events or promotions will behave like \textit{organic adopters} after adoption.

Our results have important implications for two core managerial tasks: forecasting Customer Lifetime Value (CLV) and evaluating the ROI of promotions.  We show that ignoring adoption pathways can lead to substantial bias in incremental CLV forecasts. In our setting, pooled-adopter CLV forecasts substantially overstate the value of promotion-induced adopters: pathway-aware CLV is about 60\% lower than the pooled forecast for \textit{Black Friday adopters} and about 80\% lower for \textit{Loyalty Program adopters}. Similarly, using \textit{organic adopters} as the benchmark in promotion evaluation makes online-adoption promotions appear to break even three to six times faster than they actually do. These results imply that firms should explicitly account for adoption pathways when forecasting incremental CLV, designing promotions, and setting caps on acceptable acquisition or promotional costs.


Overall, our analysis makes both substantive and managerial contributions. Substantively, we extend the multichannel shopping literature by introducing \emph{channel adoption pathways} as a key dimension of heterogeneity. Whereas prior empirical work primarily studies organic adopters or treats adopters as homogeneous (Table \ref{tab:rel_lit}), we compare organic, macro-shock-driven, and firm-induced promotional adoption, and show that these pathways are associated with systematically different post-adoption trajectories in spending, channel mix, and profitability. We also link these differences to a conceptual framework based on consumer inertia, habit persistence, and forward buying, which explains why externally induced adopters need not behave like organic adopters after adoption. Managerially, we provide a cautionary tale against naively extrapolating the behavior and profitability of one adopter group to another. We translate the pathway-specific estimates into pathway-aware CLV forecasts and promotion break-even calculations, showing that pooled-adopter or organic-adopter benchmarks can substantially misstate the value of customers acquired through promotions, and offer an implementable framework for incorporating adoption pathways into CLV forecasting, promotion design, and ROI/break-even analysis.

\section{Related Literature}
\label{sec:literature}

\begin{table}[!htp]
         \resizebox{\textwidth}{!}{%
\footnotesize
\begin{tabular}{lllll}
\toprule 
Paper & Setting & Channels & Methodology & Adoption Pathways \\ \midrule 
\textbf{Our Paper} & Pet Supplies & Physical stores only to & Diff-in-diff & Organic +  \\ 
 & &   adding website and/or & & multiple external \\ 
  & &   mobile app & & stimuli \\ 
\citet{narang2019mobile} & Video games & Physical stores + website  & Diff-in-diff& Organic only \\ 
&  and Electronics& to adding mobile app& &  \\ 
\citet{montaguti2016can} & Books  & Mail order, fax, phone,  & Cross-sectional & Organic + \\
 & (subscribers only) &   SMS, Internet, &  & mailing campaign  \\
  &  &  and bookstores &  &   \\
\citet{wang2015go}& Online grocery & Website to adding  & Diff-in-diff & Organic only \\
&  & mobile app & &  \\
\citet{kushwaha2013multichannel} & Multiple categories & Website and catalog & Cross-sectional& Organic only \\
\citet{ansari2008customer} & Apparel & Website and catalog & Within-customer& Organic only \\
\citet{venkatesan2007multichannel} & Apparel & Full price stores, & Within-customer& Organic only \\
 &  &  Discount Stores,& & \\
  &  &  and website& & \\
  \citet{kumar2005multichannel} & Technology & Salespersons, direct mail, & Cross-sectional & Organic only \\
   &  & telephone and website & & \\
     \citet{thomas2005managing} &  Retail & Catalog, physical stores, & Cross-sectional & Organic only \\
   &  & website & & \\
\bottomrule
\end{tabular}
}
\caption{Prior Empirical Literature on Multichannel Shopping in Retail Settings}
\label{tab:rel_lit}
\end{table}

Our work contributes to the empirical literature on multichannel customer management, which studies how customers' behavior changes when they purchase through multiple channels rather than a single channel. Table \ref{tab:rel_lit} summarizes closely related studies; for broader reviews see \citet{neslin2006challenges, neslin2009key, verhoef2012multichannel}.

Substantively, the existing literature overlooks heterogeneity in channel adoption pathways, implicitly assuming that the value of multichannel customers is independent of the motivations underlying their adoption decisions. As Table \ref{tab:rel_lit} shows, almost all prior work focuses on \textit{organic adopters}. \citet{montaguti2016can} is an exception, using a mailing campaign to nudge consumers into adopting multiple channels. However, they do not compare organic multichannel customers with those nudged by the campaign, whereas this comparison is central in our paper.  Moreover, our setting is a fast-moving consumer packaged goods retailer (pet supplies), whereas much of the earlier literature examines books, apparel, video games, electronics, and technology. These differ from our context along several dimensions, including purchase frequency, competitive intensity, and product type (often more hedonic than utilitarian), which has been shown to matter for multichannel effects \citep{kushwaha2013multichannel}. 

Thus, we extend the multichannel literature by (i) examining how post-adoption behavior varies with channel adoption pathways and providing theoretical explanations for these differences, and (ii) focusing on a practically important channel transition—physical store only to combined physical store and online usage—in a fast moving consumer packaged goods setting, that has not been separately studied.

From a methodological perspective, prior research has used three broad approaches to documenting the value of adopting new channels: cross-sectional, within-customer, and difference-in-differences (DID). Early work \citep{thomas2005managing, kumar2005multichannel, kushwaha2013multichannel, montaguti2016can} uses cross-sectional comparisons of multichannel versus single-channel customers and concludes that multichannel customers tend to be more valuable. However, cross-sectional comparisons lack individual-specific controls and cannot isolate how behavior \textit{changes} after multichannel adoption. A second stream \citep{venkatesan2007multichannel, ansari2008customer} employs within-customer analysis, comparing a customer's purchase behavior before versus after adopting multiple channels. This work also finds higher spending after multichannel adoption, but without an appropriate control group of non-adopters, it requires strong assumptions that no other time-varying factors affect behavior. More recent work \citep{wang2015go,narang2019mobile} uses DID with non-adopters as a control group, addressing many of these concerns. Our paper also employs a DID framework with non-adopters as a control group, but brings this methodology to a new setting and focuses on heterogeneity across adoption pathways, which has not been explored in prior empirical work.

\section{Theoretical Framework}
\label{sec:theory}


We use this section to develop a concise theoretical framework linking the \emph{reason} for online adoption to subsequent spending, channel mix, and profitability. We define a channel adoption pathway as the circumstance under which a previously offline-only customer first adopts the retailer's online channel. Conceptually, we distinguish three pathways: (1) \emph{organic adoption}, where customers voluntarily adopt because they value the convenience of an additional channel; (2) \emph{macro-environmental shocks}, such as COVID-19, nearby store closures, or severe weather events that temporarily disrupt offline shopping; and (3) \emph{firm-driven promotions}, where retailers explicitly incentivize online adoption through discounts, loyalty benefits, or online-only offers.


These pathways are not randomized treatments; they are forms of consumer self-selection into online adoption. This feature is central to the managerial problem rather than a nuisance to be eliminated. Firms can encourage adoption through promotions or observe adoption triggered by external shocks, but customers ultimately decide whether to adopt.\footnote{This logic parallels the broader literature on ``encouragement design'' in marketing and the social sciences, which distinguishes interventions assigned by firms from focal actions that consumers self-select into \citep{bradlow_1998, syrgkanis2019machine, mummalaneni2023content}. In such settings, the natural estimand is the Average Treatment Effect on the Treated (ATT), rather than the Average Treatment Effect (ATE).} Accordingly, our framework focuses on pathway-specific descriptive ATTs: how post-adoption behavior changes for customers who actually adopt through each pathway, and why those changes may differ across pathways.

We focus on two behavioral mechanisms. The first is consumer inertia and habit persistence. Organic adoption likely signals a customer's readiness or preference for online shopping. In contrast, adoption induced by a shock or incentive may reflect a temporary disruption rather than a durable preference shift. Consumer inertia and habit theory \citep{verplanken2006interventions,wood2009habitual} therefore suggest that externally induced adopters may use the newly adopted online channel less intensively after adoption and may revert more strongly toward prior offline routines. This mechanism yields two related predictions. First, externally induced adopters should allocate a larger share of post-adoption spending to the offline channel than \textit{organic adopters}. Second, this gap should be larger when the external event attracts older customers, because older consumers tend to be more entrenched in existing habits and exhibit greater inertia \citep{yang2014dynamics}. Differences in post-adoption channel mix may have profit implications for retailers. Profit margins can differ across online and offline channels because of differences in prices, costs, fulfillment, or promotional intensity. As a result, two adopter groups with similar total spend can differ in profitability if they allocate spending differently across channels. The direction of this effect depends on relative channel margins. In our empirical setting, offline margins are higher on average, so greater offline persistence can raise profitability even when total spend is similar.

The second mechanism is forward buying. Time-limited promotions can induce customers to stock up or shift purchases into the promotional period, reducing subsequent demand \citep{neslin1985consumer,hendel2006sales,van2017sales}. This mechanism is particularly relevant for promotions designed to induce online adoption. Customers may adopt online shopping to take advantage of a discount or offer, spend more during the adoption period, and then spend less afterward. Thus, promotion-induced adoption may generate lower post-adoption spending than organic adoption, even if the promotion successfully moves customers online.


This framework yields the following empirical predictions:
\squishlist
\item H1: Customers who adopt a firm's online channels and become multichannel shoppers will increase their spending post-adoption relative to \textit{offline-only customers} (consistent with prior findings on multichannel shoppers; see $\S$\ref{sec:literature}).
\item H2: Customers with different adoption pathways will exhibit different post-adoption behavior in terms of spend, profitability, and channel mix. These differences will arise from a combination of customer demographics, consumer habit theory, and forward buying. Specifically,
\squishlisttwo
\item H2a: Consumers who adopt in response to time-limited promotions may increase spending during the promotional period but reduce spending thereafter, dampening (or reversing) the spending boost from multichannel adoption in the post-promotion period.
\item H2b: Externally driven adopters (vs.\ organic adopters) may allocate a larger share of post-adoption spending to the offline channel and may exhibit slower declines in offline spending share, as predicted by consumer inertia and habit theory.
\item H2c: The difference in offline spending share between externally driven adopters and \textit{organic adopters} may be larger when external factors cause relatively older consumers to adopt
\item H2d: Differences in offline spending share may contribute to higher (lower) profitability for externally driven adopters relative to \textit{organic adopters} if offline channels have higher (lower) margins than online channels.

\squishend
\squishend

\section{Setting and Data}
\label{sec:set_data}

Recall that the key goals of the paper are: (1) to empirically document the effect of adopting online shopping on post-adoption consumption behavior, (2) to test whether different channel adoption pathways lead to different post-adoption behaviors, and (3) to verify that the differences in post-adoption behavior can be predicted using theoretical underpinnings of purchase behavior in retail settings, so as to provide generalizable insights for managers and researchers. To accomplish these goals, we need data from a setting where we can observe not only when an \textit{offline-only consumer} adopts online shopping but also the driver of this decision, i.e., the adoption pathway. Further, we need sufficient variants in channel adoption pathways to meaningfully connect the post-adoption behavior to theory. Therefore, we focus on a retail data source that satisfies these requirements. 

Our data comes from one of the largest Brazilian pet supplies retailers, which sells a wide variety of products for pets, such as pet food and snacks, medicines/pharmacy items, hygiene and grooming products, and pet toys and accessories. Pet supplies is a large and competitive industry with about \$250 billion in annual revenue globally \citep{fortune_2020}. The firm has both offline and online channels. The retailer operates more than 200 physical stores in Brazil, which serve as the offline channel, and has a website as well as a mobile app, which serve as the online channel. Moreover, our data span 2019-2024 during which there were three different events including a macro/environmental shock (COVID-19) and two different types of promotions designed to incentivize online shopping (discussed in more detail in $\S$\ref{sec:est_sum}) which led to many consumers adopting online shopping with the firm. Further, our data are quite rich, and we are able to observe when a consumer adopts online shopping (and thus can connect the timing of adoption to the channel adoption pathway) and both the pre- and post-adoption consumption behavior. For these reasons, it is an ideal setting for our study.

Our data includes transaction-level records of all individual customers from both online and offline channels during the 63-month period from January 1st, 2019, to March 31st, 2024. Each transaction record is uniquely identified by an order number with one or more purchased items. For each purchased item, we have information on the unit price, quantity purchased, brand, purchased channel (online/offline), subcategory, and category (e.g., food, hygiene). 

The firm tracks customer purchases using a National Identification Number (similar to a Social Security Number), which is unique to each individual in Brazil. Customers who purchase online (either through the website or the app) must be logged in to their profile, and therefore, all online purchases can be traced back to the National ID. Customers are also asked for their National ID when making an in-store purchase; however, this is not mandatory. Discussions with the retailer revealed that while most offline transactions are also linked with a customer profile, about a quarter of in-store transactions are not linked to a National ID. This number has remained fairly steady during the entire observation period and was not affected by COVID-19 or other marketing events\footnote{Because the number of offline transactions not linked to customer IDs remained similar through our observation period and exhibited no significant shifts in response to any of the events that spurred online adoption (including COVID-19), it is unlikely that this issue systematically biases our findings on the adoption of online shopping.}. 

A potential concern with any channel analysis is that the results may reflect assortment differences across the two channels. To ensure that this is not the case, we exclude data from four product categories that were not available in both channels for the entirety of the observation period; see Table \ref{category_remove} in Web Appendix $\S$\ref{appsec:data_cleaning} for a summary of these categories and their availability information. Together, sales from these categories account for only 0.63\% of the total sales during the observation period; as such, we do not expect these to play any significant role in the findings.

\section{Estimation Sample and Summary Statistics}
\label{sec:est_sum} 

We first discuss the sample construction process for the three focal events in $\S$\ref{ssec:sample_construction}, and present the details of the samples for each of the three focal events in $\S$\ref{ssec:focal_events},  and then present some summary statistics of the three samples of interest in $\S$\ref{ssec:summary_stats}.

\subsection{Sample Construction}
\label{ssec:sample_construction}
We now discuss how we construct the samples used in our empirical analyses and the main outcome variables of interest. 

\noindent \textbf{Focal events that triggered different pathways of adoption:}
Since we seek to connect channel adoption pathway to post-adoption behavior, we consider three events that each led to a surge in online adoption: (1) a macro-environmental shock (COVID-19), (2) a short-term promotion (Black Friday sales), and (3) a long-term loyalty program. The first event was a macroeconomic shock outside the firm's control, whereas the latter two were firm-driven marketing activities.

\noindent \textbf{Customer cohorts:} For each focal event, we first define three customer cohorts -- (a)  \textit{event adopters}, consumers who adopted online shopping during the event, (b) \textit{organic adopters}, consumers who adopted organically (without external incentives) just prior to the event, and (c) \textit{offline-only customers}, consumers, who did not adopt online shopping at all. 

\noindent \textbf{Time periods:} Next, we establish the pre-adoption, adoption, and post-adoption periods for each focal event as follows:
\squishlist
\item The pre-adoption period is defined as the common period during which all three groups of customers purchase only through physical stores (i.e., before the onset of the focal event). 
\item The adoption period is defined as the time during which adopters started purchasing online. Because each event has two sets of adopters, we have two adoption periods:
\squishlisttwo
\item Adoption period for \textit{event adopters}: This is the period during which the focal event took place, and induced customers to adopt online shopping. For instance, in the case of COVID-19, this is the first month and a half since the onset of COVID-19. Similarly, for the Black Friday event, the adoption period for \textit{event adopters} is simply the day of the Black Friday sale. 
\item Adoption period for \textit{organic adopters}: Since \textit{organic adopters} serve as a control group for \textit{event adopters}, for each event, we choose consumers who organically adopted online shopping {\it just before} the onset of the focal event, and this period varies based on the event.
\squishend
\item The post-adoption period refers to the months \textit{after} the adoption periods of both \textit{event adopters} and \textit{organic adopters}.
\squishend 
See Table \ref{tab:adoption_periods} for an overview of these periods for each of the three events. In our estimation sample for each event, we only include data from the pre- and post-adoption periods for that event. We exclude the adoption periods for both {\it event} and \textit{organic adopters} from the estimation sample because consumers are in flux during that time (where some adopters have adopted while others have not). Furthermore, by excluding these observations, we avoid capturing any abnormal purchase behaviors associated with the initial adoption of online shopping and the onset of specific events (i.e., COVID-19, Black Friday, Loyalty Program). As such, it allows us to cleanly identify the post-online-adoption difference in customers' purchase behaviors; see \citet{gu2021dark} for a similar specification. We now summarize the choice of these three periods for each event. 

\squishlist
\item COVID-19 event: We use one year of data before the adoption period as the pre-adoption period (the calendar year of 2019). For the post-adoption period, we use all the data from May 2020 till April 2023.
\item Black Friday event: We use eight months of data, starting from January 2019 to October 2020, as the pre-adoption period. For the post-adoption period, we use data from December 2019 (following the Black Friday event) until the next Black Friday.
\item Loyalty Program event: We use the data from five months before the adoption period, and all the data available (from the end of the adoption period till March 2024) as the post-adoption period.
\squishend 

\begin{table}[!ht]
\centering
\caption{Pre-adoption, Adoption, and Post-adoption Periods Under Each Event}
\label{tab:adoption_periods}
\small
\setlength{\tabcolsep}{8pt}
\begin{tabular}{@{} l  c  c  c @{}}
\toprule
 & \textbf{COVID-19} & \textbf{Black Friday} & \textbf{Loyalty Program} \\
\midrule
Pre-adoption & 
  2019-01-01 -- 2019-12-31 & 
  2019-01-01 -- 2019-10-31 & 
  2023-01-01 -- 2023-05-31 \\[6pt]

Adoption & & & \\[-1pt]
\quad Organic & 
  2020-01-01 -- 2020-02-29 & 
  2019-11-14 -- 2019-11-28 & 
  2023-06-01 -- 2023-07-31 \\[-2pt]
\quad Event-Induced & 
  2020-03-16 -- 2020-04-30 & 
  2019-11-29 & 
  2023-08-01 -- 2023-09-30 \\[6pt]

Post-adoption & 
  2020-05-01 -- 2023-04-30 & 
  2019-12-01 -- 2020-10-31 & 
  2023-10-01 -- 2024-03-31 \\
\bottomrule
\end{tabular}
\end{table}

We discuss each event and the definition of the three cohorts of consumers in detail below.

\noindent \textbf{Outcome metrics:} For all our analyses, we consider three metrics that quantify a consumer's shopping behavior in a given month: (1) Total Spend, (2) Profitability, and (3) Share of Offline Spend. Formally, we are interested in the Average Treatment Effect on Treated (ATT) of adopting online shopping on these three metrics for each of the three focal events. Each of these metrics is defined at the customer-month level. 

\subsection{Estimation Samples for the Focal Events}
\label{ssec:focal_events}
We now describe the estimation sample used in the analysis of each of the three focal events. 
\subsubsection{Event 1: COVID-19}
\label{sssec:covid_sample}

\begin{figure}[htp!]
    \centering
    \caption{First-Time Online Adoption Around Three Focal Events}\label{fig:first_time_online_adoption_events}
    \begin{subfigure}{0.30\textwidth}
        \centering
        \caption{COVID-19}
\includegraphics[width=\linewidth]{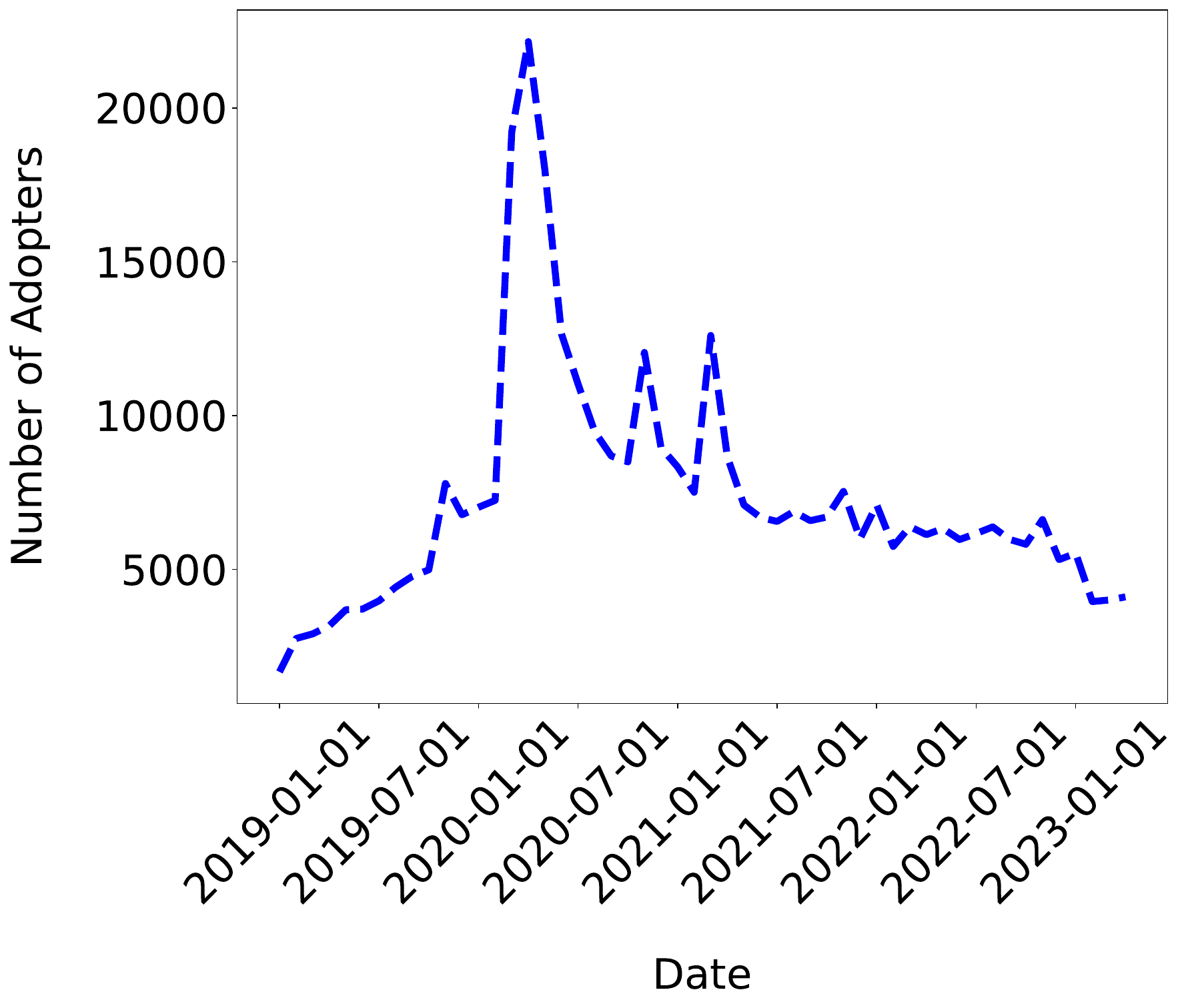}
\label{fig:covid_shock}
    \end{subfigure}%
    \hfill
    \begin{subfigure}{0.30\textwidth}
        \centering
        \caption{Black Friday}
\includegraphics[width=\linewidth]{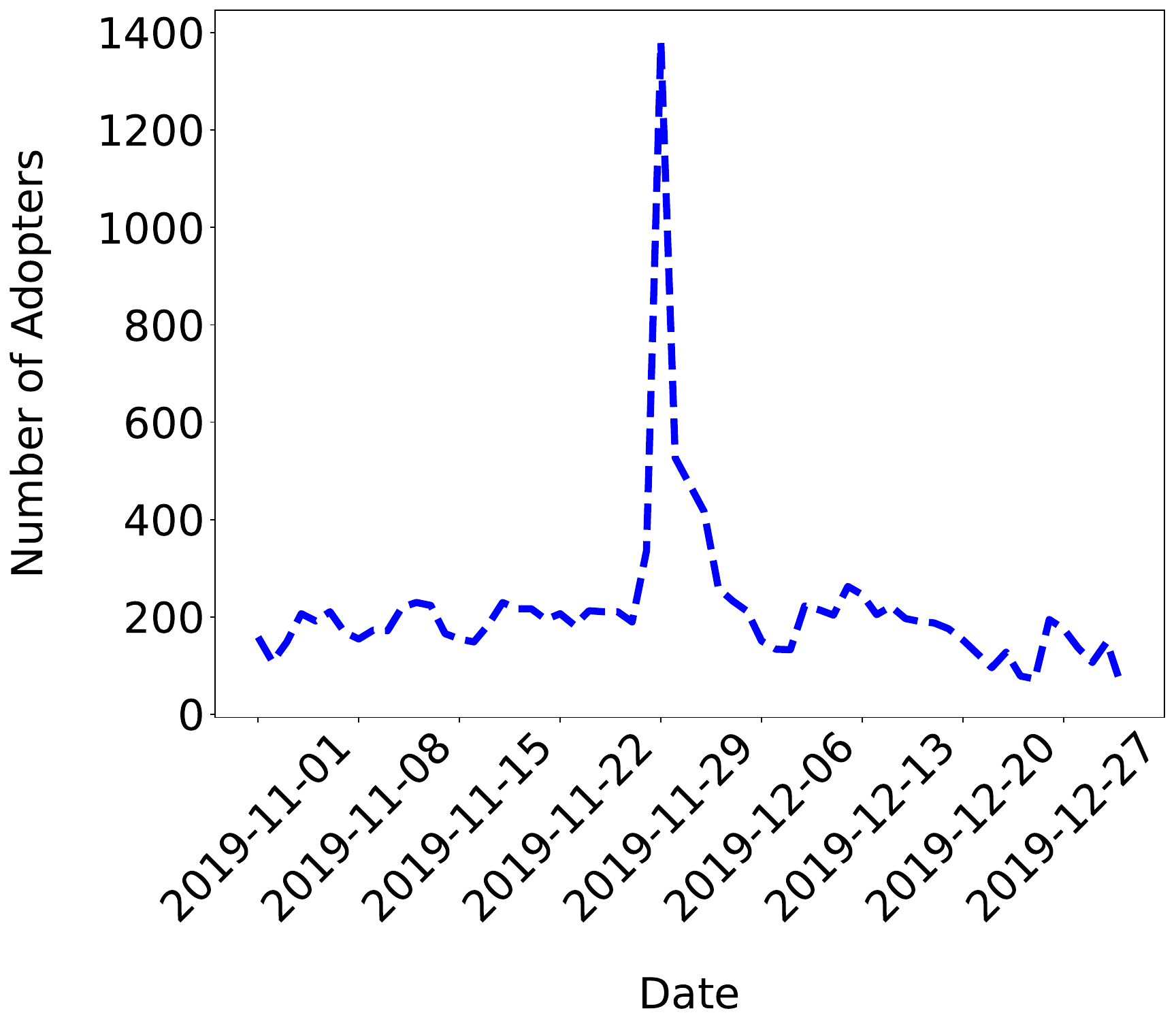}
 \label{fig:blackfriday_shock}
    \end{subfigure}%
    \hfill
    \begin{subfigure}{0.30\textwidth}
        \centering
        \caption{Loyalty Program}
\includegraphics[width=\linewidth]{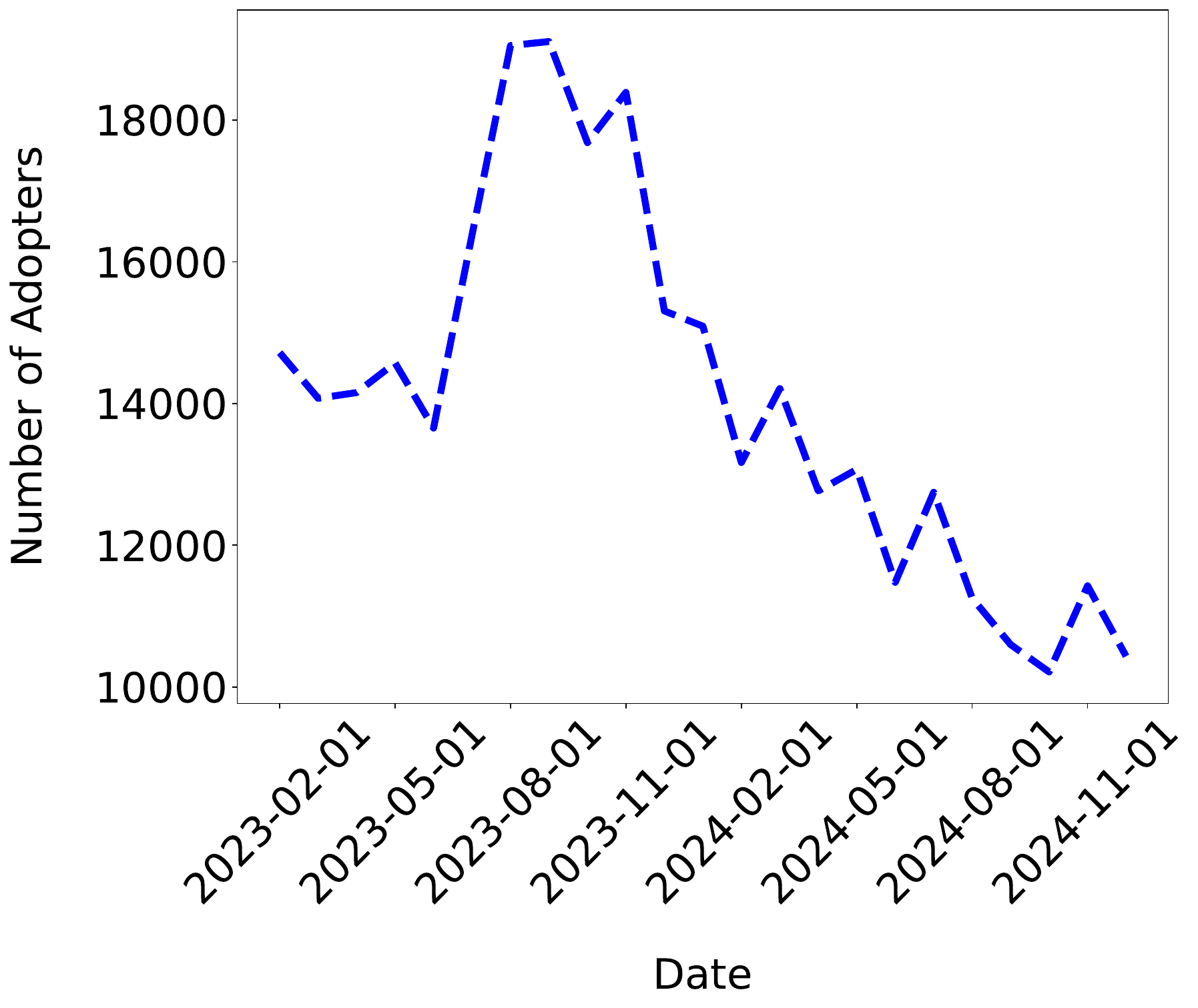}      
\label{fig:loyalty_shock}
    \end{subfigure}
\vspace{0.3em} 
\begin{minipage}{\textwidth}
\footnotesize
\textit{Notes:} This figure plots the number of previously \textit{offline-only} customers who first adopted the retailer's online channel around each focal event. Panel (a) reports adoption around the onset of COVID-19 by month, Panel (b) reports adoption around Black Friday by day, and Panel (c) reports adoption around the Loyalty Program by month.
\end{minipage}
\end{figure}

The COVID-19 pandemic was an exogenous macro-environmental shock that prompted many consumers to adopt online shopping. Figure \ref{fig:covid_shock} shows the number of customers who adopted online shopping through the retailer's website and/or mobile app (from previously only shopping at physical stores) for each month for our observation period. As we can see, during March and April 2020, there was a large peak in the number of offline customers who first switched to online shopping; the number of adopters sharply increased from approx. 7,000 in January and February 2020 to approximately 20,000 in March and April 2020. Customers who adopted online shopping during the onset of COVID-19 may have done so for reasons like health concerns, social distancing norms, etc., whereas the customers who adopted online shopping before that did so of their own volition, for other reasons (e.g., the convenience of online shopping), unrelated to COVID-19.

We consider the set of customers who made their first offline purchase before 2019-12-31 and were offline-only through the end of 2019 and continued shopping with the firm after the onset of COVID-19. Then, we categorize each consumer from this set into one of the following three groups (see second column in Table \ref{tab:adoption_periods} for exact dates):
\squishlist
    \item {\it COVID adopters}: Customers who made their first online purchase during the onset of COVID-19, i.e., from 2020-03-16 to 2020-04-30. 2020-03-16 was the date on which the Brazilian government began implementing physical distancing and confinement measures in response to COVID-19 \citep{faria2021covid}. Thus, we use this date as the start of the onset of COVID-19 and define the customers who adopted online shopping in the first one-and-a-half months as \textit{COVID adopters}.
    
    \item \textit{Organic adopters}: Customers who adopted online shopping organically in the period just before the onset of COVID-19. They made their first online purchase between 2020-01-01 and 2020-02-29, i.e., the two months before the onset of COVID-19. 
    
    \item \textit{Offline-only customers}: Customers who are active through both the pre- and post-adoption periods (i.e., at least one purchase before 2019-12-31 and after 2020-05-01), and never make online purchases during the estimation period.
\squishend

In total, we have 33,246 {\it COVID adopters}, 12,799 \textit{organic adopters}, and 573,934 \textit{offline-only customers} for this COVID-19 analysis. Note that a large fraction of consumers were able to remain offline-only during COVID-19 because the focal retailer's stores were allowed to remain open, as pet supplies retailing was deemed an essential service by the Brazilian government.

\subsubsection{Event 2: Black Friday Sales}
\label{sssec:blackfriday_sample}

Next, we discuss the retailer's Black Friday sales event in 2019, characterized by a one-day, time-limited flash sales event offering discounts exclusively for online purchases. Because of the exclusive online nature of the event, consumers interested in these deals were motivated to adopt online shopping. Figure \ref{fig:blackfriday_shock} indicates a peak in the number of adopters on 2019-11-29, which is the date that Black Friday sales took place.

For this estimation sample, we consider the cohort of customers who made their first offline purchase before 2019-10-31 and continued to be the firm's customers after the Black Friday sales date. Then, we categorize each consumer from this cohort into one of the following three groups (see third column in Table \ref{tab:adoption_periods} for exact dates):
\squishlist
    \item {\it Black Friday adopters}: Customers who made their first online purchase during the 2019 Black Friday sales.
    \item \textit{Organic adopters}: Customers who adopted online shopping organically in the two weeks preceding the 2019 Black Friday sales. 
    \item \textit{Offline-only customers}: Customers who are active through both the pre- and post-adoption periods, and never make online purchases during the estimation period. 
\squishend
In total, we have 1380 {\it Black Friday adopters}, 3061 {\it organic adopters}, and 534,871 \textit{offline-only customers} for the Black Friday analysis. 
\subsubsection{Event 3: Loyalty Program}
\label{sssec:loyalty_sample}
 In August and September 2023,\footnote{Pilot studies for the loyalty program ran in two states (Rio Grande do Sul and Paraná) as early as May. Therefore, we exclude these two states from the estimation sample for this event.} the retailer launched a loyalty program and extensively advertised it to customers through email and store marketing.  Customers who wanted to join the LP had to use the mobile app (no offline options were available for activating the loyalty program). After completing the online activation, customers could access and activate monthly special offers and receive discounts on both {\it online} and {\it offline} purchases. Note that customers had to activate each month to participate in that month's special offers. Figure \ref{fig:loyalty_shock} presents the number of adopters by month since January 2023 and shows clear peaks in the number of adopters during August and September 2023, coinciding with the launch of the loyalty program. This suggests that online-only activation of the newly available loyalty program prompted a significant number of customers to adopt online shopping during the roll-out period.

For this event, we consider the cohort of customers who made their first offline purchase before 2022-12-31 and continued to be the firm's customers during the post-online-adoption period (i.e., after 2023-10-01). Then, we categorize each consumer from this cohort into one of the following groups (see last column of Table \ref{tab:adoption_periods} for details):
\squishlist
    \item {\it Loyalty Program adopters}: Customers who made their first online purchase in August or September 2023, with their first online purchase occurring \textit{after} they activated the loyalty program online. In such cases, we can attribute the online loyalty activation as a reason for online adoption, as the activation precedes the first online purchase and both events occur within a similar time frame. 
    \item \textit{Organic adopters}: Customers who made their first online purchase in June or July 2023, prior to the launch of the loyalty program. These consumers adopt online shopping organically, but may still activate loyalty offers after their online adoption. We will take this into account in our empirical analysis.
    \item \textit{Offline-only customers}: Customers who are active through both the pre- and post-adoption periods (i.e., at least one purchase before 2022-12-31 and after 2023-10-01), and never make online purchases during the estimation period. 
\squishend
In total, we have 1567 \textit{loyalty program adopters}, 16,746 \textit{organic adopters}, and 396,396 \textit{offline-only customers} for this analysis.

\subsection{Summary Statistics}
\label{ssec:summary_stats}


For each consumer in the estimation sample of any focal event, we aggregate their transaction-level data into customer-month-level panel data over the customer's entire purchase history within the relevant period (January 2019 to March 2024), starting from their initial purchase month and ending with the last observed purchase month. For example, if a customer purchased for the first time in our data in August 2019 and the last observed purchase is in July 2020, then the panel for this customer spans 12 months, from August 2019 to July 2020. If there are intermediate months in which a customer did not make any purchase, the values for the monthly purchase-related variables are set to zero.  

\begin{table}[htp!]
\centering
\caption{Customer-Level Pre-adoption Summary Statistics Across Adoption Events}
\label{tab:pre_adoption_stacked}
\resizebox{\textwidth}{!}{
\footnotesize{
\begin{tabular}{lcccccc}
\toprule
Variable  & Adopters (Mean) & Organic (Mean) & Diff & t-stats & p-value & Offline-only (Mean) \\
\midrule

\multicolumn{7}{c}{\textit{Panel A1: COVID-19 Event — Demographics and Pre-adoption Characteristics}} \\
\midrule
Age & 43.36 & 41.44 & 1.92 & 15.65 & 0.00 & 45.62 \\
Monthly Household Income & 2432.11 & 2200.69 & 231.41 & 12.30 & 0.00 & 2033.01 \\
Tenure (Days) & 245.12 & 215.63 & 29.49 & 23.99 & 0.00 & 239.70 \\

\midrule
\multicolumn{7}{c}{\textit{Panel B1: COVID-19 Event — Pre-adoption Purchase Behaviors}} \\
\midrule
AvgSpendPerMonth & 463.46 & 453.26 & 10.20 & 1.80 & 0.07 & 284.38 \\
AvgQuantitiesPerMonth & 4.98 & 4.43 & 0.55 & 6.15 & 0.00 & 2.96 \\
AvgOrdersPerMonth & 0.95 & 0.95 & -0.00 & -0.38 & 0.71 & 0.74 \\
AvgUniqueItemsPerMonth & 2.92 & 2.89 & 0.03 & 0.97 & 0.33 & 1.91 \\
AvgUniqueBrandsPerMonth & 2.32 & 2.34 & -0.02 & -0.92 & 0.36 & 1.55 \\
AvgUniqueSubcategoriesPerMonth & 2.02 & 2.04 & -0.02 & -1.10 & 0.27 & 1.39 \\
AvgUniqueCategoriesPerMonth & 1.65 & 1.66 & -0.01 & -0.62 & 0.54 & 1.17 \\
AvgProfitPerMonth & 162.12 & 162.36 & -0.24 & -0.12 & 0.91 & 97.52 \\

\midrule
\multicolumn{7}{c}{\textit{Panel A2: Black Friday Event — Demographics and Pre-adoption Characteristics}} \\
\midrule
Age & 41.86 & 42.72 & -0.86 & -2.49 & 0.01 & 47.31 \\
Monthly Household Income & 2278.60 & 2254.31 & 24.29 & 0.42 & 0.68 & 2046.46 \\
Tenure (Days) & 203.25 & 188.93 & 14.32 & 4.51 & 0.00 & 204.45 \\

\midrule
\multicolumn{7}{c}{\textit{Panel B2: Black Friday Event — Pre-adoption Purchase Behaviors}} \\
\midrule
AvgSpendPerMonth & 484.56 & 472.50 & 12.06 & 0.67 & 0.50 & 291.87 \\
AvgQuantitiesPerMonth & 5.00 & 4.49 & 0.50 & 2.07 & 0.04 & 3.00 \\
AvgOrdersPerMonth & 1.00 & 0.99 & 0.02 & 0.55 & 0.58 & 0.76 \\
AvgUniqueItemsPerMonth & 3.06 & 2.95 & 0.10 & 0.99 & 0.32 & 1.93 \\
AvgUniqueBrandsPerMonth & 2.44 & 2.39 & 0.05 & 0.72 & 0.47 & 1.57 \\
AvgUniqueSubcategoriesPerMonth & 2.13 & 2.08 & 0.04 & 0.74 & 0.46 & 1.41 \\
AvgUniqueCategoriesPerMonth & 1.72 & 1.69 & 0.03 & 0.77 & 0.44 & 1.19 \\
AvgProfitPerMonth & 165.90 & 168.36 & -2.46 & -0.36 & 0.72 & 99.52 \\

\midrule
\multicolumn{7}{c}{\textit{Panel A3: Loyalty Program Event — Demographics and Pre-adoption Characteristics}} \\
\midrule
Age & 44.57 & 43.43 & 1.14 & 2.78 & 0.01 & 47.01 \\
Monthly Household Income & 1959.34 & 2072.83 & -113.49 & -2.24 & 0.02 & 1975.35 \\
Tenure (Days) & 786.58 & 827.54 & -40.96 & -3.09 & 0.00 & 915.96 \\

\midrule
\multicolumn{7}{c}{\textit{Panel B3: Loyalty Program Event — Pre-adoption Purchase Behaviors}} \\
\midrule
AvgSpendPerMonth & 576.42 & 444.70 & 131.71 & 8.49 & 0.00 & 332.57 \\
AvgQuantitiesPerMonth & 4.31 & 3.25 & 1.06 & 6.44 & 0.00 & 2.62 \\
AvgOrdersPerMonth & 1.04 & 0.83 & 0.20 & 9.64 & 0.00 & 0.70 \\
AvgUniqueItemsPerMonth & 2.55 & 2.05 & 0.50 & 7.96 & 0.00 & 1.64 \\
AvgUniqueBrandsPerMonth & 2.00 & 1.64 & 0.36 & 8.48 & 0.00 & 1.31 \\
AvgUniqueSubcategoriesPerMonth & 1.70 & 1.41 & 0.29 & 8.76 & 0.00 & 1.14 \\
AvgUniqueCategoriesPerMonth & 1.42 & 1.20 & 0.22 & 8.57 & 0.00 & 0.98 \\
AvgProfitPerMonth & 198.59 & 164.46 & 34.12 & 6.04 & 0.00 & 125.81 \\

\bottomrule
\end{tabular}
}
}
\end{table}

Thus, for each event (COVID-19, Black Friday, and Loyalty Program), we generate customer-month-level panel data for the pre- and post-adoption periods. These three panel datasets form the basis of our empirical analysis going forward. To preserve the confidentiality of the data provider, we mask monetary values by multiplying the true values by an undisclosed scalar here and in subsequent sections. All descriptive statistics and model estimates involving monetary outcomes in the paper are reported in Masked Currency Units (MCU). 
Table \ref{tab:pre_adoption_stacked} presents the mean statistics of demographic (age and household income) and pre-adoption purchase behaviors (spend, quantity, orders, brands, categories, etc.) for the three groups of \textit{event adopters}, their corresponding \textit{organic adopters}, and \textit{offline-only} customers. We further present the detailed summary statistics on demographics (gender and location), pre- and post- and adoption purchase behaviors for the three groups of customers (\textit{offline-only}, \textit{organic adopters}, and \textit{event adopters}) from each event (COVID-19, Black Friday, and Loyalty Program) in Web Appendix $\S$\ref{appsec:app_descriptive}.  We discuss some key takeaways here.

Across our three samples for COVID-19, Black Friday, and the Loyalty Program, we find consistent differences between the two groups of adopters (\textit{organic adopters} and \textit{event adopters}) and \textit{offline-only customers}. For instance, the adopter groups are consistently younger, higher income, and have higher spend with the retailer pre-adoption, relative to \textit{offline-only customers}. 

However, there are some important differences between the two groups of adopters (\textit{event adopters} and \textit{organic adopters}) across all three events. For instance, in the case of COVID-19, \textit{event/COVID adopters} tend to be older, have higher incomes, and have been shopping with the retailer for a longer time compared to \textit{organic adopters}. However, these two groups have similar pre-adoption monthly spending. In the case of Black Friday, {\it event/Black Friday adopters} are younger, have been shopping with the retailer for a longer time, and have lower monthly spend with the retailer in the pre-adoption period relative to \textit{organic adopters}. In contrast, in the Loyalty program case, \textit{event adopters} tend to be slightly older, have been shopping with the retailer for a shorter period, and have a higher monthly spend with the retailer in the pre-adoption period compared to \textit{organic adopters}. However, the differences in characteristics between the two groups of adopters for each event are much smaller in magnitude relative to the differences between adopters and non-adopters (i.e. \textit{offline-only customers}).

\section{Empirical Analysis and Results}
\label{sec:did}

Our goal is to estimate the ATT (Average Treatment Effect on the Treated) of online adoption on post-adoption purchase behavior for different types of events that trigger the adoption of online shopping, and then connect these ATTs estimates with the different adoption pathways. To estimate our ATTs of interest, we use the difference-in-differences approach. This approach allows us to account for unobserved heterogeneity across customers and time-specific shocks. For each event, we use customer-month-level panel data for the estimation sample corresponding to that event and include customer-month observations from the pre- and post-adoption periods. We exclude the adoption periods from the estimation for the reasons discussed in $\S$\ref{ssec:sample_construction}.

The rest of this section is organized as follows. First, in $\S$\ref{ssec:results_adopters_offline_only} we compare \textit{adopters} and \textit{offline-only customers}. Next, in $\S$\ref{ssec:results_event_organic_adopters}, we compare \textit{event adopters} and \textit{organic adopters}. Finally, in $\S$\ref{ssec:parallel_tests}, we discuss parallel pre-trend tests for our difference-in-differences specifications, and in cases where parallel pre-trends fail, we present HonestDID results to assess the sensitivity of our measured effect sizes to violations of pre-trends.

\subsection{\textit{Adopters} vs. \textit{offline-only customers}}
\label{ssec:results_adopters_offline_only}

We first estimate the ATT of online adoption by comparing \textit{adopters} with \textit{offline-only customers} for each of the three events: COVID-19, Black Friday, and the Loyalty Program. In this section, we focus on two outcomes: total spend and profitability. We do not analyze the share of offline spend here because channel mix is mechanically determined for \textit{offline-only customers}, who never adopt the online channel during the estimation period. We return to channel mix in $\S$\ref{ssec:results_event_organic_adopters}, where both comparison groups consist of online adopters.

We estimate the following two-way fixed effects (TWFE) specification with customer fixed effects ($\gamma_i$) and time (month) fixed effects ($\eta_t$):
\begin{equation}
\begin{aligned}
\label{eq:did_fe}
    y_{it} & = \gamma_{i} + \eta_{t} 
    + \beta_{0} \mathbbm{I}\{Post_{t}\} \times \mathbbm{I}\{Adopter_{i}\} + \boldsymbol{{\beta}}_{1}' \boldsymbol{{x}}_{it} + \epsilon_{it}.
\end{aligned}
\end{equation} 
Here, $y_{it}$ is the monthly outcome of interest for customer $i$ in month $t$. $\mathbbm{I}\{Adopter_i\}$ equals 1 for customers who adopt online shopping, including both \textit{organic adopters} and \textit{event adopters}, and 0 for \textit{offline-only customers}. $\mathbbm{I}\{Post_t\}$ equals 1 in the post-adoption period and 0 in the pre-adoption period. The coefficient of interest, $\beta_0$, captures the average incremental post-adoption change in the outcome for \textit{adopters}, relative to \textit{offline-only customers}. For the Loyalty Program analysis, we also include time-varying controls $\boldsymbol{x}_{it}$ that capture customers' participation in the loyalty program, such as whether they activate an offer. We describe these controls in Web Appendix $\S$\ref{appssec:control_variable_def}.

This specification addresses our first research question and tests H1: whether customers who adopt the firm's online channel increase their spending and profitability after adoption, relative to \textit{offline-only customers}.

\paragraph{Spend.}
Table \ref{tab:did_spend_adopter_offlineonly} presents the results with monthly spend as the dependent variable. Across all three events, the coefficient on $Adopter \times Post$ is positive and statistically significant. To interpret the magnitudes of the estimated ATTs, we scale each ATT by the pre-adoption monthly spend of adopters, using the corresponding estimates from the standard two-by-two DID specification reported in Web Appendix $\S$\ref{appssec:two_by_two_did_adopter_nonadopter}. The implied effect sizes are substantial: monthly spend increases by 30.6\% for COVID adopters, 16.7\% for Black Friday adopters, and 18.6\% for Loyalty Program adopters. Thus, online adoption is associated with economically meaningful increases in spend across all three events.

\begin{table}[htp!]
  \caption{DID Main Analysis -- Spend (\textit{Adopters} vs. \textit{Offline-only})}
  \centering
  \label{tab:did_spend_adopter_offlineonly}
  \begin{threeparttable}
\footnotesize{
\begin{tabular}{lccc}
\midrule\midrule
Study: & COVID-19 & Black Friday & Loyalty Program \\
\midrule
DV: & \multicolumn{3}{c}{Spend} \\
\midrule
\emph{Variables} \\

Adopter $\times$ Post 
& 140.1 (2.974) 
& 78.37 (7.078) 
& 84.89 (4.238) \\
& [0.000] & [0.000] & [0.000] \\

\midrule \midrule
\% Effect size relative to the pre-adoption baseline & 30.6\% & 16.7\% & 18.6\% \\ 
\midrule \midrule
Loyalty program controls &  &  & Yes \\
\midrule

\emph{Fixed-effects} \\
Customer & Yes & Yes & Yes \\
YearMonth & Yes & Yes & Yes \\

\midrule
\emph{Fit statistics} \\
Observations & 22,204,608 & 9,505,999 & 4,403,449 \\
R$^2$ & 0.38214 & 0.41557 & 0.48127 \\
Within R$^2$ & 0.00102 & $6.84\times10^{-5}$ & 0.00294 \\

\midrule\midrule
\multicolumn{4}{l}{\emph{Clustered (Customer) standard errors in parentheses; p-values in brackets.}} \\
\end{tabular}
}
\begin{tablenotes}[flushleft]
\footnotesize
\item \textit{Notes:} Percentage effects equal the DID estimate divided by the pre-adoption mean of adopters, reported in Table \ref{tab:did_spend_standard} in Web Appendix \ref{appssec:two_by_two_did_adopter_nonadopter}.
\end{tablenotes}
\end{threeparttable}
\end{table}

\paragraph{Profitability.}
We next examine whether the increase in spend translates into higher profitability. In our data, we observe item-level profit margins for each purchase, which allows us to construct monthly customer-level profitability. Table \ref{tab:did_profitability_adopter_offlineonly} presents the results with monthly profit as the dependent variable. The coefficient on $Adopter \times Post$ is again positive and statistically significant across all three events. To interpret the effect sizes, we scale the estimates by adopters' pre-adoption monthly profitability. We get effect sizes of 24.3\%, 11.7\%, and 11.0\%, respectively. Thus, the post-adoption increase in spend is accompanied by a corresponding increase in profitability.

\begin{table}[htp!]
   \caption{DID Main Analysis -- Profit (\textit{Adopters} vs. \textit{Offline-only})}
   \centering
   \label{tab:did_profitability_adopter_offlineonly}
     \begin{threeparttable}
\footnotesize{
\begin{tabular}{lccc}
\midrule\midrule
Study: & COVID-19 & Black Friday & Loyalty Program \\
\midrule
DV: & \multicolumn{3}{c}{Profit} \\
\midrule
\emph{Variables} \\

Adopter $\times$ Post 
& 37.37 (1.070) 
& 18.44 (2.506) 
& 18.41 (1.506) \\
& [0.000] & [0.000] & [0.000] \\ 
\midrule \midrule
\% Effect size relative to the pre-adoption baseline & 24.3\% & 11.7\% & 11.0\% \\
\midrule \midrule
Loyalty program controls &  &  & Yes \\
\midrule

\emph{Fixed-effects} \\
Customer & Yes & Yes & Yes \\
YearMonth & Yes & Yes & Yes \\

\midrule
\emph{Fit statistics} \\
Observations & 22,204,608 & 9,505,999 & 4,403,449 \\
R$^2$ & 0.31903 & 0.34750 & 0.44546 \\
Within R$^2$ & 0.00047 & $2.48\times 10^{-5}$ & 0.00193 \\

\midrule\midrule
\multicolumn{4}{l}{\emph{Clustered (Customer) standard errors in parentheses; p-values in brackets.}}
\end{tabular}
}
\begin{tablenotes}[flushleft]
\footnotesize
\item \textit{Notes:} Percentage effects equal the DID estimate divided by the pre-adoption mean of adopters, reported in Table \ref{tab:did_profit_standard} in Web Appendix \ref{appssec:two_by_two_did_adopter_nonadopter}.
\end{tablenotes}
\end{threeparttable}
\end{table}

\paragraph{Summary of key takeaways.}
The results in this subsection support H1. Across all three events, customers who adopt online shopping subsequently spend more and generate higher profit than comparable \textit{offline-only customers}. The magnitude of these effects is economically meaningful: spend increases by 16.7\% to 30.6\%, while profitability increases by 11.0\% to 24.3\%, depending on the adoption setting.

These results are consistent with prior work showing that multichannel customers are more valuable than single-channel customers \citep{neslin2009key, montaguti2016can}. However, this comparison pools \textit{organic adopters} and \textit{event adopters} into a single adopter group. As a result, it abstracts away from heterogeneity in why customers adopted online shopping. In the next subsection, we examine this heterogeneity directly by comparing each group of \textit{event adopters} with \textit{organic adopters}.

\subsection{\textit{Event adopters} vs. \textit{organic adopters}}
\label{ssec:results_event_organic_adopters}

We next examine whether post-adoption behavior differs by the reason for online adoption. Specifically, for each event, we compare the corresponding group of \textit{event adopters} with \textit{organic adopters}. This comparison isolates heterogeneity across adoption pathways among customers who all adopt online shopping. We estimate the following two-way fixed effects (TWFE) specification:
\begin{equation}
\label{eq:did_fe_2by2_FE}
y_{it} = \gamma_{i}  + \eta_{t} 
+ \beta_2 \mathbbm{I}\{Event\_Adopter_{i}\}\times  \mathbbm{I}\{Post_{t}\} 
+ \boldsymbol{\beta}_{3}' \boldsymbol{x}_{it} + \epsilon_{it}.
\end{equation}
Here, $y_{it}$ is the monthly outcome of interest for customer $i$ in month $t$. $\gamma_i$ and $\eta_t$ denote customer and month fixed effects, respectively. $\mathbbm{I}\{Event\_Adopter_i\}$ equals 1 for the relevant group of \textit{event adopters} and 0 for the corresponding group of \textit{organic adopters}. $\mathbbm{I}\{Post_t\}$ equals 1 in the post-adoption period and 0 in the pre-adoption period. The coefficient of interest, $\beta_2$, captures the difference in the post-adoption change in the outcome between \textit{event adopters} and \textit{organic adopters}. For the Loyalty Program analysis, we include the same time-varying loyalty-program controls described in Web Appendix $\S$\ref{appssec:control_variable_def}.

We report the TWFE estimates as our main estimates. We also estimate a standard two-by-two DID specification for all the outcomes of interest and report those results in Web Appendix $\S$\ref{appssec:two_by_two_did_event_adopter_spend}. The two-by-two DID results are consistent with the TWFE results and are used to construct the effect-size benchmarks reported in the main tables.

To facilitate interpretation, we report two effect-size benchmarks for spend and profitability: the DID estimate scaled by the corresponding post-adoption change for \textit{organic adopters}, and the DID estimate scaled by the pre-adoption mean of \textit{event adopters}. If the organic-adopter benchmark is not statistically significant, we report the relative effect as NA; if the DID estimate is statistically insignificant, we report the effect size as 0.0\%. When both the DID estimate and the organic-adopter post-adoption change are negative, a positive value under the first benchmark indicates an additional decline relative to the organic-adopter decline, not a positive effect. For share of offline spend, the DID coefficient is already interpretable in percentage points, so we do not rescale it.

We first present results for spend in $\S$\ref{sssec:spend}, then examine the share of offline spend in $\S$\ref{sssec:share}, and finally analyze profitability in $\S$\ref{sssec:profitability}. These analyses address our second research question and test the second set of predictions, H2a through H2d, presented in $\S$\ref{sec:theory}.

\subsubsection{Spend}
\label{sssec:spend}

\paragraph{Main DID results and economic magnitudes}
Table \ref{tab:total_spend} presents the results from estimating Equation \eqref{eq:did_fe_2by2_FE} with monthly spend as the dependent variable. The coefficient on $Event\_Adopter \times Post$ is small and statistically insignificant for \textit{COVID adopters}, indicating no detectable difference in post-adoption spend between \textit{COVID adopters} and \textit{organic adopters}. In contrast, the estimates for both promotion-driven adoption pathways are negative and statistically significant. 

\begin{table}[htp!]
   \caption{DID Main Analysis -- Spend ({\it Event Adopters} vs.\ {\it Organic Adopters})}
   \label{tab:total_spend}
   \centering
   \begin{threeparttable}
\footnotesize{
\begin{tabular}{lccc}
\midrule\midrule
Study: & COVID-19 & Black Friday & Loyalty Program \\
\midrule
DV: & \multicolumn{3}{c}{Spend} \\
\midrule
\emph{Variables}\\
Event\_Adopter $\times$ Post 
& -3.095 (6.465) & -69.88 (14.26) & -53.22 (19.42) \\
& [0.632] & [0.000] & [0.006] \\
\midrule \midrule
\makecell[l]{\% Effect size relative to the \\post-adoption change for organic adopters} 
& 0.0\% & NA & 177.9\% \\
\midrule
\midrule
\makecell[l]{\% Effect size relative to the \\ pre-adoption baseline for event adopters} 
& 0.0\% & -14.2\% & -9.2\% \\
\midrule
\midrule
Loyalty program controls &  &  & Yes \\
\midrule
\emph{Fixed-effects}\\
Customer  & Yes & Yes & Yes \\
YearMonth & Yes & Yes & Yes \\
\midrule
\emph{Fit statistics}\\
Observations & 1,783,949 & 76,842 & 197,658 \\
R$^2$        & 0.39193   & 0.41147 & 0.47126 \\
Within R$^2$ & $5.37\times10^{-7}$ & 0.00068 & 0.02692 \\
\midrule\midrule
\multicolumn{4}{l}{\emph{Clustered (Customer) standard errors in parentheses; p-values in brackets.}}
\end{tabular}
}
\begin{tablenotes}[flushleft]
\footnotesize
\item \textit{Notes:} Effect sizes are calculated using the standard two-by-two DID estimates reported in Table \ref{tab:did_spend_standard_event_adopter} in Web Appendix \ref{appssec:two_by_two_did_event_adopter_spend}. The effect-size conventions are described at the beginning of $\S$\ref{ssec:results_event_organic_adopters}.
\end{tablenotes}
\end{threeparttable}
\end{table}

The effect-size rows in Table \ref{tab:total_spend} show that the promotion-driven differences are economically meaningful. Relative to their own pre-adoption baselines, \textit{Black Friday adopters} spend 14.2\% less and \textit{Loyalty Program adopters} spend 9.2\% less than comparable \textit{organic adopters} in the post-adoption period. The relative-to-organic benchmark also shows that the Loyalty Program reduction is large: the additional decline for \textit{Loyalty Program adopters} is 177.9\% of the post-adoption decline observed among \textit{organic adopters}. For \textit{COVID adopters}, the spend difference is statistically indistinguishable from zero.

Taken together, these results show that the adoption pathway matters for post-adoption spend. Customers who adopt online shopping during COVID-19 spend similarly to \textit{organic adopters} after adoption. By contrast, customers who adopt through firm-driven promotions spend less than \textit{organic adopters} after adoption. This pattern is consistent with H2a and with theories of forward buying and demand distortion \citep{neslin1985consumer,hendel2006sales,van2017sales}: time-limited promotions can induce customers to pull future demand into the promotional period, reducing spending in the post-promotion period.

\paragraph{Forward buying mechanism.}
The negative post-adoption spend effects for \textit{Black Friday adopters} and \textit{Loyalty Program adopters} may arise because these customers increase spending during the adoption period itself and then reduce spending afterward. That is, promotion-driven adopters may pull forward future demand, generating a post-promotion dip in spending.

To examine this mechanism, we estimate an event study specification that compares each promotion-driven adopter group with the corresponding group of \textit{organic adopters} in each month around adoption:
\begin{equation}
\begin{aligned}
\label{eq:event_study}
    y_{it} & = \gamma_{i} + \eta_{t} 
    + \sum_{j} \theta_{j} \left(PRE_{it}(j)\cdot \mathbbm{I}\{Event\_Adopter_{i}\}\right) \\
    & \quad + \sum_{k} \theta_{k} \left(POST_{it}(k)\cdot \mathbbm{I}\{Event\_Adopter_i\}\right) 
    + \boldsymbol{\lambda}'\boldsymbol{x}_{it} + \epsilon_{it}.
\end{aligned}
\end{equation}
Here, $PRE_{it}(j)$ is an indicator that equals 1 if month $t$ is $j$ months before the adoption month. The coefficients on these pre-adoption indicators allow us to assess differential pre-trends between \textit{event adopters} and \textit{organic adopters}; we discuss these parallel-trend diagnostics in $\S$\ref{ssec:parallel_tests}. Similarly, $POST_{it}(k)$ is an indicator that equals 1 if month $t$ is $k$ months after adoption. The corresponding coefficients capture the dynamic differences in spend between \textit{event adopters} and \textit{organic adopters} around and after adoption. We normalize outcomes relative to the month immediately prior to adoption.

\begin{figure}[htp!]
    \centering
    \caption{Event-Study Evidence on Forward Buying}
    \label{fig:Forward_Buying}
        \begin{subfigure}{0.48\textwidth}
        \centering
        \caption{Black Friday}
        \includegraphics[scale=0.3]{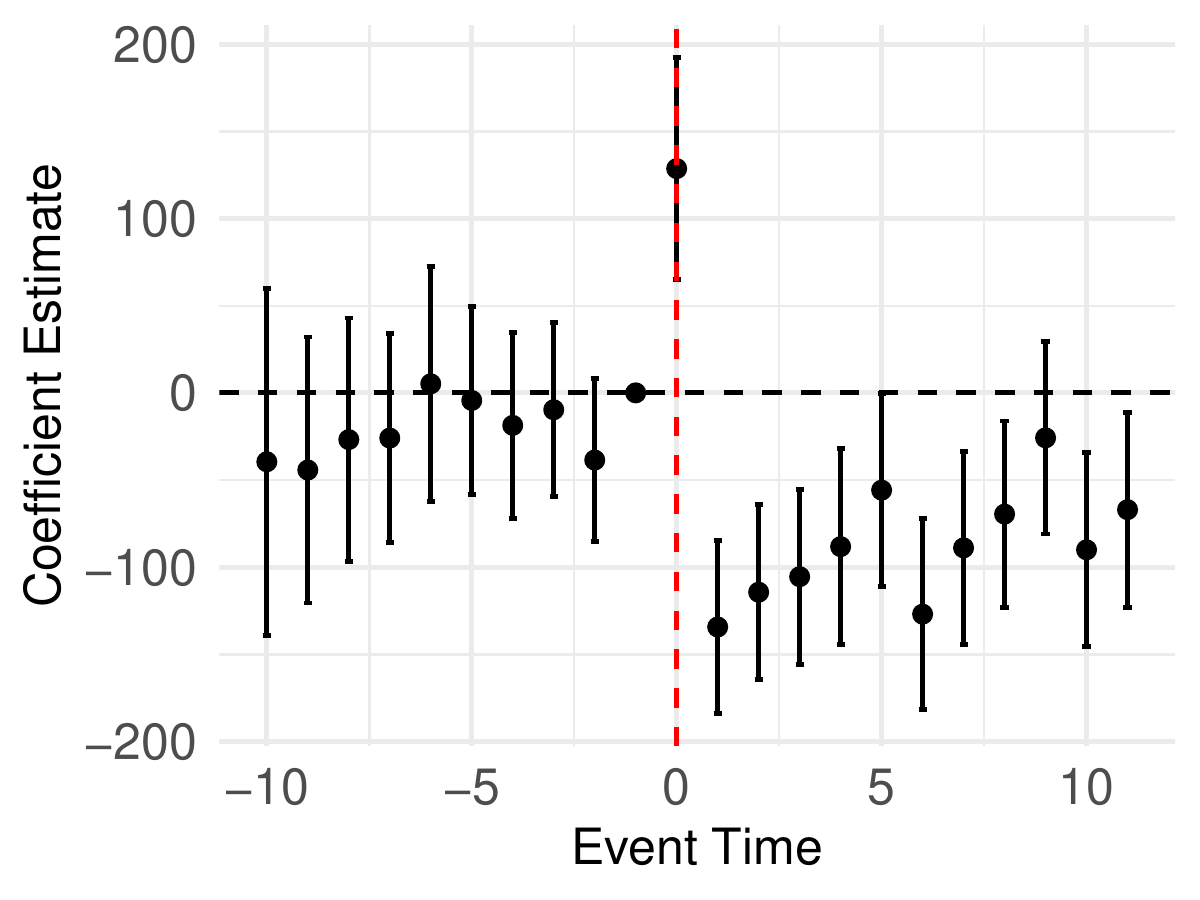}
    \label{fig:Forward_Buying_BF}
    \end{subfigure}%
    ~
    \begin{subfigure}{0.48\textwidth}
        \centering
        \caption{Loyalty Program}
        \includegraphics[scale=0.3]{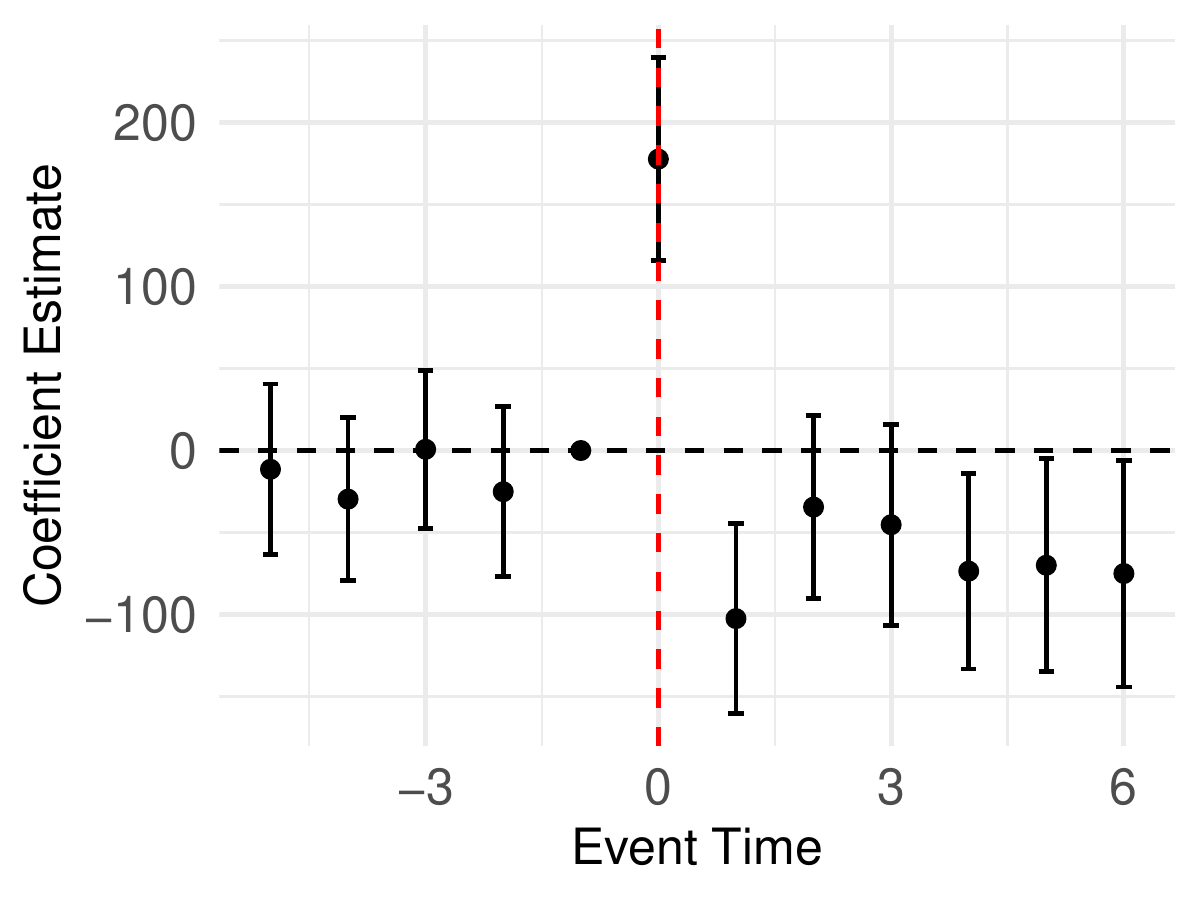}
        \label{fig:Forward_Buying_Loyalty}
    \end{subfigure}
\vspace{2em} 
\begin{minipage}{0.96\textwidth}
\footnotesize
\textit{Notes:} This figure plots event-study estimates of monthly spend for \textit{Black Friday adopters} and \textit{Loyalty Program adopters} relative to the corresponding group of \textit{organic adopters}. Estimates before the adoption month assess pre-event differences, while estimates after the adoption month show post-adoption spending patterns. Error bars = +/- 1.96 SE. Both promotion-driven adopter groups spend more during the adoption month but less in the following months, consistent with forward buying.
\end{minipage}
\end{figure}

We estimate Equation \eqref{eq:event_study} for Black Friday and the Loyalty Program and report the full estimation results in Web Appendix $\S$\ref{appssec:two_by_two_did_event_adopter_spend}. Figure \ref{fig:Forward_Buying} plots the dynamic treatment effects and 95\% confidence intervals. The evidence is consistent with forward buying for both promotion-driven adoption pathways. During the month of adoption, \textit{Black Friday adopters} and \textit{Loyalty Program adopters} spend more than \textit{organic adopters}. In the months following adoption, they spend less.\footnote{When we examine offline spending only, there are no significant differences in offline spending between \textit{Black Friday adopters} and \textit{organic adopters}; see Figure \ref{fig:Forward_Buying_OfflineSpend_BF} in Web Appendix $\S$\ref{appssec:two_by_two_did_event_adopter_spend}. This suggests that the forward buying effect for Black Friday is specific to the online channel. For the Loyalty Program, we observe the effect across both channels.} We also confirm that for COVID-19, which is not a firm-driven promotion, there is no comparable evidence of forward buying; see Figure \ref{fig:Forward_Buying_Covid} in Web Appendix $\S$\ref{appssec:two_by_two_did_event_adopter_spend}. 

We interpret these event-study patterns as evidence consistent with forward buying, rather than as a definitive mechanism test. Other explanations, such as one-time deal seeking, channel disappointment after initial online use, or differential retention among promotion-induced adopters, could also contribute to the observed post-adoption decline in spending.

\paragraph{Summary of key takeaways.}
The spend results provide three main takeaways. First, \textit{COVID adopters} do not differ meaningfully from \textit{organic adopters} in post-adoption spend. Second, promotion-driven adopters spend less than \textit{organic adopters} after adoption, with economically meaningful reductions of 14.2\% for \textit{Black Friday adopters} and 9.2\% for \textit{Loyalty Program adopters}, relative to their respective pre-adoption baselines. Third, the event study evidence shows that these post-adoption reductions are consistent with forward buying: promotion-driven adopters spend more during the adoption period but less afterward. Thus, managers should not expect promotion-driven online adopters to generate the same post-adoption spending gains as \textit{organic adopters}.

\subsubsection{Share of offline spend}
\label{sssec:share}

We next examine how adoption pathways affect post-adoption channel usage. As discussed in $\S$\ref{sec:theory}, consumer inertia and habit theory \citep{verplanken2006interventions,wood2009habitual} predict that adoption induced by external shocks or incentives may temporarily disrupt existing shopping routines, but that consumers may revert toward prior habits absent continued reinforcement. In our setting, this means that customers who adopt online shopping because of COVID-19, Black Friday, or the Loyalty Program may continue allocating a larger share of their spending to the offline channel than customers who adopt online shopping organically. This analysis tests H2b by comparing the post-adoption share of offline spend for each group of \textit{event adopters} relative to the corresponding group of \textit{organic adopters}.

The outcome variable is the share of monthly spending that occurs offline. It is defined only for customer-months with positive total spend; customer-months with zero total spend are treated as missing because the offline spending share is undefined in those months.

\paragraph{Main DID results and economic magnitude.}

Table \ref{tab:share_offline_online_spend} presents the results from estimating Equation \eqref{eq:did_fe_2by2_FE} with the share of offline spend as the dependent variable. Since this outcome is a share, the coefficient on $Event\_Adopter \times Post$ is directly interpretable as the incremental post-adoption difference in offline spending share for \textit{event adopters} relative to \textit{organic adopters}. The estimates imply that \textit{COVID adopters} allocate 6.98 percentage points more of their post-adoption spending to the offline channel, while \textit{Black Friday adopters} allocate 9.22 percentage points more. Both differences are statistically significant. In contrast, the estimate for \textit{Loyalty Program adopters} is only 0.18 percentage points and is statistically insignificant, indicating no detectable difference in post-adoption channel mix relative to \textit{organic adopters}.

\begin{table}[htp!]
   \caption{DID Main Analysis -- Share of Offline Spend ({\it Event Adopters} vs. {\it Organic Adopters})}
   \label{tab:share_offline_online_spend}
   \centering
   \begin{threeparttable}
   \footnotesize{
   \begin{tabular}{lccc}
\midrule\midrule
Study: & COVID-19 & Black Friday & Loyalty Program\\
\midrule
DV: & \multicolumn{3}{c}{Share of Offline Spend}\\
\midrule
\emph{Variables}\\
Event\_Adopter $\times$ Post
& 0.0698 (0.0042)
& 0.0922 (0.0120)
& 0.0018 (0.0114)\\
& [0.000]
& [0.000]
& [0.889]\\
\midrule \midrule
\makecell[l]{Incremental difference in share of offline spend\\relative to \textit{organic adopters} (percentage points)} 
& 6.98
& 9.22
& 0.18\\
\midrule \midrule
Loyalty program controls
& 
& 
& Yes\\
\midrule
\emph{Fixed-effects}\\
Customer
& Yes
& Yes
& Yes\\
YearMonth
& Yes
& Yes
& Yes\\
\midrule
\emph{Fit statistics}\\
Observations
& 949,448
& 42,299
& 110,339\\
R$^2$
& 0.50739
& 0.53685
& 0.57059\\
Within R$^2$
& 0.00140
& 0.00528
& 0.00904\\
\midrule\midrule
\multicolumn{4}{l}{\emph{Clustered (Customer) standard errors in parentheses; p-values in brackets.}}\\
   \end{tabular}
   }
   \begin{tablenotes}[flushleft]
\footnotesize
\item \textit{Notes:}  The percentage-point row equals the DID coefficient multiplied by 100. The effect-size conventions are described at the beginning of $\S$\ref{ssec:results_event_organic_adopters}.
\end{tablenotes}
\end{threeparttable}
\end{table}

\begin{figure}[htp!]
    \centering
    \caption{Share of Offline Spend Relative to Total Spend -- {\it Event Adopters} vs. {\it Organic Adopters}}
\label{fig:share_offline_spend_trend}
    \begin{subfigure}{0.32\textwidth}
        \centering
        \caption{COVID-19}
        \includegraphics[width=\linewidth]{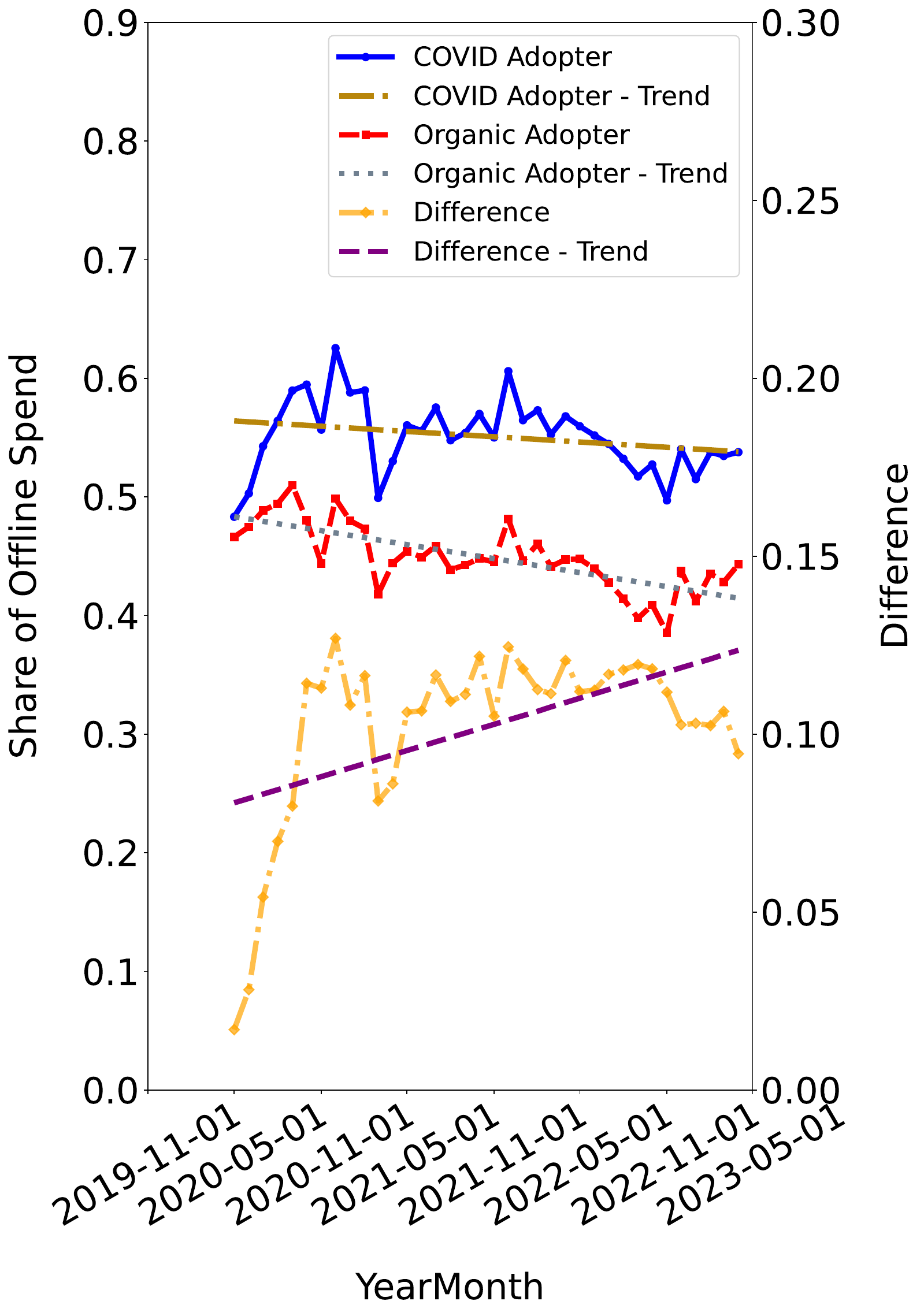}
     \label{fig:share_offline_spend_trend_covid}
    \end{subfigure}%
    \hfill
    \begin{subfigure}{0.32\textwidth}
        \centering
        \caption{Black Friday}
        \includegraphics[width=\linewidth]{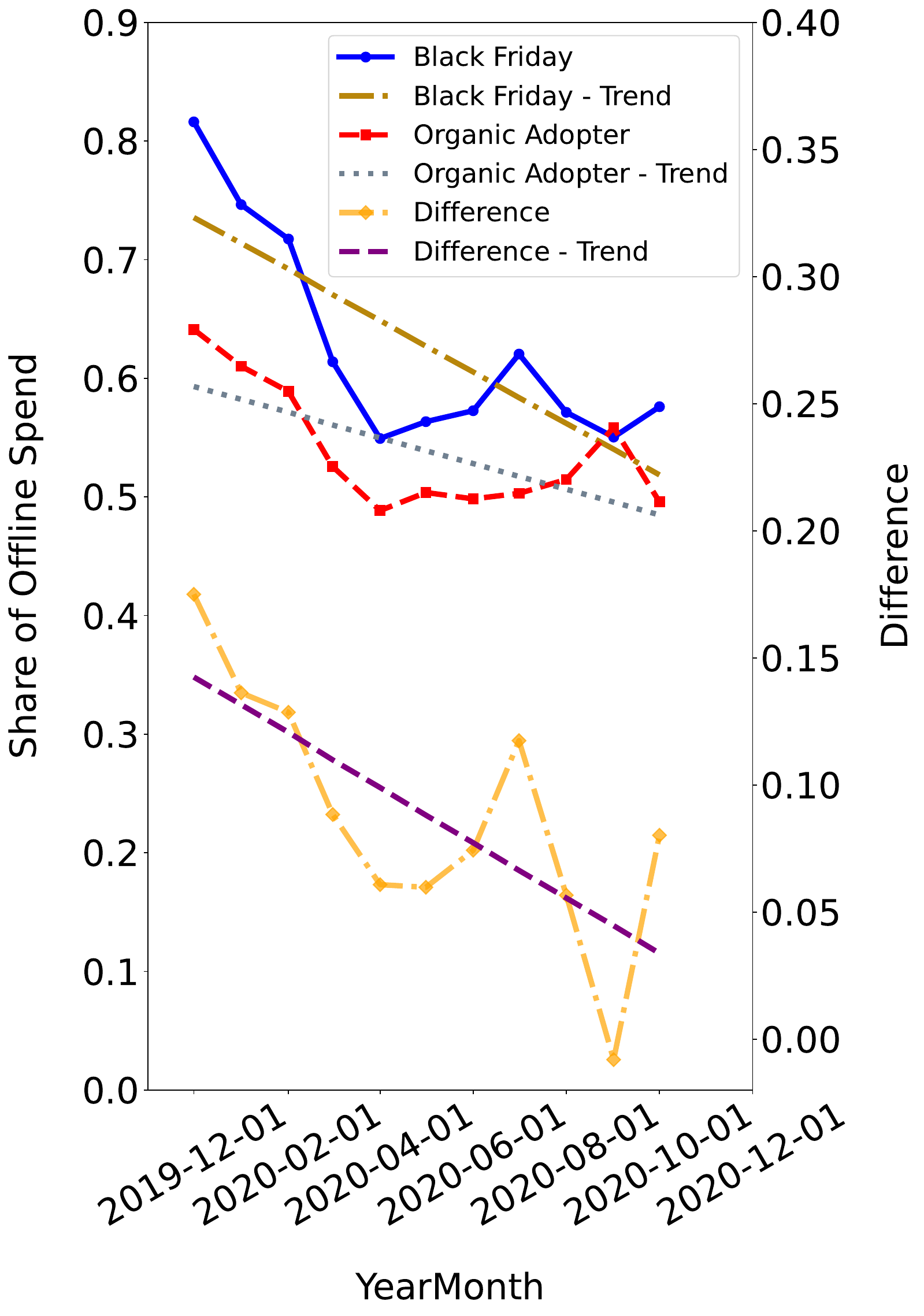}
 \label{fig:share_offline_spend_trend_blackfriday}
    \end{subfigure}%
    \hfill
    \begin{subfigure}{0.33\textwidth}
        \centering
        \caption{Loyalty Program}
        \includegraphics[width=\linewidth]{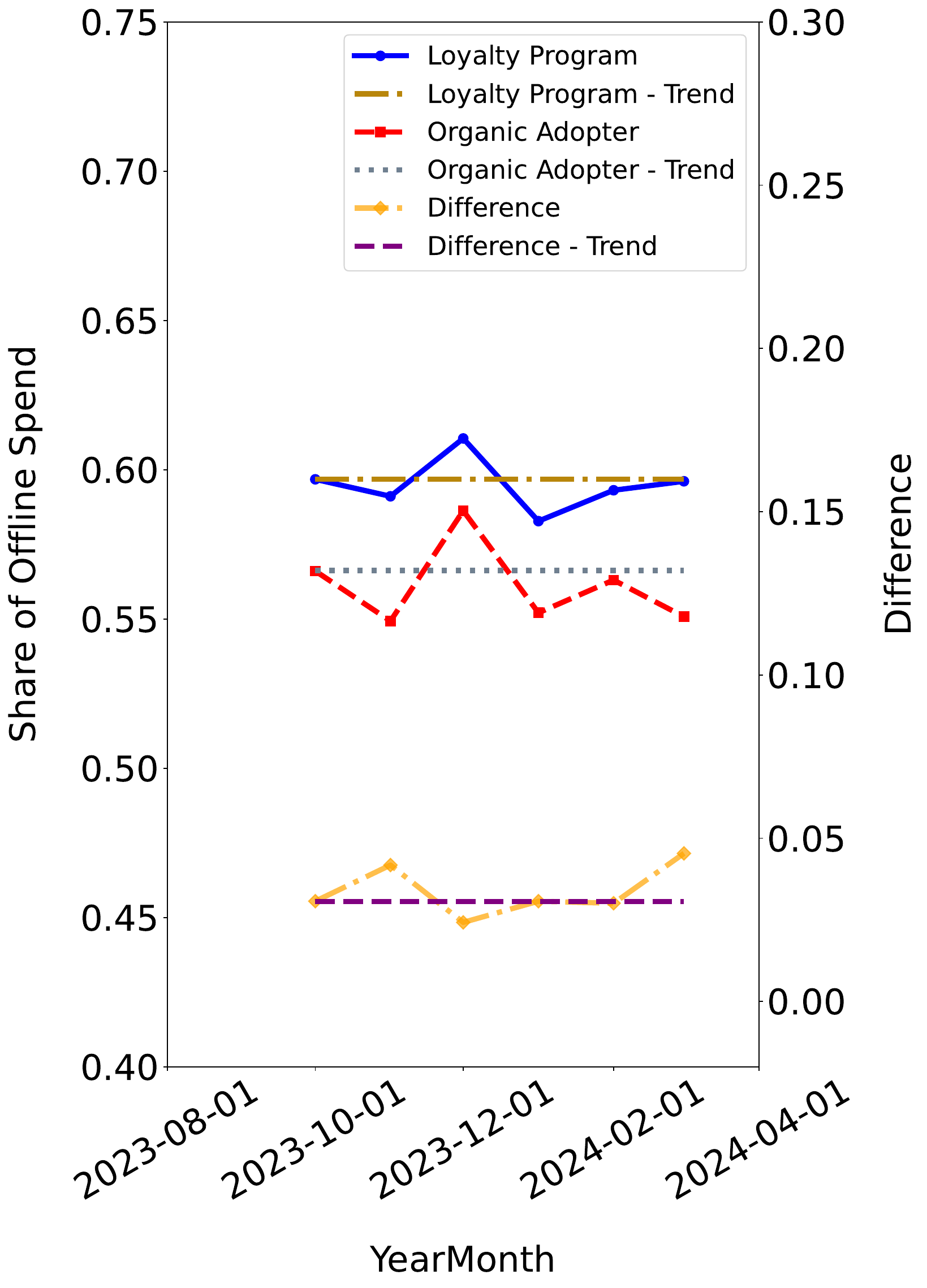}
\label{fig:share_offline_spend_trend_loyalty}
    \end{subfigure}
\end{figure}

\paragraph{Dynamic patterns and mechanism.}
The DID estimates above summarize average differences over the full post-adoption period. We next examine whether these differences persist, widen, or narrow over time. Figure \ref{fig:share_offline_spend_trend} plots the share of offline spend for each group of \textit{event adopters} and the corresponding group of \textit{organic adopters}, along with the difference between the two groups.

For COVID-19, Figure \ref{fig:share_offline_spend_trend_covid} shows that \textit{COVID adopters} not only have a higher post-adoption offline spending share than \textit{organic adopters}, but also move toward online shopping more slowly over time; see Web Appendix $\S$\ref{appssec:did_share_offline_spend} for formal tests of the magnitude of the difference. This pattern is consistent with consumer inertia and habit theory, and supports H2b.

\paragraph{Theory and mechanism.}
The COVID-19 pattern is also consistent with H2c. As shown in $\S$\ref{ssec:summary_stats}, \textit{COVID adopters} are older than \textit{organic adopters}. Older consumers are more likely to be set in their shopping habits and exhibit greater inertia \citep{yang2014dynamics}. To examine whether the slower movement toward online shopping is concentrated among older customers, we split \textit{COVID adopters} and \textit{organic adopters} by median age. We find that the difference in offline-share trends appears among older adopters, but not among younger adopters.\footnote{See Figure \ref{fig:covid_offline_fraction_purchase_age} and Table \ref{tab:fraction_offline_organic_covid_timeTrend_byAge} in Web Appendix $\S$\ref{appssec:did_share_offline_spend} for details.} This supports the interpretation that the channel-mix differences for \textit{COVID adopters} are related to consumer inertia and age-based differences in habit persistence.

For Black Friday, Figure \ref{fig:share_offline_spend_trend_blackfriday} shows that \textit{Black Friday adopters} initially allocate a higher share of spending to the offline channel than \textit{organic adopters}, consistent with the positive DID estimate in Table \ref{tab:share_offline_online_spend}. However, they also increase their online share of spend faster over time.\footnote{The p-value for the trend difference is 0.109; see Table \ref{tab:fraction_offline_organic_covid_timeTrend} in Web Appendix $\S$\ref{appssec:did_share_offline_spend}. Extending the post-adoption period further reduces the p-value.} This pattern may reflect the fact that \textit{Black Friday adopters} are younger than \textit{organic adopters}, and thus may be less entrenched in offline habits and more willing to explore the convenience of online shopping. Nevertheless, because their initial adoption was induced by an external incentive, they still begin the post-adoption period with greater offline reliance than \textit{organic adopters}.

For the Loyalty Program, the DID estimate in Table \ref{tab:share_offline_online_spend} is small and statistically insignificant, and Figure \ref{fig:share_offline_spend_trend_loyalty} shows little separation between \textit{Loyalty Program adopters} and \textit{organic adopters}. This may be because the two groups are relatively similar in demographics and because the Loyalty Program requires recurring monthly online activation, which may reinforce online channel usage after adoption.

\paragraph{Summary of key takeaways.}
The share-of-offline-spend results show that externally induced adoption does not always produce the same post-adoption channel mix as organic adoption. \textit{COVID adopters} and \textit{Black Friday adopters} allocate significantly more of their post-adoption spending to the offline channel than \textit{organic adopters}, by 6.98 and 9.22 percentage points, respectively. In contrast, \textit{Loyalty Program adopters} do not differ significantly from \textit{organic adopters} in their offline spending share. The dynamic patterns further suggest that age and reinforcement matter: older \textit{COVID adopters} move online more slowly, while younger \textit{Black Friday adopters} appear to catch up over time, and \textit{Loyalty Program adopters} show no meaningful channel-mix divergence, possibly because repeated app-based activation reinforces online use. These findings support the role of consumer inertia and habit persistence in shaping post-adoption channel usage.

\subsubsection{Profitability}
\label{sssec:profitability}

We next examine how adoption pathways affect profitability. The spend results in $\S$\ref{sssec:spend} show that promotion-driven adopters spend less than \textit{organic adopters} after adoption, while \textit{COVID adopters} spend similarly to \textit{organic adopters}. The channel-mix results in $\S$\ref{sssec:share} show that \textit{COVID adopters} and \textit{Black Friday adopters} allocate a larger share of their post-adoption spending to the offline channel. Profitability can therefore differ across adoption pathways for two reasons: differences in total spend and differences in the allocation of spend across channels.

\paragraph{Channel margin differences.}
To interpret the profitability results, we first document differences in prices and profit margins across the offline and online channels. We construct a panel at the item-channel-month level.\footnote{For instance, in the case of price, an observation in this panel is the average price of item $j$ sold through channel $c$ in month $t$, calculated as the total sales of item $j$ in channel $c$ during month $t$ divided by the total quantity of that item sold in that channel-month.} We then calculate offline-online differences at the item-month level and average these differences over time to obtain item-level average differences in price and profit margin across channels. Table \ref{tab:item_level_price_margin} reports item-level summary statistics for the 22,803 items that were sold in both offline and online channels in our data and were purchased at least once in the same month. For confidentiality, we multiply profit margins by the same undisclosed multiplier used for prices.

\begin{table}[htp!]
    \centering
    \caption{Item-Level Offline Price and Profit Margin Relative to Online}
    \label{tab:item_level_price_margin}
    \footnotesize{
\begin{tabular}{lrrrrrrr}
\toprule
{} &  mean & std & 25\% & 50\% & 75\% & (min, max) & count \\
\midrule
Offline - Online Price & 6.01 & 91.07 & -0.84 & 4.61 & 17.71 & (-7,598.46, 1,430.91) & 22,803 \\
Offline - Online Profit Margin & 5.37 & 78.49 & -1.77 & 3.47 & 14.05 & (-6,586.62, 2,163.31) & 22,803 \\
\bottomrule
\end{tabular}
}
\end{table}

On average, prices are higher in physical stores than online, and profit margins are also higher in physical stores than online.\footnote{While the retailer in our context has higher offline prices, there is considerable heterogeneity across retailers, countries, and product categories in whether online prices are lower, the same, or higher than offline prices. In the US, 69\% of multichannel retailers have identical prices online and offline, while 22\% have higher offline prices and only 8\% have higher online prices \citep{cavallo2017online}.} Thus, even when two adopter groups have similar total spend, the group with a higher share of offline spend may generate higher profit. This is the mechanism underlying H2d. At the same time, the retailer still has an incentive to promote online adoption because, as shown in $\S$\ref{ssec:results_adopters_offline_only}, multichannel adoption is associated with higher total spending and profitability relative to remaining offline-only.

\begin{table}[htp!]
   \caption{DID Main Analysis -- Profit ({\it Event Adopters} vs. {\it Organic Adopters})}
   \label{tab:profitability}
\centering
\begin{threeparttable}
\footnotesize{
\begin{tabular}{lccc}
\midrule\midrule
Study: & COVID-19 & Black Friday & Loyalty Program \\
\midrule
DV: & \multicolumn{3}{c}{Profit} \\
\midrule
\emph{Variables}\\
Event\_Adopter $\times$ Post 
& 9.079 (2.274) & -16.04 (5.094) & -15.41 (6.835) \\
& [0.000] & [0.002] & [0.024] \\
\midrule \midrule
\makecell[l]{\% Effect size relative to the \\ post-adoption change for organic adopters} 
& 44.8\% & NA & 73.2\% \\
\midrule \midrule
\makecell[l]{\% Effect size relative to the \\ pre-adoption baseline for event adopters} 
& 5.8\% & -9.9\% & -7.8\% \\
\midrule
\midrule
Loyalty program controls &  &  & Yes \\
\midrule
\emph{Fixed-effects}\\
Customer  & Yes & Yes & Yes \\
YearMonth & Yes & Yes & Yes \\
\midrule
\emph{Fit statistics}\\
Observations & 1,783,949 & 76,842 & 197,658 \\
R$^2$        & 0.29792   & 0.35806 & 0.44335 \\
Within R$^2$ & $2.79\times10^{-5}$ & 0.00025 & 0.02248 \\
\midrule\midrule
\multicolumn{4}{l}{\emph{Clustered (Customer) standard errors in parentheses; p-values in brackets.}}
\end{tabular}
}
\begin{tablenotes}[flushleft]
\footnotesize
\item \textit{Notes:} Effect sizes are calculated using the standard two-by-two DID estimates reported in Table \ref{tab:did_profit_standard_event_adopter} in Web Appendix \ref{appssec:two_by_two_did_event_adopter_profit}. The effect-size conventions are described at the beginning of $\S$\ref{ssec:results_event_organic_adopters}.
\end{tablenotes}
\end{threeparttable}
\end{table}

\paragraph{Main DID results and economic magnitude.}
Table \ref{tab:profitability} presents the results from estimating Equation \eqref{eq:did_fe_2by2_FE} with monthly profit as the dependent variable. The coefficient on $Event\_Adopter \times Post$ is positive and statistically significant for \textit{COVID adopters}, indicating that they become more profitable than \textit{organic adopters} after adoption. In contrast, the coefficients are negative and statistically significant for both \textit{Black Friday adopters} and \textit{Loyalty Program adopters}, indicating lower post-adoption profitability relative to \textit{organic adopters}.

The effect-size rows in Table \ref{tab:profitability} show that these profitability differences are economically meaningful. For \textit{COVID adopters}, monthly profitability increases by 5.8\% relative to their own pre-adoption baseline, and the incremental profitability gain is 44.8\% of the post-adoption gain observed among \textit{organic adopters}. This profitability advantage occurs despite similar total spend and is consistent with \textit{COVID adopters}' higher offline spending share, combined with the higher offline margins documented above. For \textit{Black Friday adopters} and \textit{Loyalty Program adopters}, profitability is lower than for \textit{organic adopters}. Relative to their own pre-adoption baselines, \textit{Black Friday adopters} generate 9.9\% lower monthly profit and \textit{Loyalty Program adopters} generate 7.8\% lower monthly profit. For \textit{Loyalty Program adopters}, the relative-to-organic benchmark also shows that the additional decline is sizable, amounting to 73.2\% of the post-adoption decline observed among \textit{organic adopters}.

\paragraph{Summary of key takeaways.}
The profitability results show that adoption pathways affect customer value in ways that are not captured by spend alone. \textit{COVID adopters} are more profitable than \textit{organic adopters} despite similar spending, consistent with their higher offline spending share and the higher offline margins in our setting. In contrast, \textit{Black Friday adopters} and \textit{Loyalty Program adopters} are less profitable than \textit{organic adopters}, largely because they spend less after adoption, as shown in $\S$\ref{sssec:spend}. These findings support H2d and show that firms should not assume that customers induced to adopt online shopping through external events or promotions will have the same post-adoption profitability as \textit{organic adopters}. We use Wald tests to further confirm that the effect sizes for spend and profit for both promotion pathways, Black Friday and the Loyalty Program, are statistically indistinguishable from each other, while both are statistically different from the COVID adoption pathway.


\subsection{Parallel pre-trends \& HonestDiD Sensitivity Analysis}
\label{ssec:parallel_tests}

\begin{figure}[htp!]
\centering
\caption{Parallel Trend Assessment -- COVID-19 Event}
\label{fig:parallel_trend_combined_main}

\begin{subfigure}{0.31\textwidth}
    \centering
        \caption{Visualization -- Spend}
\includegraphics[width=\linewidth]{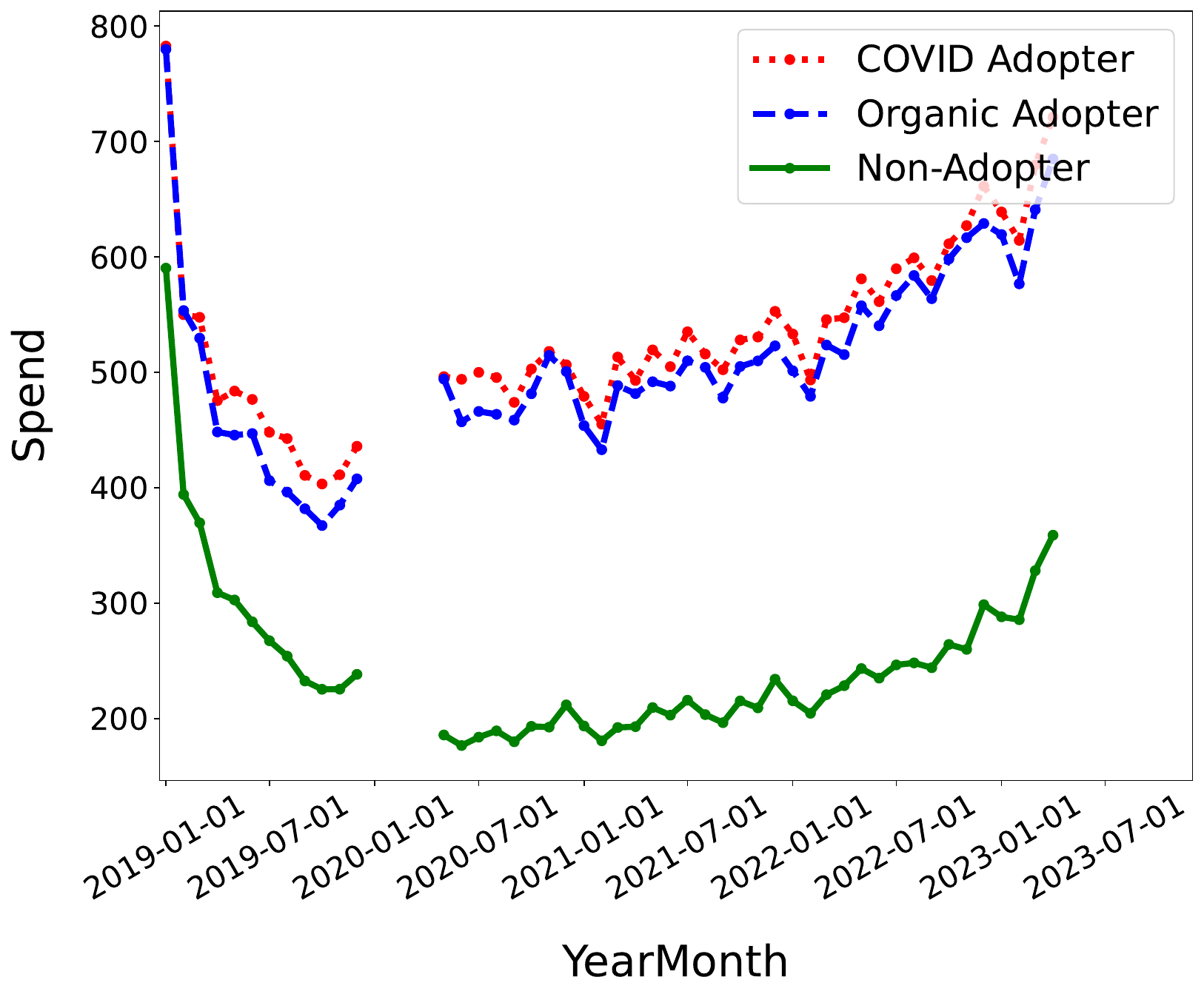}
    \label{fig:spend_sub1}
\end{subfigure}
\hfill
\begin{subfigure}{0.32\textwidth}
    \centering
      \caption{\textit{Adopters} vs. \textit{Offline-only} -- Spend}\includegraphics[width=\linewidth]{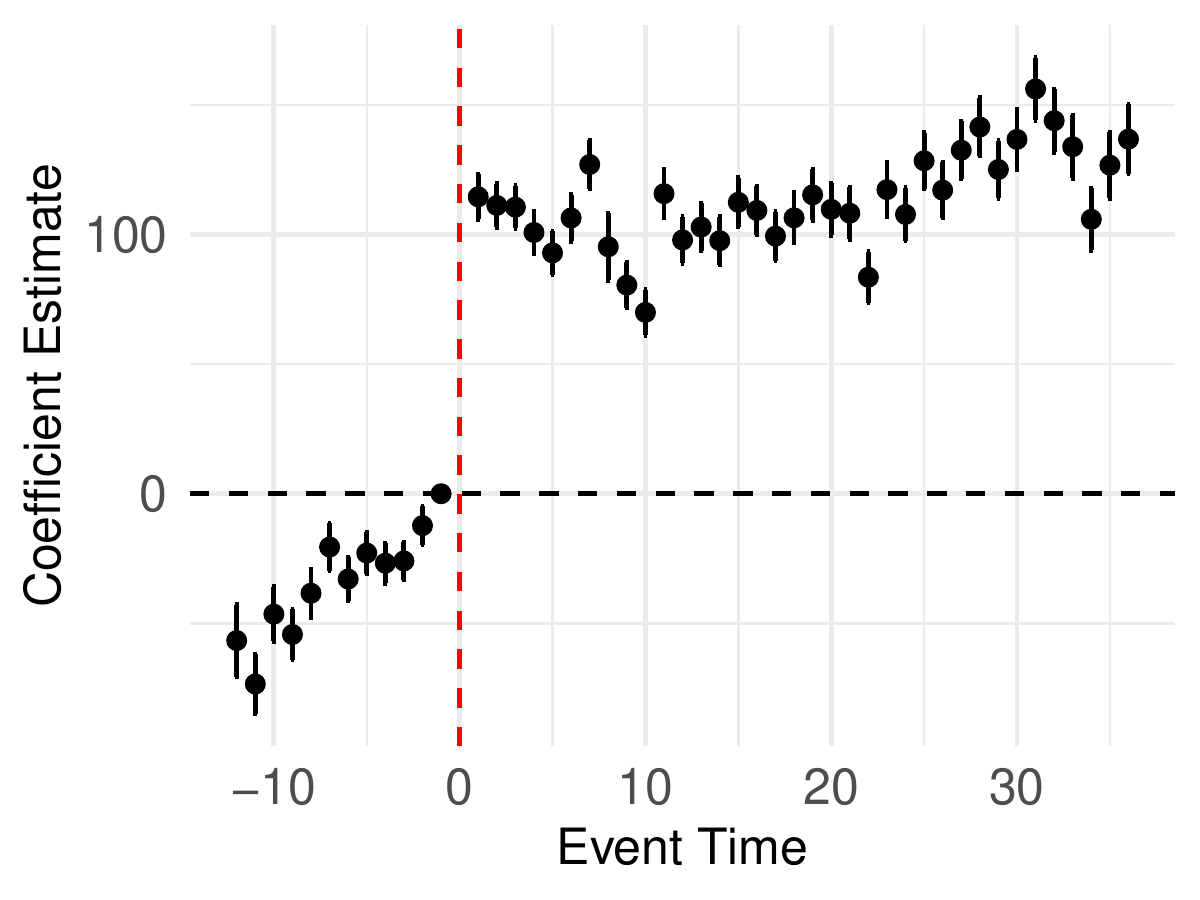}

        \label{fig:spend_sub2}
\end{subfigure}
\hfill
\begin{subfigure}{0.32\textwidth}
    \centering
        \caption{\textit{COVID} vs. \textit{Organic Adopters} -- Spend}\includegraphics[width=\linewidth]{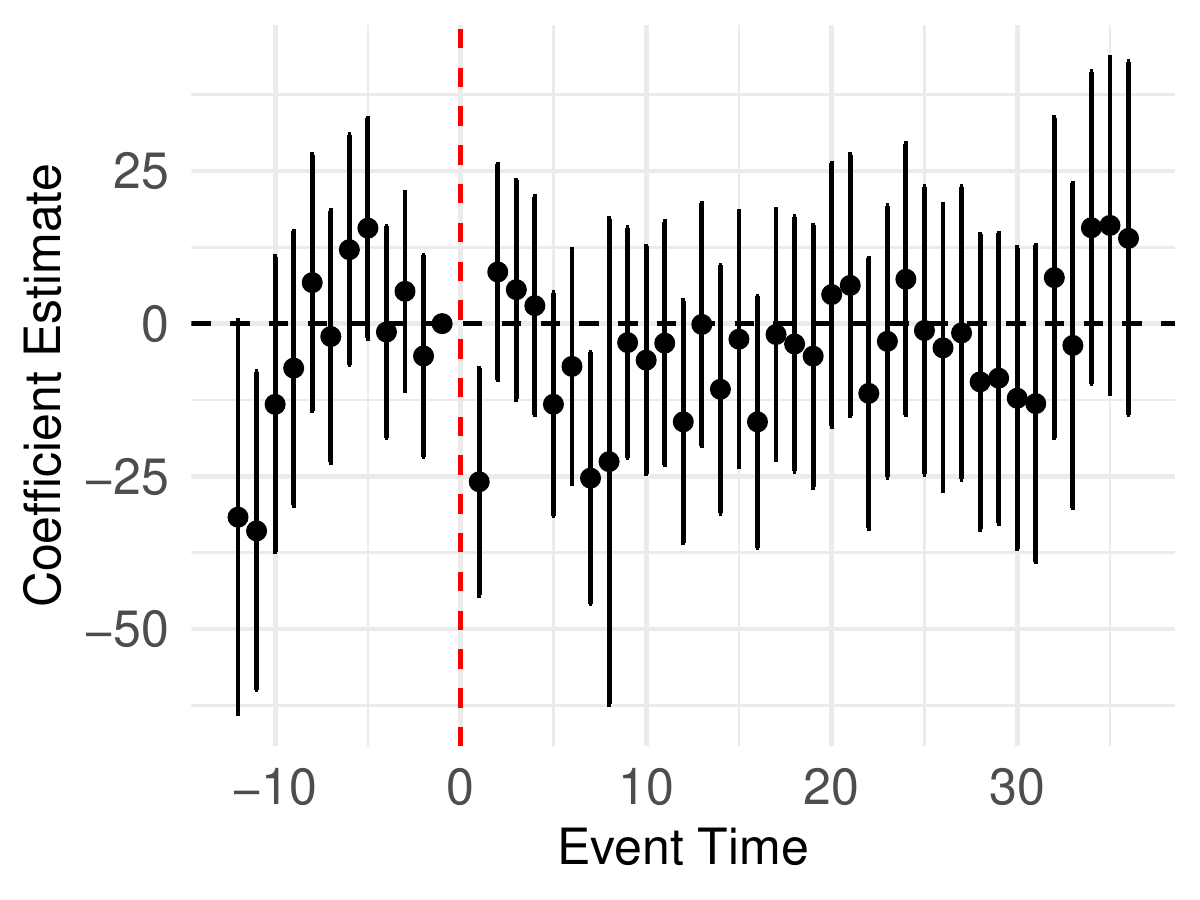}

        \label{fig:spend_sub3}
\end{subfigure}

\vspace{0.5em}

\begin{subfigure}{0.31\textwidth}
    \centering
        \caption{Visualization -- Profit}    \includegraphics[width=\linewidth]{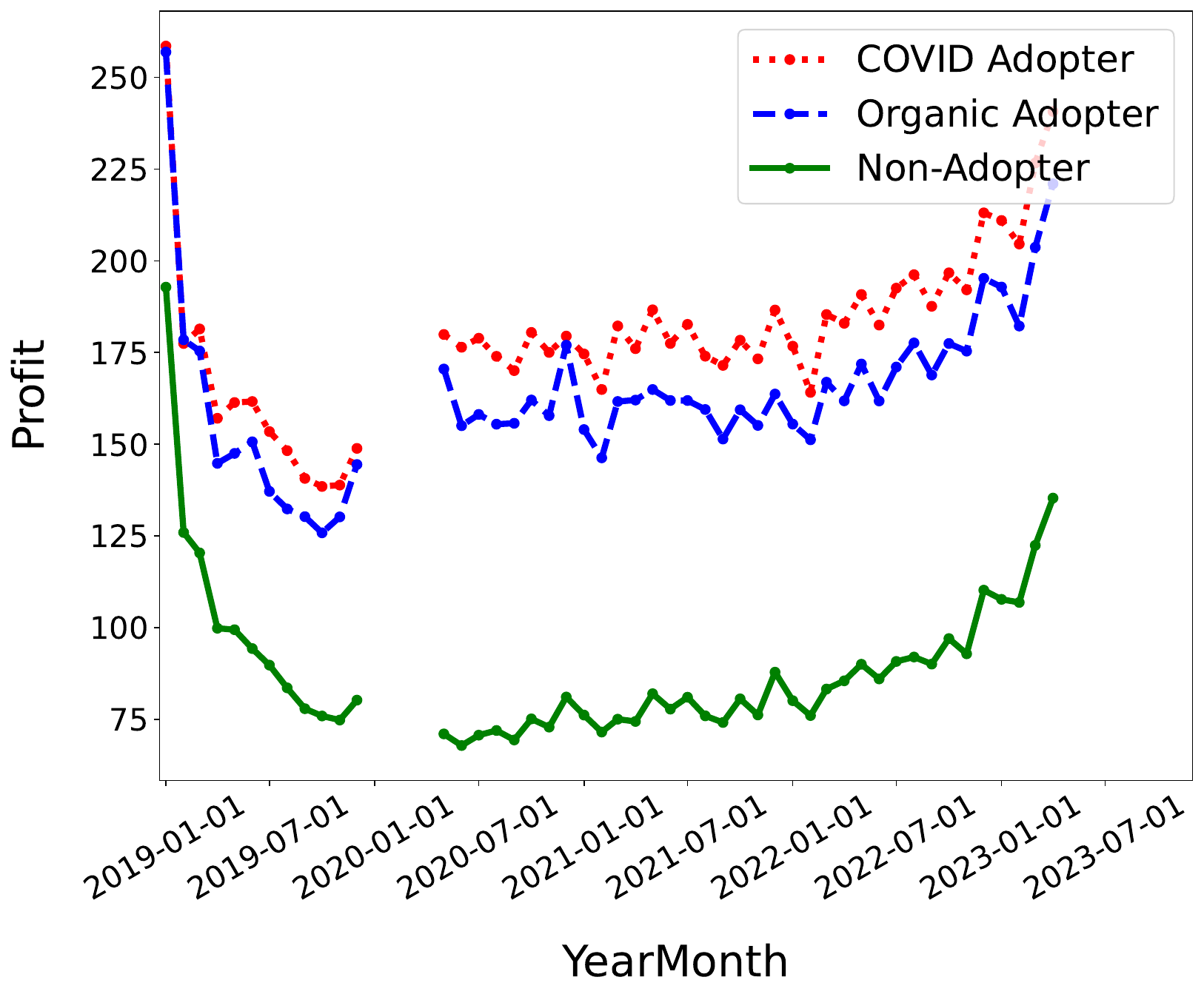}
        \label{fig:profit_sub1}
\end{subfigure}
\hfill
\begin{subfigure}{0.32\textwidth}
    \centering
       \caption{\textit{Adopters} vs. \textit{Offline-only} -- Profit}
\includegraphics[width=\linewidth]{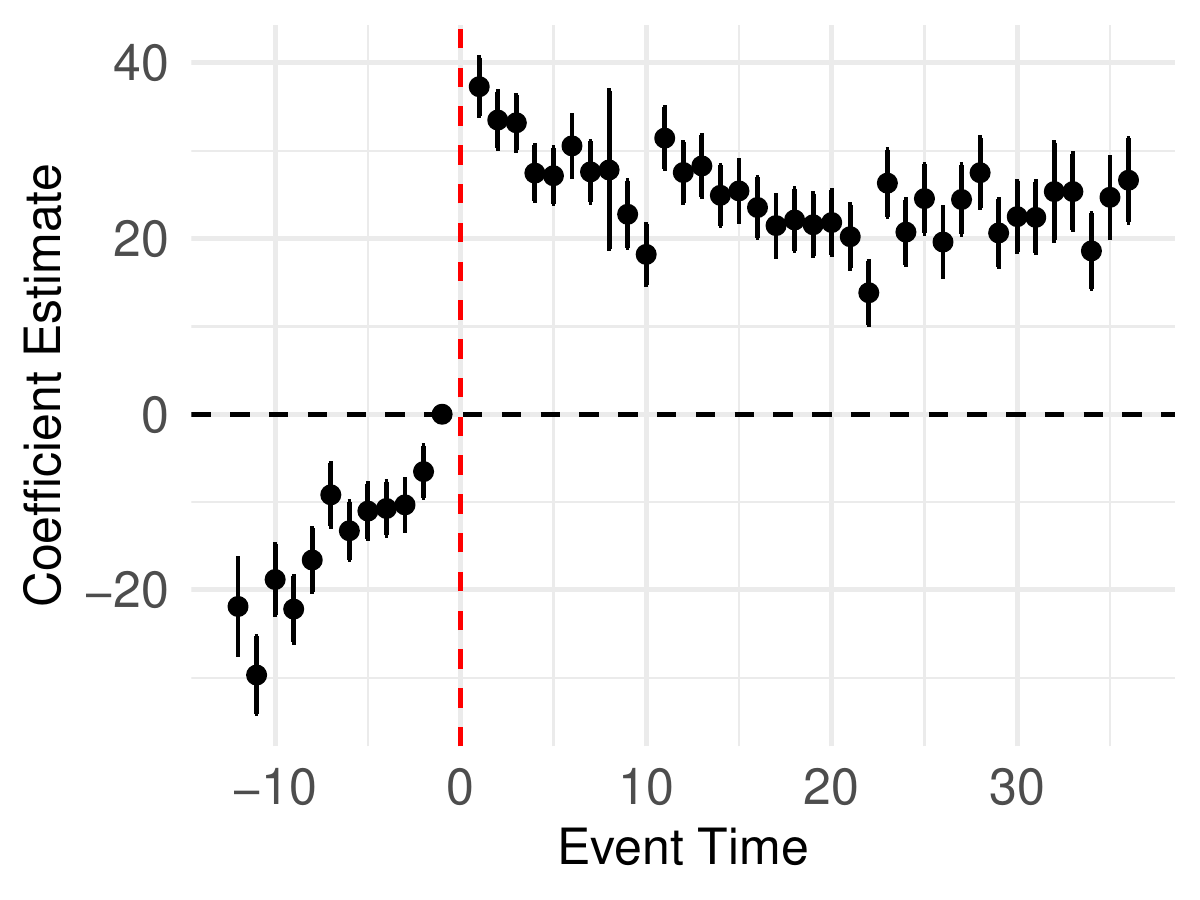}
 
            \label{fig:profit_sub2}
\end{subfigure}
\hfill
\begin{subfigure}{0.32\textwidth}
    \centering
        \caption{\textit{COVID} vs. \textit{Organic Adopters} -- Profit}
        \includegraphics[width=\linewidth]{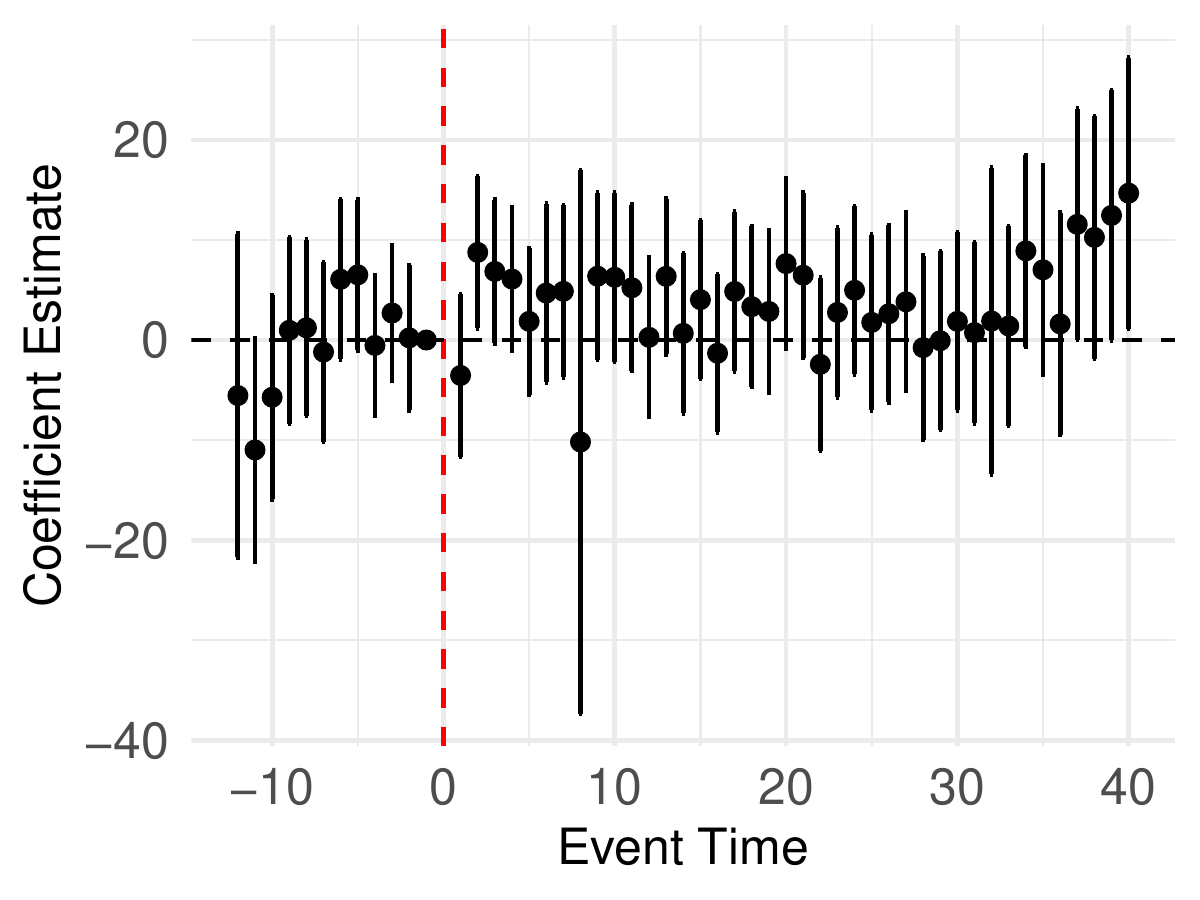}
            \label{fig:profit_sub3}
\end{subfigure}
\vspace{1em} 
\begin{minipage}{0.95\textwidth}
\footnotesize
\textit{Notes:} Panels (\ref{fig:spend_sub1},  \ref{fig:profit_sub1}) plot monthly outcomes, while panels (\ref{fig:spend_sub2},   \ref{fig:profit_sub2}) and (\ref{fig:spend_sub3}, \ref{fig:profit_sub3}) report event study estimates (based on Equation \eqref{eq:event_study}) comparing (i) \textit{adopters} to \textit{offline-only customers} and (ii) \textit{COVID} to \textit{organic adopters}, respectively.
\end{minipage}
\end{figure}

We test whether the parallel trends assumption holds for our DID analyses comparing spend and profitability between \textit{event adopters} and \textit{organic adopters}, and between adopters and \textit{offline-only customers}, using both visual inspection and more formal event study regressions \citep{autor2003outsourcing}.
The parallel trends assumption appears to be satisfied for both spend and profit when comparing \textit{event adopters} and \textit{organic adopters} for all three events. However, the parallel pre-trends assumption is not satisfied for the comparisons between adopters and \textit{offline-only customers}, across all three events. Further, specifically for the case of \textit{COVID adopters} and \textit{organic adopters}, there seems to be a difference in pre-trends visually, although the differences in pre-adoption outcomes are not statistically different for most of the pre-adoption months. Taken together, these patterns point to four settings in which the parallel trends assumption is less well supported: the comparisons of \textit{adopters} versus \textit{offline-only} customers across three events, and the comparison of \textit{COVID adopters} versus \textit{organic adopters}. Figures \ref{fig:parallel_trend_combined_main} presents COVID-19 as a representative case for the total spend (top row) and profit (bottom row), while the full set of diagnostics for other two events (Black Friday and Loyalty Program) are reported in Web Appendix $\S$\ref{appsec:honest_did}.

\begin{figure}[htp!]
\centering
\caption{HonestDiD Sensitivity Bounds for Post-Adoption Spend and Profit}
\label{fig:honest_did_all_main}
\captionsetup[subfigure]{position=top,font=footnotesize,skip=1pt}
{\scriptsize \textbf{Panel A: Organic + COVID vs. Offline-only}} \\[-0.6em]
\begin{subfigure}{0.3\textwidth}
    \centering
    \caption{Spend}
     \label{fig:honest_did_spend_organiccovid_offlineonly}
    \includegraphics[width=\linewidth]{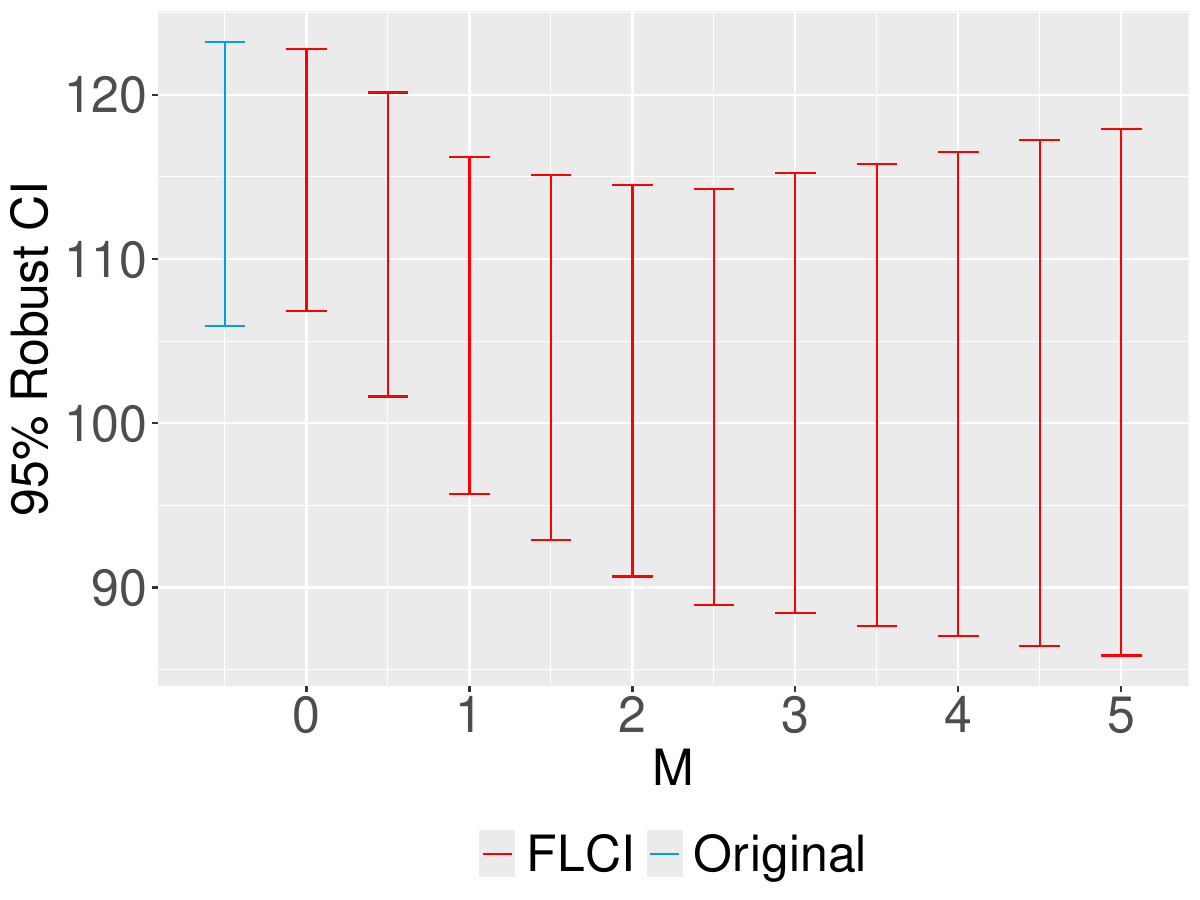}
\end{subfigure}\hspace{0.02\textwidth}
\begin{subfigure}{0.3\textwidth}
    \centering
    \caption{Profit}
    \includegraphics[width=\linewidth]{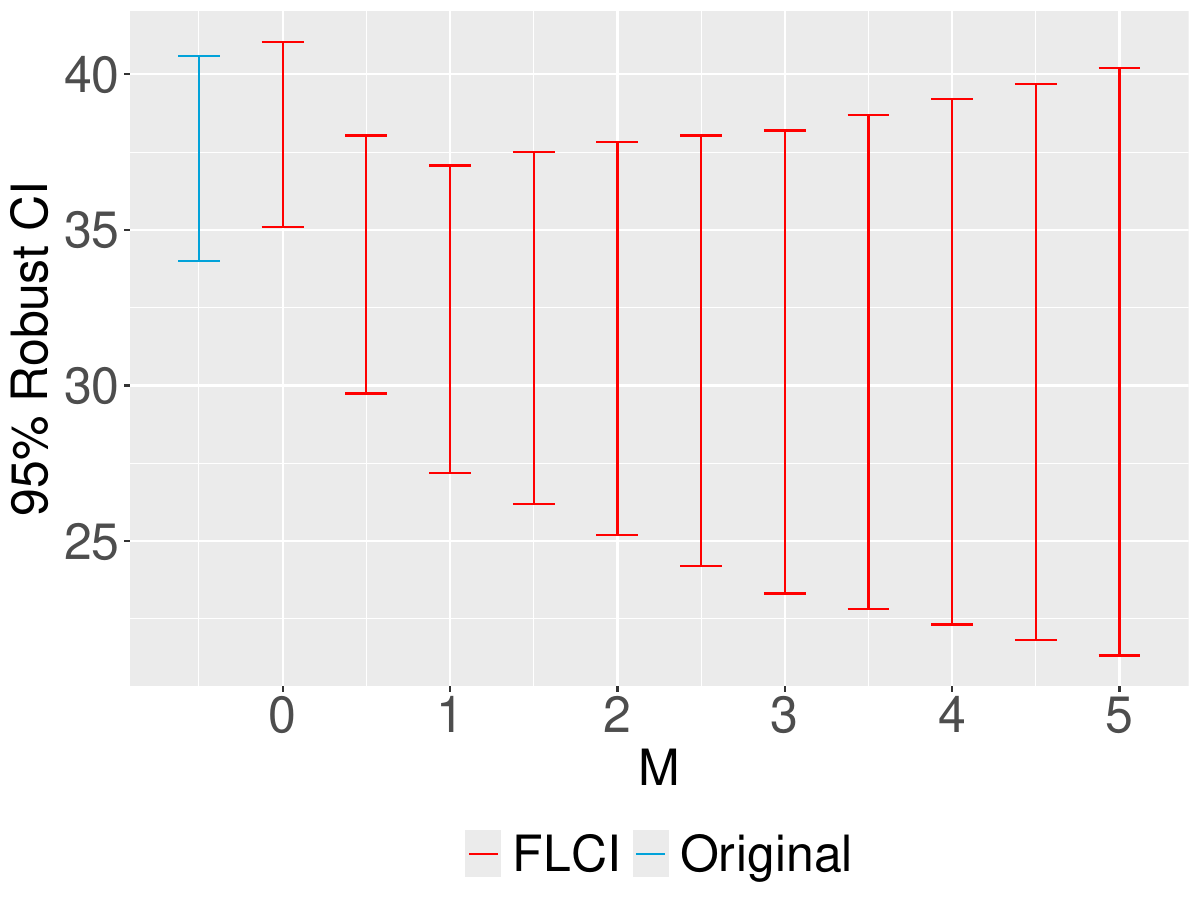}
\end{subfigure}

\vspace{0.3em}

{\scriptsize\textbf{Panel B: Organic + Black Friday vs. Offline-only}} \\[-0.6em]
\begin{subfigure}{0.3\textwidth}
    \centering
    \caption{Spend}
    \includegraphics[width=\linewidth]{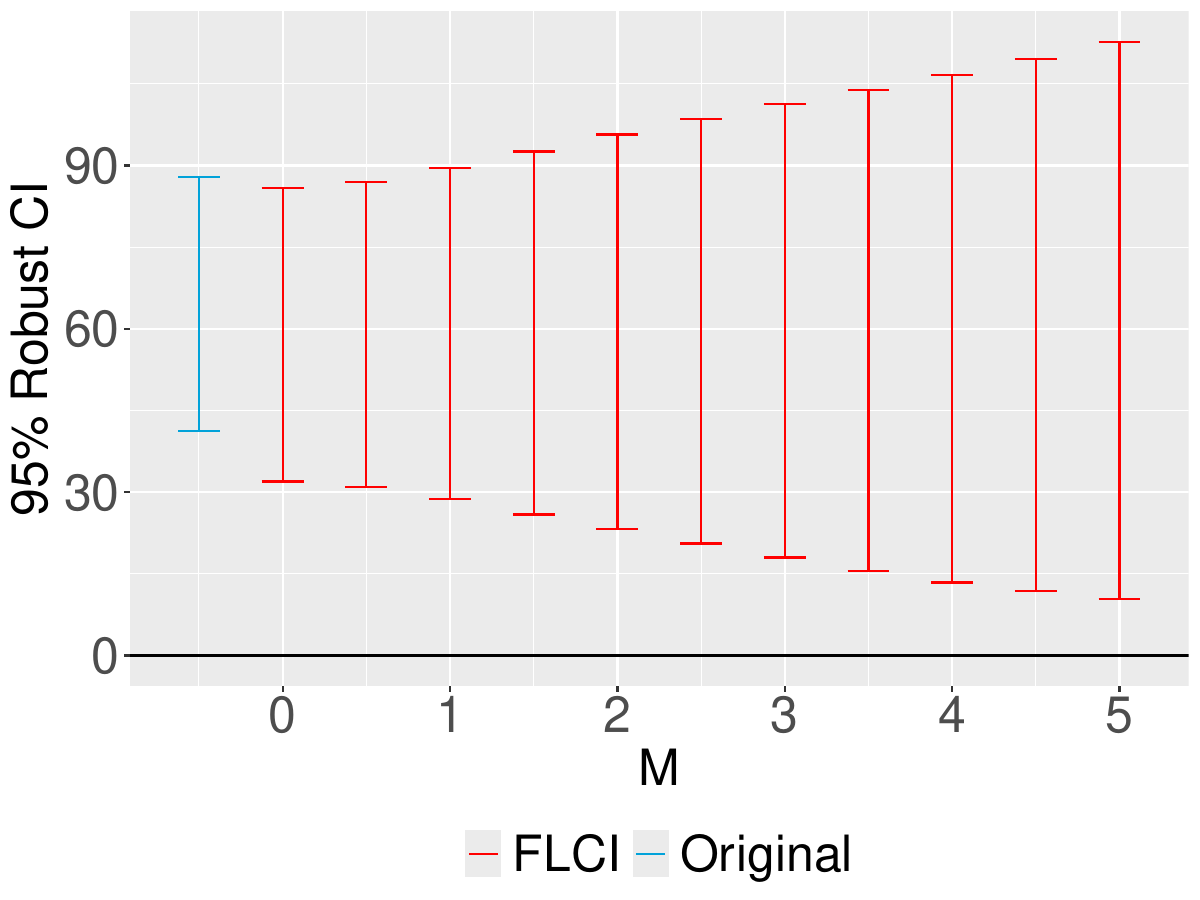}
\end{subfigure}\hspace{0.02\textwidth}
\begin{subfigure}{0.3\textwidth}
    \centering
    \caption{Profit}
    \includegraphics[width=\linewidth]{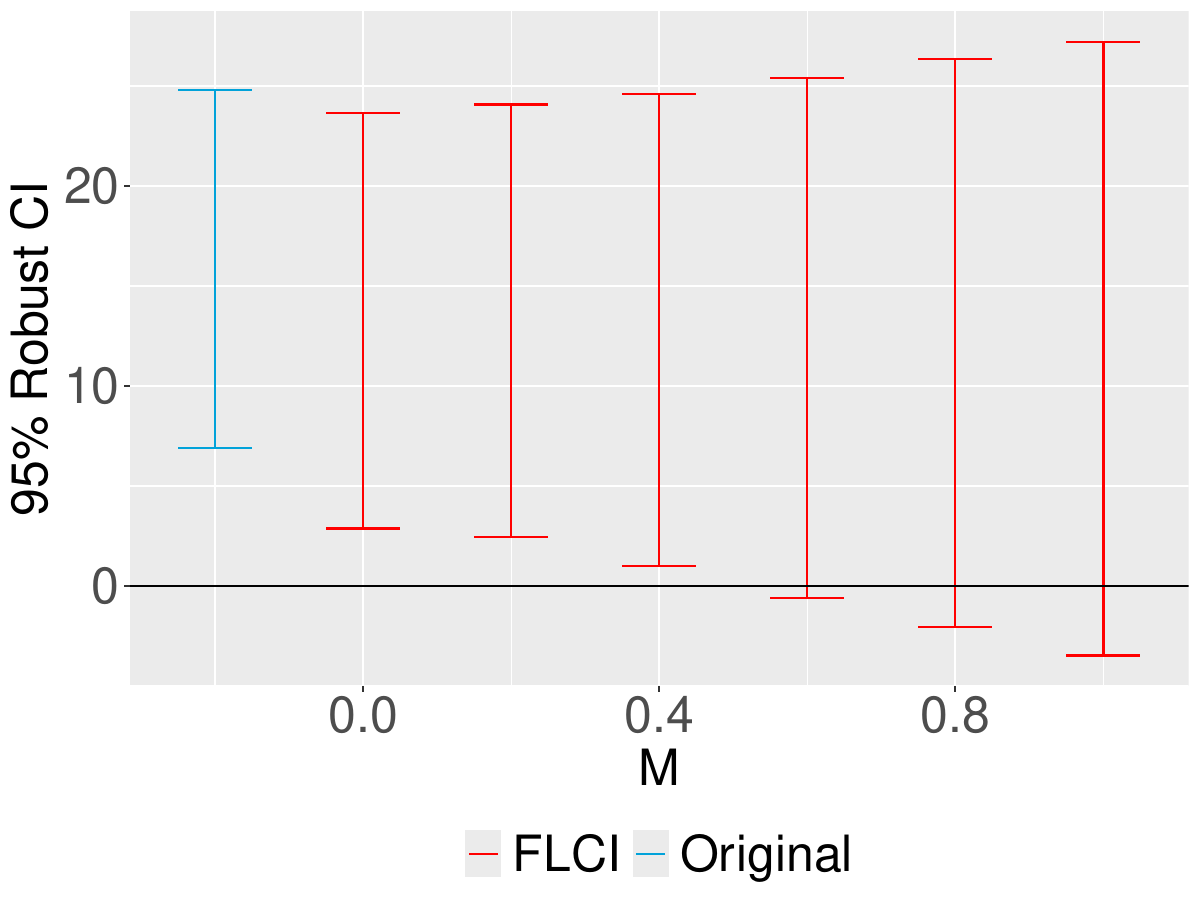}
\end{subfigure}

\vspace{0.3em}

{\scriptsize\textbf{Panel C: Organic + Loyalty Program vs. Offline-only}} \\[-0.6em]
\begin{subfigure}{0.3\textwidth}
    \centering
    \caption{Spend}
    \includegraphics[width=\linewidth]{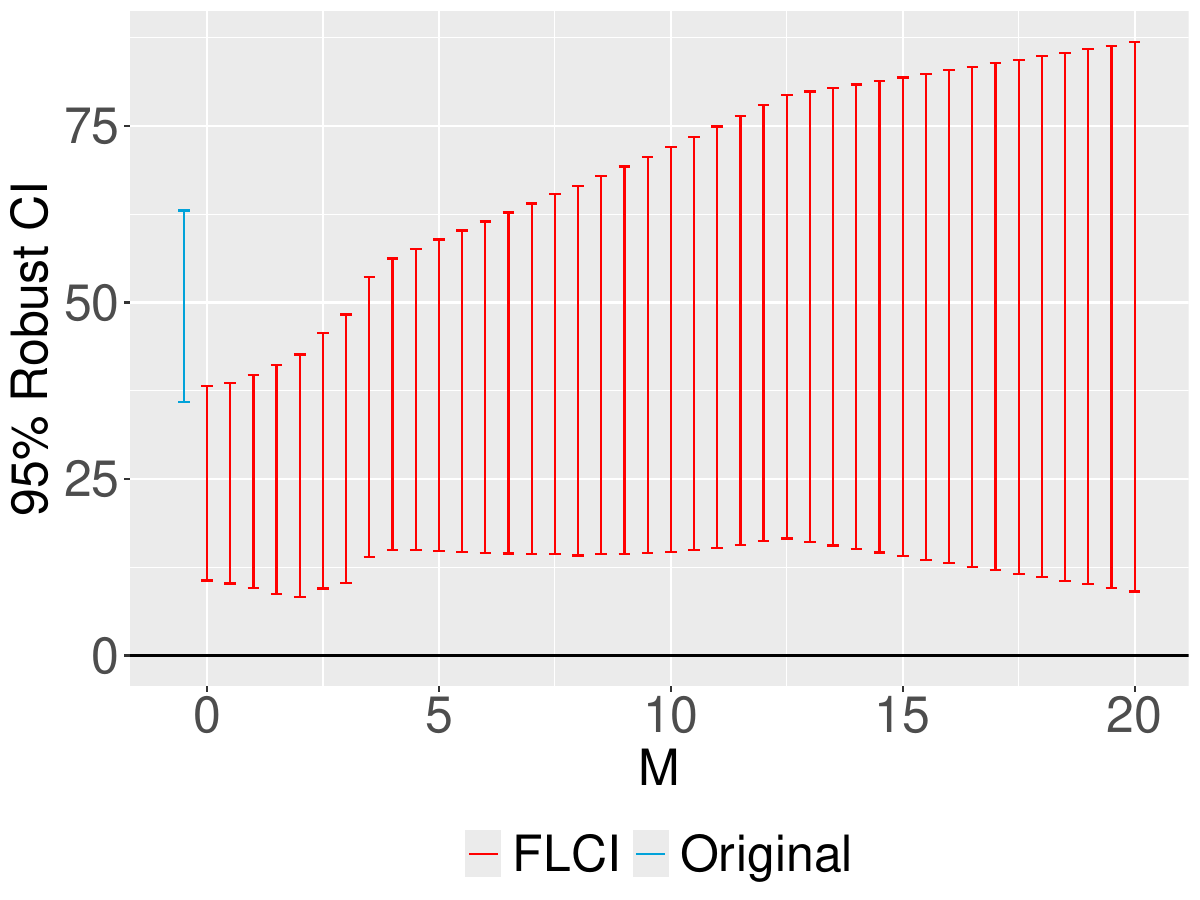}
\end{subfigure}\hspace{0.02\textwidth}
\begin{subfigure}{0.3\textwidth}
    \centering
    \caption{Profit}
    \includegraphics[width=\linewidth]{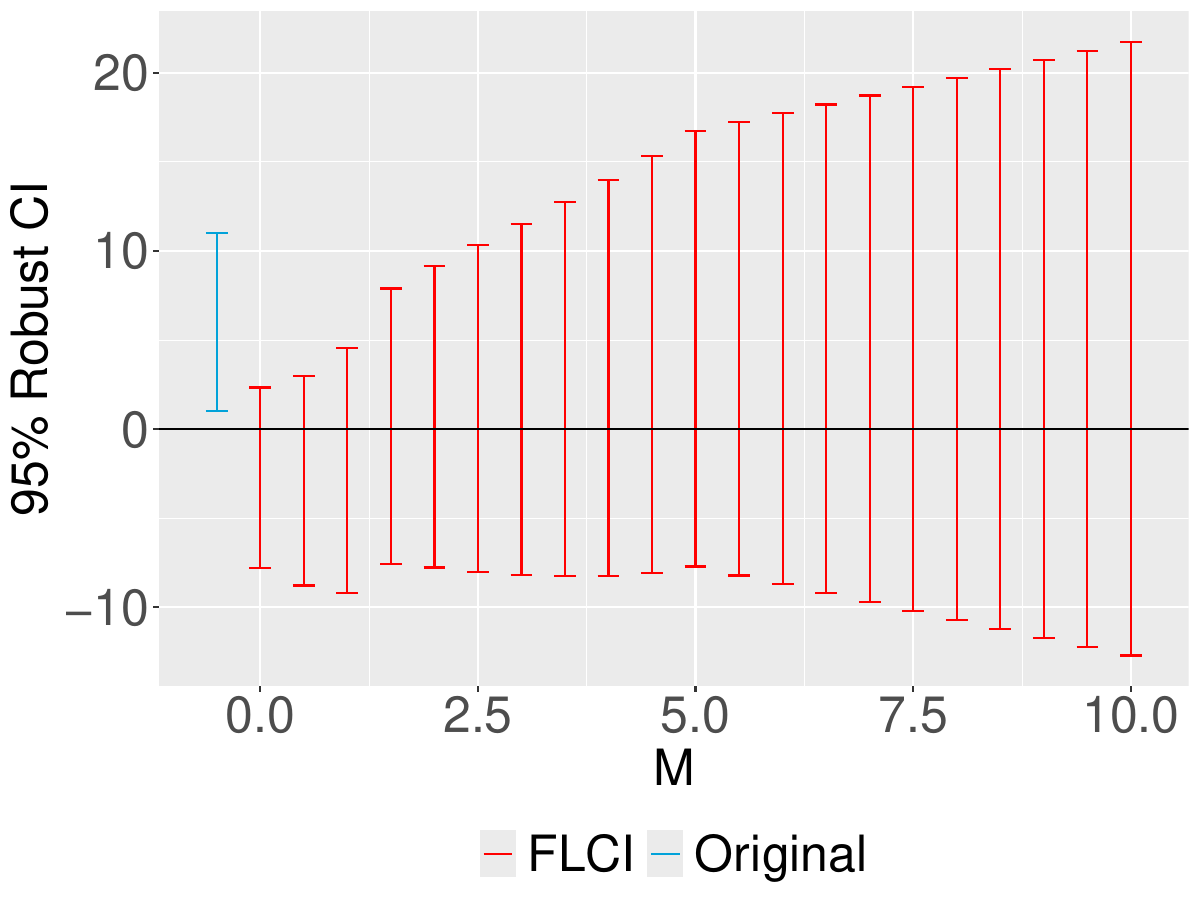}
\end{subfigure}

\vspace{0.3em}

{\scriptsize\textbf{Panel D: COVID Adopters vs. Organic Adopters}} \\[-0.6em]
\begin{subfigure}{0.3\textwidth}
    \centering
    \caption{Spend}
    \includegraphics[width=\linewidth]{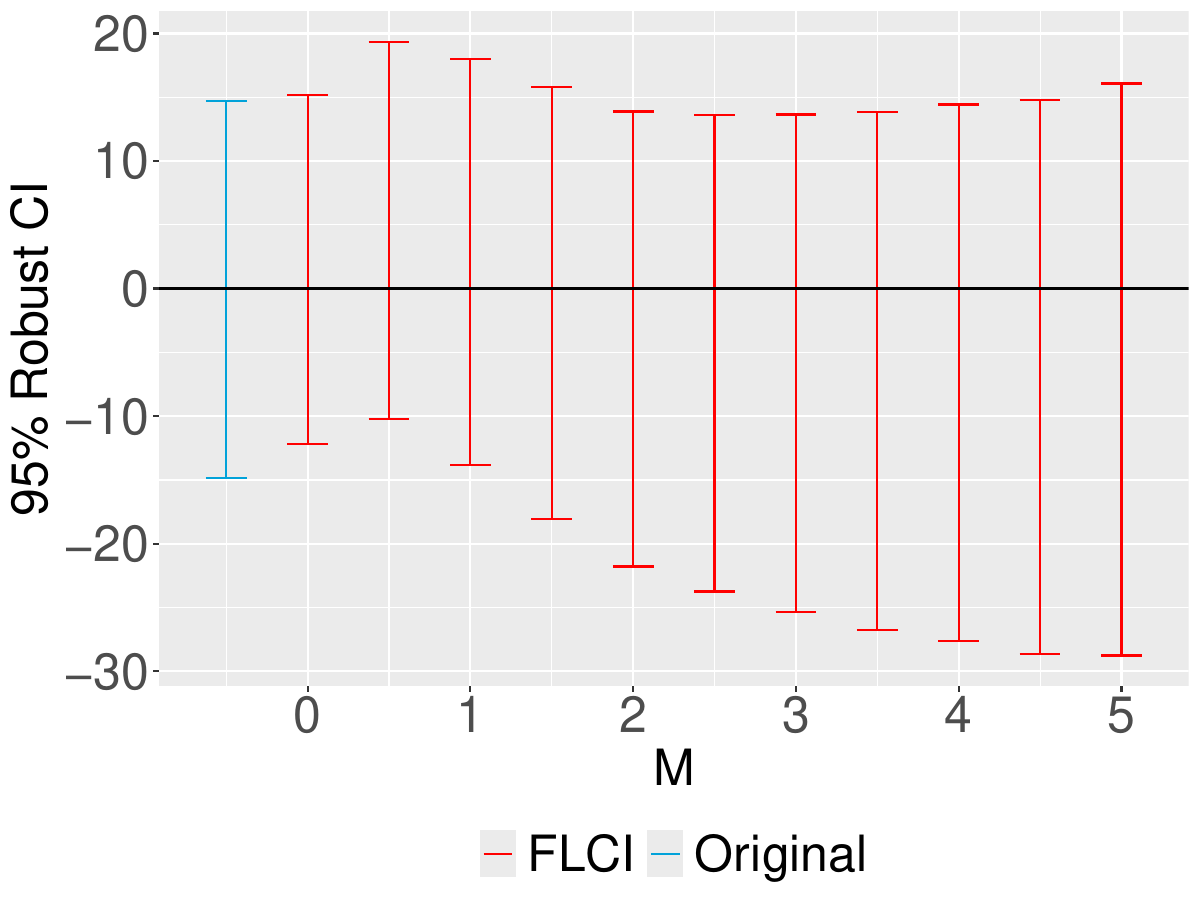}
\end{subfigure}\hspace{0.02\textwidth}
\begin{subfigure}{0.3\textwidth}
    \centering
    \caption{Profit}
    \includegraphics[width=\linewidth]{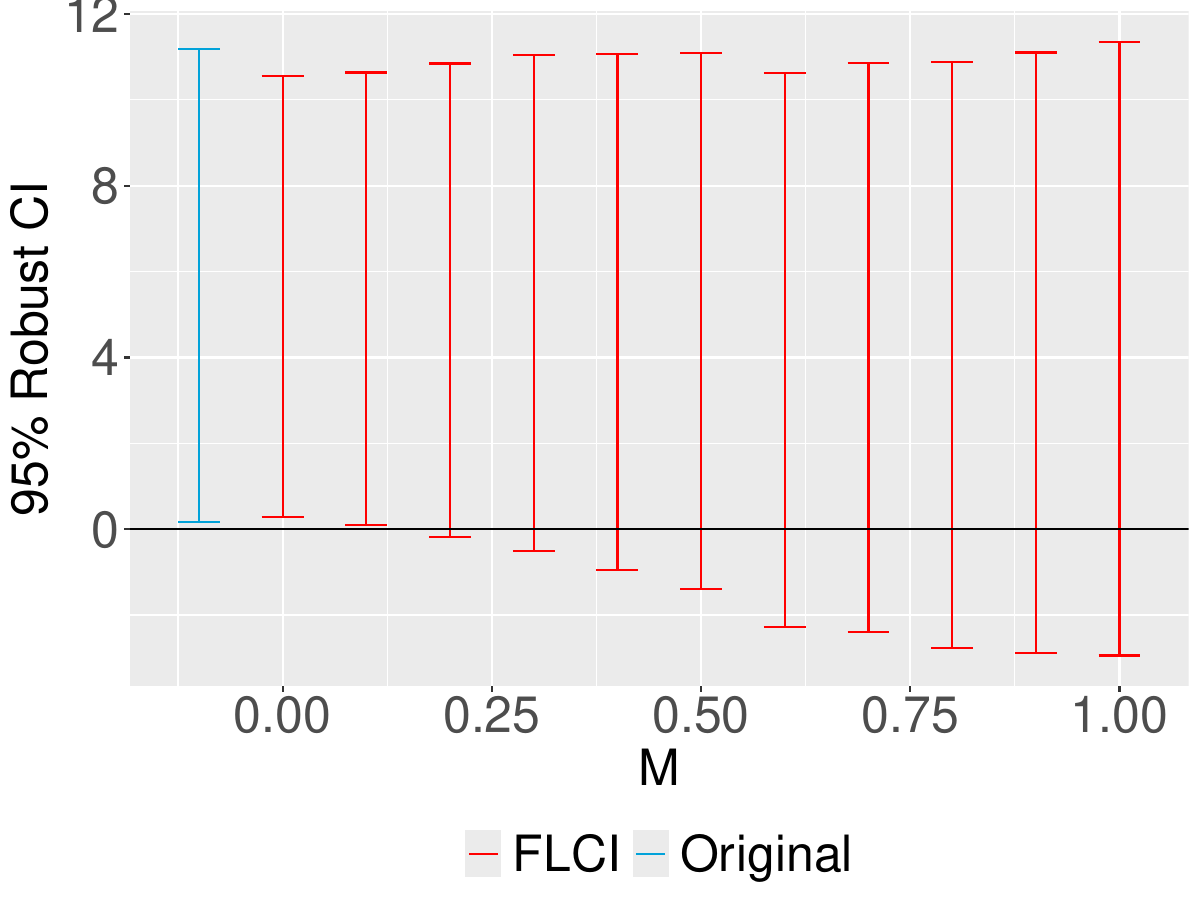}
\end{subfigure}
\par\vspace{0.6em}
\begin{minipage}{0.65\textwidth}
\footnotesize
\textit{Notes:} Each panel reports HonestDiD 95\% confidence interval bounds for the post-treatment effect (labeled as FLCI) under increasing restrictions ($M$) on deviations from parallel trends. 
\end{minipage}
\end{figure}

For these four cases (spend and profits for \textit{adopters} vs. \textit{offline-only customers} for all three events, and \textit{COVID adopters} vs. \textit{organic adopters}), we bound our treatment effects under different assumptions on the extent of the violation using the HonestDiD approach proposed by \cite{rambachan2023more}. To conduct this HonestDiD sensitivity analysis, we impose the restriction that the change in slope of the differential trend between treatment and control customers across two periods is bounded by:
\begin{equation}
    \Delta^{SD} := \{\zeta: |(\zeta_{t+1} - \zeta_{t}) - (\zeta_{t} - \zeta_{t-1})| \leq M, ~ \forall t\}
\end{equation}
Here, $\zeta_{t}$ is the difference in trends between two groups at time $t$, and $M$ is the magnitude of maximum allowable change in trend between consecutive periods. We test a wide range of $M$ values, up to and beyond the linear pre-trend estimate (see Web Appendix $\S$\ref{appssec:honest_did_linear_trend_est}). 

We present the HonestDiD 95\% confidence interval of ATT bounds under different violations $M$ for spend and profit under four cases in Figure \ref{fig:honest_did_all_main}. Our results remain statistically significant for values for $M$ larger than the estimated linear pre-trends. In all instances but one, we conclude that even if the post-adoption trend violation is twice as large as the pre-adoption trend, our substantive conclusions remain unchanged in terms of both direction and statistical significance. The only exception is profit for the Loyalty Program event. However, even for this event, the difference in post-adoption spend is large and significant.

\section{Managerial Implications}
\label{sec:managerial}


Our main findings indicate that consumers who adopt online shopping through different channel pathways (organic, macro-environmental shocks, or promotions) behave differently in the post-adoption period. These patterns are consistent with consumer inertia and forward buying, and the implications may be especially relevant in settings where similar mechanisms operate. In particular, forward buying and stockpiling may reduce the long-run spending gains from promotion-induced online adoption. To mitigate this possibility, firms may benefit from designing incentives that discourage stockpiling (e.g., phased or behavior-contingent rewards) and combining promotions with rapid follow-up nudges to encourage repeat rather than one-off behavior. For consumers pushed online by shocks such as COVID-19, habit-reinforcing promotions and personalized nudges in the immediate post-adoption window may help sustain online engagement after the initial shock subsides. Beyond these general lessons, we derive concrete implications for two common marketing tasks: (a) CLV forecasting and (b) ROI/breakeven analysis for promotions.

\subsection{Implications for CLV forecasts}
\label{ssec:clv}

Customer Lifetime Value (CLV) is a core metric for customer relationship management, targeting, and budgeting \citep{fader2005rfm, gupta2006modeling, fader2010customer}. Our results show that when \textit{offline-only customers} adopt online shopping for different reasons (e.g., Black Friday, COVID-19, loyalty program, or organically), their spending and profitability trajectories differ markedly by pathway. Ignoring this heterogeneity leads to biased forecasts of \textit{incremental} CLV from online adoption and, in turn, to suboptimal decisions, for example, overestimating the effectiveness of promotions aimed at inducing online adoption.

We quantify the \textit{average} incremental (relative to \textit{offline-only customers}) CLV from online adoption for each adopter cohort $s$. Following standard practice \citep{gupta2006modeling}, the (average) incremental CLV equals the present value of all future incremental monthly profits attributable to adoption under an infinite horizon:
\begin{equation}
\label{eq:incremental_clv}
    Incremental ~ CLV_{s} = \sum_{t = 1}^{\infty} \frac{ATT_{s} \times r_{s}^{t}}{(1 + i)^{t}} = ATT_{s}\frac{r_{s}}{1+i-r_{s}},
\end{equation}
where: (1) $ATT_{s}$ is the monthly average incremental profit (our ATT estimate) for cohort $s$ relative to \textit{offline-only customers}; (2) $i$ is the firm’s monthly cost of capital (WACC), and (3) $r_{s}$ is the (average) monthly retention rate of cohort $s$ post-adoption, defined as the probability of purchasing next month conditional on purchasing in the focal month. All quantities are monthly. The $ATT_{s}$ estimates come from our DID models for each event, and retention rates are obtained non-parametrically. Further, based on discussions with the retailer, we set WACC to 15\% annually, implying a monthly cost of capital of $i = (1 + 15\%)^{1/12} - 1 = 1.17\%$.

We next show how CLV forecasts can be biased when managers ignore adoption pathways. We consider two types of Incremental CLV estimates for each type of {\it event adopter}: (1) \textit{Pooled} forecast, where the firm ignores the adoption pathway and pools {\it event} and {\it organic adopters} and estimates a single incremental CLV for all adopters, and (2) {\it Pathway aware} forecast estimates based only on the event adopters (without including {\it organic} adopters).
The inputs as well as the results of this exercise are shown in Table \ref{tab:clv_pathway_vs_pooled}. For the COVID-19 event, for example, the pooled ATT from Table \ref{tab:did_profitability_adopter_offlineonly} is 37.30 MCU (Masked Currency Units as defined in \S \ref{ssec:summary_stats}), and the pooled monthly retention rate is 0.654. Overall, we see that pooled estimates overstate the value of promotion-induced adopters by almost 60\% (Black Friday) and 80\% (Loyalty Program), while understating the value of COVID adopters by 7.5\%.\footnote{Note that the difference for COVID adopters is mainly a result of higher offline prices (vs. online) at our focal retailer. Not all retailers have higher offline prices. In the US, 69\% of multi-channel retailers have identical prices online and offline, while 22\% have higher offline prices and only 8\% have higher online prices\citep{cavallo2017online}.} 

Among the profit estimates that enter the CLV and breakeven exercises, the Loyalty Program estimate is the one for which the HonestDiD sensitivity bounds are least robust to deviations from parallel trends; see $\S$\ref{ssec:parallel_tests}. We therefore interpret the corresponding Loyalty Program CLV and breakeven calculations as suggestive of the direction and potential magnitude of the bias from ignoring adoption pathways, rather than as precise point estimates. The Black Friday managerial calculations are more robust to this sensitivity check.



\begin{table}[t]
\centering
\caption{\textit{Incremental} CLV (relative to \textit{offline-only}, in Masked Currency Units (MCU)): Pathway-Aware (\textit{Event Adopters}) vs. Pooled Forecasts -- By Three Events (COVID-19, Black Friday, and Loyalty Program)}
\label{tab:clv_pathway_vs_pooled}
\begin{threeparttable}
\setlength{\tabcolsep}{6pt}
\renewcommand{\arraystretch}{1.15}
\footnotesize{
\begin{tabular}{lccc}
\toprule
& \multicolumn{1}{c}{COVID-19} & \multicolumn{1}{c}{Black Friday} & \multicolumn{1}{c}{Loyalty Program} \\
\midrule
\addlinespace
\multicolumn{4}{l}{\textbf{Panel A: Inputs}}\\
ATT (\textit{Event Adopters})  & 39.780 & 7.813  & 2.986 \\
ATT ({\it Pooled})                  & 37.370 & 18.440 & 18.410 \\
\addlinespace
Monthly Retention Rate (\textit{Event Adopters}) & 0.656 & 0.621 & 0.625 \\
Monthly Retention Rate ({\it Pooled})                 & 0.654 & 0.633 & 0.565 \\ \midrule
\addlinespace
\multicolumn{4}{l}{\textbf{Panel B: Incremental CLV}}\\
\textit{Incremental} CLV (\textit{Event Adopters}) & 73.486 & 12.417 & 4.831 \\
\textit{Incremental} CLV ({\it Pooled})                 & 68.332 & 30.822 & 23.301 \\
\addlinespace
\multicolumn{4}{l}{\textbf{Panel C: Forecast Bias}}\\
\makecell[l]{Diff in \textit{Incremental} CLV of \\ \textit{Pathway aware} vs. {\it Pooled} Estimates} 
& +7.5\% & -59.7\% & -79.3\% \\
\bottomrule
\end{tabular}
}
\begin{tablenotes}[flushleft]
\footnotesize
\item \textit{Notes:} \textit{Incremental} CLV is defined relative to the \textit{offline-only} counterfactual and computed using Equation \eqref{eq:incremental_clv}. ATT estimates for \textit{event adopters} are reported in Table \ref{tab:did_three_groups_profit} in Web Appendix $\S$\ref{appsec:managerial_implication}. Pooled ATT estimates are from Table \ref{tab:did_profitability_adopter_offlineonly}. Monthly retention rates for \textit{event adopters} are estimated separately by adoption pathway, while pooled retention rates are estimated from a combined sample of \textit{organic} and \textit{event adopters}. Diff in \textit{Incremental} CLV of \textit{Pathway aware} vs. Pooled Estimates is defined as $\frac{(\textrm{Incremental ~ CLV\_{event}} - \textrm{Incremental ~ CLV\_{pooled}})}{\textrm{Incremental ~ CLV\_{pooled}}}$ for a given event. Positive values indicate that the pooled forecast understates the pathway-aware incremental CLV, while negative values indicate that the pooled forecast overstates the pathway-aware incremental CLV.
\end{tablenotes}
\end{threeparttable}
\end{table}


Together, these findings suggest that failing to distinguish adoption pathways can bias CLV forecasts by overstating the value of promotion-induced adopters and making weak promotional strategies appear more attractive. A key managerial takeaway is that retailers may benefit from (a) estimating incremental CLV separately by adoption pathway and (b) setting pathway-specific caps on the cost of inducing online adoption, aligning offer depth and paid-media spend with those caps.

\subsection{Implications for ROI and Breakeven Analyses for Promotions}
\label{ssec:implications}

We next examine how ignoring adoption pathways affects ROI and breakeven analyses for promotions designed to induce online adoption. We focus on the two firm-driven promotion pathways in our setting: Black Friday and the Loyalty Program. The key managerial question is whether the incremental profits generated by promotion-induced adopters are sufficient to recover the cost of the promotion. If managers use \textit{organic adopters} as proxies for promotion-induced adopters, they may overstate the expected incremental profit from induced adoption and therefore understate the time required to break even.

For a promotion $c$ and adopter benchmark group $g$, define $Cost_c$ as the per-customer cost of campaign $c$, and $ATT^{profit}_{c,g}$ as the average monthly incremental profit from online adoption for benchmark group $g$, relative to \textit{offline-only customers}. The ROI over a $T$-month horizon is:
\begin{equation}
\label{eq:promotion_roi}
    ROI_{c,g}(T) = \frac{T \times ATT^{profit}_{c,g} - Cost_c}{Cost_c}.
\end{equation}
The breakeven time is the value of $T$ that sets Equation \eqref{eq:promotion_roi} equal to zero:
\begin{equation}
\label{eq:promotion_breakeven}
    Breakeven_{c,g} = \frac{Cost_c}{ATT^{profit}_{c,g}}.
\end{equation}
We compare two benchmarks. The first is a naive benchmark that uses the incremental profit of \textit{organic adopters}. This is the type of benchmark a manager might use when evaluating a new promotion using historical organic-adopter behavior. The second is a pathway-aware benchmark that uses the incremental profit of the actual promotion-induced adopters.

Table \ref{tab:breakeven_pathway} reports the inputs and breakeven calculations. For Black Friday, we quantify the promotion cost as the difference between actual consumer spending at promotional prices and what consumers would have paid at regular prices. The average discount is about 28\% across categories, yielding an average promotional cost of 299.18 MCU per customer. For the Loyalty Program, we use the retailer-provided cost assumption of 100 MCU per customer to set up the program and induce online activation.\footnote{This estimate is based on conversations with the retailer; we consider alternative values of 50 MCU and 150 MCU in Web Appendix $\S$\ref{appsec:managerial_implication}.}

\begin{table}[htp!]
\centering
\caption{Breakeven Analysis for Promotions Designed to Induce Online Adoption}
\label{tab:breakeven_pathway}
\begin{threeparttable}
\footnotesize
\renewcommand{\arraystretch}{1.15}
\setlength{\tabcolsep}{6pt}
\begin{tabular}{p{6.4cm}cc}
\toprule
& Black Friday & Loyalty Program \\
\midrule
\multicolumn{3}{l}{\textbf{Panel A: Inputs}} \\
Campaign cost per induced adopter, $Cost_c$ (MCU)
& 299.18
& 100.00 \\

\makecell[l]{Monthly incremental profit using \textit{organic adopters} \\ as benchmark, $ATT^{profit}_{c,organic}$ (MCU)}
& 23.57
& 18.55 \\

\makecell[l]{Monthly incremental profit using \textit{event adopters} \\ as benchmark, $ATT^{profit}_{c,event}$ (MCU)}
& 7.813
& 2.986 \\

\midrule
\multicolumn{3}{l}{\textbf{Panel B: Breakeven Calculations}} \\

\makecell[l]{Naive breakeven using \textit{organic adopters}, \\ $Cost_c/ATT^{profit}_{c,organic}$}
& \makecell[c]{$299.18/23.57$ \\ $=12.69$ months \\ $(\approx 13)$}
& \makecell[c]{$100/18.55$ \\ $=5.39$ months} \\

\makecell[l]{Pathway-aware breakeven using \textit{event adopters}, \\ $Cost_c/ATT^{profit}_{c,event}$}
& \makecell[c]{$299.18/7.813$ \\ $=38.30$ months \\ $(\approx 39)$}
& \makecell[c]{$100/2.986$ \\ $=33.49$ months} \\

\midrule
\multicolumn{3}{l}{\textbf{Panel C: Managerial Implication}} \\

\makecell[l]{Breakeven error from using \textit{organic adopters} \\ as proxies}
& Nearly 3$\times$ longer
& More than 6$\times$ longer \\

\bottomrule
\end{tabular}
\begin{tablenotes}[flushleft]
\footnotesize
\item \textit{Notes:} Breakeven time is computed using Equation \eqref{eq:promotion_breakeven}. Monthly incremental profit estimates for \textit{organic adopters} and \textit{event adopters} are from Table \ref{tab:did_three_groups_profit} in Web Appendix $\S$\ref{appsec:managerial_implication}. The Black Friday campaign cost is calculated as the difference between actual spending at promotional prices and counterfactual spending at regular prices. The Loyalty Program campaign cost is based on the retailer-provided cost estimate; alternative cost assumptions are reported in Web Appendix $\S$\ref{appsec:managerial_implication}.
\end{tablenotes}
\end{threeparttable}
\end{table}

The results show that ignoring adoption pathways can substantially distort promotion evaluation. For Black Friday, using \textit{organic adopters} as the benchmark implies a breakeven period of about 13 months. In contrast, using the incremental profit of actual \textit{Black Friday adopters} implies a breakeven period of about 39 months, nearly three times longer. For the Loyalty Program, the distortion is even larger: the naive organic-adopter benchmark implies breakeven in 5.39 months, whereas the pathway-aware estimate implies breakeven in 33.49 months, more than six times longer.\footnote{The Loyalty Program breakeven estimate should be read together with the HonestDiD sensitivity analysis in $\S$\ref{ssec:parallel_tests}.}


In sum, managers need to calibrate expectations of incremental profitability when running promotions to encourage online adoption. The key managerial concern is not merely that promotion-induced adopters have different monthly profit estimates; it is that using the wrong benchmark can make promotions appear to recover their costs much faster than suggested by pathway-aware estimates. This may lead to overly optimistic ROI assessments, excessive promotional spending, and underestimation of the time required to recoup promotional investments.

\section{Robustness Checks}
\label{ssec:validity_robustness}
We now present a series of robustness checks to provide additional credibility for our findings. 
\subsection{Effect size measurement}
\label{ssec:effect_size_measurement}

A potential concern with our ATT measurement, especially for our COVID-19 analysis, is that \textit{COVID adopters} may be ``contaminated'' with some \textit{organic adopters} -- customers who would have adopted the online channel organically during the onset of COVID-19, had COVID-19 not occurred. As a result, the treatment group may include a mix of ``true" \textit{COVID adopters} and \textit{organic adopters}. If such contamination exists, the ATT estimates reported in \S\ref{ssec:results_event_organic_adopters} would be a lower bound on the true ATT for COVID adoption (relative to organic), and we can back out the ATT for the subset of customers of interest (i.e., the ``true" \textit{COVID adopters}). The true ATT for \textit{COVID adopters} relative to \textit{organic adopters} can be obtained by dividing the estimated ATT (reported in \S\ref{ssec:results_event_organic_adopters}) by the fraction of ``true'' \textit{COVID adopters}. Assuming the monthly incidence of \textit{organic adopters} remains constant, we estimate that around 38.5\% of the \textit{COVID adopters} could actually be \textit{organic adopters} and $61.5\%$ can be the fraction of ``true'' \textit{COVID adopters}. Using this fraction, we can compute the ``true" ATT of spend for \textit{COVID adopters} relative to \textit{organic adopters} as -5.03 (std. error: 10.363, p-value: 0.627) and the ``true" ATT of profit as 15.763 (std. error: 3.638, p-value: 0.000). The details of these calculations and the block bootstrapping used for standard error computations are shown in Web Appendix $\S$\ref{appssec:effect_size_measurement}. This confirms that, even with potential contamination, the difference in spend relative to \textit{organic adopters} is small, with a confidence interval that includes zero, whereas the difference in profit is larger and has a confidence interval that does not include zero. As such, our substantive conclusions remain unchanged. 

Note that we are less concerned about contamination for the ATT estimates for the other two events for a few reasons, which we briefly discuss below.
\squishlist
\item First, since the Black Friday sale occurs only on a single day (and not over a prolonged period, unlike COVID-19), and there is a large spike in adopters on that single day compared to the days around it, the extent of potential contamination is very small. In this case, the results are very similar to those from $\S$\ref{ssec:results_event_organic_adopters} since historical data suggest that close to 85.7\% of adopters during the Black Friday event adopt because of the event (and not organically). Nevertheless, we still conduct a similar analysis by backing out the ``true'' ATT using the estimated ATT of \textit{Black Friday adopters} relative to \textit{organic adopters} as above (details in Web Appendix $\S$\ref{appssec:effect_size_measurement}) to confirm that even with such potential contamination, \textit{Black Friday adopters} have lower spend and profit relative to \textit{organic adopters}.

\item Next, recall that the \textit{Loyalty Program adopters} are customers who specifically went online for the first time to activate the loyalty program on their mobile app and only made their first online purchase after activation, so contamination with \textit{organic adopters} (those adopted online shopping before the loyalty program was launched and then activated the loyalty program after it was launched) is unlikely. 
\squishend

\subsection{Doubly robust DID, Propensity Score Matching and IPW approaches}
\label{ssec:psm}

In $\S$\ref{ssec:summary_stats}, we saw that there are some differences in the demographics and other pre-adoption variables between (1) adopters and \textit{offline-only} customers, and (2) \textit{organic adopters} and {\it event adopters}. Therefore, to provide additional credibility to our findings and ensure that they are robust to any factors correlated with observable characteristics that may affect the propensity to adopt online shopping organically versus spurred by external events, we also consider doubly robust DID \citep{sant2020doubly} and propensity score-based approaches \citep{rosenbaum1985constructing,stuart2011matchit}. These methods provide better comparability between groups by improving covariate balance by discarding poorly matched customers. 
    We compute propensity scores using a logistic regression model between (1) adopters vs. \textit{offline-only} customers, and (2) \textit{organic adopters} vs. \textit{event adopters}, with variables including demographics (e.g., gender, age, and household income) and pre-adoption purchase variables (i.e., average monthly spend, number of unique orders, unique brands, and unique categories). With the estimated propensity scores, we apply our DID estimator with both IPW and PSM methods and also run a doubly robust DID estimator. The details of the analysis and the results are shown in Web Appendices $\S$\ref{appssec:did_psm_ipw} for \textit{organic adopters} vs. \textit{event adopters} and $\S$\ref{appssec:did_psm_ipw_adopter_nonadopter} for \textit{adopters} vs. \textit{offline-only} customers. We find that our substantive conclusions remain the same with these approaches. 


\subsection{Alternate definitions for the \textit{organic adopter} and \textit{COVID adopter} cohorts}
\label{ssec:redefine_organic_adopters}
To check whether our results are sensitive to the particular months we chose to define our \textit{COVID adopter} and \textit{organic adopter} cohorts in our COVID-19 analysis, and to test whether our conclusions would change if we used ``later" \textit{COVID adopters} vs. ``early" \textit{COVID adopters}, we repeat our analysis with alternative definitions for both cohorts and present results in Web Appendix $\S$\ref{appsec:alternative_definition_covid_organic}. To start with, we redefine \textit{organic adopters} as those customers who were offline-only before 2019-10-31 and made their first online purchase between 2019-11-01 and 2020-02-29 (i.e., we extend the period we previously used to define \textit{organic adopters}). We re-run the same set of DID models and find that all results are consistent with our main analysis. Next, we redefine the cohort of \textit{COVID adopters} as those customers who were offline-only before 2019-12-31 and made their first online purchase between 2020-05-01 and 2020-06-30 (i.e., two months later than we had used in our main analysis). Again, we find that all the results are consistent with the main analysis.
Finally, we compare ``early" \textit{COVID adopters} (who adopted online shopping between 2020-03-16 and 2020-04-30) and ``later" \textit{COVID adopters} (who adopted online shopping between 2020-05-01 and 2020-06-30). Again, we run the same set of DID models and do not find any significant differences except that the ``later" \textit{COVID adopters} spend slightly higher by 2.48\%.


These analyses also address another potential concern with our main analysis, which is that \textit{organic adopters} adopt for two extra months relative to \textit{COVID adopters} before our common post-adoption period, and this may drive some of the results. The fact that our conclusions don't change when we change the set of months used to define our cohorts also allays this concern. Finally, our definitions for \textit{Black Friday adopters} and \textit{Loyalty Program adopters} are much more precise than for {\it COVID adopters}, so we don't repeat this robustness check for those analyses.

\subsection{Alternate operationalizations of dependent variables}

To test whether the positive skewness in purchase outcomes affects our results, we re-estimate all DID models using a natural log transformation for the dependent variables ($\ln(y+1)$) and report the results in Web Appendix $\S$\ref{appsec:log_dependent_variables}. Across all specifications, the results remain consistent with our main findings. Another potential concern in our setting is that many customer-month spending observations are zero in our data. Therefore, we consider an alternative specification that accommodates zero-inflated dependent variables. Specifically, we estimate a two-part (hurdle) model at the customer-month level \citep{cameron2005microeconometrics}. This approach involves: 
(1) a linear probability model with a DID specification to estimate changes in the likelihood of making a purchase during the post-adoption periods\footnote{Here, the dependent variable of linear probability model is $\mathbbm{I}\{Spend_{it}\}$, a binary indicator for whether customer $i$ made purchase during month $t$, and the DID specification is Equations \eqref{eq:did_fe} for comparing \textit{adopters} vs. \textit{offline-only} customers, and Equation \eqref{eq:did_fe_2by2_FE} for comparing \textit{organic adopters} vs. \textit{event adopters})}; (2) a DID model that estimates the changes in spend conditional on making a purchase.\footnote{We estimate the same equations as the main analysis, but use the customer-month observations with non-zero spend.} Overall, we find that the results from these specifications are consistent with our main results; see Web Appendix $\S$\ref{appsec:zero_inflated_dependent_variables} for detailed estimation results. 

\section{Conclusions}
\label{sec:conclusion}

We study how the adoption of a retailer's online channels by previously \textit{offline-only customers} affects post-adoption behavior, and whether these effects depend on the \emph{reason} for adoption. Prior work shows that multichannel customers are more valuable than single-channel customers, but largely treats channel adoption as homogeneous. We show that this masks important heterogeneity. Using transaction data from a large Brazilian pet supplies retailer from 2019--2024, we compare customers who adopt online shopping organically with those induced to adopt through three external pathways: COVID-19, Black Friday promotions, and a mobile-app-activated Loyalty Program.

\paragraph{Main findings.} We document three main findings. First, online adoption is associated with higher spend and profitability relative to remaining offline-only, consistent with the standard multichannel-customer result. Second, conditional on adoption, post-adoption behavior differs sharply across adoption pathways. \textit{COVID adopters} spend similarly to \textit{organic adopters} but are more profitable, because they continue allocating more of their spending to the higher-margin offline channel. This pattern is consistent with consumer inertia and habit persistence, especially among older customers. Third, promotion-induced adopters behave differently: \textit{Black Friday adopters} and \textit{Loyalty Program adopters} spend less and are less profitable than \textit{organic adopters} after adoption. Event-study evidence shows that these customers spend more around the adoption period but less afterward, consistent with forward buying and post-promotion demand distortion.

The channel-mix results further show that external adoption does not always translate into sustained online usage. \textit{COVID adopters} move toward online shopping more slowly than \textit{organic adopters}. \textit{Black Friday adopters} initially retain a higher offline share, but move online faster over time, consistent with their younger profile. \textit{Loyalty Program adopters} do not differ meaningfully from \textit{organic adopters} in offline spending share, likely because recurring app-based activation reinforces online engagement. Thus, adoption pathways affect not only how much customers spend, but also where they spend and how profitable they are.

\paragraph{Managerial implications.} The results suggest that firms may misvalue online adopters when they rely on a common multichannel benchmark. In our setting, ignoring adoption pathways substantially biases incremental CLV forecasts and promotion ROI calculations. In particular, pooling adopters or using \textit{organic adopters} as proxies can substantially overstate the long-run value of promotion-induced adopters and make online-adoption promotions appear to break even much faster than suggested by pathway-aware estimates. Firms may therefore benefit from estimating incremental CLV separately by adoption pathway, setting pathway-specific caps on acquisition and promotional costs, and designing online-adoption campaigns that account for post-adoption behavior. For promotion-induced adopters, this may mean limiting stockpiling and encouraging repeat engagement after the promotion. For shock-induced adopters, it may mean using habit-reinforcing interventions to sustain online engagement after the initial shock subsides.
 

\paragraph{Limitations and future research.} Our study has three limitations that suggest opportunities for future work. First, we study one macro shock and two firm-driven promotions. Future research could examine other adoption triggers, such as store closures, weather shocks, localized disruptions, or different promotion designs. Second, our setting is a single retailer in one category and country; studying other categories and retail environments would help assess the generalizability of the mechanisms we document. Relatedly, while our evidence is consistent with consumer inertia, habit persistence, and forward buying, additional diagnostics are needed to distinguish these mechanisms from other explanations, such as channel disappointment, one-time deal-seeking, or differential retention. Future research could use surveys, randomized encouragement, or richer behavioral diagnostics to more directly distinguish among these explanations. Third, our estimates are descriptive ATTs that capture the post-adoption value of customers who actually adopt through each pathway. This is the estimand most relevant for managerial forecasting and campaign evaluation, but future work could complement our analysis by separately estimating causal ATEs of online adoption using randomized encouragements, instruments, or other identification strategies. 


\section*{Funding and Competing Interests}
The authors were not paid by the data provider. A preliminary version of this manuscript was reviewed by the data provider to ensure that the data was accurately depicted, but they did not review/comment on the content of the manuscript.


\singlespacing



\newpage
\setcounter{table}{0}
\setcounter{figure}{0}
\setcounter{equation}{0}
\setcounter{page}{1}
\renewcommand{\thetable}{A\arabic{table}}
\renewcommand{\thefigure}{A\arabic{figure}}
\renewcommand{\theequation}{A\arabic{equation}}
\renewcommand{\thepage}{\roman{page}}
\singlespacing
\renewcommand{\thesection}{\Alph{section}}
\pagenumbering{roman} 
\

\begin{appendices}

\section{Data cleaning}
\label{appsec:data_cleaning}

\begin{table}[htp!]
\centering
\caption{Item categories with substantially different availability time via online and offline channels (YY-MM-DD)}
\label{category_remove}
\footnotesize{
\begin{tabular}{lll}
\toprule
      Item Category & First Offline Purchase Date & First Online Purchase Date \\
\midrule
Large size food &                  2020-09-12 &                 2020-02-28 \\
Human food &                  2019-01-02 &                 2021-02-12 \\
Large size pharmacy &                  2021-02-03 &                 2020-12-09 \\
Pet Cleaning Supplies  &                  2019-01-03 &                 2022-03-13 \\
\bottomrule
\end{tabular}
}
\end{table}

There are four product categories with significant discrepancies in their initial availability time on the offline and online channels. For example, the product category \textit{Large size food} (e.g. pack size greater than 10KG) was available online in February 2020 but only became available offline in September 2020. In addition, the product category Pet Cleaning Supplies was only available offline before COVID-19 but became available online in August 2022. 

Including these food categories in our analysis can be problematic since it also reflects retailers' assortment and the channel-specific effects (i.e. consumers can only buy some products exclusively from one channel). Therefore, we exclude these four categories from our analysis. The fraction of sales of these four categories only accounts for 0.63\% of the total revenue during the sample period, and as such unlikely to have any meaningful impact on our findings.

\clearpage

\section{Appendix to Summary Statistics}
\label{appsec:app_descriptive}

In this section, we summarize the demographics, pre-adoption, and post-adoption purchase behaviors of three customer groups for each focal event: \textit{event adopters}, \textit{organic adopters}, and \textit{offline-only customers}. 

To characterize the pre-online-adoption purchase behavior of each group of customers, we compute several consumer-level metrics, such as \textit{AvgSpendPerMonth}, \textit{AvgOrdersPerMonth}, \textit{AvgUniqueCategoriesPerMonth}, using data from the pre-adoption period specific to each case study. We summarize these variables for each group -- {\it event adopters}, \textit{organic adopters}, and \textit{offline-only} -- by calculating the group mean as follows: $\frac{1}{N_g}\sum_{i=1}^{N_g} (\frac{1}{T_i^{pre}}\sum_{t=1}^{T_i^{pre}} X_{it})$, where $i$ denotes a customer belonging to group $g$, $t$ denotes the year-month index, $T_i^{pre}$ denotes the length of the consumer $i$'s panel during the pre-adoption period, and $N_g$ is the number of customers in group $g$. We next summarize the post-online adoption behaviors in a similar way using the post-online adoption data under each case study. The post-online adoption means for each group $g$ -- {\it event adopters}, \textit{organic adopters}, and \textit{offline-only} -- are calculated as follows: $\frac{1}{N_g}\sum_{i=1}^{N_g} (\frac{1}{T_i^{post}}\sum_{t=1}^{T_i^{post}} X_{it})$, where $i$ denotes a customer belonging to group $g$, $t$ denotes the year-month index, $T_i^{post}$ denotes the length of the consumer $i$'s panel during the post-adoption period, and $N_g$ is the number of customers in group $g$.

\subsection{Event 1: COVID-19}
\label{appssec:covid}

In this section, we present the descriptive statistics for customers in the sample for the COVID-19, supplementing those in \S\ref{ssec:summary_stats}. We first summarize demographics in Table \ref{tab:covid_gender_state_percentage}. We also report customer-level summary statistics for average purchase variables during the post-adoption period in Tables \ref{tab:covid_post_adoption_summary_stat}, complementing the pre-adoption statistics shown in Table \ref{tab:pre_adoption_stacked} in \S\ref{ssec:summary_stats}. Finally, in Table \ref{tab:covid_post_adoption_quartile}, we present more detailed summary statistics (e.g., median and quartiles) for all three groups during the post-online-adoption periods (i.e., between May 1st, 2020 and April 30, 2023). Note that the counts of \textit{organic adopters} and {\it COVID adopters} from Table \ref{tab:covid_post_adoption_quartile} slightly differ from the raw count based on our cohorts' definition because 469 (3.66\% of) \textit{organic adopters} and 933 (2.81\% of) {\it COVID adopters} churned before 2020-05-01, and therefore are not observed in the post-COVID-19 panel. Including these customers does not affect our identification of post-online-adoption ATT, and the estimation results remain the same.

\begin{table}[htp!]
    \centering
\caption{Customer Demographics -- Summary Statistics of Gender (Panel A) and Geographic State (Panel B). }
\label{tab:covid_gender_state_percentage}
\footnotesize{
\begin{tabular}{lrrr}
\toprule
       Categories &  \textit{COVID adopter} (\%) &  \textit{Organic adopter} (\%)  &  \textit{Offline-only} (\%)  \\
\midrule
\multicolumn{4}{c}{Panel A: Gender} \\ \midrule
No Information & 12.09 & 13.52 & 15.63 \\
Female & 59.59 & 57.36 & 46.51 \\
Male & 28.31 & 29.13 & 37.86 \\ \midrule
Total & 100.00 & 100.00 & 100.00 \\
\midrule
\multicolumn{4}{c}{Panel B: Geographic State}\\ \midrule
São Paulo & 58.78 & 49.68 & 63.70 \\
Minas Gerais & 8.58 & 12.13 & 4.81 \\
Rio de Janeiro & 6.69 & 6.95 & 4.30 \\
Distrito Federal & 6.21 & 5.48 & 4.26 \\
Zip\_State\_Missing & 4.07 & 4.28 & 10.34 \\
Santa Catarina & 3.33 & 3.46 & 2.90 \\
Goiás & 2.37 & 3.36 & 2.13 \\
Bahia & 1.74 & 3.31 & 0.69 \\
Rio Grande do Sul & 3.21 & 2.77 & 2.20 \\
Espírito Santo & 1.27 & 2.43 & 0.91 \\
Mato Grosso do Sul & 0.93 & 1.88 & 1.05 \\
Paraná & 1.68 & 1.86 & 2.31 \\
Other States & 1.14 & 2.41 & 0.39 \\ \midrule
Total & 100.00 & 100.00 & 100.00 \\
\bottomrule
\end{tabular}
}
\end{table}

\begin{table}[htp!] 
\centering
\caption{Customer-Level Summary Statistics of \textbf{Post-online-adoption} Purchase Behaviors between May 1st 2020 and April 30th 2023 (Panel A). Columns 1 - 2 describe statistics for {\it COVID adopters} and \textit{Organic adopters}, respectively. Column 3 calculates the difference between {\it COVID adopters} and \textit{Organic adopters}, and columns 4 and 5 show the two-sample t-test and p-value results.}
\label{tab:covid_post_adoption_summary_stat}
 \resizebox{\textwidth}{!}{
\footnotesize{
\begin{tabular}{lcccccc}
  \hline
Variable  &\makecell{{\it COVID adopters} \\ (Mean)}  & \makecell{ {\it Organic adopters} \\ (Mean)} & Diff & t-stats & p-value & \makecell{{\it Offline-only} \\ (Mean)} \\ \midrule
\multicolumn{6}{c}{Panel A: Post-online-adoption Monthly Purchase Behaviors} \\ \midrule
\textit{AvgSpendPerMonth} & 514.47 & 489.47 & 25.00 & 3.75 & 0.00 & 202.29 \\
\textit{AvgQuantitiesPerMonth} & 4.88 & 4.11 & 0.77 & 7.58 & 0.00 & 1.85 \\
\textit{AvgOrdersPerMonth} & 0.89 & 0.89 & -0.00 & -0.08 & 0.93 & 0.45 \\
\textit{AvgUniqueItemsPerMonth} & 2.28 & 2.04 & 0.23 & 9.46 & 0.00 & 1.13 \\
\textit{AvgUniqueBrandsPerMonth} & 1.78 & 1.63 & 0.15 & 9.18 & 0.00 & 0.91 \\
\textit{AvgUniqueSubcategoriesPerMonth} & 1.52 & 1.39 & 0.13 & 10.35 & 0.00 & 0.80 \\
\textit{AvgUniqueCategoriesPerMonth} & 1.27 & 1.16 & 0.11 & 10.82 & 0.00 & 0.68 \\
\textit{AvgProfitPerMonth} & 175.95 & 157.42 & 18.53 & 8.18 & 0.00 & 76.73 \\
\bottomrule
\end{tabular}
}
}
\end{table}

\begin{table}[htp!]
    \centering
\caption{Customer-Level Summary Statistics for All Three Groups -- {\it COVID adopters}, \textit{Organic adopters}, and \textit{Offline-only} customers for post-online-adoption period (i.e., between May 1st, 2020 and April 30th, 2023). }
\label{tab:covid_post_adoption_quartile}
 \resizebox{\textwidth}{!}{
\footnotesize{
\begin{tabular}{lrrrrrrrl}
\toprule
Variables & Mean & Std. & Min & 25\% & 50\% & 75\% & Max & Count \\
\midrule
\multicolumn{9}{c}{{\it COVID adopters}} \\ \midrule
\textit{AvgSpendPerMonth} & 514.47 & 647.83 & 0.00 & 155.32 & 338.44 & 655.25 & 40153.38 & 32,313 \\
\textit{AvgQuantitiesPerMonth} & 4.88 & 10.11 & 0.00 & 0.97 & 2.18 & 4.89 & 414.11 & 32,313 \\
\textit{AvgOrdersPerMonth} & 0.89 & 0.81 & 0.00 & 0.36 & 0.69 & 1.17 & 17.81 & 32,313 \\
\textit{AvgUniqueItemsPerMonth} & 2.28 & 2.41 & 0.00 & 0.78 & 1.57 & 2.97 & 58.62 & 32,313 \\
\textit{AvgUniqueBrandsPerMonth} & 1.78 & 1.61 & 0.00 & 0.67 & 1.33 & 2.41 & 18.61 & 32,313 \\
\textit{AvgUniqueSubcategoriesPerMonth} & 1.52 & 1.23 & 0.00 & 0.62 & 1.21 & 2.11 & 12.86 & 32,313 \\
\textit{AvgUniqueCategoriesPerMonth} & 1.27 & 0.93 & 0.00 & 0.56 & 1.06 & 1.78 & 8.81 & 32,313 \\
\textit{AvgProfitPerMonth} & 175.95 & 222.05 & -95.36 & 54.56 & 116.51 & 224.52 & 15880.26 & 32,313 \\
\midrule
\multicolumn{9}{c}{{\it Organic adopters}} \\ \midrule
\textit{AvgSpendPerMonth} & 489.47 & 582.43 & 0.00 & 141.99 & 317.95 & 629.33 & 10709.68 & 12,330 \\
\textit{AvgQuantitiesPerMonth} & 4.11 & 8.10 & 0.00 & 0.83 & 1.89 & 4.22 & 245.86 & 12,330 \\
\textit{AvgOrdersPerMonth} & 0.89 & 0.83 & 0.00 & 0.33 & 0.67 & 1.19 & 11.03 & 12,330 \\
\textit{AvgUniqueItemsPerMonth} & 2.04 & 2.15 & 0.00 & 0.67 & 1.39 & 2.67 & 30.58 & 12,330 \\
\textit{AvgUniqueBrandsPerMonth} & 1.63 & 1.51 & 0.00 & 0.58 & 1.19 & 2.19 & 16.31 & 12,330 \\
\textit{AvgUniqueSubcategoriesPerMonth} & 1.39 & 1.16 & 0.00 & 0.54 & 1.08 & 1.92 & 9.47 & 12,330 \\
\textit{AvgUniqueCategoriesPerMonth} & 1.16 & 0.88 & 0.00 & 0.50 & 0.94 & 1.61 & 6.81 & 12,330 \\
\textit{AvgProfitPerMonth} & 157.42 & 191.05 & -24.34 & 46.69 & 101.13 & 203.75 & 5028.99 & 12,330 \\
\midrule
\multicolumn{9}{c}{\textit{Offline-only} customers} \\ \midrule
\textit{AvgSpendPerMonth} & 202.29 & 330.96 & 0.00 & 35.98 & 99.68 & 242.19 & 24573.87 & 573,934 \\
\textit{AvgQuantitiesPerMonth} & 1.85 & 4.59 & 0.00 & 0.30 & 0.77 & 1.85 & 487.17 & 573,934 \\
\textit{AvgOrdersPerMonth} & 0.45 & 0.53 & 0.00 & 0.12 & 0.28 & 0.58 & 40.11 & 573,934 \\
\textit{AvgUniqueItemsPerMonth} & 1.13 & 1.56 & 0.00 & 0.25 & 0.63 & 1.39 & 123.22 & 573,934 \\
\textit{AvgUniqueBrandsPerMonth} & 0.91 & 1.09 & 0.00 & 0.23 & 0.55 & 1.17 & 48.44 & 573,934 \\
\textit{AvgUniqueSubcategoriesPerMonth} & 0.80 & 0.87 & 0.00 & 0.21 & 0.50 & 1.06 & 19.90 & 573,934 \\
\textit{AvgUniqueCategoriesPerMonth} & 0.68 & 0.70 & 0.00 & 0.19 & 0.44 & 0.94 & 10.00 & 573,934 \\
\textit{AvgProfitPerMonth} & 76.73 & 120.02 & -491.73 & 14.17 & 38.99 & 93.21 & 9834.37 & 573,934 \\ \midrule
\bottomrule
\end{tabular}
}
}
\end{table}

\clearpage

\subsection{Event 2: Black Friday}
\label{appssec:black_friday}

In this section, we present the descriptive statistics for customers in the sample for the Black Friday event, supplementing those in \S\ref{ssec:summary_stats}. We first summarize demographics in Table \ref{tab:black_friday_gender_state_percentage}. We also report customer-level summary statistics for average purchase variables during the post-adoption period in Table \ref{tab:black_friday_post_adoption_summary_stat}, complementing the pre-adoption statistics shown in Table \ref{tab:pre_adoption_stacked} in \S\ref{ssec:summary_stats}. Finally, in Table \ref{tab:black_friday_post_adoption_quartile}, we present more detailed summary statistics (e.g., median and quartiles) for all three groups during the post-online-adoption periods (i.e., between December 1st, 2019 and October 31, 2020). Note that the counts of \textit{organic adopters} and {\it Black Friday adopters} from Table \ref{tab:black_friday_post_adoption_quartile} slightly differ from the raw count based on our cohorts' definition because 79 (2.58\% of) \textit{organic adopters} and 24 (1.74\% of) {\it Black Friday adopters} churned before 2019-12-01, and therefore are not observed in the post-Black Friday panel. Including these customers does not affect our identification of post-online-adoption ATT, and the estimation results remain the same.

\begin{table}[htp!]
    \centering
\caption{Customer Demographics -- Summary Statistics of Gender (Panel A) and Geographic State (Panel B).}
\label{tab:black_friday_gender_state_percentage}
\footnotesize{
\begin{tabular}{lrrr}
\toprule
       Categories &  \textit{Black Friday Adopter} (\%) &  \textit{Organic Adopter} (\%)  &  \textit{Offline-only} (\%)  \\
\midrule
\multicolumn{4}{c}{Panel A: Gender} \\ \midrule
No Information & 14.13 & 13.59 & 15.42 \\
Female & 59.42 & 57.89 & 46.46 \\
Male & 26.45 & 28.52 & 38.12 \\ \midrule
Total & 100.00 & 100.00 & 100.00 \\
\midrule
\multicolumn{4}{c}{Panel B: Geographic State}\\ \midrule
São Paulo & 57.97 & 51.91 & 64.83 \\
Minas Gerais & 8.48 & 13.10 & 4.69 \\
Rio de Janeiro & 7.90 & 6.99 & 4.50 \\
Distrito Federal & 4.78 & 4.64 & 4.38 \\
Zip\_State\_Missing & 4.93 & 3.76 & 9.90 \\
Santa Catarina & 2.54 & 3.40 & 2.60 \\
Bahia & 2.61 & 3.20 & 0.50 \\
Goiás & 2.83 & 3.17 & 2.19 \\
Rio Grande do Sul & 2.17 & 3.17 & 2.10 \\
Espírito Santo & 1.96 & 2.52 & 0.73 \\
Paraná & 1.74 & 1.99 & 2.39 \\
Mato Grosso do Sul & 1.52 & 1.83 & 1.10 \\
Other States & 0.58 & 0.33 & 0.09 \\ \midrule
Total & 100.00 & 100.00 & 100.00 \\ \midrule
\bottomrule
\end{tabular}
}
\end{table}

\begin{table}[htp!] 
\centering
\caption{Customer-Level Summary Statistics of \textbf{Post-online-adoption} Purchase Behaviors between December 1st 2019 and October 31st 2020 (Panel A). Columns 1 - 2 describe statistics for {\it Black Friday adopters} and \textit{Organic adopters}, respectively. Column 3 calculates the difference between {\it Black Friday adopters} and \textit{Organic adopters}, and columns 4 and 5 show the two-sample t-test and p-value results.}
\label{tab:black_friday_post_adoption_summary_stat}
 \resizebox{\textwidth}{!}{
\footnotesize{
\begin{tabular}{lcccccc}
  \hline
Variable  &\makecell{{\it \makecell{Black Friday\\ Adopters}} \\ (Mean)}  & \makecell{ {\it Organic Adopters} \\ (Mean)} & Diff & t-stats & p-value & \makecell{{\it Offline-only} \\ (Mean)} \\ \midrule
\multicolumn{6}{c}{Panel A: Post-online-adoption Monthly Purchase Behaviors} \\ \midrule
\textit{AvgSpendPerMonth} & 417.53 & 468.23 & -50.70 & -2.94 & 0.00 & 195.21 \\
\textit{AvgQuantitiesPerMonth} & 4.37 & 4.55 & -0.17 & -0.59 & 0.56 & 2.06 \\
\textit{AvgOrdersPerMonth} & 0.88 & 0.97 & -0.10 & -3.36 & 0.00 & 0.48 \\
\textit{AvgUniqueItemsPerMonth} & 2.28 & 2.39 & -0.11 & -1.33 & 0.18 & 1.24 \\
\textit{AvgUniqueBrandsPerMonth} & 1.83 & 1.90 & -0.08 & -1.25 & 0.21 & 0.99 \\
\textit{AvgUniqueSubcategoriesPerMonth} & 1.56 & 1.63 & -0.07 & -1.55 & 0.12 & 0.87 \\
\textit{AvgUniqueCategoriesPerMonth} & 1.29 & 1.35 & -0.06 & -1.66 & 0.10 & 0.74 \\
\textit{AvgProfitPerMonth} & 144.17 & 157.64 & -13.47 & -2.31 & 0.02 & 73.46 \\
\bottomrule
\end{tabular}
}
}
\end{table}

\begin{table}[htp!]
    \centering
\caption{Customer-Level Summary Statistics for All Three Groups -- {\it Black Friday adopters}, \textit{Organic Adopters}, and \textit{Offline-only} customers for post-online-adoption period (i.e., between December 1st, 2019 and October 31st, 2020). }
\label{tab:black_friday_post_adoption_quartile}
 \resizebox{\textwidth}{!}{
\footnotesize{
\begin{tabular}{lrrrrrrrl}
\toprule
Variables & Mean & Std. & Min & 25\% & 50\% & 75\% & Max & Count \\
\midrule
\multicolumn{9}{c}{{\it Black Friday adopters}} \\ \midrule
\textit{AvgSpendPerMonth} & 417.53 & 485.46 & 0.00 & 118.34 & 280.31 & 536.91 & 6029.03 & 1,356 \\
\textit{AvgQuantitiesPerMonth} & 4.37 & 10.24 & 0.00 & 0.82 & 2.00 & 4.64 & 219.18 & 1,356 \\
\textit{AvgOrdersPerMonth} & 0.88 & 0.82 & 0.00 & 0.27 & 0.64 & 1.18 & 6.91 & 1,356 \\
\textit{AvgUniqueItemsPerMonth} & 2.28 & 2.51 & 0.00 & 0.64 & 1.50 & 3.09 & 20.27 & 1,356 \\
\textit{AvgUniqueBrandsPerMonth} & 1.83 & 1.82 & 0.00 & 0.55 & 1.27 & 2.55 & 13.82 & 1,356 \\
\textit{AvgUniqueSubcategoriesPerMonth} & 1.56 & 1.40 & 0.00 & 0.55 & 1.18 & 2.18 & 8.91 & 1,356 \\
\textit{AvgUniqueCategoriesPerMonth} & 1.29 & 1.06 & 0.00 & 0.45 & 1.00 & 1.91 & 7.00 & 1,356 \\
\textit{AvgProfitPerMonth} & 144.17 & 167.61 & -56.75 & 40.64 & 96.10 & 184.10 & 2184.19 & 1,356 \\
\midrule
\multicolumn{9}{c}{ {\it Organic Adopters}} \\ \midrule
\textit{AvgSpendPerMonth} & 468.23 & 543.57 & 0.00 & 127.57 & 315.86 & 605.36 & 6608.86 & 2,982 \\
\textit{AvgQuantitiesPerMonth} & 4.55 & 8.55 & 0.00 & 0.82 & 2.09 & 4.91 & 162.82 & 2,982 \\
\textit{AvgOrdersPerMonth} & 0.97 & 0.93 & 0.00 & 0.36 & 0.73 & 1.27 & 7.82 & 2,982 \\
\textit{AvgUniqueItemsPerMonth} & 2.39 & 2.66 & 0.00 & 0.67 & 1.64 & 3.09 & 27.00 & 2,982 \\
\textit{AvgUniqueBrandsPerMonth} & 1.90 & 1.88 & 0.00 & 0.64 & 1.36 & 2.55 & 17.82 & 2,982 \\
\textit{AvgUniqueSubcategoriesPerMonth} & 1.63 & 1.45 & 0.00 & 0.55 & 1.27 & 2.27 & 12.45 & 2,982 \\
\textit{AvgUniqueCategoriesPerMonth} & 1.35 & 1.09 & 0.00 & 0.55 & 1.09 & 1.91 & 7.73 & 2,982 \\
\textit{AvgProfitPerMonth} & 157.64 & 182.93 & -105.01 & 42.74 & 105.73 & 203.51 & 2349.98 & 2,982 \\
\midrule
\multicolumn{9}{c}{ {\it Offline-only} Customers} \\ \midrule
\textit{AvgSpendPerMonth} & 195.21 & 330.14 & 0.00 & 14.31 & 86.21 & 243.99 & 16616.77 & 534,871 \\
\textit{AvgQuantitiesPerMonth} & 2.06 & 5.51 & 0.00 & 0.18 & 0.73 & 2.09 & 522.36 & 534,871 \\
\textit{AvgOrdersPerMonth} & 0.48 & 0.62 & 0.00 & 0.09 & 0.27 & 0.64 & 19.73 & 534,871 \\
\textit{AvgUniqueItemsPerMonth} & 1.24 & 1.84 & 0.00 & 0.18 & 0.64 & 1.55 & 52.91 & 534,871 \\
\textit{AvgUniqueBrandsPerMonth} & 0.99 & 1.31 & 0.00 & 0.12 & 0.55 & 1.36 & 23.18 & 534,871 \\
\textit{AvgUniqueSubcategoriesPerMonth} & 0.87 & 1.06 & 0.00 & 0.09 & 0.55 & 1.25 & 16.09 & 534,871 \\
\textit{AvgUniqueCategoriesPerMonth} & 0.74 & 0.85 & 0.00 & 0.09 & 0.45 & 1.09 & 9.82 & 534,871 \\
\textit{AvgProfitPerMonth} & 73.46 & 120.07 & -331.15 & 5.58 & 33.33 & 93.25 & 6313.29 & 534,871 \\
\midrule
\bottomrule
\end{tabular}
}
}
\end{table}

\clearpage

\newpage

\subsection{Event 3: Loyalty Program}
\label{appssec:loyalty}

In this section, we present the descriptive statistics for customers in the sample for the loyalty program analysis, supplementing those in \S\ref{ssec:summary_stats}. Table \ref{tab:loyalty_gender_state_percentage} summarizes customer demographics. We also report customer-level summary statistics for average purchase variables during the post-adoption period in Table \ref{tab:loyalty_post_adoption_summary_stat}, complementing the pre-adoption statistics shown in Table \ref{tab:pre_adoption_stacked} in \S\ref{ssec:summary_stats}. Finally, in Table \ref{tab:loyalty_post_adoption_quartile}, we present more detailed summary statistics (e.g., median and quartiles) for all three groups during the post-online-adoption periods (i.e., between October 1st, 2023 and March 31, 2024)
\begin{table}[htp!]
    \centering
\caption{Customer Demographics -- Summary Statistics of Gender (Panel A) and Geographic State (Panel B). Our data providers label around 45\% of overall customers as no gender information in this study. }
\label{tab:loyalty_gender_state_percentage}
\footnotesize{
\begin{tabular}{lrrr}
\toprule
       Categories &  \textit{Loyalty Program Adopter} (\%) &  \textit{Organic Adopter} (\%)  &  \textit{Offline-only} (\%)  \\
\midrule
\multicolumn{4}{c}{Panel A: Gender} \\ \midrule
No Information & 50.16 & 48.30 & 42.85 \\
Female & 30.57 & 32.52 & 31.02 \\
Male & 19.27 & 19.17 & 26.13 \\ \midrule
Total & 100.00 & 100.00 & 100.00 \\
\midrule
\multicolumn{4}{c}{Panel B: Geographic State} \\ \midrule
São Paulo & 35.99 & 43.29 & 55.28 \\
Zip\_State\_Missing & 32.99 & 19.92 & 21.93 \\
Minas Gerais & 5.30 & 6.69 & 3.84 \\
Rio de Janeiro & 3.89 & 4.39 & 3.15 \\
Santa Catarina & 6.25 & 4.14 & 3.04 \\
Distrito Federal & 2.62 & 3.86 & 3.09 \\
Espírito Santo & 2.23 & 2.59 & 1.28 \\
Ceará & 2.04 & 2.37 & 0.83 \\
Pernambuco & 1.08 & 2.10 & 0.94 \\
Bahia & 1.21 & 1.99 & 1.12 \\
Goiás & 1.34 & 1.83 & 1.63 \\
Mato Grosso do Sul & 0.70 & 1.22 & 0.95 \\
Other States & 4.34 & 5.62 & 2.93 \\ \midrule
Total & 100.00 & 100.00 & 100.00 \\ \midrule
\bottomrule
\end{tabular}
}
\end{table}

\begin{table}[htp!] 
\centering
\caption{Customer-Level Summary Statistics of \textbf{Post-online-adoption} Purchase Behaviors between October 1st, 2023 and March 31st, 2024 (Panel A). Columns 1 - 2 describe statistics for {\it Loyalty Program Adopters} and \textit{Organic adopters}, respectively. Column 3 calculates the difference between {\it Loyalty Program Adopters} and \textit{Organic adopters}, and columns 4 and 5 show the two-sample t-test and p-value results.}
\label{tab:loyalty_post_adoption_summary_stat}
 \resizebox{\textwidth}{!}{
\footnotesize{
\begin{tabular}{lcccccc}
  \hline
Variable  &\makecell{{\it \makecell{Loyalty Program\\ Adopters}} \\ (Mean)}  & \makecell{\textit{organic adopters} \\ (Mean)} & Diff & t-stats & p-value & \makecell{{\it Offline-only} \\ (Mean)} \\ \midrule
\multicolumn{7}{c}{Panel A: Post-online-adoption Monthly Purchase Behaviors} \\ \midrule
\textit{AvgSpendPerMonth} & 696.00 & 524.96 & 171.04 & 9.42 & 0.00 & 284.45 \\
\textit{AvgQuantitiesPerMonth} & 5.22 & 3.58 & 1.64 & 7.70 & 0.00 & 2.18 \\
\textit{AvgOrdersPerMonth} & 1.36 & 1.02 & 0.34 & 12.57 & 0.00 & 0.59 \\
\textit{AvgUniqueItemsPerMonth} & 2.75 & 2.05 & 0.69 & 10.81 & 0.00 & 1.36 \\
\textit{AvgUniqueBrandsPerMonth} & 2.16 & 1.65 & 0.51 & 11.62 & 0.00 & 1.09 \\
\textit{AvgUniqueSubcategoriesPerMonth} & 1.84 & 1.42 & 0.41 & 12.33 & 0.00 & 0.96 \\
\textit{AvgUniqueCategoriesPerMonth} & 1.54 & 1.22 & 0.32 & 12.27 & 0.00 & 0.83 \\
\textit{AvgProfitPerMonth} & 223.51 & 180.19 & 43.32 & 6.86 & 0.00 & 110.25 \\
\bottomrule
\end{tabular}
}
}
\end{table}

\begin{table}[htp!]
    \centering
\caption{Customer-Level Summary Statistics for All Three Groups -- {\it Loyalty Program Adopters}, \textit{Organic Adopters}, and \textit{Offline-only Customers} for post-online-adoption period (i.e., between October 1st, 2023 and March 31st, 2024). }
\label{tab:loyalty_post_adoption_quartile}
 \resizebox{\textwidth}{!}{
\footnotesize{
\begin{tabular}{lrrrrrrrl}
\toprule
Variables & Mean & Std. & Min & 25\% & 50\% & 75\% & Max & Count \\
\midrule
\multicolumn{9}{c}{{\it Loyalty Program Adopters}} \\ \midrule
\textit{AvgSpendPerMonth} & 696.00 & 810.94 & 0.00 & 210.82 & 452.64 & 876.14 & 9712.41 & 1,567 \\
\textit{AvgQuantitiesPerMonth} & 5.22 & 11.23 & 0.00 & 1.00 & 2.67 & 5.45 & 262.33 & 1,567 \\
\textit{AvgOrdersPerMonth} & 1.36 & 1.31 & 0.00 & 0.50 & 1.00 & 1.83 & 20.50 & 1,567 \\
\textit{AvgUniqueItemsPerMonth} & 2.75 & 2.89 & 0.00 & 1.00 & 2.00 & 3.67 & 43.33 & 1,567 \\
\textit{AvgUniqueBrandsPerMonth} & 2.16 & 1.97 & 0.00 & 0.83 & 1.67 & 2.83 & 27.67 & 1,567 \\
\textit{AvgUniqueSubcategoriesPerMonth} & 1.84 & 1.48 & 0.00 & 0.83 & 1.50 & 2.67 & 15.17 & 1,567 \\
\textit{AvgUniqueCategoriesPerMonth} & 1.54 & 1.11 & 0.00 & 0.67 & 1.33 & 2.17 & 9.00 & 1,567 \\
\textit{AvgProfitPerMonth} & 223.51 & 272.37 & -44.82 & 66.43 & 144.49 & 280.82 & 3800.20 & 1,567 \\
\midrule
\multicolumn{9}{c}{ {\it Organic Adopters}} \\ \midrule
\textit{AvgSpendPerMonth} & 524.96 & 674.20 & 0.00 & 132.98 & 336.48 & 678.11 & 21482.62 & 16,746 \\
\textit{AvgQuantitiesPerMonth} & 3.58 & 7.69 & 0.00 & 0.67 & 1.67 & 3.83 & 358.33 & 16,746 \\
\textit{AvgOrdersPerMonth} & 1.02 & 0.99 & 0.00 & 0.33 & 0.83 & 1.33 & 15.17 & 16,746 \\
\textit{AvgUniqueItemsPerMonth} & 2.05 & 2.38 & 0.00 & 0.67 & 1.33 & 2.67 & 41.67 & 16,746 \\
\textit{AvgUniqueBrandsPerMonth} & 1.65 & 1.64 & 0.00 & 0.50 & 1.17 & 2.17 & 22.33 & 16,746 \\
\textit{AvgUniqueSubcategoriesPerMonth} & 1.42 & 1.25 & 0.00 & 0.50 & 1.17 & 2.00 & 14.33 & 16,746 \\
\textit{AvgUniqueCategoriesPerMonth} & 1.22 & 0.97 & 0.00 & 0.50 & 1.00 & 1.67 & 9.00 & 16,746 \\
\textit{AvgProfitPerMonth} & 180.19 & 235.65 & -1078.75 & 45.07 & 114.10 & 233.52 & 7693.62 & 16,746 \\
\midrule
\multicolumn{9}{c}{ {\it Offline-only} Customers} \\ \midrule
\textit{AvgSpendPerMonth} & 284.45 & 484.92 & 0.00 & 36.70 & 146.34 & 356.93 & 37203.48 & 396,396 \\
\textit{AvgQuantitiesPerMonth} & 2.18 & 5.39 & 0.00 & 0.33 & 1.00 & 2.17 & 859.33 & 396,396 \\
\textit{AvgOrdersPerMonth} & 0.59 & 0.69 & 0.00 & 0.17 & 0.40 & 0.83 & 50.67 & 396,396 \\
\textit{AvgUniqueItemsPerMonth} & 1.36 & 1.90 & 0.00 & 0.33 & 0.83 & 1.80 & 114.83 & 396,396 \\
\textit{AvgUniqueBrandsPerMonth} & 1.09 & 1.29 & 0.00 & 0.25 & 0.67 & 1.50 & 56.00 & 396,396 \\
\textit{AvgUniqueSubcategoriesPerMonth} & 0.96 & 1.02 & 0.00 & 0.20 & 0.67 & 1.33 & 22.67 & 396,396 \\
\textit{AvgUniqueCategoriesPerMonth} & 0.83 & 0.82 & 0.00 & 0.17 & 0.67 & 1.17 & 10.67 & 396,396 \\
\textit{AvgProfitPerMonth} & 110.25 & 180.70 & -690.21 & 14.02 & 56.76 & 138.84 & 11025.70 & 396,396 \\
\midrule
\bottomrule
\end{tabular}
}
}
\end{table}

\clearpage

\section{Appendix to DID results for Adopters vs. Offline-only customers}

\subsection{Loyalty Program Control Variables}
\label{appssec:control_variable_def}

Below, we provide the definition of time-varying control variables related to customers' participation in the loyalty program. These variables ($\boldsymbol{x}_{it}$) are included in the DID and event study analysis for the loyalty program study.

\squishlist
    \item $isActivated_{it}$: 1 if customer i activates the offer during month $t$ and 0 otherwise. Use to compare within-activators, how much spend differs between months with/without activation.

\item $isFirstActivation_{it}$: A binary variable equal to 1 if customer $i$ activates the offer for the first time in month $t$, and 0 otherwise. This variable isolates the novelty (first-time) in the month of initial activation, relative to subsequent activation months. 

\item $Months\_Since\_First\_Activation_{it}$: A non-negative integer indicating the number of months elapsed since customer $i$'s first activation of the offer, as of month $t$; equals 0 for months prior to the first activation. For example, if the first activation occurs in August 2023, then \\$Months\_Since\_First\_Activation_{i,September\_2023} = 1$, and so on.

\item $\%Month\_WithActivation_{it}$: The proportion of months in which customer $i$ has activated the offer up to (but not including) month $t$, calculated as the total number of activation months up to $t$ divided by the number of months since first activation (excluding $t$ itself). For the first month of activation, this variable is defined as 0 (since the denominator is 0); all first-month effects are captured by $isFirstActivation_{it}$. Starting from the second month post-activation, this variable reflects the historical activation rate prior to month $t$.

\squishend

\subsection{Standard DiD Estimation Results -- \textit{Adopters} vs. \textit{Offline-only} customers}
\label{appssec:two_by_two_did_adopter_nonadopter}

In the main text, we only considered the TWFE model. We now present the estimation results for the standard two-by-two DID model specified below:
\begin{equation}
\begin{aligned}
    y_{it} & = \alpha_{0} + \alpha_{1}\mathbbm{I}\{Adopter_i\}  +\alpha_{2} \mathbbm{I}\{Post_{t}\} 
+ \alpha_{3}\mathbbm{I}\{Adopter_{i}\} \times \mathbbm{I}\{Post_{t}\} + \boldsymbol{{\alpha}}_{4}' \boldsymbol{{x}}_{it} + \epsilon_{it}
\end{aligned}
\end{equation} 

Table \ref{tab:did_spend_standard} and \ref{tab:did_profit_standard} are estimation results for the standard two-by-two DID model for spend and profit between \textit{adopters} vs. \textit{offline-only} customers, respectively. 

\begin{table}[htp!]
\caption{DID Results — Spend (\textit{Adopters} vs. \textit{Offline-only}) (Standard 2$\times$2 DID)}
\centering
\label{tab:did_spend_standard}
\footnotesize{
\begin{tabular}{lccc}
\midrule\midrule
Study: & COVID-19 & Black Friday & Loyalty Program \\
\midrule
DV: & \multicolumn{3}{c}{Spend} \\
Model: & (1) & (2) & (3) \\
\midrule
\emph{Variables} \\

Constant 
& 280.1 (0.557) 
& 291.7 (0.586) 
& 332.6 (0.800) \\
& [0.000] & [0.000] & [0.000] \\

Adopter 
& 177.2 (2.885) 
& 177.8 (9.177) 
& 123.4 (4.42) \\
& [0.000] & [0.000] & [0.000] \\

Post 
& -62.90 (0.496) 
& -96.95 (0.426) 
& -51.36 (0.632) \\
& [0.000] & [0.000] & [0.000] \\

Adopter $\times$ Post 
& 140.4 (2.843) 
& 81.29 (7.543) 
& 10.02 (5.191) \\
& [0.000] & [0.000] & [0.053] \\

\midrule
Loyalty program controls &  &  & Yes \\

\midrule
\emph{Fit statistics} \\
Observations & 22,204,608 & 9,505,999 & 4,403,449 \\
R$^2$ & 0.01905 & 0.00980 & 0.00918 \\
\midrule\midrule
\multicolumn{4}{l}{\emph{Clustered (Customer) standard errors in parentheses; p-values in brackets.}}
\end{tabular}
}
\end{table}
\begin{table}[H]
\caption{DID Results — Profit (\textit{Adopters} vs. \textit{Offline-only}) (Standard 2$\times$2 DID)}
\centering
\label{tab:did_profit_standard}
\footnotesize{
\begin{tabular}{lccc}
\midrule\midrule
Study: & COVID-19 & Black Friday & Loyalty Program \\
\midrule
DV: & \multicolumn{3}{c}{Profit} \\
Model: & (1) & (2) & (3) \\
\midrule
\emph{Variables} \\

Constant 
& 92.57 (0.177) 
& 96.13 (0.187) 
& 125.8 (0.288) \\
& [0.000] & [0.000] & [0.000] \\

Adopter 
& 61.35 (0.925) 
& 60.94 (3.041) 
& 41.57 (1.607) \\
& [0.000] & [0.000] & [0.000] \\

Post 
& -10.56 (0.174) 
& -22.83 (0.152) 
& -16.81 (0.238) \\
& [0.000] & [0.000] & [0.000] \\

Adopter $\times$ Post 
& 35.87 (0.974) 
& 19.61 (2.547) 
& -7.231 (1.820) \\
& [0.000] & [0.000] & [0.000] \\

\midrule
Loyalty program controls &  &  & Yes \\

\midrule
\emph{Fit statistics} \\
Observations & 22,204,608 & 9,505,999 & 4,403,449 \\
R$^2$ & 0.01228 & 0.00450 & 0.00651 \\

\midrule\midrule
\multicolumn{4}{l}{\emph{Clustered (Customer) standard errors in parentheses; p-values in brackets.}}
\end{tabular}
}
\end{table}

\section{Appendix to DID Results for {\it Event Adopters} vs. {\it Organic Adopters}}
\label{appsec:additional_did}


\subsection{Appendix to DID results for spend}
\label{appssec:two_by_two_did_event_adopter_spend}

We now run the standard two-by-two DID as follows:
\begin{equation}
y_{it} = \delta_0  + \delta_1 \mathbbm{1}\{Event\_Adopter_{i}\} + \delta_2 \mathbbm{1}\{Post_{t}\} + \delta_3 \mathbbm{1}\{Event\_Adopter_{i}\}\times  \mathbbm{1}\{Post_{t}\} + \boldsymbol{{\delta_{4}}}' \boldsymbol{{x}}_{it} + \epsilon_{it}
\end{equation}
Table \ref{tab:did_spend_standard_event_adopter} shows the estimation results for the standard two-by-two DID model for spend between {\it event Adopters} vs. {\it organic Adopters}. 

\begin{table}[htp!]
   \caption{DID Main Analysis -- Spend ({\it Event Adopters} vs. {\it Organic Adopters}) (Standard 2$\times$2 DID)}
   \label{tab:did_spend_standard_event_adopter}
   \centering
   \footnotesize{
   \begin{tabular}{lccc}
       \midrule \midrule
Study: & COVID-19 & Black Friday & Loyalty Program\\
\midrule
DV: & \multicolumn{3}{c}{Spend}\\
Model: & (1) & (2) & (3)\\  
\midrule
\emph{Variables}\\
Organic\_Adopter
    & 432.8 (5.325) & 459.1 (10.48) & 444.7 (4.415) \\
    & [0.000] & [0.000] & [0.000] \\

Event\_Adopter
    & 32.84 (6.281) & 32.01 (20.72) & 131.7 (19.08) \\
    & [0.000] & [0.122] & [0.000] \\

Post
    & 85.85 (5.34) & 10.54 (9.334) & -29.92 (5.041) \\
    & [0.000] & [0.257] & [0.000] \\

Event\_Adopter $\times$ Post
    & -10.64 (6.271) & -82.25 (15.47) & -254.7 (20.82) \\
    & [0.089] & [0.000] & [0.000] \\
\midrule
Loyalty program controls
    &  &  & Yes \\
\midrule
\emph{Fit statistics}\\
Observations
    & 1,783,949 & 76,842 & 197,658 \\
R$^2$
    & 0.00131 & 0.00081 & 0.05635 \\
\midrule \midrule
\multicolumn{4}{l}{\emph{Clustered (Customer) standard errors in parentheses; p-values in brackets.}}\\
   \end{tabular}
   }
\end{table}

\subsection*{Discussion of Forward buying}


Now, we discuss how to set up the event study regression for testing the forward buying under Black Friday and loyalty program analysis.  For the Black Friday analysis, we define event time 0 as November 2019, given that Black Friday occurred on 2019-11-29. The pre-adoption period is from 2019-01-01 to 2019-10-31 (event time -10 to -1), and the post-adoption period is from 2019-12-01 to 2020-10-31 (event time 1 to 11). By estimating the event study regression in Equation \eqref{eq:event_study}, we present the estimation result in Table \ref{tab:1_BlackFriday_Event_Study} and show that \textit{Black Friday} adopters spend significantly higher than \textit{organic adopters} during the adoption month, but spend significantly lower during the post-adoption periods, which indicates a pattern of forward buying. 

\begin{table}[htp!]
   \caption{Event Study for Spend -- \textit{Black Friday Adopters} vs. \textit{Organic Adopters}}
   \label{tab:1_BlackFriday_Event_Study}
   \centering
   \footnotesize{
   \begin{tabular}{lc}
\midrule\midrule
DV:                            & Spend       \\  
Model:                                         & (1)         \\  
\midrule
\emph{Variables}\\
BlackFriday\_Adopter $\times$ Lag –10          & –39.54 (50.73)  \\
                                               & [0.436]         \\
BlackFriday\_Adopter $\times$ Lag –9           & –44.26 (38.91)  \\
                                               & [0.255]         \\
BlackFriday\_Adopter $\times$ Lag –8           & –26.81 (35.61)  \\
                                               & [0.452]         \\
BlackFriday\_Adopter $\times$ Lag –7           & –25.93 (30.61)  \\
                                               & [0.397]         \\
BlackFriday\_Adopter $\times$ Lag –6           &   5.206 (34.45) \\
                                               & [0.880]         \\
BlackFriday\_Adopter $\times$ Lag –5           &  –4.364 (27.55) \\
                                               & [0.874]         \\
BlackFriday\_Adopter $\times$ Lag –4           & –18.65 (27.25)  \\
                                               & [0.494]         \\
BlackFriday\_Adopter $\times$ Lag –3           &  –9.695 (25.51) \\
                                               & [0.704]         \\
BlackFriday\_Adopter $\times$ Lag –2           & –38.55 (23.94)  \\
                                               & [0.107]         \\
BlackFriday\_Adopter $\times$ Adoption Month   & 128.6 (32.46)   \\
                                               & [0.000]         \\
BlackFriday\_Adopter $\times$ Lead 1           & –134.3 (25.36)  \\
                                               & [0.000]         \\
BlackFriday\_Adopter $\times$ Lead 2           & –114.4 (25.65)  \\
                                               & [0.000]         \\
BlackFriday\_Adopter $\times$ Lead 3           & –105.5 (25.58)  \\
                                               & [0.000]         \\
BlackFriday\_Adopter $\times$ Lead 4           &  –88.21 (28.63) \\
                                               & [0.002]         \\
BlackFriday\_Adopter $\times$ Lead 5           &  –55.82 (28.28) \\
                                               & [0.048]         \\
BlackFriday\_Adopter $\times$ Lead 6           & –126.9 (27.95)  \\
                                               & [0.000]         \\
BlackFriday\_Adopter $\times$ Lead 7           &  –88.91 (28.21) \\
                                               & [0.002]         \\
BlackFriday\_Adopter $\times$ Lead 8           &  –69.56 (27.24) \\
                                               & [0.011]         \\
BlackFriday\_Adopter $\times$ Lead 9           &  –25.79 (28.13) \\
                                               & [0.359]         \\
BlackFriday\_Adopter $\times$ Lead 10          &  –90.03 (28.42) \\
                                               & [0.002]         \\
BlackFriday\_Adopter $\times$ Lead 11          &  –67.00 (28.53) \\
                                               & [0.019]         \\
\midrule
\emph{Fixed-effects}\\
Customer                                       & Yes            \\  
YearMonth                                      & Yes            \\  
\midrule
\emph{Fit statistics}\\
Observations                                   & 81,283         \\  
R$^2$                                          & 0.42086        \\  
Within R$^2$                                   & 0.00211        \\  
\midrule\midrule
      \multicolumn{2}{l}{\emph{Clustered (Customer) standard-errors are presented in standard brackets,}}\\
            \multicolumn{2}{l}{\emph{while p‑values are shown in square brackets.}}\\
   \end{tabular}
   }
\end{table}

Regarding the loyalty program analysis, for both \textit{Loyalty Program adopters} and \textit{organic adopters}, the month of adoption is labeled as event time 0. For example, event time 0 is June 2023 for \textit{organic adopters} who adopted online shopping in that month, and September 2023 for \textit{Loyalty Program adopters} who adopted online shopping in that month. 

We compare each group’s spending at event time 0 to their most recent pre-adoption outcome, specifically those observed in May 2023. This period serves as a common baseline for pre-treatment behaviors across all adopter groups. As defined in \S\ref{sssec:loyalty_sample}, the post-adoption periods span from October 2023 and March 2024, and we label October 2023 as the event time 1 for both \textit{organic} and \textit{Loyalty Program adopters} and label all the following months until March 2024 as event time 6. The event study coefficient at time 1 captures the difference in the monthly spend at October 2023 between two groups relative to the pre-adoption difference in the reference period May 2023. 

We first compare the spend of \textit{organic} and \textit{Loyalty Program adopters} during their adoption month relative to the pre-adoption reference period (i.e., spend in May 2023), and Table \ref{tab:Loyalty_Immediate_Adoption_Effect} shows that \textit{Loyalty Program adopters} tend to spend 177.7 MCU more than \textit{organic adopters} during the adoption month. Next, we notice that both \textit{Loyalty Program adopters} and \textit{organic adopters} can activate the monthly loyalty offer during the post-adoption period. To test the post-adoption differences in purchase behavior, we estimate the event study regression in Equation \eqref{eq:event_study} with control variables for the customers' activations. Table \ref{tab:Loyalty_Post_Adoption_Effect} shows that \textit{Loyalty Program adopters} spend significantly less than \textit{organic adopters} in most of the post-adoption periods. 

\begin{table}[htp!]
   \caption{Event Study for Spend during the Adoption Month - \textit{Loyalty Program Adopters} vs. \textit{Organic Adopters}}
   \label{tab:Loyalty_Immediate_Adoption_Effect}
   \centering
   \footnotesize{
\begin{tabular}{lc}
 \midrule \midrule
      DV:                        & Spend\\  
      Model:                                     & (1)\\  
      \midrule
      \emph{Variables}\\
      Loyalty\_Adopter $\times$ Lag -5           & -11.43 (26.52)\\   
                                                 & [0.672]\\   
      Loyalty\_Adopter $\times$ Lag -4           & -29.64 (25.37)\\   
                                                 & [0.245]\\   
      Loyalty\_Adopter $\times$ Lag -3           & 0.7290 (24.62)\\   
                                                 & [0.976]\\   
      Loyalty\_Adopter $\times$ Lag -2           & -25.12 (26.40)\\   
                                                 & [0.340]\\   
      Loyalty\_Adopter $\times$ Adoption Month   & 177.7 (31.54)\\   
                                                 & [0.000]\\   
      \midrule
      \emph{Fixed-effects}\\
      Customer                                   & Yes\\  
      Event\_Time                                & Yes\\  
      \midrule
      \emph{Fit statistics}\\
      Observations                               & 109,857\\  
      R$^2$                                      & 0.53391\\  
      Within R$^2$                               & 0.00111\\  
      \midrule \midrule
      \multicolumn{2}{l}{\emph{Clustered (Customer) standard-errors are presented in standard brackets,}}\\
            \multicolumn{2}{l}{\emph{while p‑values are shown in square brackets.}}\\
   \end{tabular}
   }
\end{table}

\begin{table}[htp!]
   \caption{Event Study for Spend during the Post-Online-Adoption Periods -- \textit{Loyalty Program Adopters} vs. \textit{Organic Adopters}}
   \label{tab:Loyalty_Post_Adoption_Effect}
   \centering
   \footnotesize{
\begin{tabular}{lc}
 \midrule \midrule
      DV:                        & Spend\\  
      Model:                                     & (1)\\  
      \midrule
      \emph{Variables}\\
      Loyalty\_Adopter $\times$ Lag -5           & -11.45 (26.52)\\   
                                                 & [0.670]\\   
      Loyalty\_Adopter $\times$ Lag -4           & -29.64 (25.37)\\   
                                                 & [0.245]\\   
      Loyalty\_Adopter $\times$ Lag -3           & 0.7526 (24.62)\\   
                                                 & [0.975]\\   
      Loyalty\_Adopter $\times$ Lag -2           & -25.10 (26.40)\\   
                                                 & [0.340]\\   
      Loyalty\_Adopter $\times$ Lead 1           & -102.5 (29.56)\\   
                                                 & [0.001]\\   
      Loyalty\_Adopter $\times$ Lead 2           & -34.48 (28.49)\\   
                                                 & [0.224]\\   
      Loyalty\_Adopter $\times$ Lead 3           & -45.26 (31.24)\\   
                                                 & [0.144]\\   
      Loyalty\_Adopter $\times$ Lead 4           & -73.56 (30.45)\\   
                                                 & [0.015]\\   
      Loyalty\_Adopter $\times$ Lead 5           & -69.94 (33.10)\\   
                                                 & [0.037]\\   
      Loyalty\_Adopter $\times$ Lead 6           & -75.00 (35.29)\\   
                                                 & [0.032]\\   
      \midrule
      Loyalty program controls                   & Yes\\  
      \midrule
      \emph{Fixed-effects}\\
      Customer                                   & Yes\\  
      YearMonth                                  & Yes\\  
      \midrule
      \emph{Fit statistics}\\
      Observations                               & 197,658\\  
      R$^2$                                      & 0.47130\\  
      Within R$^2$                               & 0.02699\\  
      \midrule \midrule
      \multicolumn{2}{l}{\emph{Clustered (Customer) standard-errors are presented in standard brackets,}}\\
            \multicolumn{2}{l}{\emph{while p‑values are shown in square brackets.}}\\
   \end{tabular}
   }
\end{table}

\begin{figure}[htp!]
    \centering
    \caption{Event Study for Offline Spend}
    \label{fig:Forward_Buying_OfflineSpend}
        \begin{subfigure}{0.49\textwidth}
        \centering
        \caption{Black Friday}
        \includegraphics[scale=0.30]{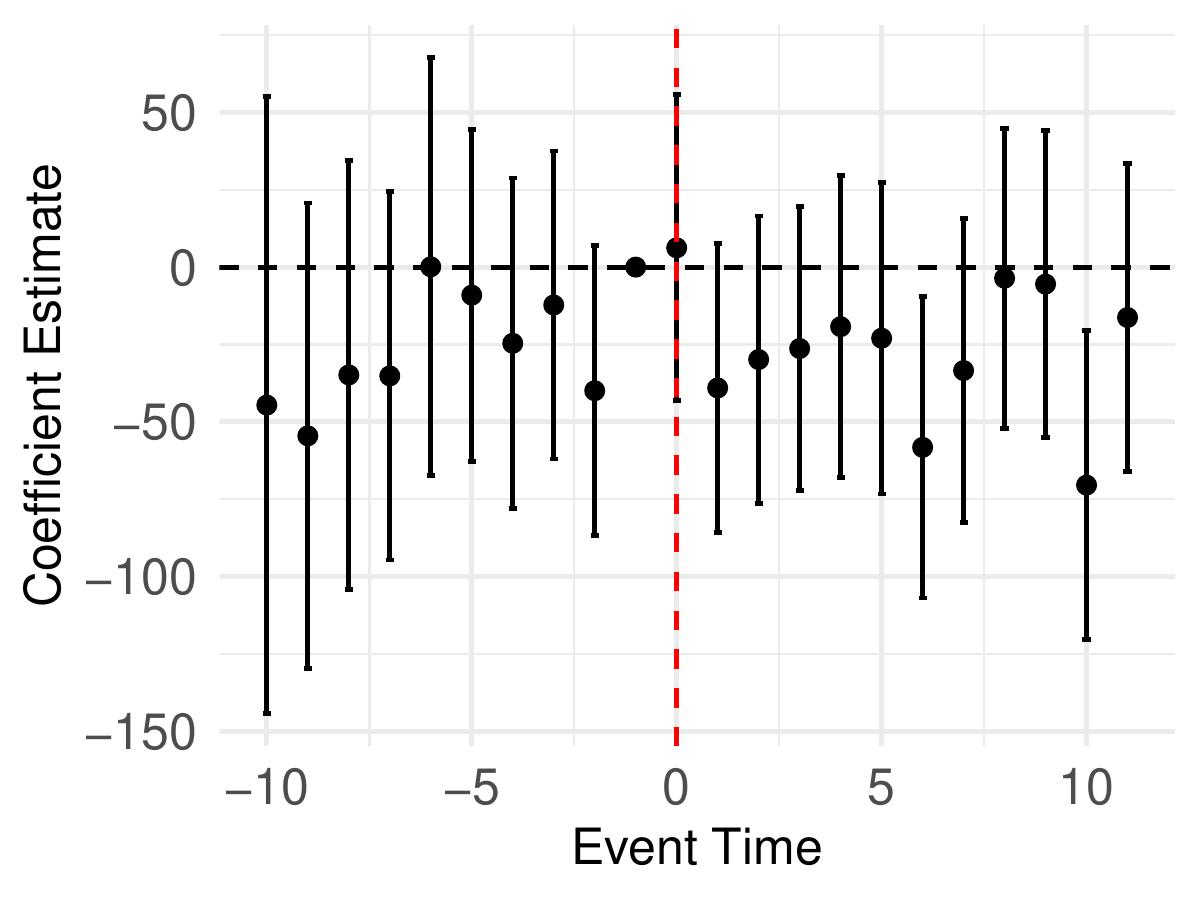}
        \label{fig:Forward_Buying_OfflineSpend_BF}
    \end{subfigure}%
    ~ 
    \begin{subfigure}{0.49\textwidth}
        \centering
        \caption{Loyalty Program}
        \includegraphics[scale=0.30]{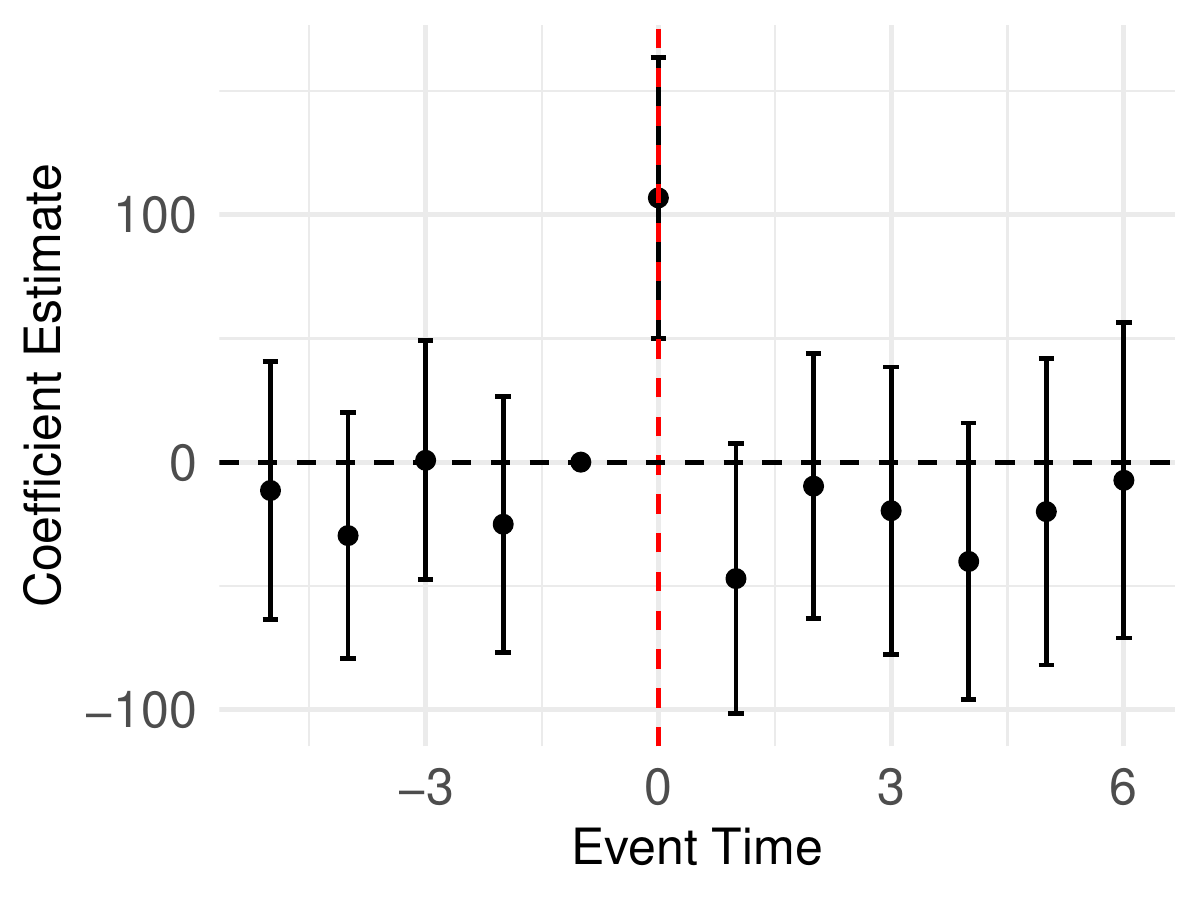}
\label{fig:Forward_Buying_OfflineSpend_Loyalty}
    \end{subfigure}
\end{figure}

Next, we run the event study regression for offline spend, and Figure \ref{fig:Forward_Buying_OfflineSpend} presents the event study figure for the offline spend of Black Friday and loyalty program study. 

\begin{figure}[htp!]
    \centering
    \caption{Event Study for Spend -- \textit{COVID adopters} vs. \textit{Organic Adopters} during the month of adoption}
    \label{fig:Forward_Buying_Covid}
        \centering
        \includegraphics[scale=0.30]{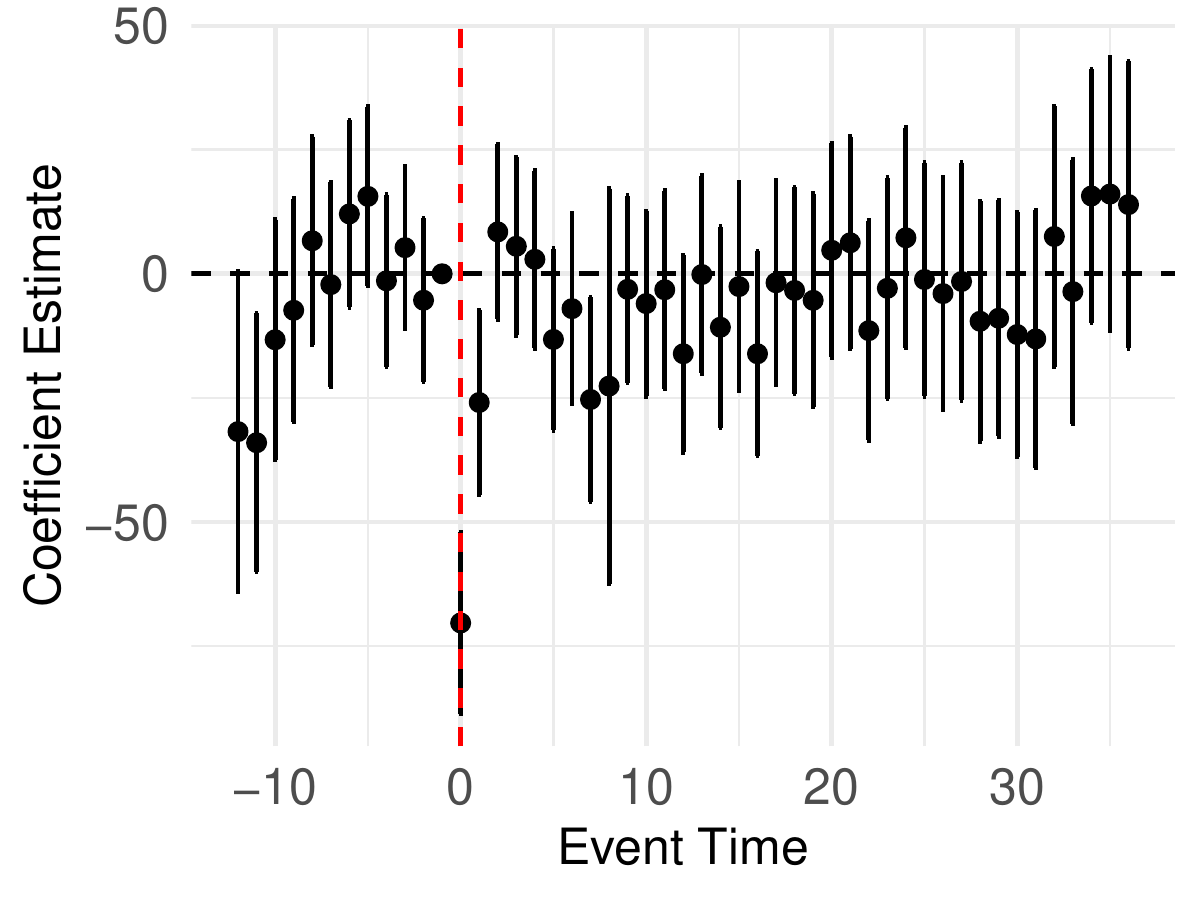}
\end{figure}

Finally, we follow a similar approach described above to test for \textit{COVID adopters}. Figure \ref{fig:Forward_Buying_Covid} shows that there is no forward buying for \textit{COVID adopters} relative to \textit{organic adopters}. 

\clearpage

\subsection{Appendix to DID results for the Share of Offline Spend}
\label{appssec:did_share_offline_spend}

Table \ref{tab:fraction_offline_organic_covid_timeTrend} presents the estimation results for the linear trend model by comparing a type of \textit{event adopters} with \textit{organic adopters}. Table \ref{tab:fraction_offline_organic_covid_timeTrend_byAge} presents the changes in share of offline spend by different age groups, showing senior adopters tend to have a slower rate of shifting their spend to the online channel.

\begin{table}[htp!]
   \caption{Differential Trends in the Share of Offline Spend (Estimation Results for Equation \eqref{eq:linear_trend}) -- {\it Event Adopters} vs. {\it Organic Adopters}}
   \label{tab:fraction_offline_organic_covid_timeTrend}
   \centering
   \footnotesize{
   \begin{tabular}{lccc}
\midrule\midrule
      DV:                       & \multicolumn{3}{c}{Share of Offline Spend}                           \\
Model:                        & COVID-19            & Black Friday      & Loyalty Program   \\  
\midrule
\emph{Variables}\\
Organic\_Adopter                      & 0.4833 (0.0084)  & 0.5930 (0.0285)   & 0.5662 (0.0093)   \\
                              & [0.000]          & [0.000]           & [0.000]           \\[0.5em]
Event\_Adopter                       & 0.0807 (0.0119)  & 0.1424 (0.0404)   & 0.0306 (0.0132)   \\
                              & [0.000]          & [0.000]           & [0.020]           \\[0.5em]
t                             & -0.0020 (0.0004) & -0.0108 (0.0048)  & -0.0020 (0.0031)  \\
                              & [0.000]          & [0.025]           & [0.518]           \\[0.5em]
Event\_Adopter $\times$ t            & 0.0012 (0.0006)  & -0.0109 (0.0068)  & 0.0013 (0.0043)   \\
                              & [0.046]          & [0.109]           & [0.763]           \\  
\midrule
\emph{Fit statistics}\\
Observations                  & 72               & 22                & 12                \\  
R$^2$                         & 0.81922          & 0.69953           & 0.72620           \\  
Adjusted R$^2$                & 0.81124          & 0.64945           & 0.62353           \\  
\midrule\midrule
\multicolumn{4}{l}{\emph{Standard errors in parentheses; p‑values in square brackets.}}
   \end{tabular}
   }
\end{table}

\begin{table}[htp!]
   \caption{Differential Trend in the Share of Offline Spend by Age -- \textit{COVID adopters} vs. \textit{Organic Adopters}}
   \label{tab:fraction_offline_organic_covid_timeTrend_byAge}
   \centering
   \footnotesize{
   \begin{tabular}{lcc}
 \midrule \midrule
      DV: & \multicolumn{2}{c}{Share of Offline Spend}\\
      Model: & (1) Senior Adopters & (2) Young Adopters \\  
      \midrule
      \emph{Variables}\\
      Organic\_Adopter       & 0.4947 (0.0092)   & 0.4728 (0.0081)\\   
                     & [0.000]           & [0.000]\\   
      COVID\_Adopter        & 0.0706 (0.0130)   & 0.0882 (0.0114)\\   
                     & [0.000]           & [0.000]\\   
      t              & -0.0019 (0.0005)  & -0.0022 (0.0004)\\   
                     & [0.001]           & [0.000]\\   
      COVID\_Adopter $\times$ t    & 0.0020 (0.0006)   & 0.0004 (0.0006)\\   
                     & [0.002]           & [0.523]\\   
      \midrule
      \emph{Fit statistics}\\
      Observations   & 72                & 72\\  
      R$^2$          & 0.79964           & 0.82429\\  
      Adjusted R$^2$ & 0.79080           & 0.81654\\  
      \midrule \midrule
      \multicolumn{3}{l}{\emph{Standard errors are shown in parentheses; p-values are shown in square brackets.}}\\
   \end{tabular}
   }
\end{table}

\begin{figure}[H]
    \centering
    \caption{Differential Trend Visualization by Age -- \textit{COVID adopters} vs. \textit{Organic adopters}}
   \label{fig:covid_offline_fraction_purchase_age}
    \begin{subfigure}{0.49\textwidth}
        \centering
        \caption{Senior Adopters (Age $\geq 41$)}
        \includegraphics[scale=0.15]{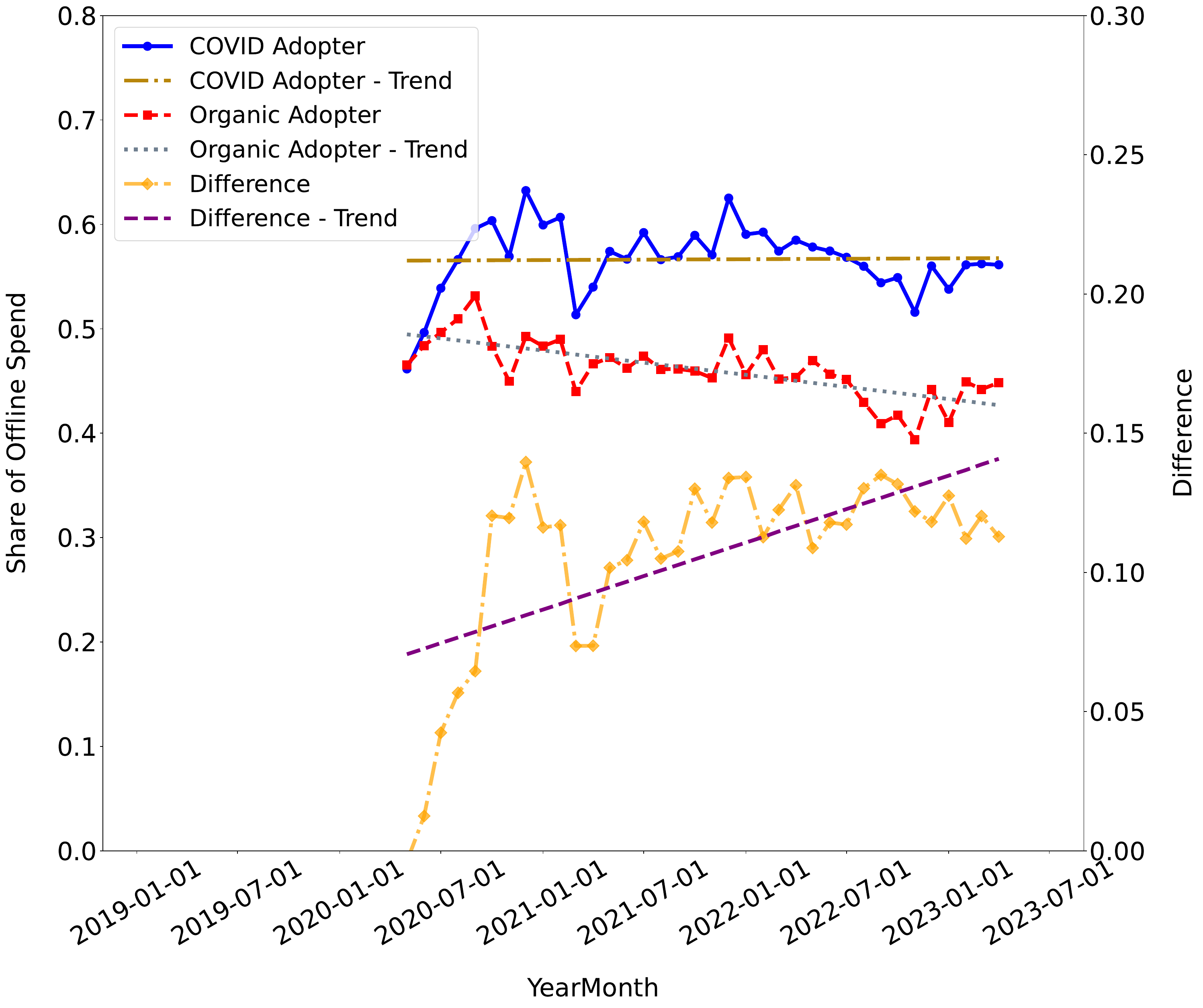}
        \label{fig:_senior_sub1}
    \end{subfigure}%
    ~ 
    \begin{subfigure}{0.49\textwidth}
        \centering
        \caption{Young Adopters (Age $< 41$)}
        \includegraphics[scale=0.15]{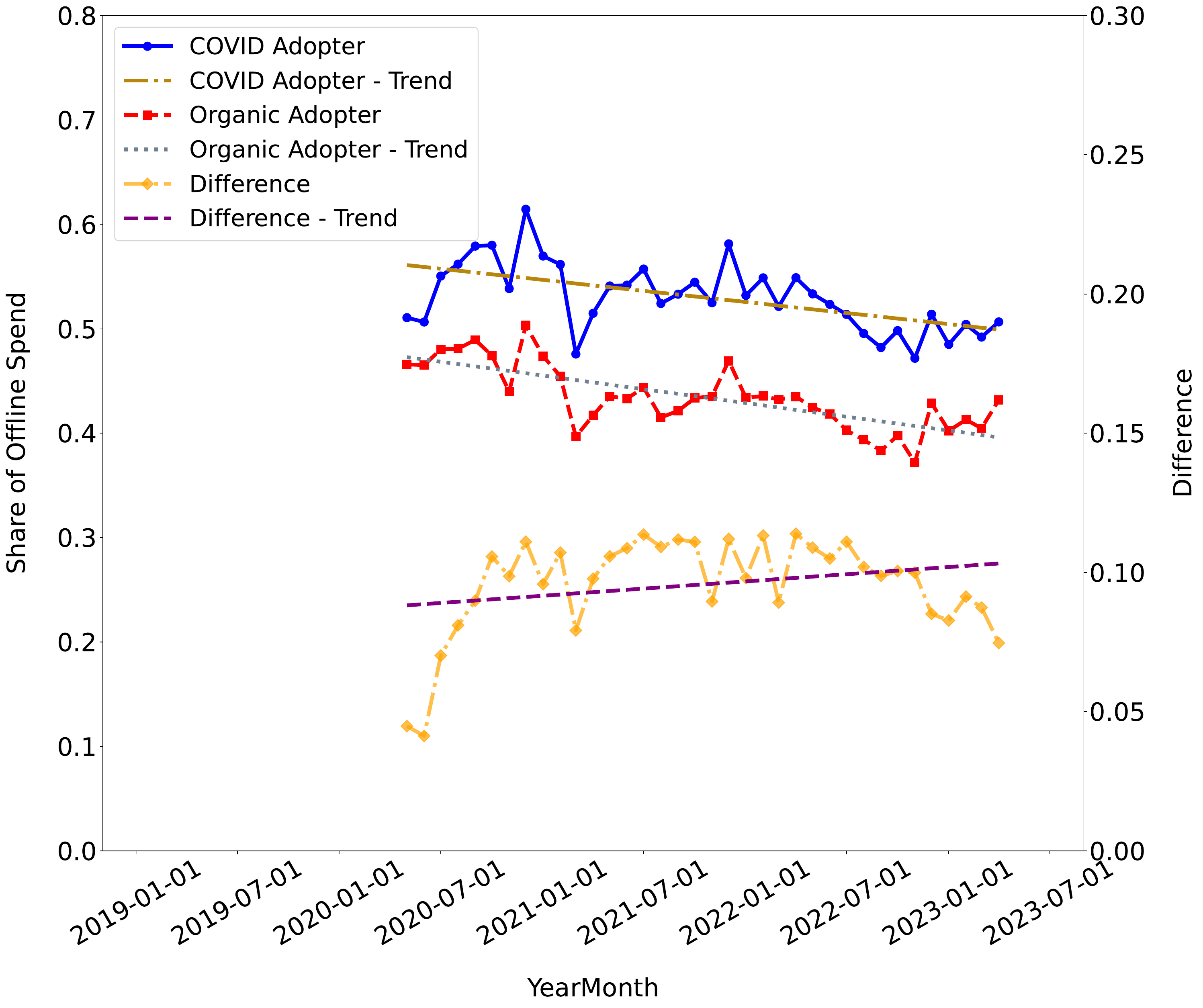}
        \label{fig:non_senior_sub2}
    \end{subfigure}
    \vspace{1ex} 
\end{figure}

In \S\ref{sssec:share}, we find that the share of offline spending is persistently higher for \textit{COVID adopters} compared to {\it organic adopters} in the post-adoption period, and the downward trend is steeper for \textit{organic adopters}, indicating that \textit{COVID adopters} shift their spending to online channels slower than {\it organic adopters}. To formally test whether these differential trends in share of offline spend are systematic, we estimate the following model with linear time trend on the post-adoption monthly panel data for both groups:
\begin{equation}
\label{eq:linear_trend}
    y_{gt} = \zeta + \rho_0 t + \rho_1\mathbbm{I}\{Event\_Adopter_g\} + \rho_2 \mathbbm{I}\{Event\_Adopter_g\}\times t +\epsilon_{gt}.
\end{equation}
Here, $y_{gt}$ is the fraction of the offline spend relative to the total spend (ranging from 0 to 1) for group $g$ in period $t$. $t$ is the numerical time index, and we set the first post-adoption month of each study as $t=0$. The results for COVID-19, Black Friday and loyalty program studies from this regression are shown in Table \ref{tab:fraction_offline_organic_covid_timeTrend}. The estimate of the difference in trend ($\rho_2$) between \textit{COVID adopters} and \textit{organic adopters} is 0.0012 (or 0.12\%). This finding suggests that although the different reasons (i.e., organic vs. exogenous shock) for adopting online shopping do not seem to have much impact on {\it how much} consumers spend with the firm in total, they do cause differences in {\it where} they spend it. Figure \ref{fig:covid_offline_fraction_purchase_age} plots the fraction of offline spending in the post-treatment period separately for older and younger adopters.

\subsection{Appendix to DID results for Profit}
\label{appssec:two_by_two_did_event_adopter_profit}

Table \ref{tab:did_profit_standard_event_adopter} shows the estimation result for the standard two-by-two DID model for the profit between {\it event Adopters} vs. {\it organic Adopters}.

\begin{table}[htp!]
   \caption{DID Main Analysis -- Profit ({\it Event Adopters} vs. {\it Organic Adopters}) (Standard 2$\times$2 DID)}
   \label{tab:did_profit_standard_event_adopter}
   \centering
   \footnotesize{
   \begin{tabular}{lccc}
       \midrule \midrule
Study: & COVID-19 & Black Friday & Loyalty Program\\
\midrule
DV: & \multicolumn{3}{c}{Profit}\\
Model: & (1) & (2) & (3)\\  
\midrule
\emph{Variables}\\
Organic\_Adopter
    & 145.7 (1.698) & 154.7 (3.519) & 164.5 (1.627) \\
    & [0.000] & [0.000] & [0.000] \\

Event\_Adopter
    & 10.96 (2.007) & 7.338 (6.79) & 34.13 (6.413) \\
    & [0.000] & [0.28] & [0.000] \\

Post
    & 20.27 (1.788) & 3.315 (3.163) & -21.05 (1.788) \\
    & [0.000] & [0.295] & [0.000] \\

Event\_Adopter $\times$ Post
    & 7.248 (2.117) & -20.52 (5.228) & -86.83 (7.352) \\
    & [0.000] & [0.000] & [0.000] \\
\midrule
Loyalty program controls
    &  &  & Yes \\
\midrule
\emph{Fit statistics}\\
Observations
    & 1,783,949 & 76,842 & 197,658 \\
R$^2$
    & 0.00134 & 0.00039 & 0.04568 \\
\midrule \midrule
\multicolumn{4}{l}{\emph{Clustered (Customer) standard errors in parentheses; p-values in brackets.}}\\
   \end{tabular}
   }
\end{table}

\clearpage


\section{Appendix to HonestDiD and Linear Trend Estimation}
\label{appsec:honest_did}

\subsection{Parallel Trends for Black Friday Analysis}
\label{appssec:parallel_tests_BlackFriday}

\begin{figure}[htp!]
\centering
\caption{Parallel Trend Assessment -- Black Friday. Panels (\ref{fig:spend_sub1_BF},  \ref{fig:profit_sub1_BF}) plot monthly outcomes, while panels (\ref{fig:spend_sub2_BF},   \ref{fig:profit_sub2_BF}) and (\ref{fig:spend_sub3_BF}, \ref{fig:profit_sub3_BF}) report event study estimates (based on Equation \eqref{eq:event_study}) comparing (i) \textit{adopters} (both \textit{Black Friday} and \textit{organic}) to \textit{offline-only customers} and (ii) \textit{Black Friday} to \textit{organic adopters}, respectively.}
\label{fig:parallel_trend_combined_main_BF}
\begin{subfigure}{0.31\textwidth}
    \centering
        \caption{Visualization -- Spend}
\includegraphics[width=\linewidth]{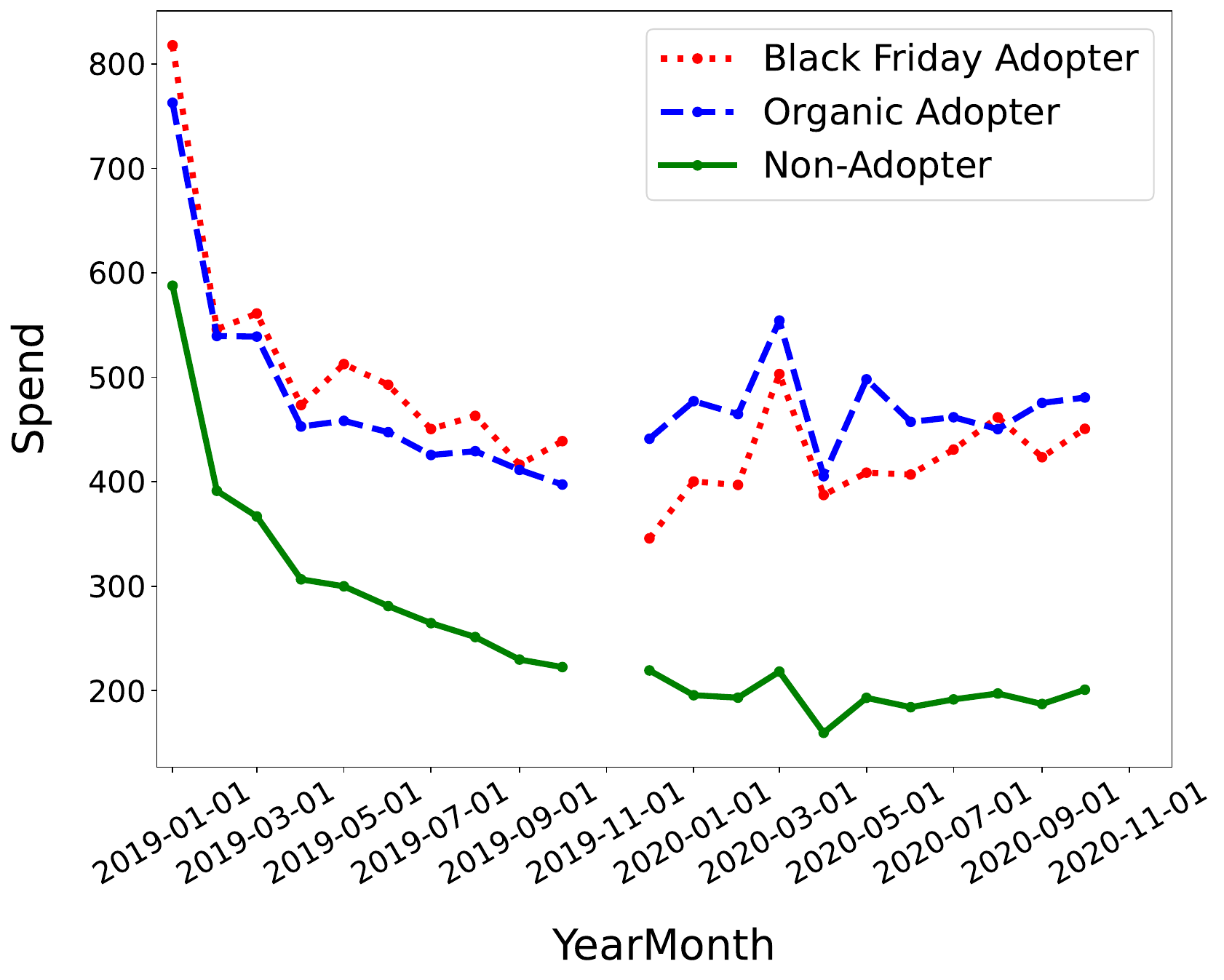}
    \label{fig:spend_sub1_BF}
\end{subfigure}
\hfill
\begin{subfigure}{0.32\textwidth}
    \centering
      \caption{\textit{Adopters} vs. \textit{Offline-only} -- Spend}\includegraphics[width=\linewidth]{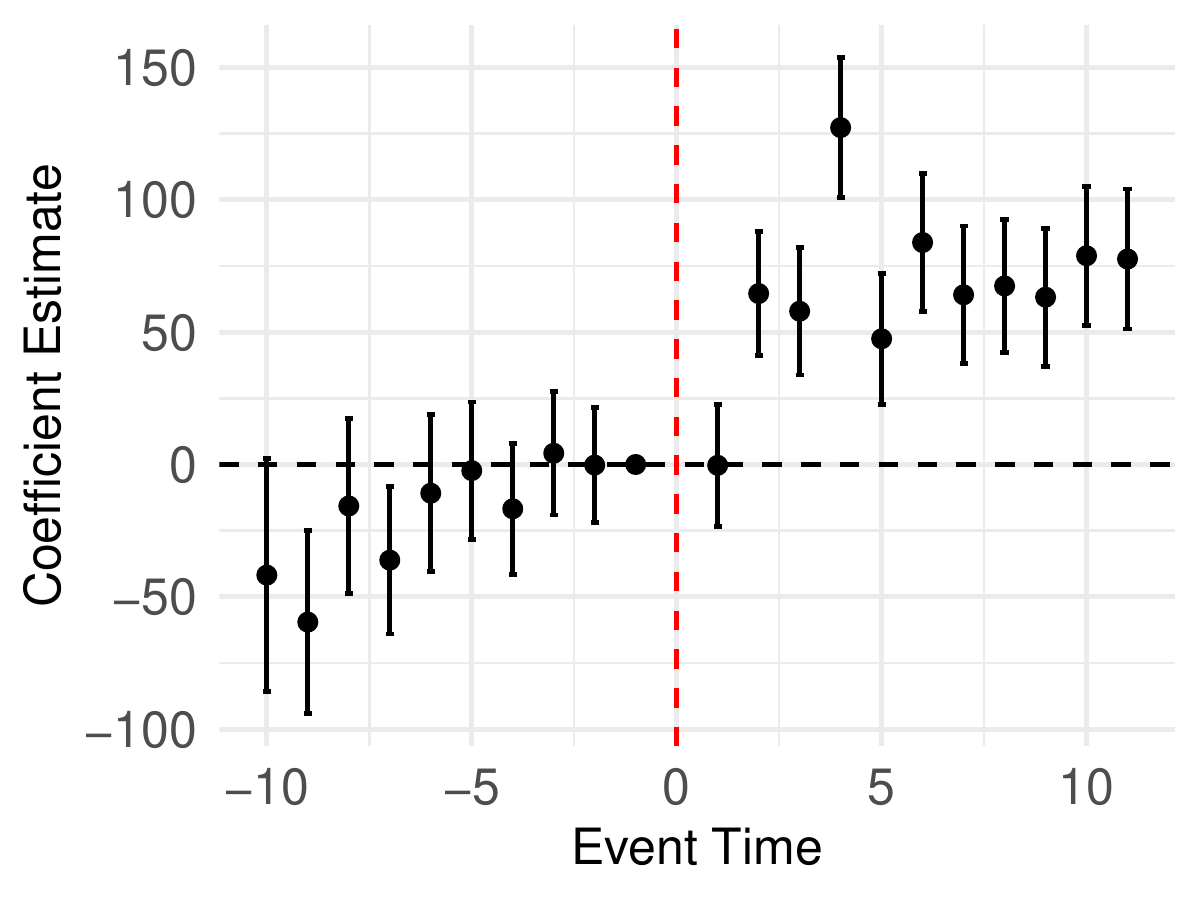}

        \label{fig:spend_sub2_BF}
\end{subfigure}
\hfill
\begin{subfigure}{0.32\textwidth}
    \centering
        \caption{\textit{Black Friday} vs. \textit{Organic Adopters} -- Spend}\includegraphics[width=\linewidth]{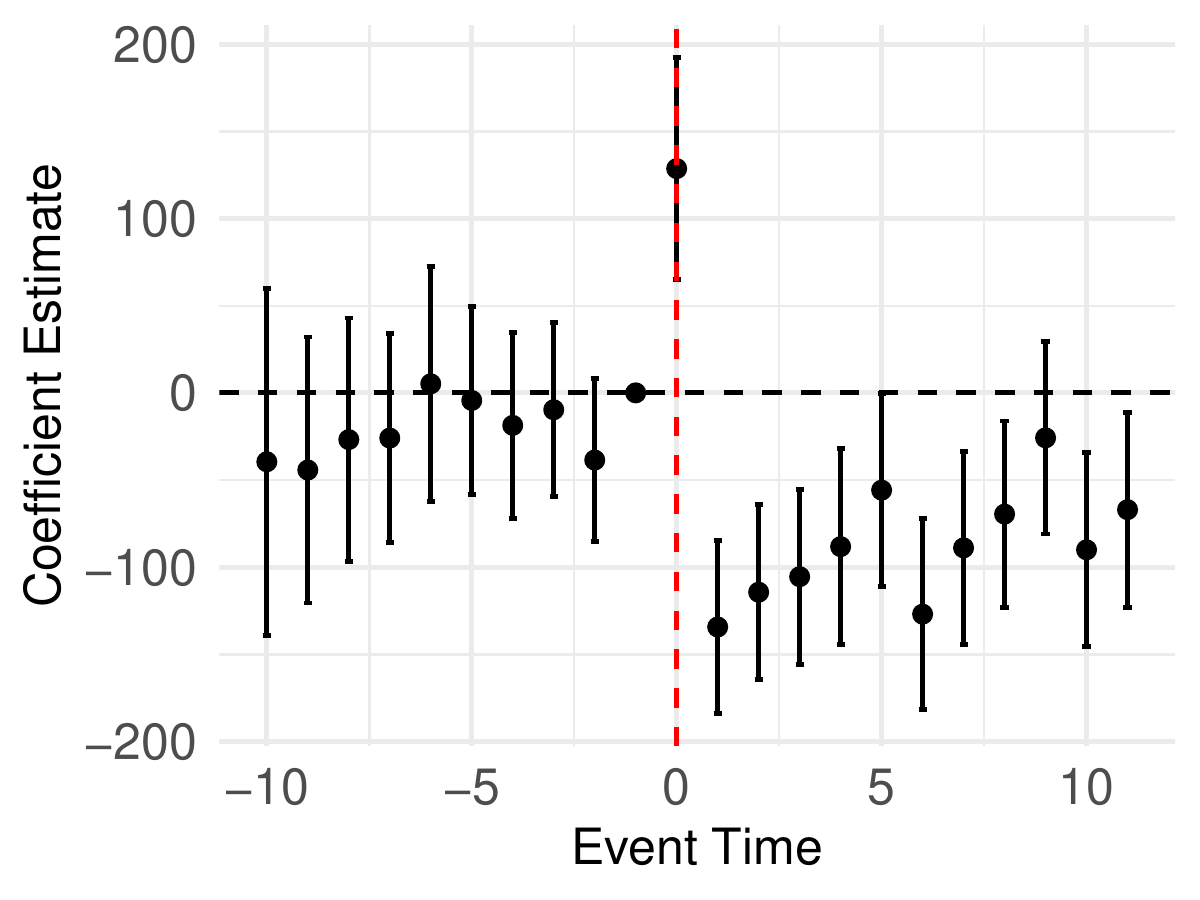}

        \label{fig:spend_sub3_BF}
\end{subfigure}
\vspace{0.5em}
\begin{subfigure}{0.31\textwidth}
    \centering
        \caption{Visualization -- Profit}    \includegraphics[width=\linewidth]{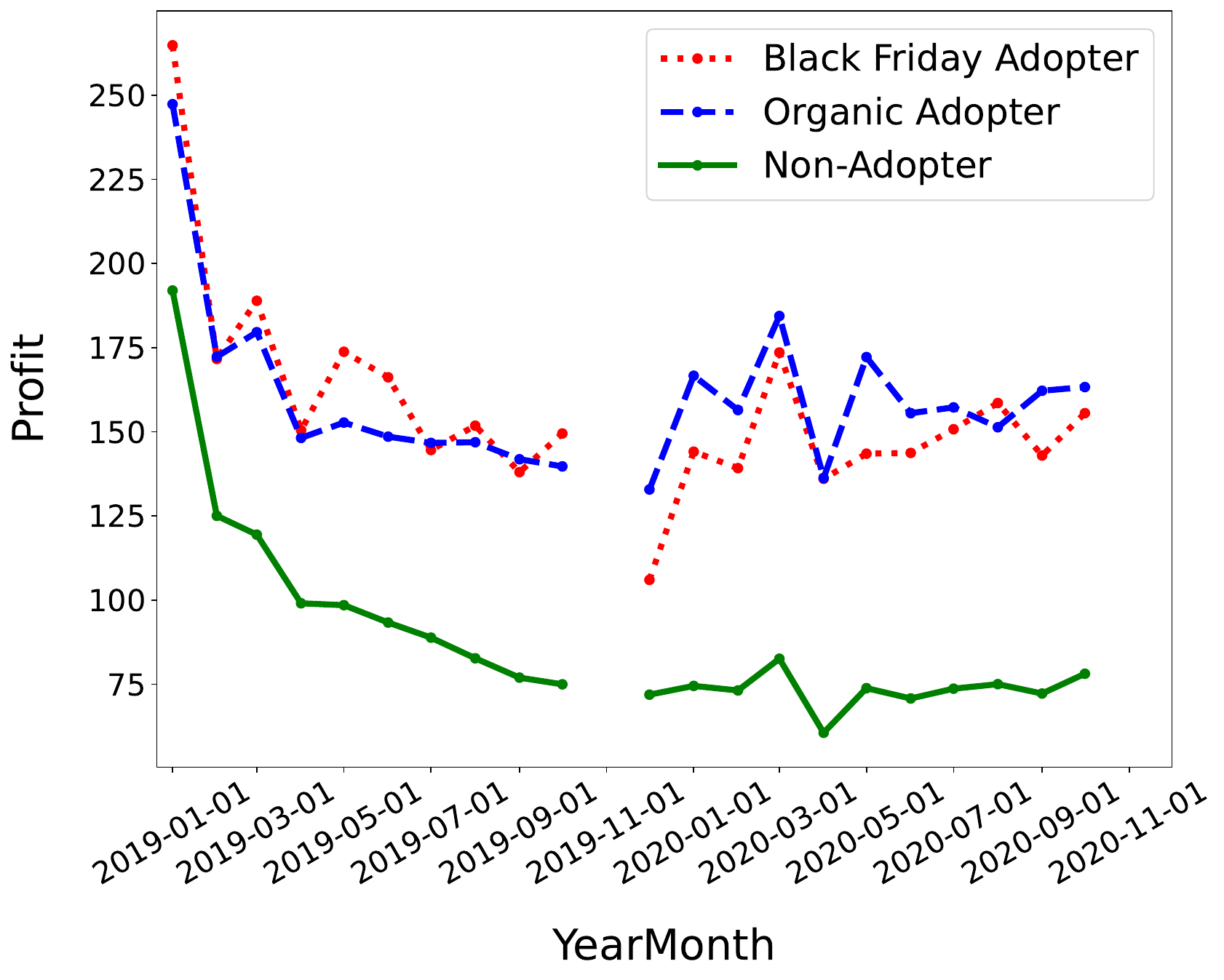}
        \label{fig:profit_sub1_BF}
\end{subfigure}
\hfill
\begin{subfigure}{0.32\textwidth}
    \centering
       \caption{\textit{Adopters} vs. \textit{Offline-only} -- Profit}
\includegraphics[width=\linewidth]{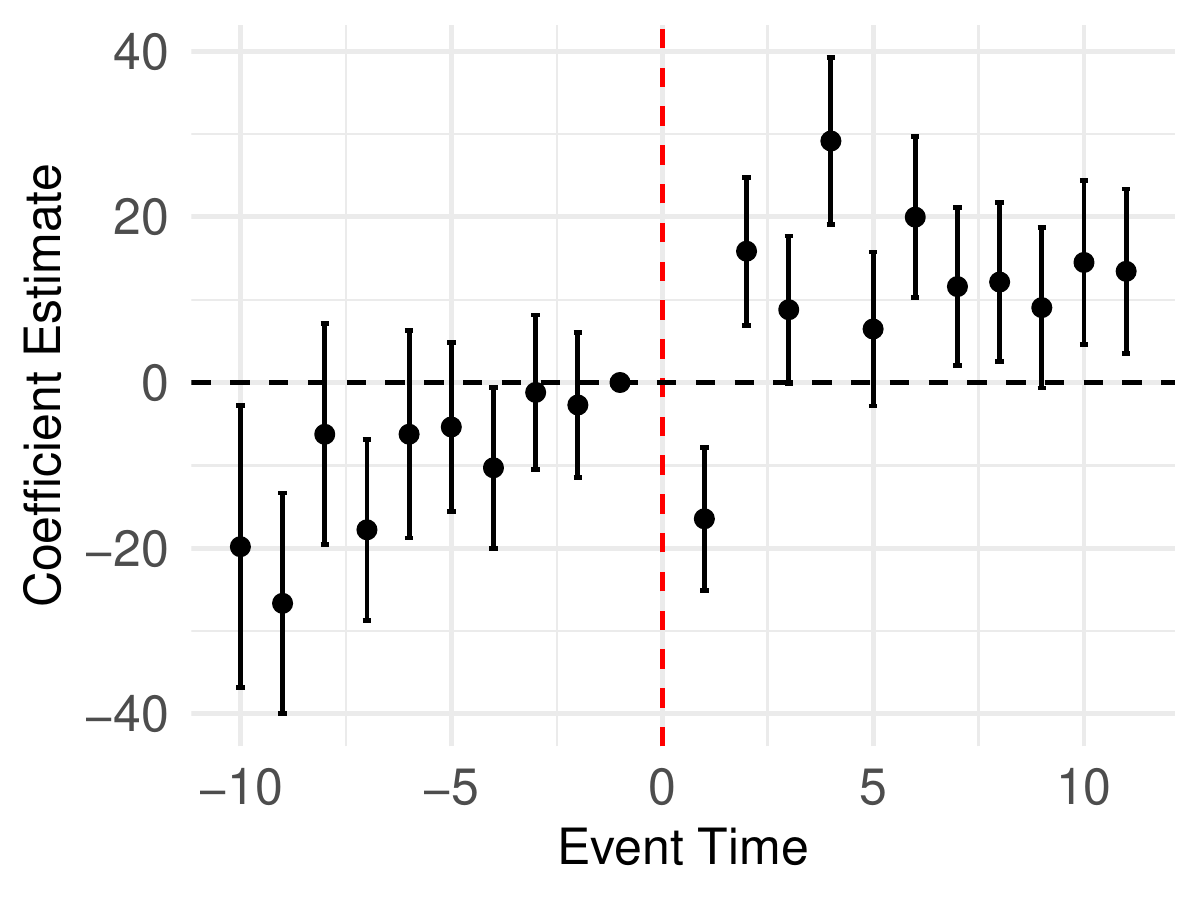}
 \label{fig:profit_sub2_BF}
\end{subfigure}
\hfill
\begin{subfigure}{0.32\textwidth}
    \centering
        \caption{\textit{Black Friday} vs. \textit{Organic Adopters} -- Profit}
        \includegraphics[width=\linewidth]{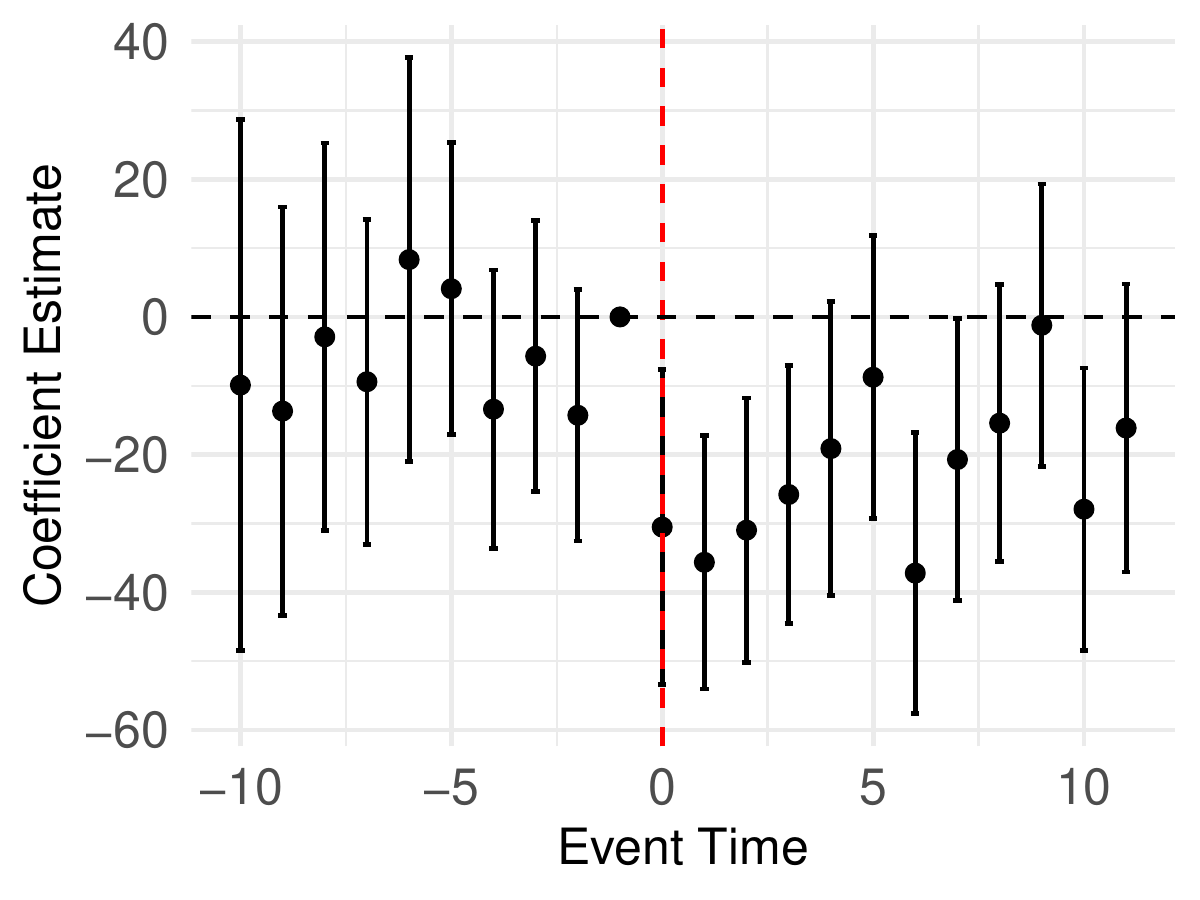}
            \label{fig:profit_sub3_BF}
\end{subfigure}
\end{figure}

Figure \ref{fig:spend_sub1_BF} and \ref{fig:profit_sub1_BF} visualize the monthly average total spend and profit values for \textit{Black Friday adopters}, \textit{organic adopters} and \textit{offline-only} customers. Then, we provide a pre-trend test for two main metrics (i.e., spend and profit) between \textit{adopters} (including both \textit{Black Friday adopters} and \textit{organic adopters}) relative to \textit{offline-only} customers. In total, we have 10 pre-treatment periods (i.e., between January 2019 and October 2019), labeled as -10 to -1, and 11 post-adoption periods (i.e., between December 2019 and October 2020), labeled as 1 to 11. The base period is October 2019, and the event time 0 refers to the adoption month (i.e., November 2019). Figure \ref{fig:spend_sub2_BF} and \ref{fig:profit_sub2_BF} exhibit a minor violation during the early stage of the pre-adoption period, but most pre-adoption coefficients are significant. To show the robustness of our main finding in the presence of violation of parallel pre-trend, we show the bounded ATT in \S\ref{ssec:parallel_tests} and confirm that the incremental spend and profit are still positive and significant in the presence of violation of parallel trend.

Finally, we provide a pre-trend test between \textit{Black Friday adopters} and \textit{organic adopters}. Figure \ref{fig:spend_sub3_BF} and \ref{fig:profit_sub3_BF} presents the event study regression estimates for the parallel pre-trend of the Black Friday study. 

\newpage

\subsection{Parallel Trends for Loyalty Program Analysis}
\label{appssec:parallel_tests_loyalty}

\begin{figure}[htp!]
\centering
\caption{Parallel Trend Assessment -- Loyalty Program. Panels (\ref{fig:spend_sub1_loyalty},  \ref{fig:profit_sub1_loyalty}) plot monthly outcomes, while panels (\ref{fig:spend_sub2_loyalty},   \ref{fig:profit_sub2_loyalty}) and (\ref{fig:spend_sub3_loyalty}, \ref{fig:profit_sub3_loyalty}) report event study estimates (based on Equation \eqref{eq:event_study}) comparing (i) \textit{adopters} (both \textit{Loyalty Program} and \textit{organic}) to \textit{offline-only customers} and (ii) \textit{Loyalty Program} to \textit{organic adopters}, respectively.}
\label{fig:parallel_trend_combined_main_loyalty}
\begin{subfigure}{0.31\textwidth}
    \centering
        \caption{Visualization -- Spend}
\includegraphics[width=\linewidth]{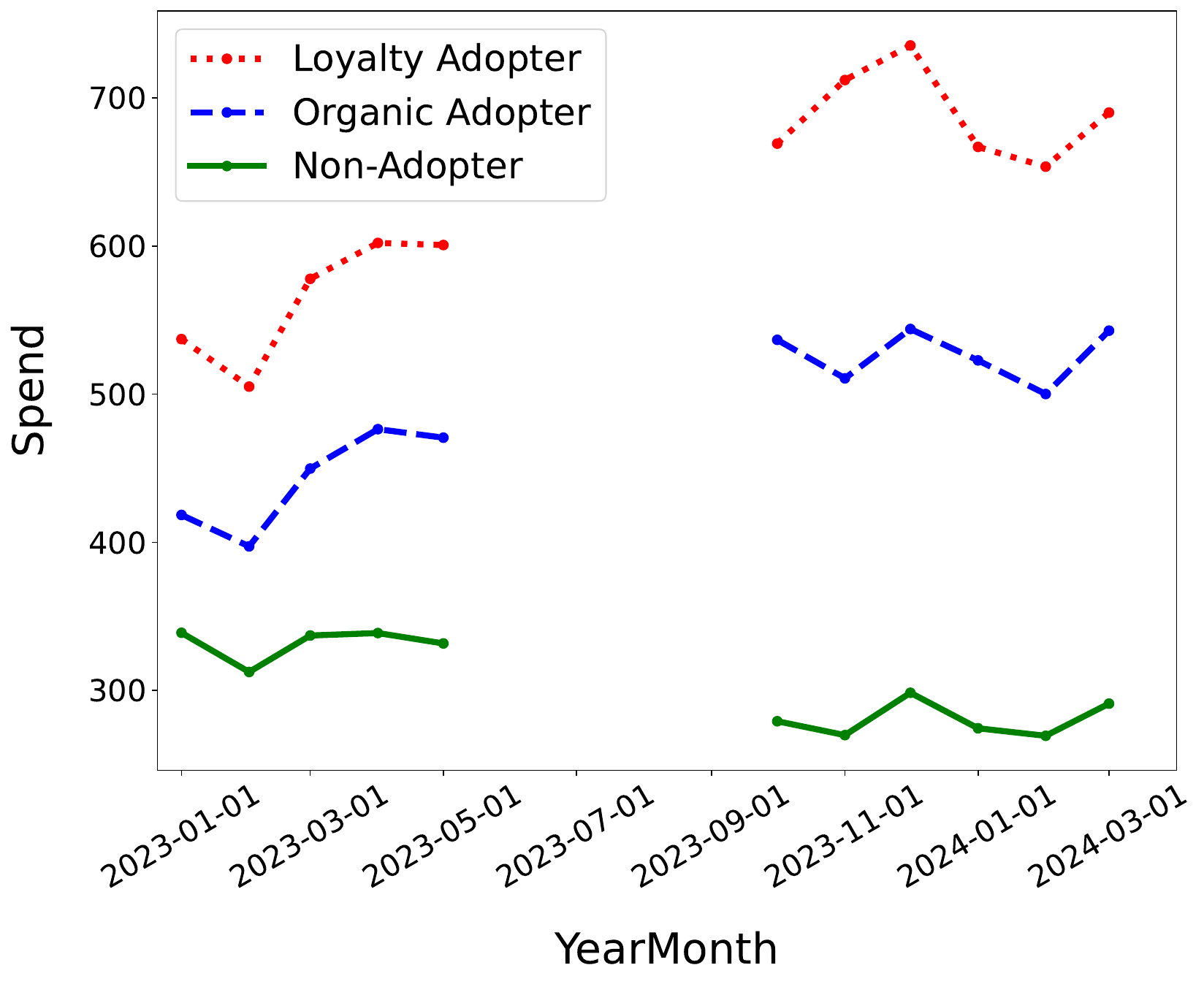}
    \label{fig:spend_sub1_loyalty}
\end{subfigure}
\hfill
\begin{subfigure}{0.32\textwidth}
    \centering
      \caption{\textit{Adopters} vs. \textit{Offline-only} -- Spend}\includegraphics[width=\linewidth]{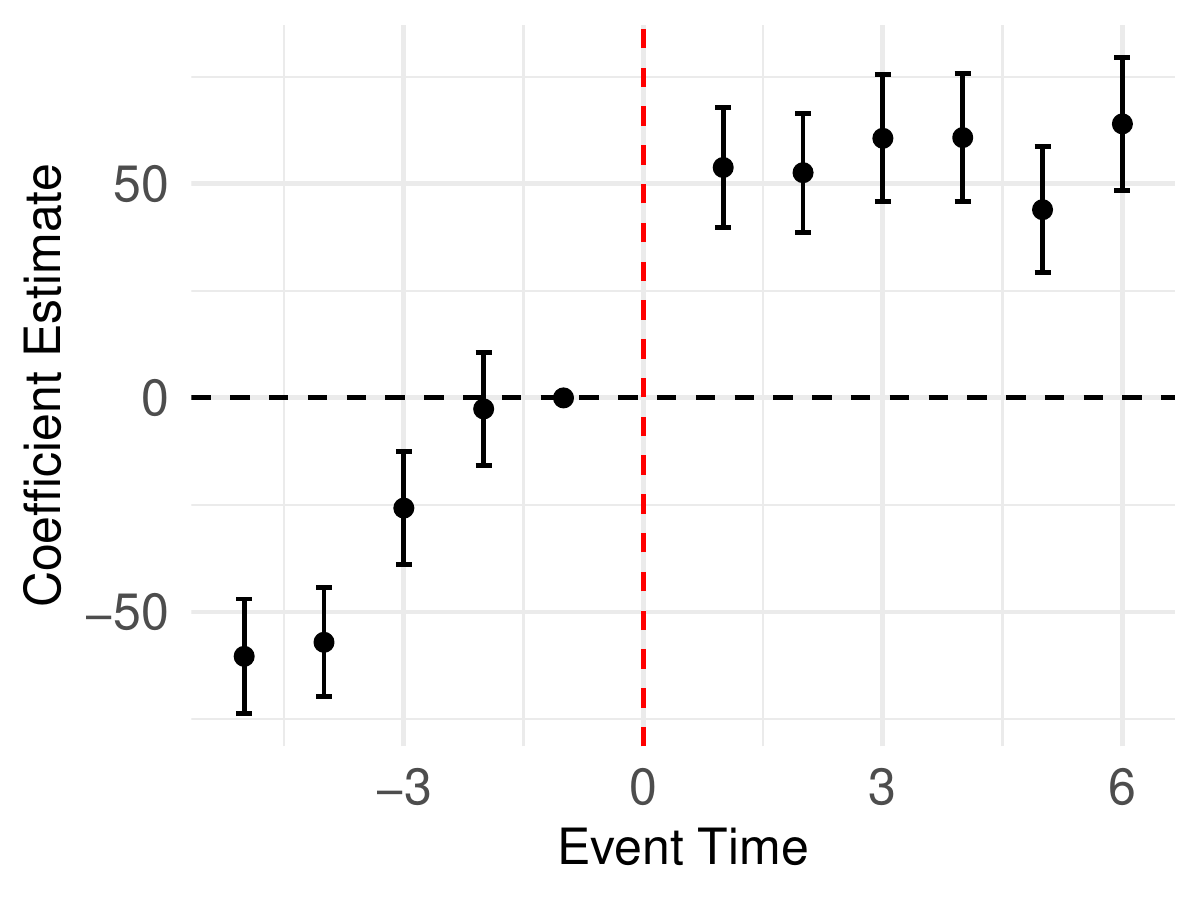}

        \label{fig:spend_sub2_loyalty}
\end{subfigure}
\hfill
\begin{subfigure}{0.32\textwidth}
    \centering
        \caption{\textit{Loyalty Program} vs. \textit{Organic Adopters} -- Spend}\includegraphics[width=\linewidth]{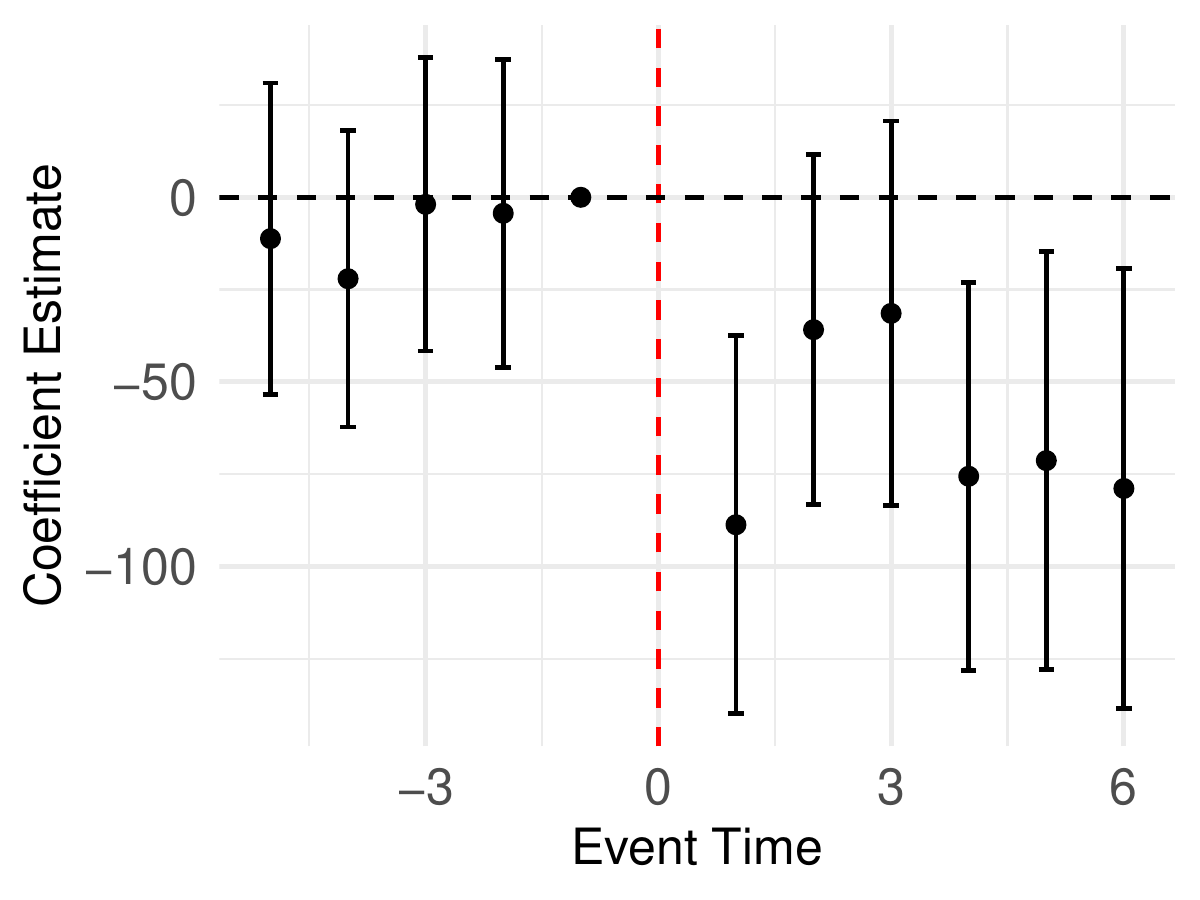}

        \label{fig:spend_sub3_loyalty}
\end{subfigure}
\vspace{0.5em}
\begin{subfigure}{0.31\textwidth}
    \centering
        \caption{Visualization -- Profit}    \includegraphics[width=\linewidth]{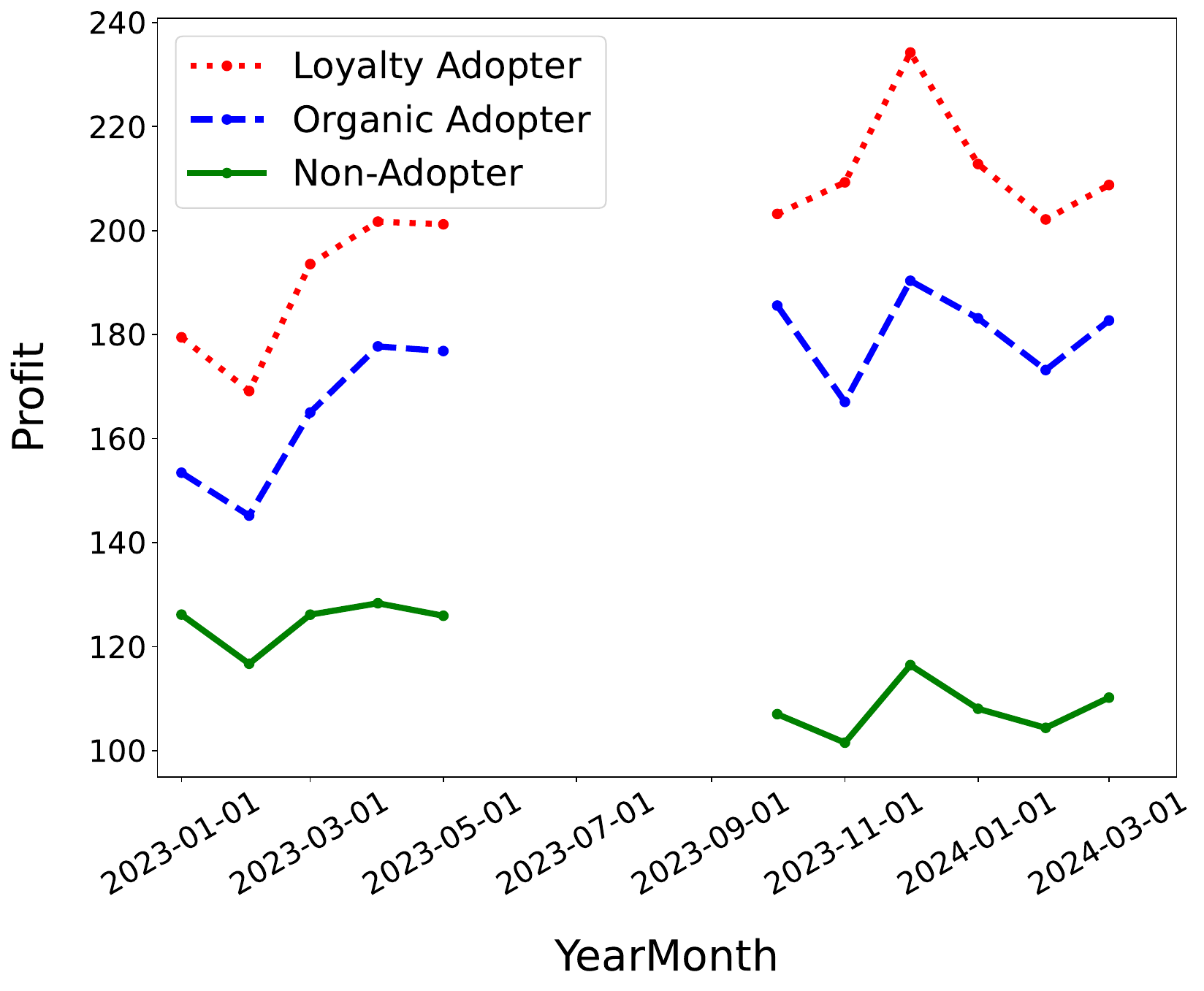}
        \label{fig:profit_sub1_loyalty}
\end{subfigure}
\hfill
\begin{subfigure}{0.32\textwidth}
    \centering
       \caption{\textit{Adopters} vs. \textit{Offline-only} -- Profit}
\includegraphics[width=\linewidth]{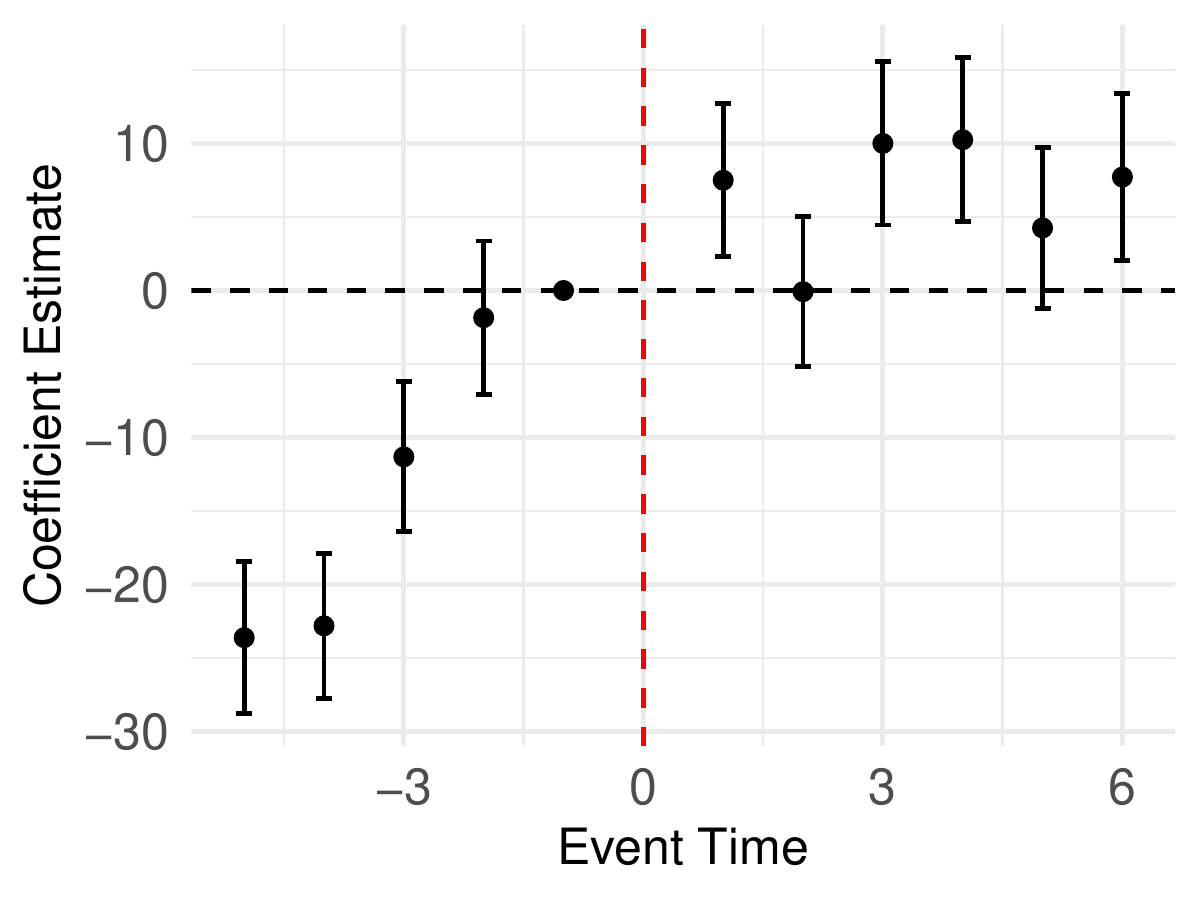}
\label{fig:profit_sub2_loyalty}
\end{subfigure}
\hfill
\begin{subfigure}{0.32\textwidth}
    \centering
        \caption{\textit{Loyalty Program} vs. \textit{Organic Adopters} -- Profit}
        \includegraphics[width=\linewidth]{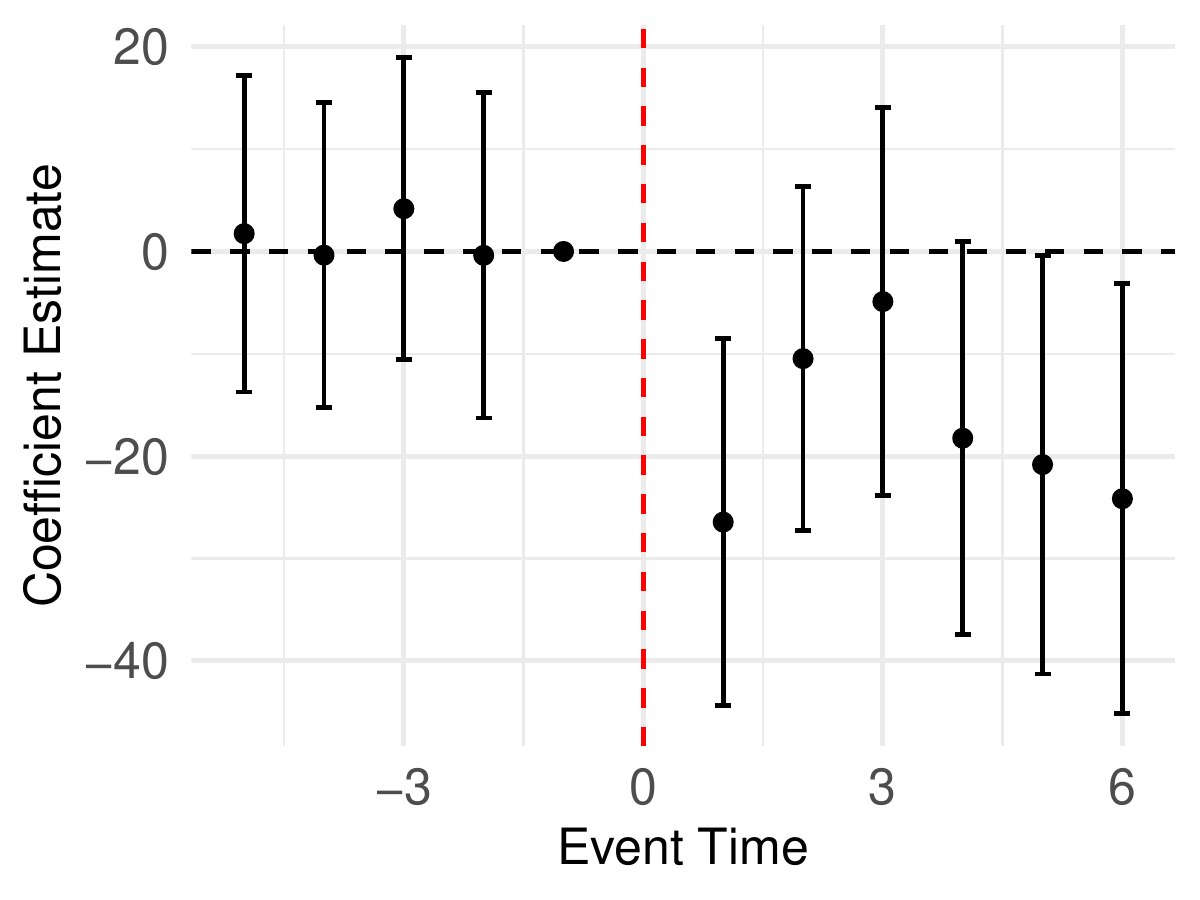}
            \label{fig:profit_sub3_loyalty}
\end{subfigure}
\end{figure}

Figure \ref{fig:spend_sub1_loyalty} and \ref{fig:profit_sub1_loyalty} visualize the monthly average total spend and profit values for \textit{Loyalty Program adopters}, \textit{organic adopters} and \textit{offline-only} customers.

Next, we provide a pre-trend test for two main metrics (i.e., spend and profit) between \textit{adopters} (including both \textit{Loyalty Program adopters} and \textit{Organic adopters}) relative to \textit{offline-only} customers and present the event study estimates (see Figure \ref{fig:spend_sub2_loyalty} and \ref{fig:profit_sub2_loyalty} for total spend and profit, respectively). We exhibit a violation of the parallel trend assumption. To show the robustness of our main finding in the presence of violation of parallel pre-trend, we show the bounded ATT in \S\ref{ssec:parallel_tests} and confirm that the incremental spend is still positive and significant in the presence of violation of parallel trend.

Finally, Figure \ref{fig:spend_sub3_loyalty} and \ref{fig:profit_sub3_loyalty} present the event study regression estimates for the parallel pre-trend of the loyalty program online activation study. 

\newpage

\subsection{Linear Trend Model Estimates}
\label{appssec:honest_did_linear_trend_est}

To quantify the magnitude of the differential pre-trends, we estimate a linear trend model using the entire pre-treatment data to test for any differential time trends in the average monthly spend and profit between the treatment and control customers before the adoption. 
\begin{equation}
    y_{it} = \gamma_{i} + \eta_{t} + \phi_{e,v}  \mathbbm{I}\{Treatment\}_{i} \times t + \epsilon_{it}
\end{equation}
Here, $\mathbbm{I}\{Treatment\}_{i}$ refers to the treatment customers. When we compare \textit{adopters} with \textit{offline-only customers}, $\mathbbm{I}\{Treatment\}_{i}=1$ is customer $i$ is an adopter. When we compare \textit{COVID adopters} with \textit{organic adopters}, $\mathbbm{I}\{Treatment\}_{i}=1$ if customer $i$ is a \textit{COVID adopter}. $t$ is the year-month number for the pre-adoption period, and $\phi_{e, v}$ is the linear trend coefficient for event $e$ variable $v$. If no linear trend exists during the pre-adoption period, the coefficient $\phi_{e, v}$ should be statistically insignificant. The results are shown in Table \ref{tab:linear_trend_consolidated}. We see some significant trend coefficients, except for one insignificant coefficient for pre-adoption monthly profit between adopters (Black Friday and organic) and offline-only customers (see column 4 in Table \ref{tab:linear_trend_consolidated}).  

\begin{table}[htp!]
\centering
\caption{Linear Trend Tests Across Adoption Pathways}
\label{tab:linear_trend_consolidated}
\footnotesize
\setlength{\tabcolsep}{4pt}
\begin{tabular}{lcccccccc}
\midrule\midrule
& \multicolumn{6}{c}{\textbf{Adopters vs. Offline-only}} & \multicolumn{2}{c}{\textbf{Within Adopters}} \\
\cmidrule(lr){2-7}\cmidrule(lr){8-9}
& \multicolumn{2}{c}{\textbf{COVID}} 
& \multicolumn{2}{c}{\textbf{ Black Friday}} 
& \multicolumn{2}{c}{\textbf{ Loyalty Program}} 
& \multicolumn{2}{c}{\textbf{COVID vs. Organic}} \\
\cmidrule(lr){2-3}\cmidrule(lr){4-5}\cmidrule(lr){6-7}\cmidrule(lr){8-9}
DVs: 
& Spend & Profit 
& Spend & Profit 
& Spend & Profit 
& Spend & Profit \\
Model: 
& (1) & (2) 
& (3) & (4) 
& (5) & (6) 
& (7) & (8) \\
\midrule
\emph{Variables} \\
Adopter $\times$ t
& 4.170 & 1.128 
& 4.244 & 1.081 
& 17.51 & 6.811 
& 4.393 & 1.415 \\
& (0.377) & (0.141) 
& (1.571) & (0.6015) 
& (1.504) & (0.578) 
& (0.8605) & (0.3183) \\
& [0.000] & [0.000] 
& [0.007] & [0.072] 
& [0.000] & [0.000] 
& [0.000] & [0.000] \\
\midrule
\emph{Fixed-effects} \\
Customer
& Yes & Yes 
& Yes & Yes 
& Yes & Yes 
& Yes & Yes \\
YearMonth
& Yes & Yes 
& Yes & Yes 
& Yes & Yes 
& Yes & Yes \\
\midrule
\emph{Fit statistics} \\
Observations
& 5,146,359 & 5,146,359 
& 3,855,908 & 3,855,908 
& 2,073,524 & 2,073,524 
& 378,387 & 378,387 \\
R$^2$
& 0.48244 & 0.39923 
& 0.48754 & 0.40597 
& 0.55397 & 0.51055 
& 0.51305 & 0.42830 \\
Within R$^2$
& $6.28\times10^{-5}$ & $3.05\times10^{-5}$ 
& $5.24\times10^{-6}$ & $2.25\times10^{-6}$ 
& 0.00012 & 0.00012 
& 0.00012 & $7.71\times10^{-5}$ \\
\midrule\midrule
\multicolumn{9}{l}{\emph{Clustered (Customer) standard errors are reported in parentheses; p-values are reported in square brackets.}} \\
\multicolumn{9}{l}{\emph{Columns (1)--(6) compare adopters in each pathway group to offline-only customers.}} \\
\multicolumn{9}{l}{\emph{Columns (7)--(8) compare COVID adopters directly to Organic adopters.}} \\
\end{tabular}
\end{table}

\clearpage

\section{Appendix to Managerial Implications}
\label{appsec:managerial_implication}

Table \ref{tab:did_three_groups_profit} presents an additional DID result for profit, which are inputs for the discussion of managerial implications. Instead of estimating Equation \eqref{eq:did_fe}, with a single indicator $\mathbbm{I}\{Adopter_i\}$, we now include two separate dummies -- $\mathbbm{I}\{Event\_Adopter_i\}$ and $\mathbbm{I}\{Organic\_Adopter_i\}$ -- each interacted with the post-adoption indicator. This allows us to estimate the ATT of online adoption on profit for \textit{organic adopters} and \textit{event adopters}, relative to \textit{offline-only} customers.

\begin{table}[htp!]
  \caption{DID Results (All Three Groups) – Profit}
  \label{tab:did_three_groups_profit}
  \centering
  \footnotesize{
  \begin{tabular}{lccc}
  \toprule
  Study:          & COVID-19 & Black Friday & Loyalty Program \\
  \midrule
  Dependent Var.: & \multicolumn{3}{c}{Profit} \\
  Model:          & (1) & (2) & (3) \\
  \midrule
  \emph{Variables} \\
  Organic\_Adopter $\times$ Post 
  & 30.28 (1.903)
  & 23.57 (3.160)
  & 18.55 (1.509)\\
  & [0.000]
  & [0.000]
  & [0.000]\\

  Event\_Adopter $\times$ Post
  & 39.78 (1.272)
  & 7.813 (4.014)
  & 2.986 (6.798)\\
  & [0.000]
  & [0.051]
  & [0.292]\\
  \midrule
  Loyalty program controls 
  & 
  & 
  & Yes \\
  \midrule
  \emph{Fixed-effects} \\
  Customer           & Yes & Yes & Yes \\
  YearMonth          & Yes & Yes & Yes \\
  \midrule
  \emph{Fit statistics} \\
  Observations       & 22,204,608 & 9,505,999 & 4,403,449 \\
  R$^2$              & 0.31903 & 0.34750 & 0.44546 \\
  Within R$^2$       & 0.00047 & $2.88\times10^{-5}$ & 0.00194 \\
  \midrule\midrule
\multicolumn{4}{l}{\emph{Clustered (Customer) standard errors in parentheses; p-values in brackets.}}\\
  \end{tabular}
  }
\end{table}

\subsubsection*{Additional Analysis for ROI/breakeven analyses for promotion:} 

We also perform a similar calculation for the Loyalty Program, with the cost per customer to set up the loyalty program and to convince customers to activate it online is 50 MCU and 150 MCU. Using our DID estimates in Web Appendix $\S$\ref{appsec:managerial_implication}, the average monthly incremental profit from online adoption is 18.55 MCU for \textit{organic adopters} and 2.986 MCU for \textit{Loyalty Program adopters}. 

When the cost per customer is 50 MCU, if managers naively used the incremental profit from \textit{organic adopters}, they would compute a breakeven time of 2.70 months (i.e., 50/18.55), whereas the correct breakeven period using \textit{Loyalty Program adopters}’ incremental profit is 16.74 months (i.e., 50/2.986) -- more than six times longer. Therefore, managers should appropriately calibrate their expectations of increased consumer spending when running promotions that encourage the adoption of online shopping. 

When the cost per customer is 150 MCU, if managers naively used the incremental profit from \textit{organic adopters}, they would compute a breakeven time of 8.09 months (i.e., 150/18.55), whereas the correct breakeven period using \textit{Loyalty Program adopters}’ incremental profit is 50.23 months (i.e., 150/2.986) -- more than six times longer. Therefore, managers should appropriately calibrate their expectations for increased consumer spending when running promotions that encourage online shopping adoption.

\clearpage
\section{Appendix to Validity and Robustness Checks}
\label{appsec:validity_robust}

\subsection{Appendix to Effect Size Measurement}
\label{appssec:effect_size_measurement}
We can compute the true ATT for COVID adopters relative to organic adopters Equation \eqref{eq:att_bound}:
\begin{equation}
\label{eq:att_bound}
    \textrm{True} ~ ATT_{COVID,organic} = \frac{\textrm{Estimated} ~ ATT_{COVID, organic}}{w_{COVID}} 
\end{equation}
where $w_{COVID}$ is the fraction of ``true'' \textit{COVID adopters} in the sample. 

Assuming the monthly incidence of \textit{organic adopters} remains constant, we estimate that around 38.5\% of the \textit{COVID adopters} could actually be \textit{organic adopters} and $61.5\%$ can be the fraction of ``true'' \textit{COVID adopters}, $w_{COVID}$.

Here, since both the $\textrm{Estimated} ~ ATT_{COVID, organic}$ and $w_{COVID}$ are estimated from the sample and thus contain sampling variability, we compute the standard errors of $\textrm{True} ~ ATT_{COVID, organic}$ using block bootstrapping with 500 replications, which treat individual customers as the resampling units. 
Accordingly, the ``true" ATT of spend for \textit{COVID adopters} relative to \textit{organic adopters} would be $\frac{-3.095}{0.615} = -5.033$ (std. error: 10.363, p-value: 0.627), and the ``true" ATT of profit would be $\frac{-9.079}{0.615} = 14.763$ (std. error: 3.638, p-value: 0.000). The ATT point estimates are taken from Tables \ref{tab:total_spend} and \ref{tab:profitability}. The standard errors are obtained from 500 bootstrap replications. The bootstrap distribution of $\textrm{True} ~ ATT_{COVID, organic}$ is presented in Figure \ref{fig:true_ATT_covid_bootstrapped} below. In each replication, we resample the same number of customers (with replacement) as in the original data, re-estimate the DID, calculate the fraction of \textit{COVID adopters}, and obtain the bootstrapped standard errors for $\textrm{True} ~ ATT_{COVID, organic}$ \citep{cameron2005microeconometrics}. 

\begin{figure}[htp!]
    \centering
    \caption{True ATT -- \textit{Black Friday Adopters} vs. \textit{Organic Adopters}}
   \label{fig:true_ATT_BF_bootstrapped}
    \begin{subfigure}{0.49\textwidth}
        \centering
        \caption{Spend}
        \includegraphics[scale=0.45]{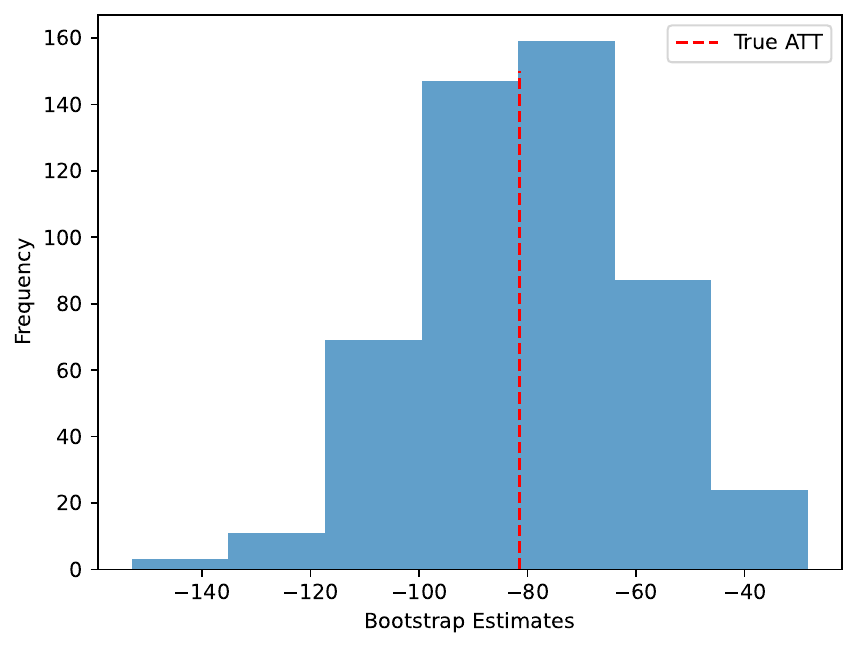}
    \end{subfigure}%
    ~ 
    \begin{subfigure}{0.49\textwidth}
        \centering
        \caption{Profit}
        \includegraphics[scale=0.45]{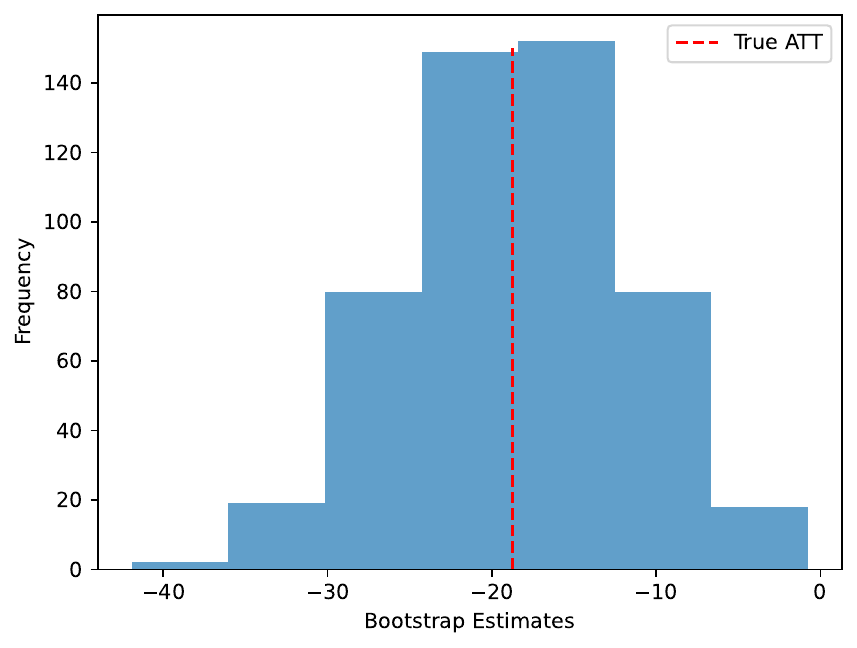}
    \end{subfigure}
    \vspace{1ex} 
\end{figure}

For the \textit{Black Friday adopters}, we can similarly compute the ``true" ATT relative to \textit{organic adopters} by using $w_{black\_friday} = 85.7\%$. The ``true" ATT of spend for \textit{Black Friday adopters} relative to \textit{organic adopters} would be $\frac{-69.88}{0.857} = -81.54$ (std. error: 20.236, p-value: 0.000), and the ATT of profit would be $\frac{-16.04}{0.857} = -18.716$ (std. error: 6.775, p-value: 0.006).The ATT point estimates are taken from Tables \ref{tab:total_spend} and \ref{tab:profitability}. The standard errors are obtained from 500 bootstrap replications. The bootstrap distribution of ``true" ATT of \textit{Black Friday adopters} relative to \textit{organic adopters} is presented in Figure \ref{fig:true_ATT_BF_bootstrapped}. 
\begin{figure}[htp!]
    \centering
    \caption{True ATT -- \textit{COVID adopters} vs. \textit{Organic Adopters}}
   \label{fig:true_ATT_covid_bootstrapped}
    \begin{subfigure}{0.49\textwidth}
        \centering
        \caption{Spend}
        \includegraphics[scale=0.45]{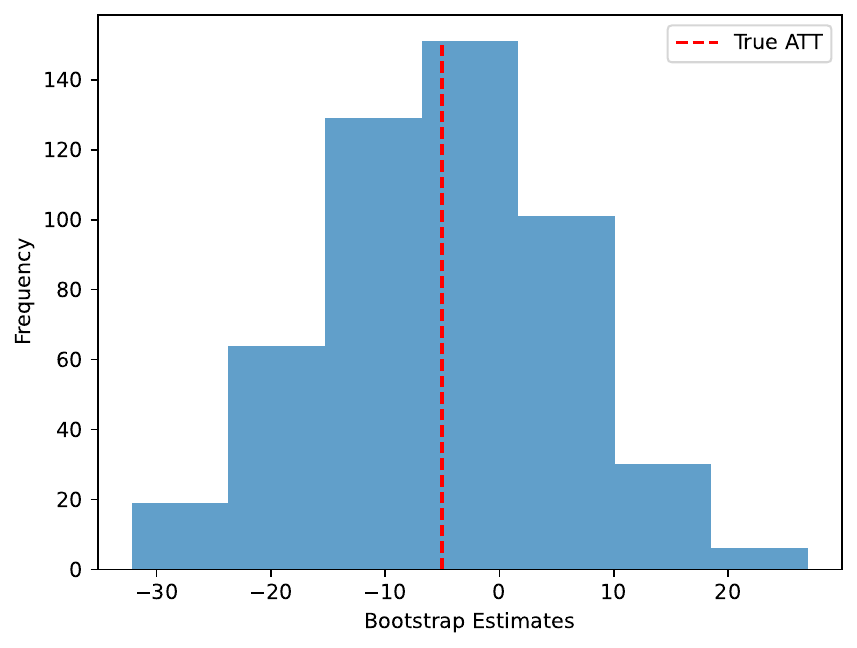}
    \end{subfigure}%
    ~ 
    \begin{subfigure}{0.49\textwidth}
        \centering
        \caption{Profit}
        \includegraphics[scale=0.45]{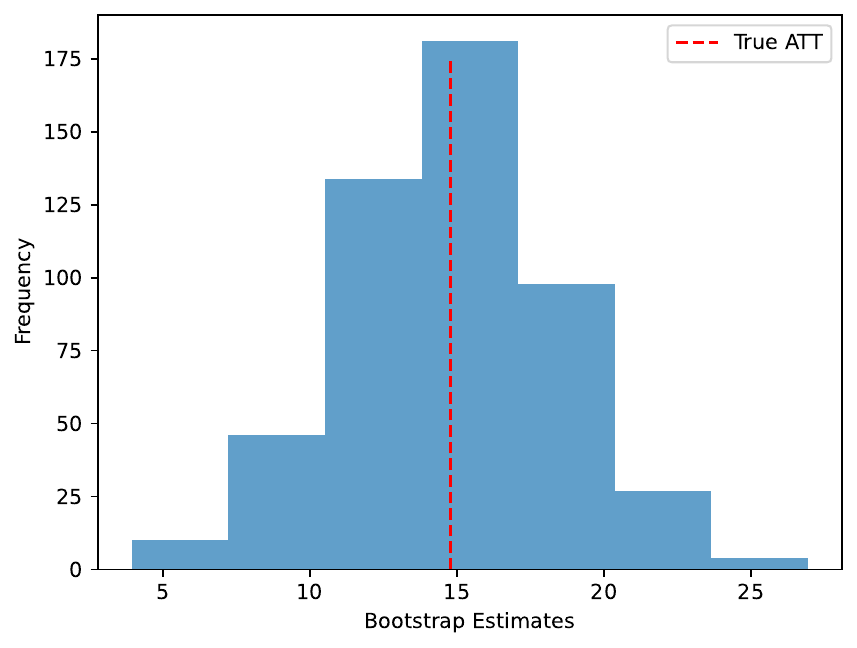}
    \end{subfigure}
    \vspace{1ex} 
\end{figure}

\subsection{Appendix to DID Models with Propensity Score Matching and Inverse Propensity Score Weighting -- \textit{Organic Adopters} vs. \textit{Event Adopters}}
\label{appssec:did_psm_ipw}

In this section, we provide a set of robustness checks under PSM and IPTW. To conduct this analysis, we label the {\it event adopters} as 1 and \textit{organic adopters} as 0 and model the propensity score of customer $i$ being an \textit{event adopter}. The propensity score model is specified as follows: 
\begin{equation}
    PropensityScore_{i} = f(X_{i}^{T}\theta),
\end{equation}
where $f(\cdot)$ is the logit link function where $f(X_{i}^{T}\theta) = \frac{1}{1+\exp(-X_{i}^{T}\theta)}$. $X_{i}\in\mathcal{R} ^ {M \times 1}$ is a column vector with $M$ covariates of customer $i$. These covariates include $\textit{Gender}_{i}, \textit{Age}_{i}, \textit{Tenure(Days)}_{i}$, $\textit{AverageSpendPerMonth}_{i}$, etc.
$\theta\in \mathcal{R}^{M\times 1}$ is a column vector of coefficients. Table \ref{tab:psm_model_est} in Web Appendix \ref{appsssec:psm_model} presents the propensity score model estimation for the COVID-19, Black Friday, and Loyalty Program analyses, respectively. Figure \ref{fig:psm_prediction_before_matching} presents the distribution of propensity scores before matching, and we observe some differences between \textit{event adopters} and \textit{organic adopters}. After matching, Figure \ref{fig:psm_prediction_after_matching} suggests that \textit{event adopters} and \textit{organic adopters} become more overlapped. 

Next, we use these propensity scores to augment our DID analysis in two ways.

\squishlist
\item First, we perform a customer-level propensity score matching and then re-run the DID analysis. Matching based on propensity score is much more efficient than matching based on a set of covariates \citep{rosenbaum1985constructing}. Based on the predicted propensity score of each customer $i$, we use \textit{MatchIt} package in R \citep{stuart2011matchit} to perform nearest-neighbor matching for each \textit{event adopter}. We then run the DID analysis on the newly matched sample (see Table \ref{tab:covid_psm_att}, \ref{tab:black_friday_psm_att}, and \ref{tab:loyalty_psm_att} in Web Appendix \ref{appsssec:psm_att_est}). We find that the results from this exercise are consistent with the main findings.

\item Next, we use a slightly different approach to control for any potential selection issues. Specifically, we use an inverse propensity of treatment weight-adjusted (IPTW) DID regression. Here, the sample is the same as the one from the main analysis, but each observation is inversely weighted by its propensity score. We then perform a two-sample t-test on all the pre-treatment variables based on the weighted sample and confirm that the two samples are statistically insignificant on all the variables (see Table \ref{tab:summary.stat.after.matching} for the weighted sample of \textit{organic adopters} and {\it event adopters}, respectively in the Web Appendix \ref{appsssec:ipw_att_est}). The DID results of {\it event adopters} and \textit{organic adopters} from this analysis are shown in Table \ref{tab:covid_ipw_att}, \ref{tab:black_friday_ipw_att} and \ref{tab:loyalty_ipw_att} for the COVID-19, Black Friday, and Loyalty Program analyses, respectively in Web Appendix \ref{appsssec:ipw_att_est}. Again, we find that the results and conclusions are quite similar to those in the main section. 

\squishend

\subsubsection{Propensity Score Model Estimation}
\label{appsssec:psm_model}

\begin{table}[!htp]
  \centering
  \begin{threeparttable}
  \caption{Propensity Score Logit Model – \textit{Event Adopters} vs. \textit{Organic Adopters}}
  \label{tab:psm_model_est}
  \footnotesize{
  \begin{tabular}{@{\extracolsep{5pt}}lccc}
    \toprule
    & \multicolumn{3}{c}{\textit{DVs:}} \\ 
    \cline{2-4}
    & COVID\_Adopter & BlackFriday\_Adopter & LoyaltyProgram\_Adopter \\ 
    \midrule
    Gender\_NoInformation 
      & -0.044 (0.027) & 0.095 (0.088) & -0.031 (0.042) \\
      & [0.103]        & [0.278]       & [0.459]       \\

    Gender\_Female 
      & 0.052 (0.018)  & 0.098 (0.062)  & -0.050 (0.035) \\
      & [0.004]        & [0.114]        & [0.152]       \\

    LogTenureDays 
      & 0.243 (0.009)  & 0.175 (0.031)  & -0.087 (0.026) \\
      & [0.000]        & [0.000]        & [0.001]       \\

    AvgSpendPerMonth 
      & -0.0001 (0.0001) & 0.0003 (0.0002) & 0.002 (0.0001) \\
      & [0.317]          & [0.133]         & [0.000]       \\

    AvgQuantityPerMonth 
      & 0.010 (0.002)  & 0.008 (0.006)  & 0.0001 (0.003) \\
      & [0.000]        & [0.183]        & [0.974]       \\

    AvgUniqueOrdersPerMonth 
      & -0.147 (0.018) & -0.121 (0.063) & 0.174 (0.028) \\
      & [0.000]        & [0.055]        & [0.000]       \\

    AvgUniqueItemsPerMonth 
      & 0.036 (0.013)  & 0.022 (0.043)  & -0.066 (0.021) \\
      & [0.006]        & [0.610]        & [0.002]       \\

    AvgUniqueBrandsPerMonth 
      & -0.086 (0.022) & -0.031 (0.070) & -0.0001 (0.042) \\
      & [0.000]        & [0.659]        & [0.998]       \\

    AvgUniqueSubcategoriesPerMonth 
      & 0.003 (0.025)  & 0.081 (0.080)  & 0.315 (0.055) \\
      & [0.905]        & [0.311]        & [0.000]       \\

    AvgUniqueCategoriesPerMonth 
      & 0.088 (0.026)  & -0.040 (0.089) & -0.096 (0.059) \\
      & [0.001]        & [0.655]        & [0.103]       \\

    AvgUniqueVisitedStoresPerMonth 
      & 0.289 (0.041)  & 0.284 (0.137)  & -0.074 (0.067) \\
      & [0.000]        & [0.038]        & [0.272]       \\

    AvgProfitPerMonth 
      & 0.0002 (0.0001) & -0.001 (0.0005) & -0.006 (0.0003) \\
      & [0.045]         & [0.045]         & [0.000]       \\

    Constant 
      & -1.063 (0.410) & -1.191 (0.203) & -0.079 (0.189) \\
      & [0.010]        & [0.000]        & [0.672]       \\

    \midrule
    Age & Yes & Yes & Yes \\
    Household Income & Yes & Yes & Yes \\

    \midrule
    Observations  & 46,045   & 4,441    & 18,313   \\
    Log Likelihood & -48,790.450 & -3,983.461 & -19,285.020 \\
    Akaike Inf. Crit. & 97,622.890 & 8,006.922 & 38,610.040 \\
    \midrule\midrule
    \multicolumn{4}{l}{\emph{Standard‐errors in standard brackets; p‑values in square brackets.}}
  \end{tabular}
  }
\begin{tablenotes}
      \small
      \item Note: \textit{AvgUniqueVisitedStoresPerMonth} measures the average number of distinct stores a customer visits per month during the pre-adoption period under each study. 
    \end{tablenotes}
\end{threeparttable}
\end{table}

\begin{figure}[htp!]
    \centering
    \caption{Propensity Score Model Predictions - Before Matching}
    \label{fig:psm_prediction_before_matching}
    \begin{subfigure}{0.32\textwidth}
        \centering
        \caption{COVID-19}
        \includegraphics[width=\linewidth]{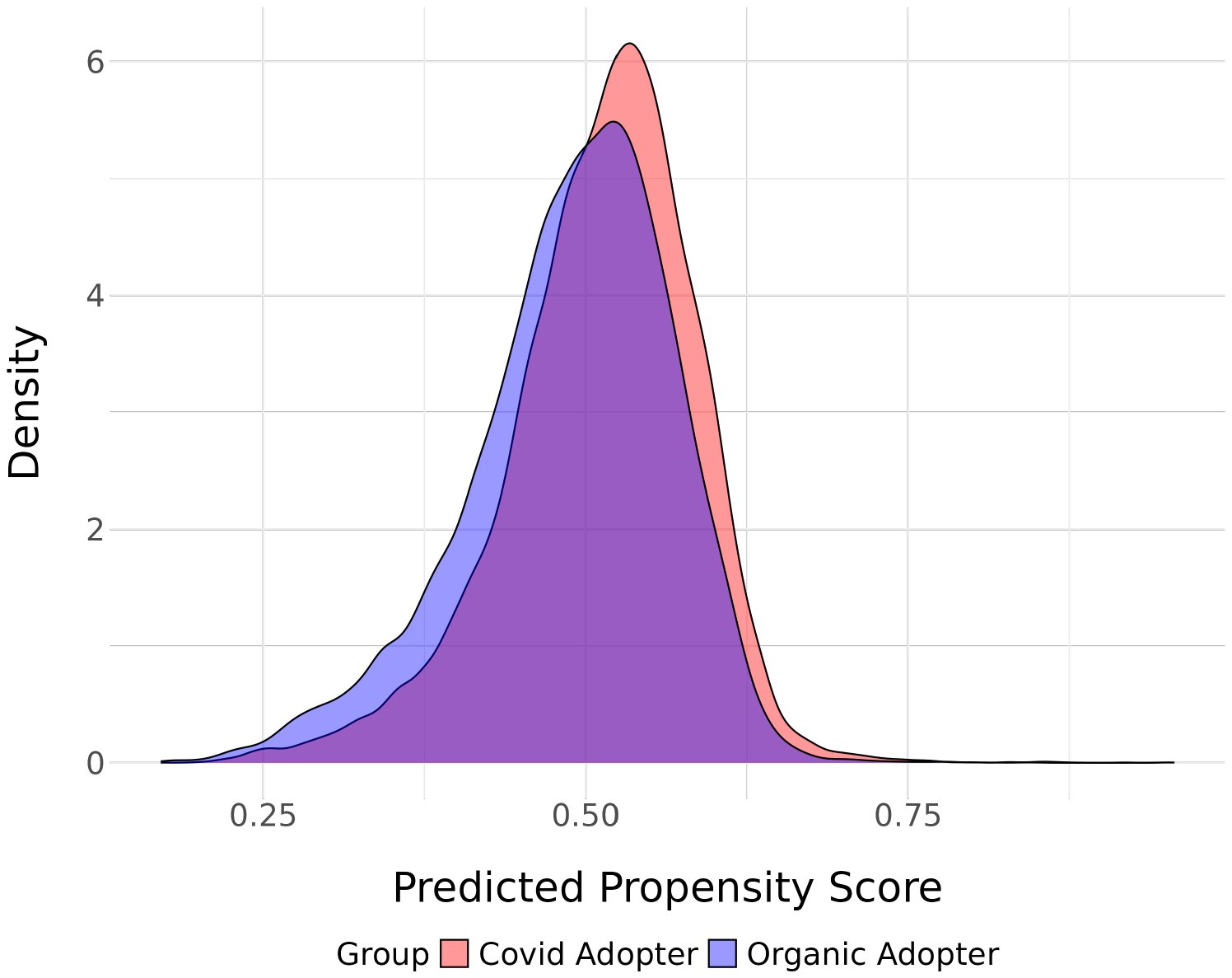}
    \end{subfigure}%
    \hfill
    \begin{subfigure}{0.32\textwidth}
        \centering
        \caption{Black Friday}
        \includegraphics[width=\linewidth]{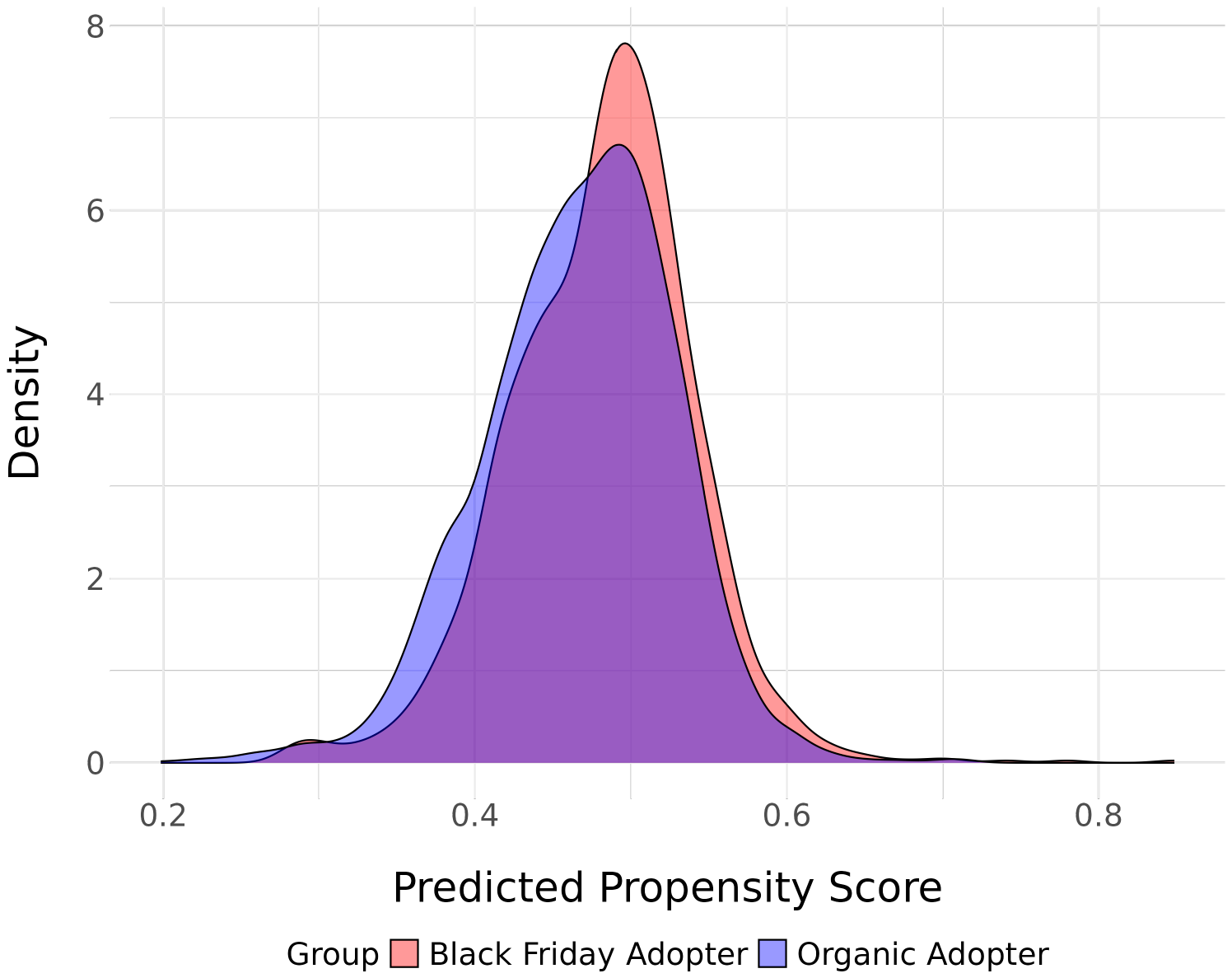}
    \end{subfigure}%
    \hfill
    \begin{subfigure}{0.32\textwidth}
        \centering
        \caption{Loyalty Program}
        \includegraphics[width=\linewidth]{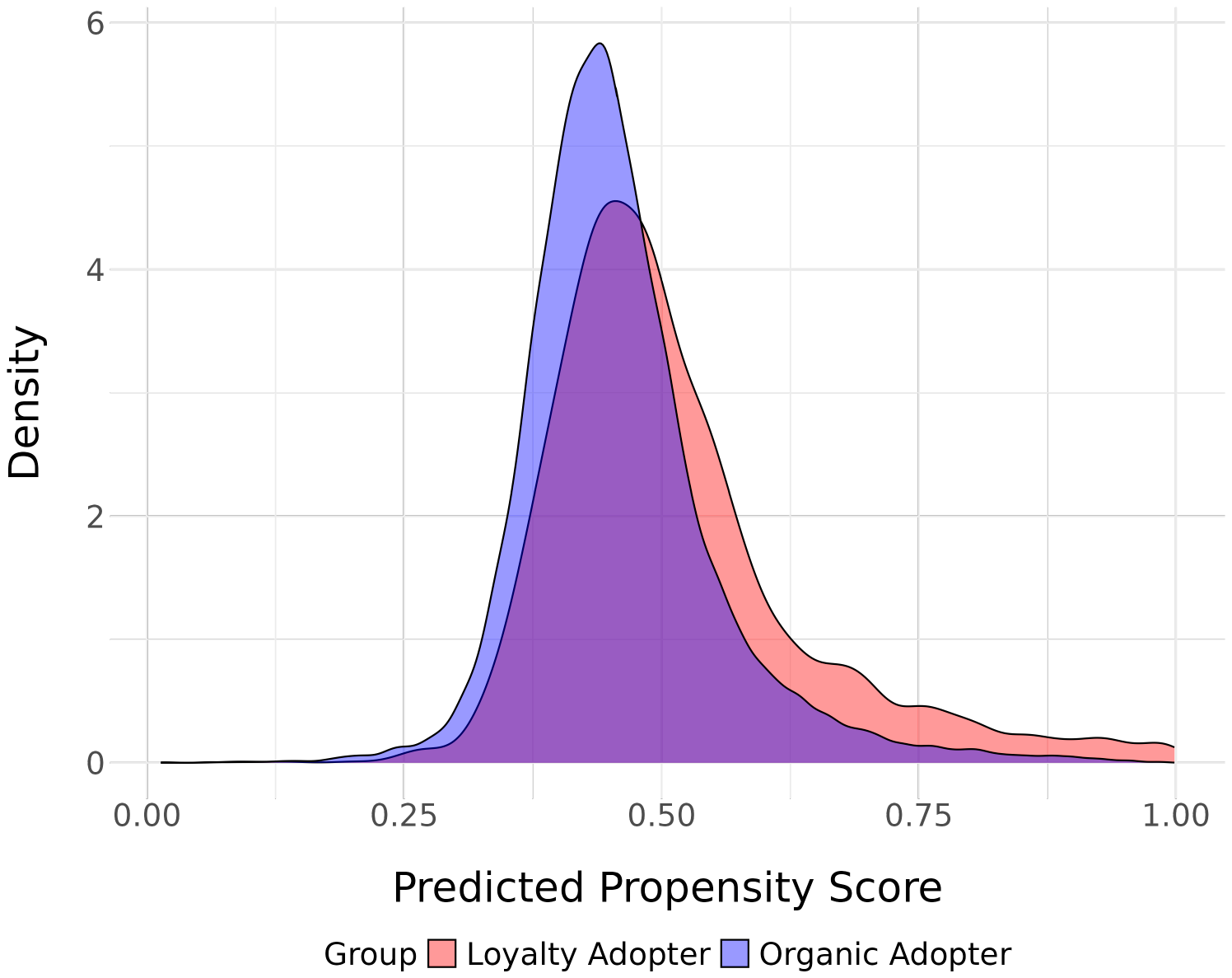}
    \end{subfigure}
\end{figure}

\begin{figure}[htp!]
    \centering
    \caption{Propensity Score Model Predictions - After Matching}
    \label{fig:psm_prediction_after_matching}
    \begin{subfigure}{0.32\textwidth}
        \centering
        \caption{COVID-19}
        \includegraphics[width=\linewidth]{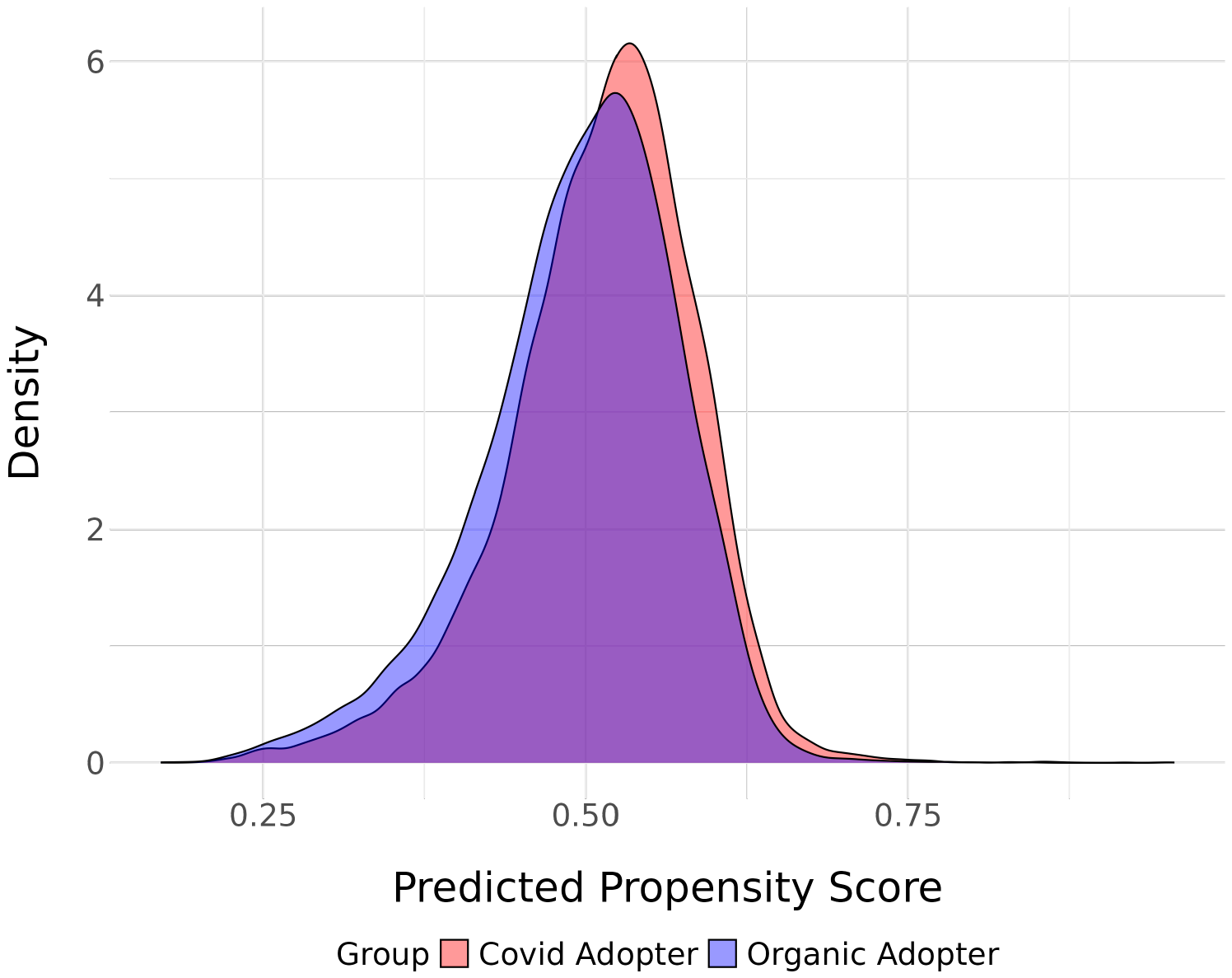}
    \end{subfigure}%
    \hfill
    \begin{subfigure}{0.32\textwidth}
        \centering
        \caption{Black Friday}
        \includegraphics[width=\linewidth]{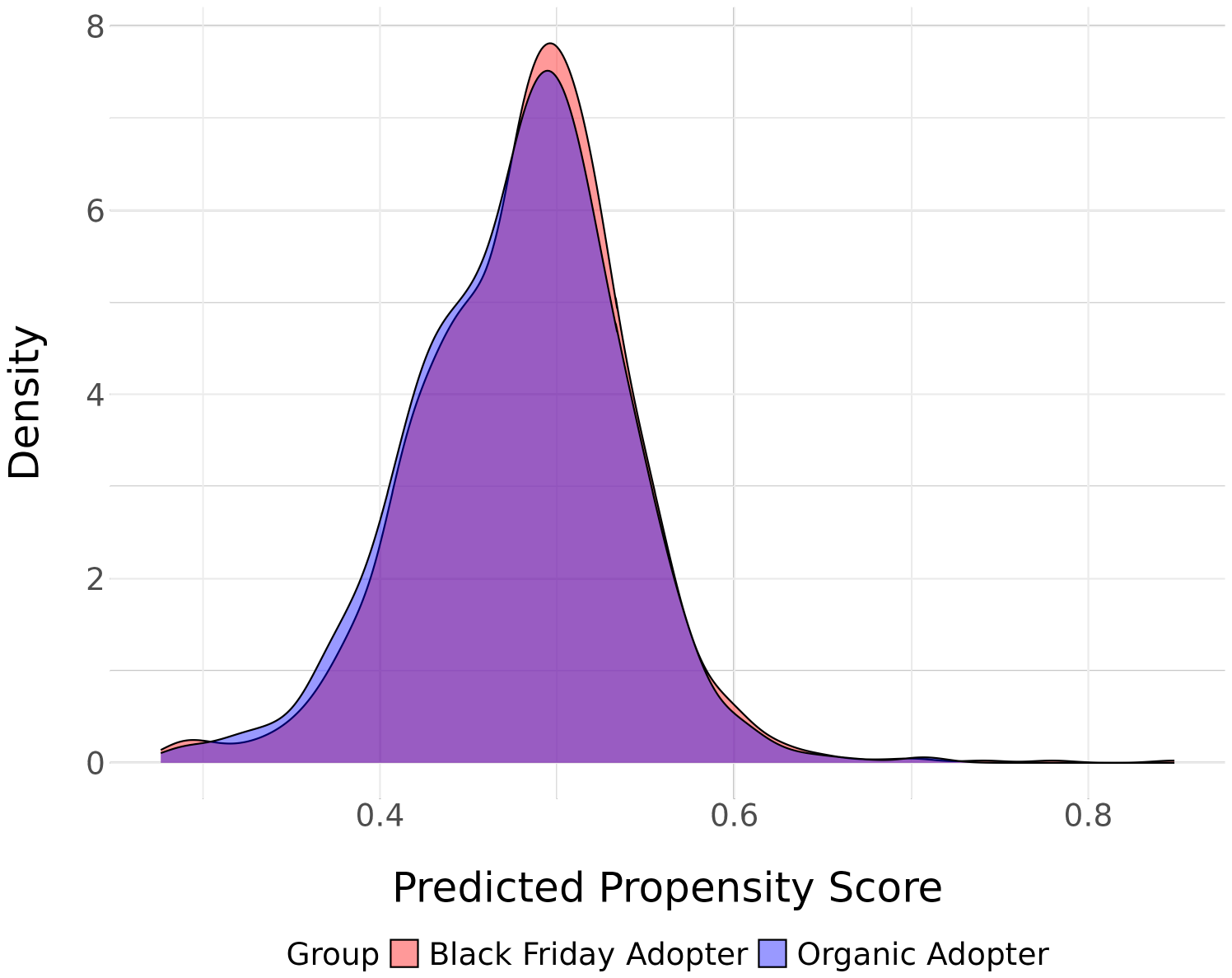}
    \end{subfigure}%
    \hfill
    \begin{subfigure}{0.32\textwidth}
        \centering
        \caption{Loyalty Program}
        \includegraphics[width=\linewidth]{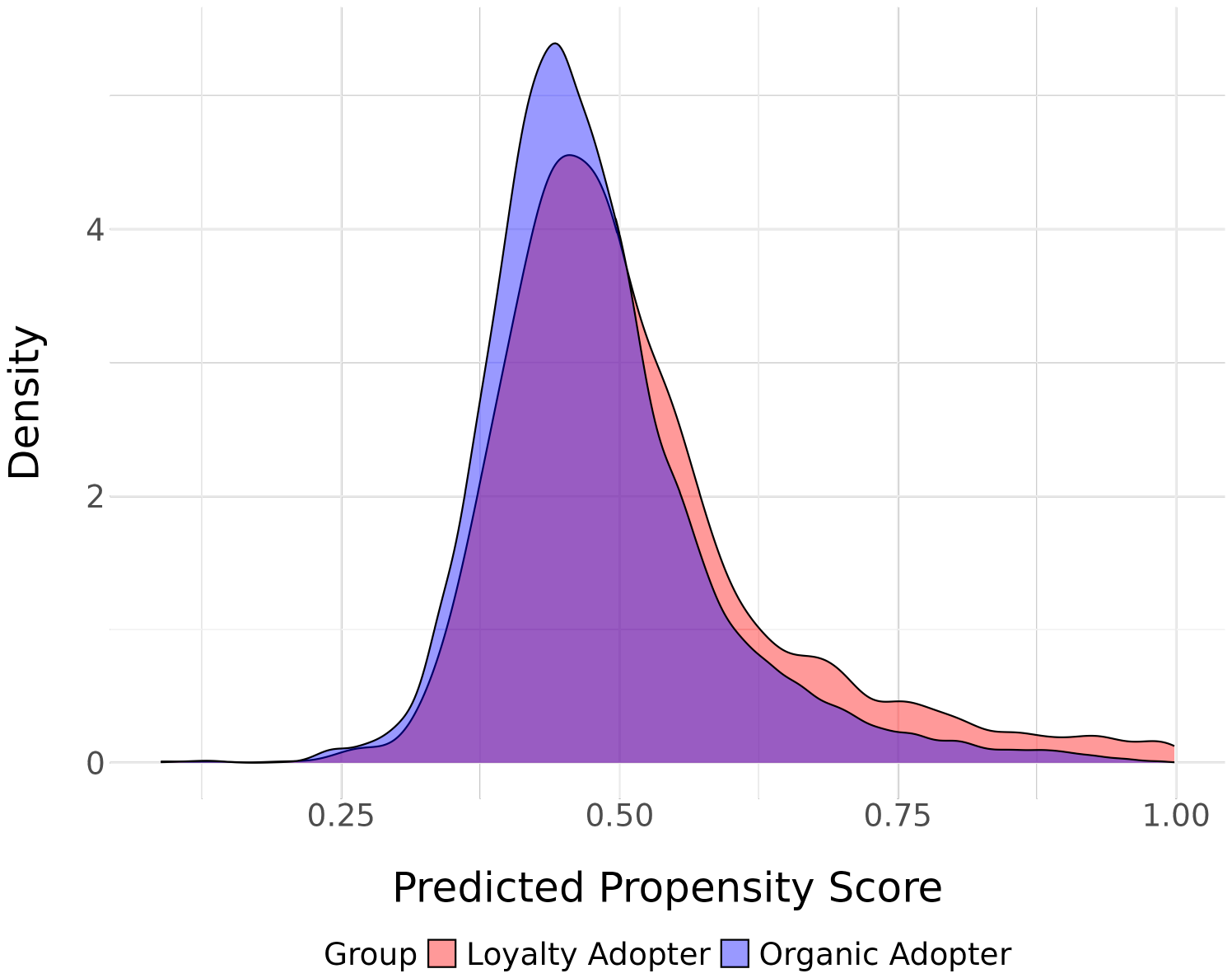}
    \end{subfigure}
\end{figure}

Table \ref{tab:psm_model_est} presents the propensity score model estimates for \textit{COVID adopters}, \textit{Black Friday adopters}, and \textit{Loyalty Program adopters} versus \textit{organic adopters}. Figure \ref{fig:psm_prediction_before_matching} presents the distribution of propensity scores of \textit{event adopters} and \textit{organic adopters} under three studies. After performing nearest-neighbor matching, \ref{fig:psm_prediction_after_matching} shows that the distribution of propensity scores between \textit{event adopters} and \textit{organic adopters} becomes more overlapped. 

\begin{table}[htp!]
\centering
\caption{Customer-level Summary Statistics of Pre-Adoption Variables for \textit{Event Adopters} and \textit{Organic Adopters}: Re-weighted Sample Based on Inverse Propensity Score. Columns 2 and 3 show the weighted mean of variables of \textit{event adopters} and \textit{organic adopters}, with weights being the inverse of propensity.} 
\label{tab:summary.stat.after.matching}
\footnotesize{
\begin{tabular}{lccccc}
  \hline
  Variable & \makecell{\textit{Event Adopters}\\
mean}  & \makecell{\textit{Organic adopters} \\ mean} & Diff & T-stats & P-value \\\hline
    \multicolumn{6}{c}{Panel A: COVID-19 Analysis} \\ \midrule 
  \textit{LogTenure(Days)} & 5.12 & 5.12 & -0.00 & -0.40 & 0.69 \\ 
    \textit{AvgSpendPerMonth} & 459.22 & 461.68 & -2.47 & -0.42 & 0.67 \\ 
    \textit{AvgQuantitiesPerMonth} & 4.72 & 4.91 & -0.19 & -1.67 & 0.09 \\ 
    \textit{AvgOrdersPerMonth} & 0.95 & 0.95 & 0.00 & 0.04 & 0.97 \\ 
    \textit{AvgUniqueItemsPerMonth} & 2.90 & 2.90 & 0.00 & 0.01 & 1.00 \\ 
    \textit{AvgUniqueBrandsPerMonth} & 2.32 & 2.33 & -0.00 & -0.04 & 0.97 \\ 
    \textit{AvgUniqueSubcategoriesPerMonth} & 2.03 & 2.03 & -0.00 & -0.10 & 0.92 \\ 
    \textit{AvgUniqueCategoriesPerMonth} & 1.65 & 1.65 & -0.00 & -0.09 & 0.93 \\ 
    \textit{AvgProfitPerMonth} & 162.25 & 162.72 & -0.47 & -0.22 & 0.83 \\ 
   \hline
    \multicolumn{6}{c}{Panel B: Black Friday Analysis} \\ \midrule
  \textit{LogTenure(Days)} & 5.01 & 5.01 & 0.00 & 0.02 & 0.99 \\ 
    \textit{AvgSpendPerMonth} & 477.58 & 477.41 & 0.17 & 0.01 & 0.99 \\ 
    \textit{AvgQuantitiesPerMonth} & 4.75 & 4.70 & 0.05 & 0.21 & 0.83 \\ 
    \textit{AvgOrdersPerMonth} & 1.00 & 1.00 & 0.00 & 0.01 & 1.00 \\ 
    \textit{AvgUniqueItemsPerMonth} & 3.00 & 3.00 & 0.00 & 0.01 & 0.99 \\ 
    \textit{AvgUniqueBrandsPerMonth} & 2.42 & 2.42 & -0.00 & -0.00 & 1.00 \\ 
    \textit{AvgUniqueSubcategoriesPerMonth} & 2.10 & 2.10 & -0.00 & -0.01 & 0.99 \\ 
    \textit{AvgUniqueCategoriesPerMonth} & 1.70 & 1.70 & -0.00 & -0.00 & 1.00 \\ 
    \textit{AvgProfitPerMonth} & 166.97 & 167.08 & -0.11 & -0.02 & 0.99 \\ \midrule
        \multicolumn{6}{c}{Panel C: Loyalty Program Analysis} \\ \midrule
  \textit{LogTenure(Days)} & 6.45 & 6.45 & 0.00 & 0.14 & 0.89 \\ 
    \textit{AvgSpendPerMonth} & 514.74 & 535.10 & -20.36 & -1.38 & 0.17 \\ 
    \textit{AvgQuantitiesPerMonth} & 3.73 & 3.85 & -0.12 & -0.70 & 0.48 \\ 
    \textit{AvgOrdersPerMonth} & 0.93 & 0.94 & -0.01 & -0.42 & 0.67 \\ 
    \textit{AvgUniqueItemsPerMonth} & 2.28 & 2.29 & -0.02 & -0.27 & 0.78 \\ 
    \textit{AvgUniqueBrandsPerMonth} & 1.80 & 1.81 & -0.01 & -0.37 & 0.71 \\ 
    \textit{AvgUniqueSubcategoriesPerMonth} & 1.54 & 1.55 & -0.01 & -0.39 & 0.70 \\ 
    \textit{AvgUniqueCategoriesPerMonth} & 1.30 & 1.31 & -0.01 & -0.38 & 0.70 \\ 
    \textit{AvgProfitPerMonth} & 180.94 & 184.54 & -3.60 & -0.70 & 0.49 \\ 
   \hline
\end{tabular}
}
\end{table}

\clearpage

\subsubsection{DID Results with PSM}
\label{appsssec:psm_att_est}

Table \ref{tab:covid_psm_att}, \ref{tab:black_friday_psm_att} and \ref{tab:loyalty_psm_att} present the DID estimates of Equation \eqref{eq:did_fe_2by2_FE} of key purchase metrics based on the matched sample under the COVID-19, Black Friday, and Loyalty Program analyses. All results are consistent with the main findings.

\begin{table}[htp!]
  \caption{\textit{COVID adopters} vs. \textit{Organic Adopters} – PSM + DID}
  \label{tab:covid_psm_att}
  \centering
  \footnotesize{
  \begin{tabular}{lcccc}
    \toprule
    DVs: & Spend & Offline Spend & Share of Offline Spend & Profit \\
    Model: & (1) & (2) & (3) & (4) \\
    \midrule
    \emph{Variables}\\
    COVID\_Adopter $\times$ Post 
    & -1.842 (6.910)
    & 38.08 (5.813)
    & 0.0693 (0.0046)
    & 9.258 (2.423)\\
    & [0.790]
    & [0.000]
    & [0.000]
    & [0.000]\\
    \midrule
    \emph{Fixed-effects}\\
    Customer & Yes & Yes & Yes & Yes\\
    YearMonth & Yes & Yes & Yes & Yes\\
    \midrule
    \emph{Fit statistics}\\
    Observations & 1,689,232 & 1,689,232 & 901,177 & 1,689,232\\
    R$^2$ & 0.39081 & 0.36431 & 0.50579 & 0.29503\\
    Within R$^2$ & 1.72e-07 & 0.00012 & 0.00125 & 2.59e-05\\
    \midrule\midrule
\multicolumn{5}{l}{\emph{Clustered (Customer) standard-errors in parentheses; p-values in brackets.}}
  \end{tabular}
  }
\end{table}
\begin{table}[htp!]
  \caption{\textit{Black Friday Adopters} vs. \textit{Organic Adopters} – PSM + DID}
  \label{tab:black_friday_psm_att}
  \centering
  \footnotesize{
  \begin{tabular}{lcccc}
    \toprule
    DVs: & Spend & Offline Spend & Share of Offline Spend & Profit \\
    Model: & (1) & (2) & (3) & (4) \\
    \midrule
    \emph{Variables}\\
    BlackFriday\_Adopter $\times$ Post 
    & -78.26 (17.94)
    & -16.68 (17.37)
    & 0.0959 (0.0159)
    & -22.46 (6.431)\\
    & [0.000]
    & [0.337]
    & [0.000]
    & [0.001]\\
    \midrule
    \emph{Fixed-effects}\\
    Customer & Yes & Yes & Yes & Yes\\
    YearMonth & Yes & Yes & Yes & Yes\\
    \midrule
    \emph{Fit statistics}\\
    Observations & 42,329 & 42,329 & 23,221 & 42,329\\
    R$^2$ & 0.41012 & 0.41414 & 0.52090 & 0.34764\\
    Within R$^2$ & 0.00099 & 5.74e-05 & 0.00671 & 0.00058\\
    \midrule\midrule
\multicolumn{5}{l}{\emph{Clustered (Customer) standard-errors in parentheses; p-values in brackets.}}
  \end{tabular}
  }
\end{table}
\begin{table}[H]
  \caption{\textit{Loyalty Program Adopters} vs. \textit{Organic Adopters} – PSM + DID}
  \label{tab:loyalty_psm_att}
  \centering
  \footnotesize{
  \begin{tabular}{lcccc}
    \toprule
    DVs: & Spend & Offline Spend & Share of Offline Spend & Profit \\
    Model: & (1) & (2) & (3) & (4) \\
    \midrule
    \emph{Variables}\\
    Loyalty\_Adopter $\times$ Post 
    & -52.62 (20.60)
    & -8.541 (19.41)
    & 0.0013 (0.0117)
    & -16.32 (7.228)\\
    & [0.011]
    & [0.660]
    & [0.912]
    & [0.024]\\
    \midrule
    Loyalty program controls & Yes & Yes & Yes & Yes\\
    \midrule
    \emph{Fixed-effects}\\
    Customer & Yes & Yes & Yes & Yes\\
    YearMonth & Yes & Yes & Yes & Yes\\
    \midrule
    \emph{Fit statistics}\\
    Observations & 159,333 & 159,333 & 90,778 & 159,333\\
    R$^2$ & 0.47539 & 0.44214 & 0.56883 & 0.44853\\
    Within R$^2$ & 0.02722 & 0.01915 & 0.00922 & 0.02278\\
    \midrule\midrule
\multicolumn{5}{l}{\emph{Clustered (Customer) standard-errors in parentheses; p-values in brackets.}}
  \end{tabular}
  }
\end{table}

\subsubsection{DID Results with IPTW}
\label{appsssec:ipw_att_est}

Table \ref{tab:covid_ipw_att}, \ref{tab:black_friday_ipw_att} and \ref{tab:loyalty_ipw_att} present the DID estimates of Equation \eqref{eq:did_fe_2by2_FE} of key purchase metrics with inverse propensity score weighting under COVID-19, Black Friday, and Loyalty Program analyses. All results are consistent with the main findings. 

\begin{table}[htp!]
   \caption{\textit{COVID adopters} vs. \textit{Organic Adopters} - IPTW + DID}
   \label{tab:covid_ipw_att}
   \centering
   \footnotesize{
\begin{tabular}{lcccc}
\midrule \midrule
DVs & Spend & Offline Spend & Share of Offline Spend & Profit\\
Model: & (1) & (2) & (3) & (4)\\
\midrule
\emph{Variables}\\
COVID\_Adopter $\times$ Post 
& -6.305 (6.550) 
& 37.96 (5.465) 
& 0.0651 (0.0043) 
& 6.651 (2.307)\\
& [0.315] 
& [0.000] 
& [0.000] 
& [0.008]\\
\midrule
\emph{Fixed-effects}\\
Customer & Yes & Yes & Yes & Yes\\
YearMonth & Yes & Yes & Yes & Yes\\
\midrule
\emph{Fit statistics}\\
Observations & 1,783,949 & 1,783,949 & 949,448 & 1,783,949\\
R$^2$ & 0.38696 & 0.36110 & 0.50768 & 0.29041\\
Within R$^2$ & $2.31\times 10^{-6}$ & 0.00014 & 0.00125 & $1.53\times 10^{-5}$\\
\midrule \midrule
\multicolumn{5}{l}{\emph{Clustered (Customer) standard-errors in parentheses; p-values in brackets.}}\\
\end{tabular}
}
\end{table}

\begin{table}[htp!]
   \caption{\textit{Black Friday Adopters} vs. \textit{Organic Adopters} - IPTW + DID}
   \label{tab:black_friday_ipw_att}
   \centering
   \footnotesize{
\begin{tabular}{lcccc}
\midrule \midrule
DVs: & Spend & Offline Spend & Share of Offline Spend & Profit\\
Model: & (1) & (2) & (3) & (4)\\
\midrule
\emph{Variables}\\
BlackFriday\_Adopter $\times$ Post 
& -68.89 (14.30) 
& -4.429 (14.12) 
& 0.0848 (0.0122) 
& -20.64 (5.231)\\
& [0.000] 
& [0.765] 
& [0.000] 
& [0.000]\\
\midrule
\emph{Fixed-effects}\\
Customer & Yes & Yes & Yes & Yes\\
YearMonth & Yes & Yes & Yes & Yes\\
\midrule
\emph{Fit statistics}\\
Observations & 76,842 & 76,842 & 42,299 & 76,842\\
R$^2$ & 0.40586 & 0.39667 & 0.53426 & 0.34948\\
Within R$^2$ & 0.00067 & $3.58\times 10^{-6}$ & 0.00455 & 0.00042\\
\midrule \midrule
\multicolumn{5}{l}{\emph{Clustered (Customer) standard-errors in parentheses; p-values in brackets.}}\\
\end{tabular}
}
\end{table}

\begin{table}[H]
   \caption{\textit{Loyalty Program Adopters} vs. \textit{Organic Adopters} - IPTW + DID}
   \label{tab:loyalty_ipw_att}
   \centering
   \footnotesize{
\begin{tabular}{lcccc}
\midrule \midrule
DVs & Spend & Offline Spend & Share of Offline Spend & Profit\\
Model: & (1) & (2) & (3) & (4)\\
\midrule
\emph{Variables}\\
Loyalty\_Adopter $\times$ Post 
& -52.53 (19.43) 
& -11.43 (18.60) 
& 0.0017 (0.0114) 
& -15.22 (6.835)\\
& [0.009] 
& [0.546] 
& [0.888] 
& [0.024]\\
\midrule
Loyalty program controls & Yes & Yes & Yes & Yes\\
\midrule
\emph{Fixed-effects}\\
Customer & Yes & Yes & Yes & Yes\\
YearMonth & Yes & Yes & Yes & Yes\\
\midrule
\emph{Fit statistics}\\
Observations & 197,658 & 197,658 & 110,339 & 197,658\\
R$^2$ & 0.47155 & 0.43591 & 0.57046 & 0.44366\\
Within R$^2$ & 0.02694 & 0.01910 & 0.00903 & 0.02249\\
\midrule \midrule
\multicolumn{5}{l}{\emph{Clustered (Customer) standard-errors in parentheses; p-values in brackets.}}\\
\end{tabular}
}
\end{table}

\subsubsection{Doubly Robust DID for the COVID-19 Analysis}
\label{appsssec:doubly_robust_did}

Given the potential concern of parallel trend assumption under COVID-19 analysis, we seek an alternative identification assumption and conduct a doubly robust DID for the COVID-19 analysis. We now show that the main conclusions remain consistent by using the doubly robust DID \citep{sant2020doubly}. We provide the ATT estimates for each metric based on the package \textit{did} in R provided by \cite{callaway2021difference}. Table \ref{tab:drdid_other_metrics} presents the doubly robust DID estimates for all metrics. Figure \ref{fig:doubly_robust_did_spend_profit} below presents the event study figure under doubly robust DID results for main metrics (i.e., spend and profit). We can see that the parallel pre-trends are also well satisfied, conditional on customer demographics and pre-adoption purchase variables. The estimation results are consistent with the main findings.
\begin{table}[htp!]
    \centering
    \caption{Doubly Robust DID -- \textit{COVID adopters} vs. \textit{Organic Adopters}}
    \label{tab:drdid_other_metrics}
    \footnotesize{
\begin{tabular}{lcccccc}
\toprule
\midrule
DVs  & ATT & SE & T-stat & P-value & CI- & CI+\\
\midrule
Spend &  -5.963 &  7.814 & -0.763  & 0.445 & -21.279 &  9.353\\
  Share of Offline Spend & 0.083 & 0.004& 20.550 & 0.000 & 0.075 & 0.091 \\
Profit & 11.623 & 2.942 & 3.951 & 0.000&  5.856 &17.390 \\ \midrule 
\bottomrule
\end{tabular}
}
\end{table}
\begin{figure}[htp!]
    \centering
    \caption{Doubly Robust Event Study for the Main Metrics -- \textit{COVID adopters} vs. \textit{Organic Adopters}}
    \label{fig:doubly_robust_did_spend_profit}
        \begin{subfigure}{0.49\textwidth}
        \centering
        \caption{Spend}
        \includegraphics[scale=0.30]{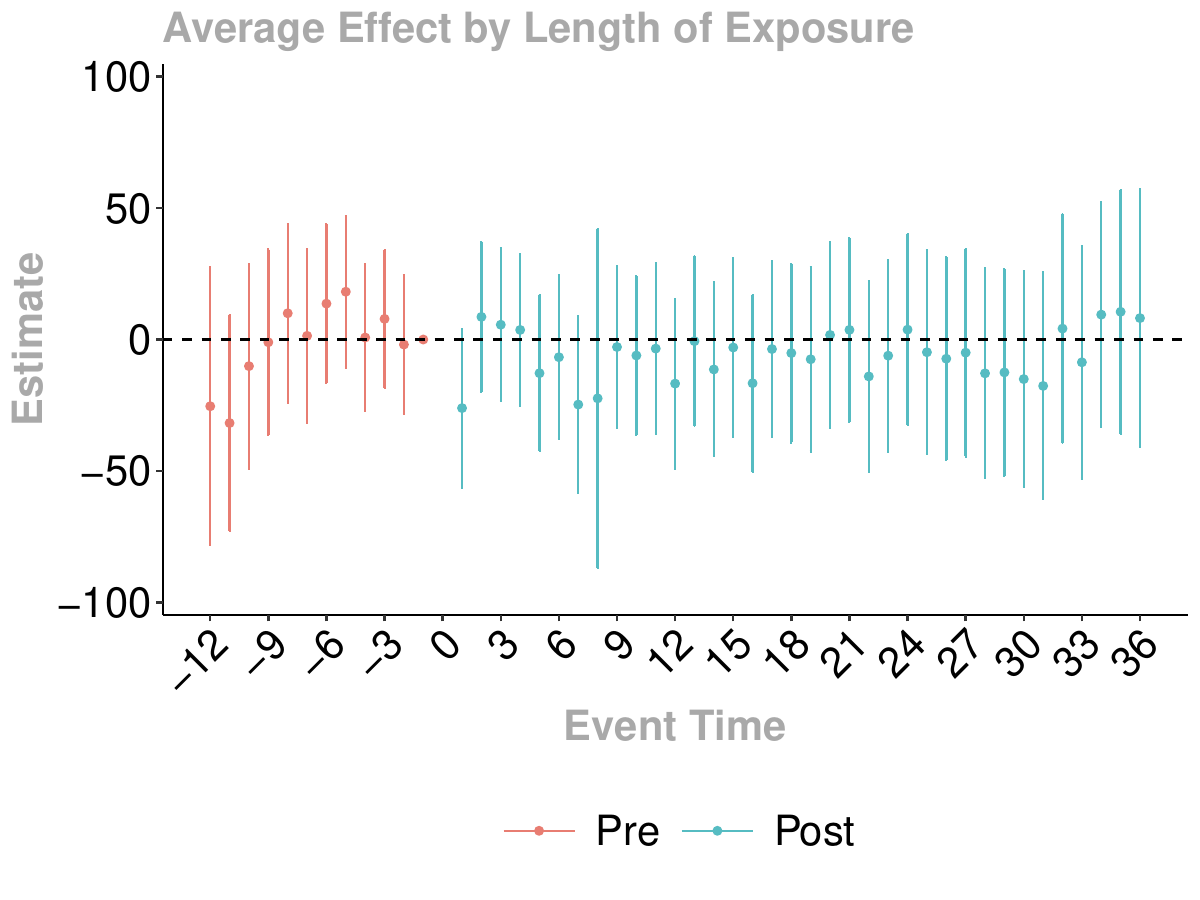}
        \label{fig:dr_did_spend_covid}
    \end{subfigure}%
    ~ 
    \begin{subfigure}{0.49\textwidth}
        \centering
        \caption{Profit}
        \includegraphics[scale=0.30]{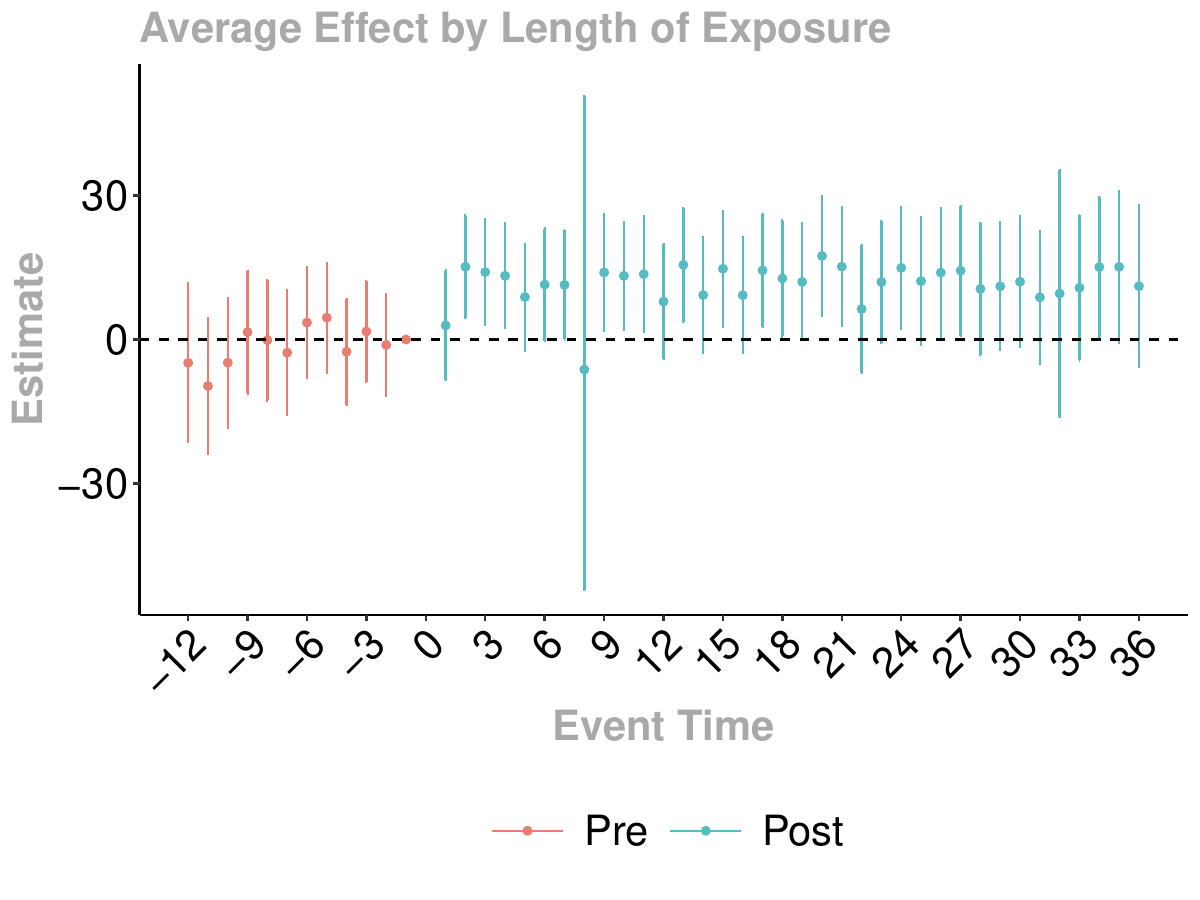}
\label{fig:dr_did_profit_covid}
    \end{subfigure}
\end{figure}

\clearpage

\subsection{Appendix to DID Models with Propensity Score Matching and Inverse Propensity Score Weighting -- \textit{Adopters} vs. \textit{Offline-only} Customers}
\label{appssec:did_psm_ipw_adopter_nonadopter}

In this section, we label the {\it adopters} (both \textit{event adopters} and \textit{organic adopters}) as 1 and \textit{offline-only} customers as 0 and model the propensity score of customer $i$ being an \textit{adopter}. The propensity score model is specified as follows: 
\begin{equation}
    PropensityScore_{i} = f(X_{i}^{T}\theta),
\end{equation}
where $f(\cdot)$ is the logit link function where $f(X_{i}^{T}\theta) = \frac{1}{1+\exp(-X_{i}^{T}\theta)}$. $X_{i}\in\mathcal{R} ^ {M \times 1}$ is a column vector with $M$ covariates of customer $i$. These covariates include $\textit{Gender}_{i}, \textit{Age}_{i}, \textit{Tenure(Days)}_{i}$, $\textit{AverageSpendPerMonth}_{i}$, etc.
$\theta\in \mathcal{R}^{M\times 1}$ is a column vector of coefficients. Table \ref{tab:psm_model_est_adopter_nonadopter} in Web Appendix \ref{appssec:did_psm_ipw_adopter_nonadopter} presents the propensity score model estimation for COVID-19, Black Friday, and Loyalty Program analyses, respectively. Figure \ref{fig:psm_prediction_before_matching_adopter_nonadopter} presents the distribution of propensity scores before matching, and we observe some differences between \textit{ adopters} and \textit{offline-only} customers. After matching, Figure \ref{fig:psm_prediction_after_matching_adopter_nonadopter} suggests that \textit{adopters} and \textit{offline-only} customers become more overlapped. 

Next, we use these propensity scores of being adopters to augment our DID analysis in two ways.

\squishlist
\item First, we perform a customer-level propensity score matching and then re-run the DID analysis. Matching based on propensity score is much more efficient than matching based on a set of covariates \citep{rosenbaum1985constructing}. Based on the predicted propensity score of each customer $i$, we use \textit{MatchIt} package in R \citep{stuart2011matchit} to perform nearest-neighbor matching for each \textit{adopter}. We then run the DID analysis on the newly matched sample and find that the results from this exercise are consistent with the main findings (see Table \ref{tab:psm_att_est_adopter_nonadopter_main}).

\item Next, we use a slightly different approach to control for any potential selection issues. Specifically, we use an inverse propensity of treatment weight-adjusted (IPTW) DID regression. Here, the sample is the same as the one from the main analysis, but each observation is inversely weighted by its propensity score. The DID results of {\it adopters} and \textit{offline-only} customers from this analysis are shown in Table \ref{tab:iptw_att_est_adopter_nonadopter_main}. Again, we find that the results and conclusions are quite similar to those in the main section.


\squishend

\subsubsection{Propensity Score Model Estimation}
\label{appsssec:psm_model_adopter_nonadopter}

\begin{table}[!htp]
  \centering
  \begin{threeparttable}
  \caption{Propensity Score Logit Model -- Adopters vs. Non-Adopters}
  \label{tab:psm_model_est_adopter_nonadopter}
  \footnotesize{
  \begin{tabular}{@{\extracolsep{5pt}}lccc}
    \toprule
    & \multicolumn{3}{c}{\textit{DV:}} \\ 
    \cline{2-4}
    & COVID-19 & Black Friday & Loyalty Program \\ 
    \midrule
    Gender\_NoInformation 
      & -0.003 (0.006) & 0.123 (0.016) & 0.290 (0.010) \\
      & [0.617]        & [0.000]       & [0.000]       \\

    Gender\_Female 
      & 0.494 (0.004) & 0.544 (0.012) & 0.355 (0.008) \\
      & [0.000]       & [0.000]       & [0.000]       \\

    LogTenureDays 
      & 0.051 (0.002) & -0.039 (0.006) & -0.323 (0.006) \\
      & [0.000]       & [0.000]        & [0.000]        \\

    AvgSpendPerMonth 
      & 0.001 (0.00002) & 0.001 (0.00003) & 0.001 (0.00003) \\
      & [0.000]         & [0.000]         & [0.000]         \\

    AvgQuantityPerMonth 
      & 0.009 (0.001) & -0.003 (0.001) & -0.014 (0.001) \\
      & [0.000]       & [0.003]        & [0.000]        \\

    AvgUniqueOrdersPerMonth 
      & -0.290 (0.005) & -0.179 (0.012) & -0.031 (0.008) \\
      & [0.000]        & [0.000]        & [0.000]        \\

    AvgUniqueItemsPerMonth 
      & -0.050 (0.004) & -0.059 (0.008) & -0.023 (0.006) \\
      & [0.000]        & [0.000]        & [0.000]        \\

    AvgUniqueBrandsPerMonth 
      & 0.085 (0.007) & 0.166 (0.014) & -0.024 (0.011) \\
      & [0.000]       & [0.000]       & [0.029]        \\

    AvgUniqueSubcategoriesPerMonth 
      & -0.055 (0.008) & -0.052 (0.016) & 0.157 (0.015) \\
      & [0.000]        & [0.001]        & [0.000]       \\

    AvgUniqueCategoriesPerMonth 
      & 0.253 (0.008) & 0.138 (0.017) & 0.123 (0.015) \\
      & [0.000]       & [0.000]       & [0.000]       \\

    AvgUniqueVisitedStoresPerMonth 
      & 0.430 (0.012) & 0.605 (0.026) & 0.159 (0.017) \\
      & [0.000]       & [0.000]       & [0.000]       \\

    AvgProfitPerMonth 
      & 0.0002 (0.00005) & -0.001 (0.0001) & -0.003 (0.0001) \\
      & [0.000]          & [0.000]         & [0.000]         \\

    Constant 
      & -2.464 (0.079) & -5.123 (0.321) & 1.166 (0.045) \\
      & [0.000]        & [0.000]        & [0.000]       \\

    \midrule
    Age & Yes & Yes & Yes \\
    HouseholdIncome & Yes & Yes & Yes \\

    \midrule
    Observations & 619,979 & 539,905 & 414,709 \\
    Log Likelihood & -720,390.800 & -145,457.300 & -349,382.200 \\
    Akaike Inf. Crit. & 1,440,824.000 & 290,956.700 & 698,804.300 \\
    \midrule\midrule
    \multicolumn{4}{l}{\emph{Standard errors in parentheses; p-values in square brackets.}} \\
  \end{tabular}
  }
  \begin{tablenotes}
      \small
      \item Note: \textit{AvgUniqueVisitedStoresPerMonth} measures the average number of distinct stores a customer visits per month during the pre-adoption period under each study.
  \end{tablenotes}
  \end{threeparttable}
\end{table}

\begin{figure}[htp!]
    \centering
    \caption{Propensity Score Model Predictions - Before Matching}
\label{fig:psm_prediction_before_matching_adopter_nonadopter}
    \begin{subfigure}{0.32\textwidth}
        \centering
        \caption{COVID-19}
        \includegraphics[width=\linewidth]{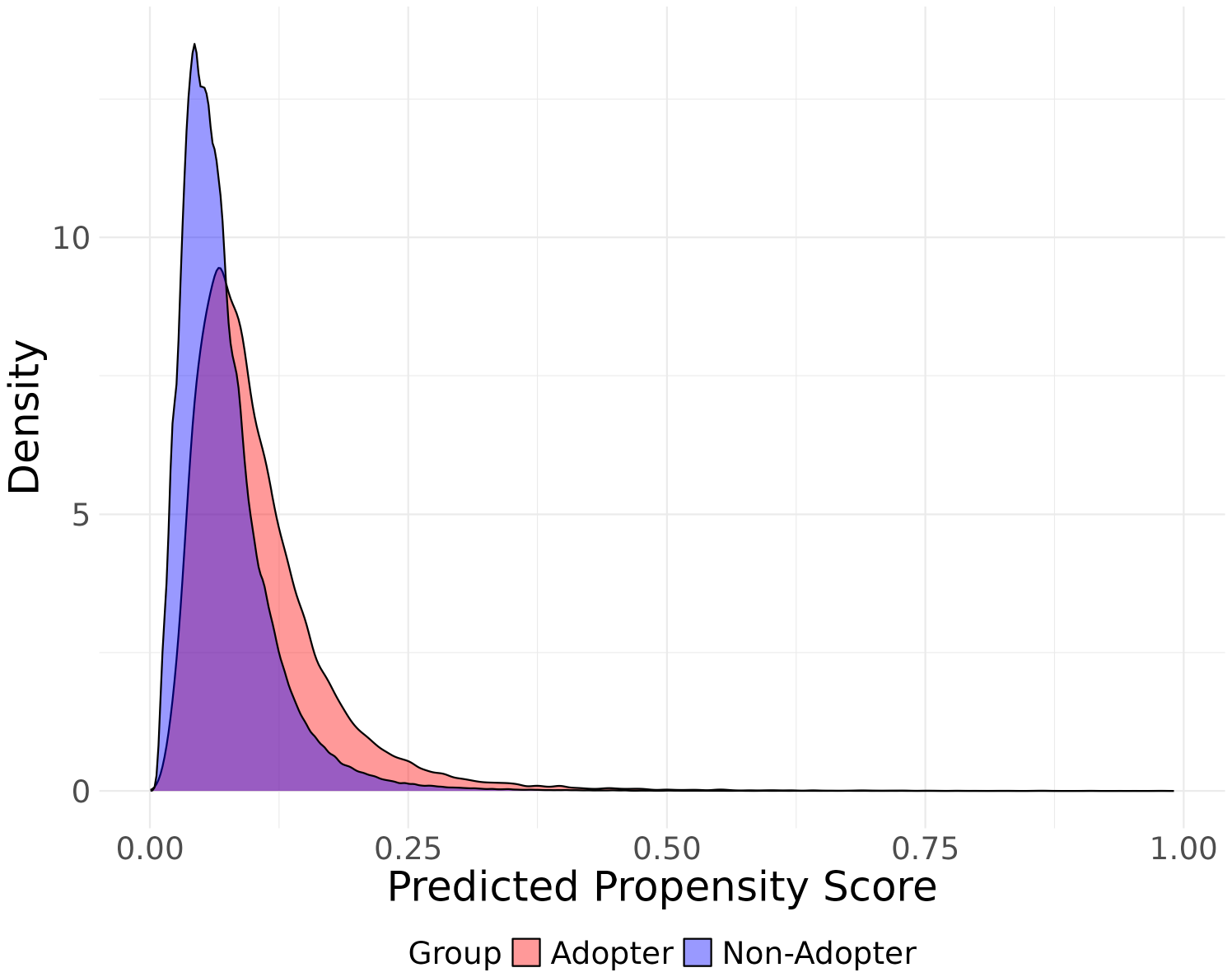}
    \end{subfigure}%
    \hfill
    \begin{subfigure}{0.32\textwidth}
        \centering
        \caption{Black Friday}
        \includegraphics[width=\linewidth]{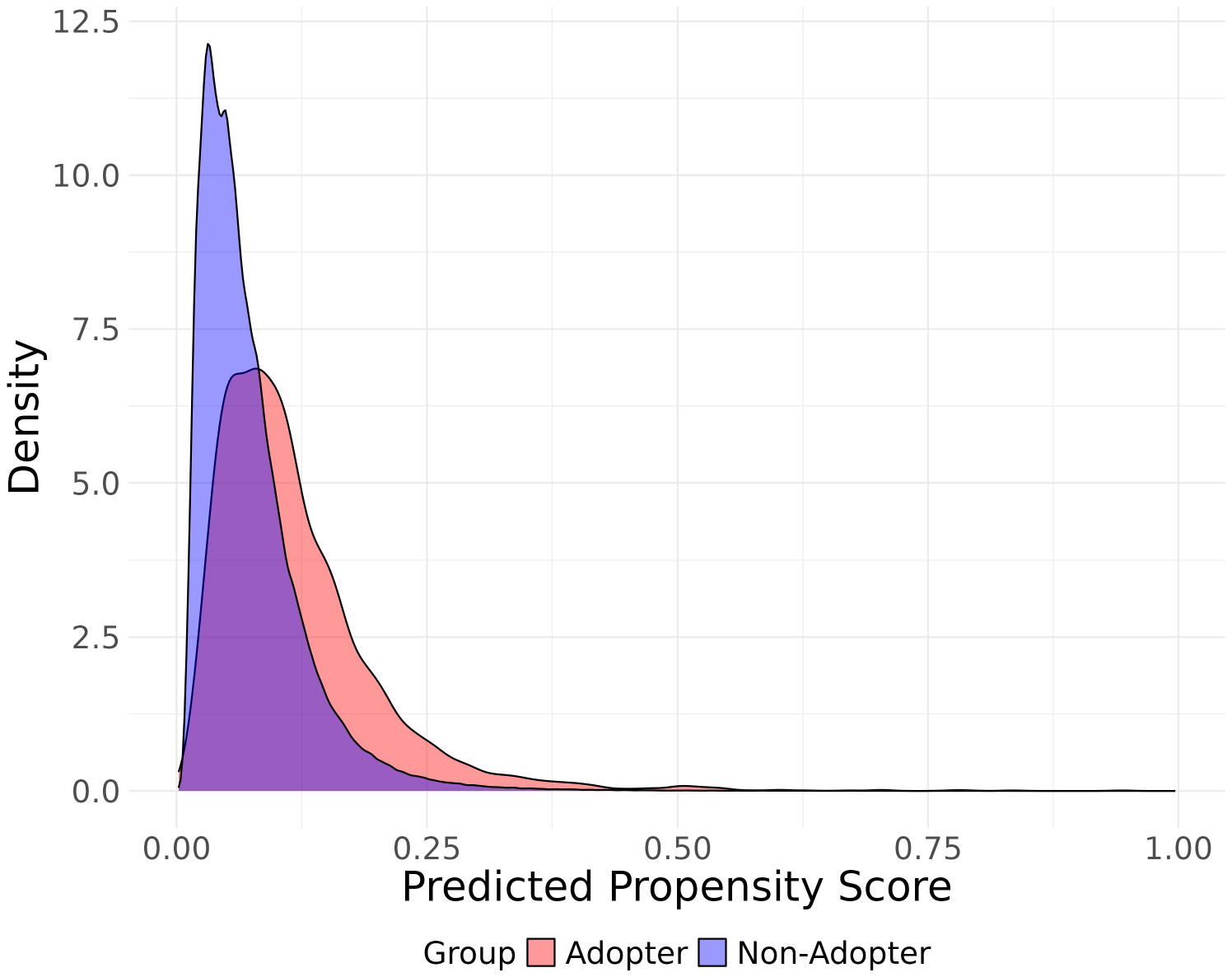}
    \end{subfigure}%
    \hfill
    \begin{subfigure}{0.32\textwidth}
        \centering
        \caption{Loyalty Program}
        \includegraphics[width=\linewidth]{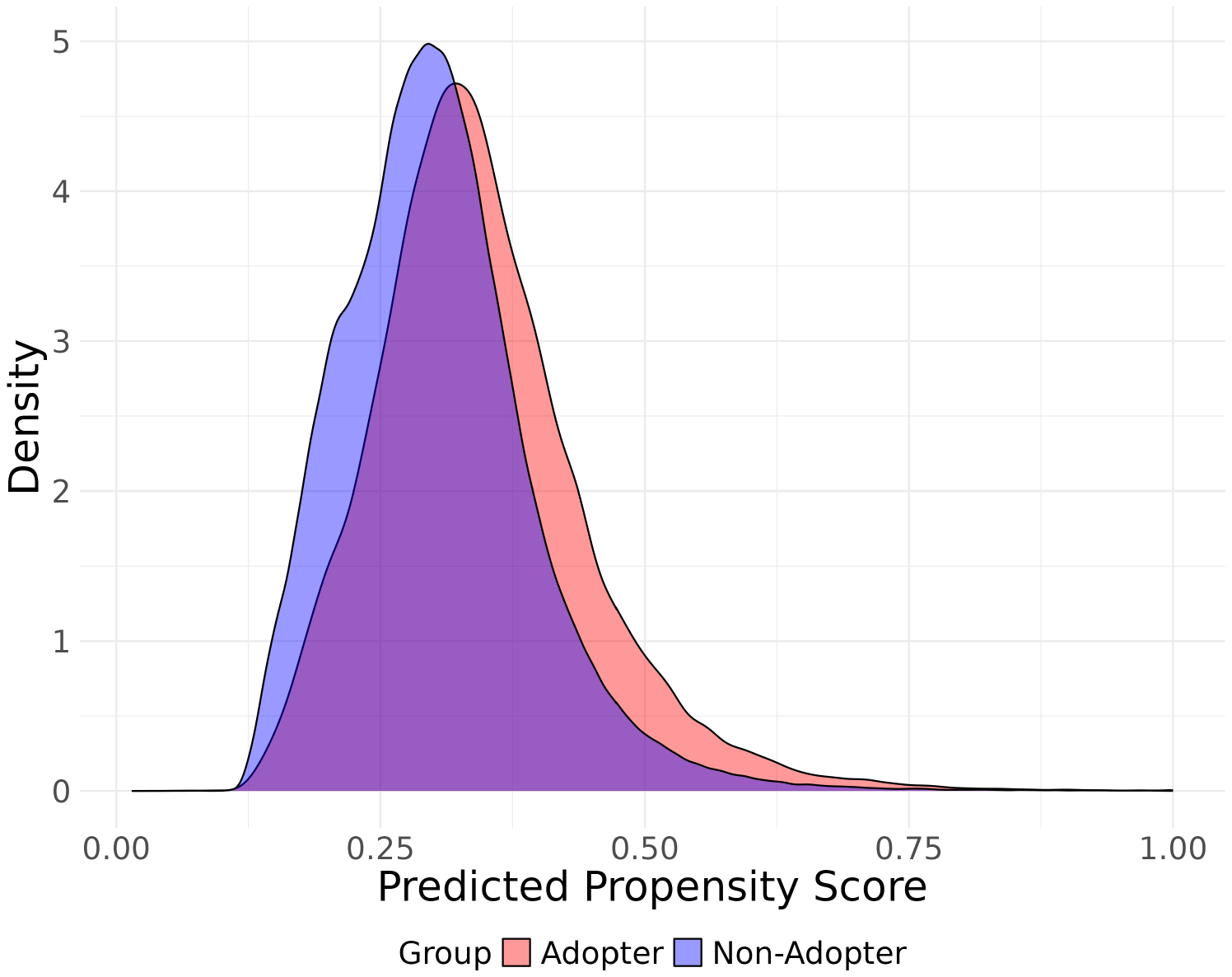}
    \end{subfigure}
\end{figure}

\begin{figure}[htp!]
    \centering
    \caption{Propensity Score Model Predictions - After Matching}  \label{fig:psm_prediction_after_matching_adopter_nonadopter}
    \begin{subfigure}{0.32\textwidth}
        \centering
        \caption{COVID-19}
        \includegraphics[width=\linewidth]{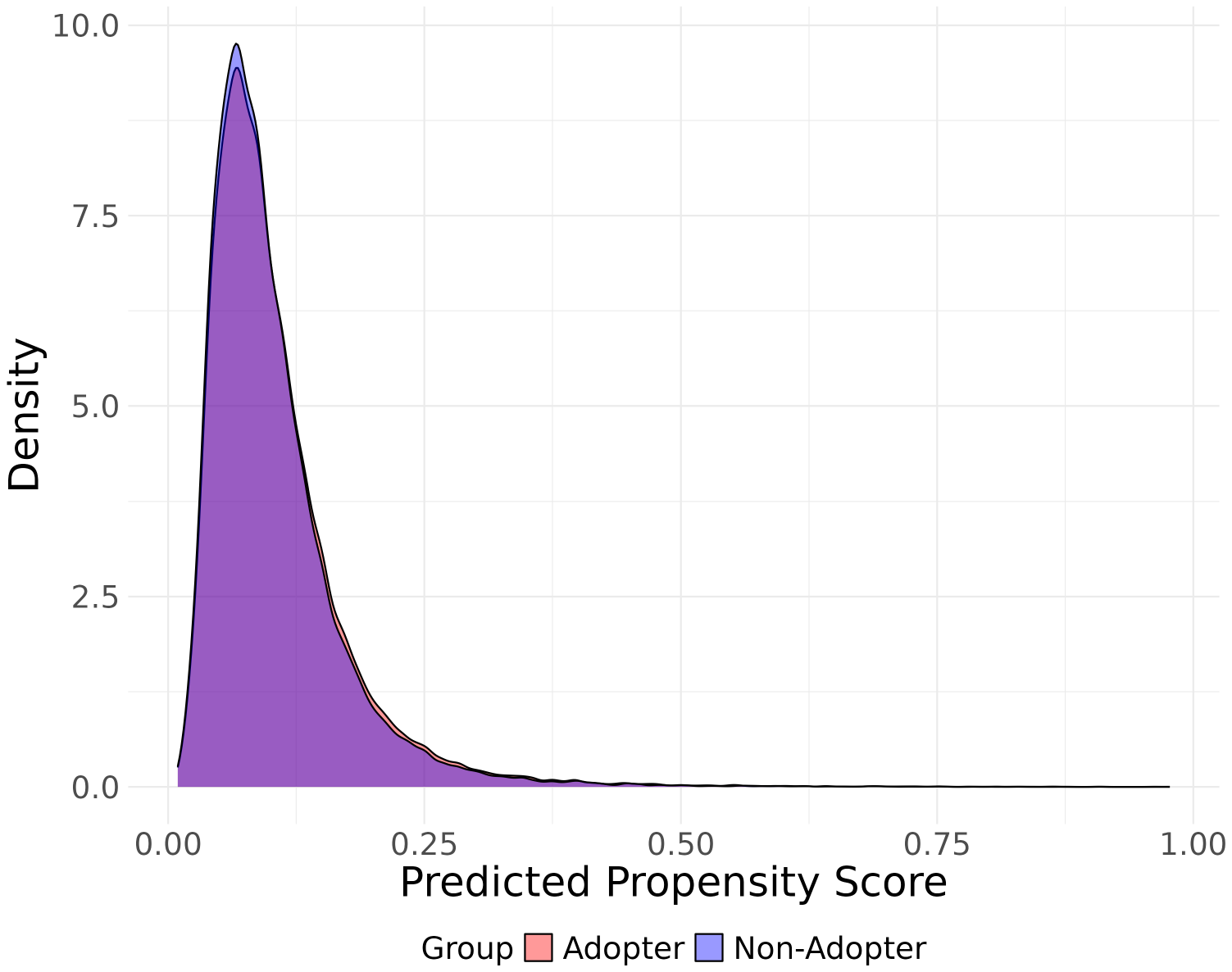}
    \end{subfigure}%
    \hfill
    \begin{subfigure}{0.32\textwidth}
        \centering
        \caption{Black Friday}
\includegraphics[width=\linewidth]{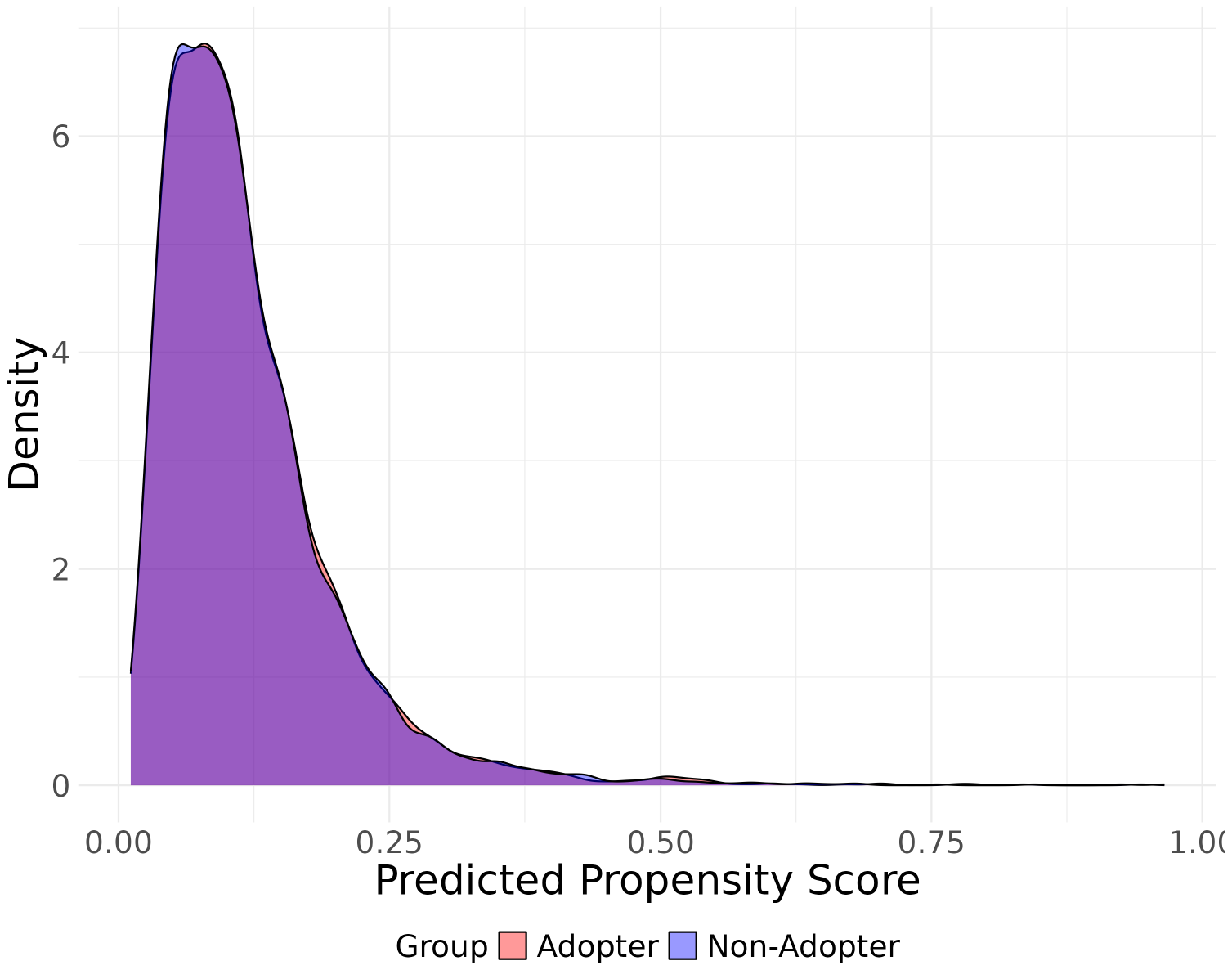}
    \end{subfigure}%
    \hfill
    \begin{subfigure}{0.32\textwidth}
        \centering
        \caption{Loyalty Program}
\includegraphics[width=\linewidth]{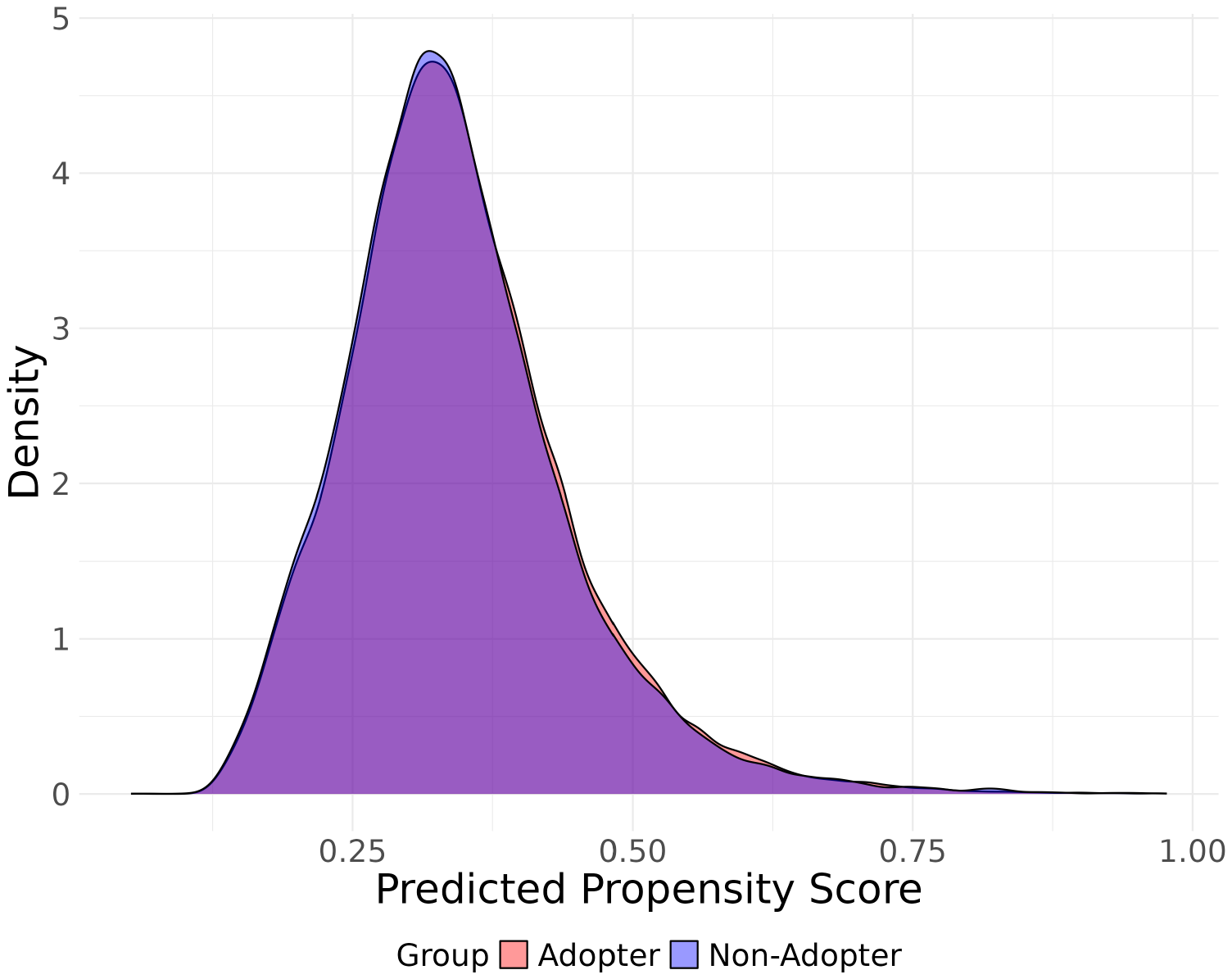}
    \end{subfigure}
\end{figure}

Table \ref{tab:psm_model_est_adopter_nonadopter} presents the propensity score model estimates for being an \textit{adopter} under three studies. Figure \ref{fig:psm_prediction_before_matching_adopter_nonadopter} presents the distribution of propensity scores of being an \textit{adopter} under three studies. After performing nearest-neighbor matching, \ref{fig:psm_prediction_after_matching_adopter_nonadopter} shows that the distribution of propensity scores of being an \textit{adopter} becomes more overlapped. 

\clearpage
\subsubsection{DID Results with PSM/IPTW}
\label{appsssec:psm_att_est_adopter_nonadopter}

Table \ref{tab:psm_att_est_adopter_nonadopter_main} and Table \ref{tab:iptw_att_est_adopter_nonadopter_main} present the DID estimates from two-way fixed effects regression based on the matched samples and with inverse propensity weighting, respectively. We see the findings are consistent with the main results.

\begin{table}[!htp]
  \centering
  \caption{PSM + DID Estimates of Adopters vs. Non-Adopters}
  \label{tab:psm_att_est_adopter_nonadopter_main}
  \resizebox{\textwidth}{!}{%
  \footnotesize{
  \begin{tabular}{@{\extracolsep{4pt}}lcccccc}
    \toprule
    & \multicolumn{2}{c}{COVID-19} & \multicolumn{2}{c}{Black Friday} & \multicolumn{2}{c}{Loyalty Program} \\
    \cline{2-3}\cline{4-5}\cline{6-7}
    Variable 
      & Spend & Profit & Spend & Profit & Spend & Profit \\
    \midrule

    Adopter $\times$ Post 
      & 191.8 (3.701) & 53.37 (1.335) & 140.4 (10.03) & 38.22 (3.553) & 127.5 (5.336) & 32.38 (1.933) \\
      & [0.000]       & [0.000]       & [0.000]       & [0.000]       & [0.000]       & [0.000]       \\

    \midrule
    Loyalty program controls 
      &  &  &  &  & Yes & Yes \\

    \midrule
    \textit{Fixed-effects} \\

    Customer  
      & Yes & Yes & Yes & Yes & Yes & Yes \\

    YearMonth 
      & Yes & Yes & Yes & Yes & Yes & Yes \\

    \midrule
    \textit{Fit statistics} \\

    Observations 
      & 3,310,659 & 3,310,659 & 153,086 & 153,086 & 384,412 & 384,412 \\

    R$^2$ 
      & 0.4107 & 0.3262 & 0.4228 & 0.3681 & 0.4948 & 0.4656 \\

    Within R$^2$ 
      & 0.00351 & 0.00169 & 0.00344 & 0.00180 & 0.02170 & 0.01570 \\

    \midrule\midrule
    \multicolumn{7}{l}{\emph{Clustered (Customer) standard-errors are presented in parentheses, while p-values are shown in square brackets.}}
  \end{tabular}
  }
  }
\end{table}

\begin{table}[!htp]
  \centering
  \caption{IPTW + DID Estimates of Adopters vs. Non-Adopters}
  \label{tab:iptw_att_est_adopter_nonadopter_main}
  \resizebox{\textwidth}{!}{%
  \footnotesize{
  \begin{tabular}{@{\extracolsep{4pt}}lcccccc}
    \toprule
    & \multicolumn{2}{c}{COVID-19} & \multicolumn{2}{c}{Black Friday} & \multicolumn{2}{c}{Loyalty Program} \\
    \cline{2-3}\cline{4-5}\cline{6-7}
    Variable 
      & Spend & Profit & Spend & Profit & Spend & Profit \\
    \midrule

  Adopter $\times$ Post
      & 200.0 (2.758) & 55.99 (0.9605) & 135.7 (7.784) & 36.64 (2.443) & 144.2 (5.584) & 36.62 (1.795) \\
      & [0.000]       & [0.000]        & [0.000]       & [0.000]       & [0.000]       & [0.000]       \\

    \midrule
    Loyalty program controls 
      &  &  &  &  & Yes & Yes \\

    \midrule
    \textit{Fixed-effects} \\

    Customer  
      & Yes & Yes & Yes & Yes & Yes & Yes \\

    YearMonth 
      & Yes & Yes & Yes & Yes & Yes & Yes \\

    \midrule
    \textit{Fit statistics} \\

    Observations 
      & 22,204,608 & 22,204,608 & 9,505,999 & 9,505,999 & 4,403,449 & 4,403,449 \\

    R$^2$ 
      & 0.4067 & 0.3394 & 0.4442 & 0.3625 & 0.6956 & 0.6298 \\

    Within R$^2$ 
      & 0.00165 & 0.00083 & 0.00160 & 0.00078 & 0.00419 & 0.00308 \\

    \midrule\midrule
    \multicolumn{7}{l}{\emph{Clustered (Customer) standard-errors are presented in parentheses, while p-values are shown in square brackets.}}
  \end{tabular}
  }
  }
\end{table}

\clearpage

\subsection{Appendix to Alternate definitions for the \textit{organic adopter} and \textit{COVID adopter} cohorts}
\label{appsec:alternative_definition_covid_organic}

\subsubsection{Appendix to the COVID-19 Analysis - Re-defining the \textit{Organic Adopter} Cohort}
\label{appsec:validity_robust_Redefine}

In this section, we provide a robustness check by redefining \textit{organic adopters} as those customers who were offline-only before 2019-10-31 and made their first online purchase between 2019-11-01 to 2020-02-29. The definition of {\it COVID adopters} remains the same. Under this new definition, we continue to have 33,246 {\it COVID adopters} and a larger number of 25,958 \textit{organic adopters}. As before, we exclude the periods used for defining these two cohorts (2019-11-01 to 2020-04-30) and re-run the same set of DID models as in the main analysis. Table \ref{tab:organic_covid_Robust_2_value} presents the DID results between {\it COVID adopters} and \textit{organic adopters}. We find that all the results are consistent with the main analysis.

\begin{table}[htp!]
  \caption{\textit{COVID Adopters vs. Organic Adopters} – Redefining Organic Adopter Group}
  \label{tab:organic_covid_Robust_2_value}
  \centering
  \footnotesize{
  \begin{tabular}{lcccc}
    \toprule
    DVs: & Spend & Offline Spend & Share of Offline Spend & Profit \\
    Model: & (1) & (2) & (3) & (4) \\
    \midrule
    \emph{Variables}\\
    COVID\_Adopter $\times$ Post 
    & 10.03 (5.944)
    & 44.76 (4.773)
    & 0.0580 (0.0034)
    & 13.20 (2.082) \\
    & [0.091]
    & [0.000]
    & [0.000]
    & [0.000] \\
    \midrule
    \emph{Fixed-effects}\\
    Customer & Yes & Yes & Yes & Yes \\
    YearMonth & Yes & Yes & Yes & Yes \\
    \midrule
    \emph{Fit statistics}\\
    Observations & 2,178,690 & 2,178,690 & 1,153,061 & 2,178,690 \\
    R$^2$ & 0.39039 & 0.36518 & 0.50935 & 0.30924 \\
    Within R$^2$ & 5.66e-06 & 0.00021 & 0.00108 & 6.31e-05 \\
    \midrule\midrule
    \multicolumn{5}{l}{\emph{Clustered (Customer) standard-errors in parentheses; p-values in brackets.}}\\
  \end{tabular}
  }
\end{table}

\newpage

\subsubsection{Appendix to the COVID-19 Analysis - Re-defining the \textit{COVID adopter} Cohort}
\label{appsec:redefine_covid_adopter}

In this section, we provide an alternative definition of the \textit{COVID adopters} cohort for the COVID-19 analysis. We provide a robustness check by redefining \textit{COVID adopters} as those customers who were offline-only before 2019-12-31 and made their first online purchase between 2020-05-01 and 2020-06-30 (i.e., two months later than the exact onset of COVID-19). The definition of \textit{organic adopters} remains the same as the definition in \S\ref{sssec:covid_sample}. Under this definition, we continue to have 12,799 \textit{organic adopters} but now have 27,448 \textit{COVID adopters}. As before, we exclude the periods used for defining these two cohorts (i.e., 2020-01-01 to 2020-06-30), re-run the same set of DID models as in the main analysis, and present the results in Table \ref{tab:did_redefine_covid_adopters_main}. We find that all results are consistent with the main analysis. 

\begin{table}[htp!]
  \caption{\textit{COVID adopters} vs. \textit{Organic adopters} – Re‐defining \textit{COVID adopters} Group}
  \label{tab:did_redefine_covid_adopters_main}
  \centering
  \footnotesize{
  \begin{tabular}{lcccc}
    \toprule
    DVs: & Spend & Offline Spend & Share of Offline Spend & Profit \\
    Model: & (1) & (2) & (3) & (4) \\
    \midrule
    \emph{Variables}\\
    COVID\_Adopter $\times$ Post
    & 7.16 (6.69)
    & 70.2 (5.50)
    & 0.078 (0.004)
    & 9.36 (2.31) \\
    & [0.286]
    & [0.000]
    & [0.000]
    & [0.000] \\
    \midrule
    \emph{Fixed‐effects}\\
    Customer & Yes & Yes & Yes & Yes \\
    YearMonth & Yes & Yes & Yes & Yes \\
    \midrule
    \emph{Fit statistics}\\
    Observations & 1,461,440 & 1,461,440 & 744,927 & 1,461,440 \\
    R$^2$ & 0.37338 & 0.36248 & 0.51561 & 0.27316 \\
    Within R$^2$ & 3.38e-06 & 0.00055 & 0.00202 & 3.41e-05 \\
    \midrule\midrule
 \multicolumn{5}{l}{\emph{Clustered (Customer) standard-errors in parentheses; p-values in brackets.}}\\
  \end{tabular}
  }
\end{table}

\newpage

\subsubsection{Appendix to the COVID-19 Analysis - Comparing \textit{COVID adopters (March/April 2020) vs. Later COVID adopters (May/June 2020)}}

\label{appsssec:covid_adopter_later_covid_adopter}

Table \ref{tab:covid_later_covid_comparison} compares \textit{COVID adopters} (i.e., adopt during March/April 2020) and \textit{later COVID adopters} (i.e., adopt during May/June 2020). Although the \textit{later COVID adopters} spend more than the \textit{COVID adopters} by 9.96 MCU, it is not large in magnitude with a p-value of 0.048. So we conclude there is no major difference in post-adoption spend, share of offline spend (i.e., channel utilization), and profit.

\begin{table}[htp!]
  \caption{ \textit{COVID adopters} (March/April) vs. \textit{Later COVID adopters} (May/June)}
  \label{tab:covid_later_covid_comparison}
  \centering
  \footnotesize{
  \begin{tabular}{lcccc}
    \toprule
    DVs: & Spend & Offline Spend & Share of Offline Spend & Profit \\
    Model: & (1) & (2) & (3) & (4) \\
    \midrule
    \emph{Variables}\\
    Later\_COVID\_Adopter $\times$ Post
    & 9.96 (5.040)
    & 30.4 (4.270)
    & -0.0002 (0.003)
    & 0.070 (1.770) \\
    & [0.048]
    & [0.000]
    & [0.947]
    & [0.968] \\
    \midrule
    \emph{Fixed-effects}\\
    Customer & Yes & Yes & Yes & Yes \\
    YearMonth & Yes & Yes & Yes & Yes \\
    \midrule
    \emph{Fit statistics}\\
    Observations & 2,240,950 & 2,240,950 & 1,170,421 & 2,240,950 \\
    R$^2$ & 0.39921 & 0.36673 & 0.50555 & 0.35185 \\
    Within R$^2$ & 8.12e-06 & 0.00011 & 9.28e-09 & 3.00e-09 \\
    \midrule\midrule
  \multicolumn{4}{l}{\emph{Clustered (Customer) standard-errors in parentheses; p-values in brackets.}}\\
  \end{tabular}
  }
\end{table}

\newpage

\subsection{Appendix to Using Natural Log Scale Dependent Variables}
\label{appsec:log_dependent_variables}
In this section, we provide a robustness check for using natural log transformed scale of the dependent variables (i.e., ln(dependent variable + 1)), and conclude that the direction of effects are consistent with the main finding. Note that the profit variable can take both positive and negative values, so we do not consider natural log transformation of profit. First, we present the result for spend between \textit{adopters} and \textit{offline-only customers} in Table \ref{tab:did_spend_adopter_offlineonly_log}, and show that the results are consistent with Table \ref{tab:did_spend_adopter_offlineonly}.

\begin{table}[htp!]
   \caption{DID Analysis - Spend (in Natural Log Scale) (\textit{Adopters} vs. \textit{Offline-only})}
   \centering
   \label{tab:did_spend_adopter_offlineonly_log}
   \footnotesize{
   \begin{tabular}{lccc}
\midrule \midrule
       Study: & COVID-19 & Black Friday & Loyalty Program\\ 
\midrule 
      DV & \multicolumn{3}{c}{ln(Spend + 1)}\\
      Model: & (1) & (2) & (3)\\  
\midrule
\emph{Variables}\\
Adopter $\times$ Post 
& 0.6883 (0.0104) 
& 0.5742 (0.0302) 
& 0.6099 (0.0190)\\
& [0.000] 
& [0.000] 
& [0.000]\\
\midrule
Loyalty program controls &  &  & Yes \\ 
\midrule
\emph{Fixed-effects}\\
Customer   & Yes & Yes & Yes\\  
YearMonth  & Yes & Yes & Yes\\  
\midrule
\emph{Fit statistics}\\
Observations   & 22,204,608 & 9,505,999 & 4,403,449\\  
R$^2$          & 0.30744 & 0.36627 & 0.29925\\  
Within R$^2$   & 0.00087 & 0.00010 & 0.00208\\  
\midrule \midrule
\multicolumn{4}{l}{\emph{Clustered (Customer) standard-errors in parentheses; p-values in brackets.}}\\
   \end{tabular}
   }
\end{table}

Next, Table \ref{tab:did_spend_log_event_organic} presents DID results for the total spend of comparing \textit{event adopters} and \textit{organic adopters} by three events. Note that the profit variable can take both positive and negative values, so we do not consider natural log transformation of profit. We show that the results in Table \ref{tab:did_spend_log_event_organic} are consistent with Table \ref{tab:total_spend}.

\begin{table}[H]
\caption{DID Analysis -- Spend (in Natural Log Scale) (\textit{Event Adopters} vs. \textit{Organic Adopters})}
\label{tab:did_spend_log_event_organic}
\centering
\footnotesize
\begin{tabular}{lccc}
\midrule\midrule
Study: & COVID-19 & Black Friday & Loyalty Program \\
\midrule
DV & \multicolumn{3}{c}{ln(Spend + 1)} \\
Model: & (1) & (2) & (3) \\
\midrule
\emph{Variables}\\
Event\_Adopter $\times$ Post 
& 0.008 (0.023) 
& -0.268 (0.062) 
& -0.151 (0.064) \\
& [0.728] 
& [0.000] 
& [0.017] \\
\midrule
Loyalty program controls 
&  &  & Yes \\
\midrule
\emph{Fixed-effects}\\
Customer 
& Yes & Yes & Yes \\
YearMonth 
& Yes & Yes & Yes \\
\midrule
\emph{Fit statistics}\\
Observations 
& 1,783,949 
& 76,842 
& 197,658 \\
R$^2$ 
& 0.28677 
& 0.32002 
& 0.31155 \\
Within R$^2$ 
& $2.16\times10^{-7}$ 
& 0.00049 
& 0.02144 \\
\midrule\midrule
\multicolumn{4}{l}{\emph{Clustered (Customer) standard-errors in parentheses; p-values in brackets.}}
\end{tabular}
\end{table}

\clearpage

\subsection{Appendix to Handling Zero-Inflated Dependent Variables}
\label{appsec:zero_inflated_dependent_variables}

In this section, we outline the procedure and results of the two-part hurdle model and the estimation results. The two-part hurdle model includes two separate estimation steps: 
\squishlist
\item Step 1: Estimate a linear probability model with outcome variable $\mathbbm{I}\{Spend_{it}\}$ with all independent variables specified in Equation \eqref{eq:did_fe} when comparing \textit{adopters} vs. \textit{offline-only}, and 
independent variables specified in Equation \eqref{eq:did_fe_2by2_FE} when comparing \textit{event adopters} vs. \textit{organic adopters}; 
\item Step 2: Estimate Equation \eqref{eq:did_fe} conditional on customers making non-zero spend when comparing \textit{adopters} vs. \textit{offline-only}, and 
independent variables specified in Equation \eqref{eq:did_fe_2by2_FE} when comparing \textit{event adopters} vs. \textit{organic adopters}. 
\squishend

Next, we present the estimation results for \textit{adopters} vs. \textit{offline-only} customers and for each \textit{event Adopters} vs. \textit{organic adopters}. 

\subsubsection*{ \textit{Adopters} vs. \textit{Offline-only}}

Table \ref{tab:lpm_adopter_offlineonly} presents the estimation result of linear probability model of $\mathbbm{I}\{Spend_{it}\}$ between \textit{adopters} and \textit{offline-only} customers.  Table \ref{tab:lm_adopter_offlineonly} presents the DID estimation results conditional on customer-month with non-zero spend between \textit{adopters} and \textit{offline-only} customers. 

\begin{table}[htp!]
   \caption{Linear Probability Model with DID Specifications -- \textit{Adopters} vs. \textit{Offline-only}}
   \centering
   \label{tab:lpm_adopter_offlineonly}
   \footnotesize{
   \begin{tabular}{lccc}
\midrule \midrule
Study: & COVID-19 & Black Friday & Loyalty Program\\ 
\midrule 
DV & \multicolumn{3}{c}{$\mathbbm{I}\{Spend_{it}\}$}\\
Model: & (1) & (2) & (3)\\  
\midrule
\emph{Variables}\\
Adopter $\times$ Post 
& 0.1033 (0.0016) 
& 0.0910 (0.0046) 
& 0.0933 (0.0029)\\
& [0.000] 
& [0.000] 
& [0.000]\\
\midrule
Loyalty program controls &  &  & Yes\\
\midrule
\emph{Fixed-effects}\\
Customer & Yes & Yes & Yes\\
YearMonth & Yes & Yes & Yes\\
\midrule
\emph{Fit statistics}\\
Observations & 22,204,608 & 9,505,999 & 4,403,449\\  
R$^2$        & 0.28000 & 0.34101 & 0.26334\\  
Within R$^2$ & 0.00073 & $9.3\times 10^{-5}$ & 0.00156\\  
\midrule \midrule
  \multicolumn{4}{l}{\emph{Clustered (Customer) standard-errors in parentheses; p-values in brackets.}}\\
   \end{tabular}
   }
\end{table}

\begin{table}[htp!]
   \caption{DID Estimates for Spend (Conditional on Purchase) -- \textit{Adopters} vs. \textit{Offline-only}}
   \centering
   \label{tab:lm_adopter_offlineonly}
   \footnotesize{
   \begin{tabular}{lccc}
\midrule \midrule
Study: & COVID-19 & Black Friday & Loyalty Program\\ 
\midrule 
DV & \multicolumn{3}{c}{Spend}\\
Model: & (1) & (2) & (3)\\  
\midrule
\emph{Variables}\\
Adopter $\times$ Post 
& 135.4 (3.953) 
& 65.40 (9.385) 
& 50.28 (6.334)\\
& [0.000] 
& [0.000] 
& [0.000]\\
\midrule
Loyalty program controls &  &  & Yes\\
\midrule
\emph{Fixed-effects}\\
Customer & Yes & Yes & Yes\\
YearMonth & Yes & Yes & Yes\\
\midrule
\emph{Fit statistics}\\
Observations & 8,187,536 & 3,744,951 & 1,939,763\\  
R$^2$ & 0.47201 & 0.50834 & 0.62202\\  
Within R$^2$ & 0.00084 & $4.37\times 10^{-5}$ & 0.00150\\  
\midrule \midrule
  \multicolumn{4}{l}{\emph{Clustered (Customer) standard-errors in parentheses; p-values in brackets.}}\\
   \end{tabular}
   }
\end{table}

\newpage

\subsubsection*{ \textit{Event Adopters} vs. \textit{Organic Adopters}}

Table \ref{tab:lpm_event_organic_adopter} presents the estimation result of linear probability model of $\mathbbm{I}\{Spend_{it}\}$ between \textit{event adopters} and \textit{organic adopters}. Table \ref{tab:lm_event_organic_adopter} presents the DID estimation results conditional on customer-month with non-zero spend between \textit{event adopters} and \textit{organic adopters}. 

\begin{table}[htp!]
   \caption{Linear Probability Model with DID Specifications -- \textit{Event Adopters} vs. \textit{Organic Adopters}}
   \centering
   \label{tab:lpm_event_organic_adopter}
   \footnotesize{
   \begin{tabular}{lccc}
\midrule \midrule
Study: & COVID-19 & Black Friday & Loyalty Program\\ 
\midrule 
DVs: & \multicolumn{3}{c}{$\mathbbm{I}\{Spend_{it}\}$}\\
Model: & (1) & (2) & (3)\\  
\midrule
\emph{Variables}\\
Event\_Adopter $\times$ Post 
& 0.0039 (0.0034) 
& -0.0361 (0.0095) 
& -0.0202 (0.0095) \\
& [0.251] 
& [0.000] 
& [0.032] \\
\midrule
Loyalty program controls &  &  & Yes\\
\midrule
\emph{Fixed-effects}\\
Customer & Yes & Yes & Yes\\
YearMonth & Yes & Yes & Yes\\
\midrule
\emph{Fit statistics}\\
Observations & 1,783,949 & 76,842 & 197,658\\  
R$^2$ & 0.25272 & 0.28908 & 0.27038\\  
Within R$^2$ & $2.36\times 10^{-6}$ & 0.00036 & 0.01574\\  
\midrule \midrule
  \multicolumn{4}{l}{\emph{Clustered (Customer) standard-errors in parentheses; p-values in brackets.}}\\
   \end{tabular}
   }
\end{table}

\begin{table}[htp!]
   \caption{DID Estimates for Spend (Conditional on Purchase)  -- \textit{Event Adopters} vs. \textit{Organic Adopters}}
   \centering
   \label{tab:lm_event_organic_adopter}
   \footnotesize{
   \begin{tabular}{lccc}
\midrule \midrule
Study: & COVID-19 & Black Friday & Loyalty Program\\ 
\midrule 
DVs: & \multicolumn{3}{c}{Spend}\\
Model: & (1) & (2) & (3)\\  
\midrule
\emph{Variables}\\
Event\_Adopter $\times$ Post 
& -3.982 (8.373) 
& -64.19 (19.68) 
& -36.03 (23.92) \\
& [0.623] 
& [0.001] 
& [0.126] \\
\midrule
Loyalty program controls &  &  & Yes\\
\midrule
\emph{Fixed-effects}\\
Customer & Yes & Yes & Yes\\
YearMonth & Yes & Yes & Yes\\
\midrule
\emph{Fit statistics}\\
Observations & 949,448 & 42,299 & 110,339\\  
R$^2$ & 0.44326 & 0.45643 & 0.56886\\  
Within R$^2$ & $7.04\times 10^{-7}$ & 0.00047 & 0.01280\\  
\midrule \midrule
  \multicolumn{4}{l}{\emph{Clustered (Customer) standard-errors in parentheses; p-values in brackets.}}\\
   \end{tabular}
   }
\end{table}



\end{appendices}
\end{document}